\documentclass[10pt]{article} 

\usepackage[numbers,sort&compress]{natbib}

\usepackage{amsmath,amsfonts,ifthen}

\usepackage{geometry} 
\usepackage[format=hang, font=footnotesize, labelfont = bf, margin=10pt]{caption}
\usepackage[format=hang, font=footnotesize, labelfont = bf, margin=10pt]{subcaption}

\title{A mixed-mode dependent interface and phase-field damage model for  solids with inhomogeneities}

\author{Roman Vodička\\
\small Technical University of Ko\v{s}ice, Faculty of Civil Engineering, Vysoko\v{s}kolsk\'{a} 4, 042 00 Ko\v{s}ice, Slovakia\\
\tt{\small roman.vodicka@tuke.sk}}
\date{} 

\usepackage{tikz}

\def\slant#1#2{%
  \tikz[baseline=(X.base), xslant=tan(#1)]
    \node[inner sep=0pt, xslant=tan(#1)](X){#2};%
}

\newcommand{\OM}{\ensuremath{\mathit\Omega}}
\newcommand{\GM}{\ensuremath{\mathit\Gamma}}
\newcommand{\D}{\ensuremath{_{\mathrm D}}}
\newcommand{\N}{\ensuremath{_{\mathrm N}}}
\newcommand{\I}{\ensuremath{_{\mathrm i}}}
\newcommand{\C}{\ensuremath{_{\mathrm c}}}
\newcommand\Gc[1][]{\ensuremath{G_{\text c}^{\hpt[0.5]\text{#1}}}}
\newcommand\scri[1][]{\ensuremath{\sigma_{\text crit}^{\hpt[0.5]\text{#1}}}}
\newcommand\pcri[1][]{\ensuremath{p_{\sfn\,\text{crit}}^{\hpt[0.5]\text{#1}}}}
\newcommand\kG{\ensuremath{\kappa_{\text G}}}
\newcommand\kk{\ensuremath{\kappa}}
\newcommand\Kp{\ensuremath{K_{\text{p}}}}
\newcommand\sfn{{\sf n}}

\newcommand\sfs{{\sf s}}
\newcommand{\dd}{{\mathrm d}}

\newcommand{\hpt}[1][1]{\hspace{#1pt}}
\newcommand{\uppartial}[1][13.5]{\text{\slant{-#1}{$\partial$}}}

\newcommand{\Fun}[1]{\ensuremath{\mathit #1}}
\newcommand{\Vector}[1]{\ensuremath{\text{\bfseries\itshape #1}}}
\newcommand{\Matrix}[1]{\ensuremath{\mathbf #1}}
\newcommand{\Functional}[1]{\ensuremath{\mathfrak #1}}
\newcommand{\Tensor}[1]{\ensuremath{\mathsf #1}}

\newcommand\JUMP[3][]{\ensuremath{\mathchoice
                  {\big[\hspace*{-.3em}\big[#2\big]\hspace*{-.3em}\big]_{#3}^{#1}}
                   {[\hspace*{-.15em}[#2]\hspace*{-.15em}]_{#3}^{#1}}
                   {[\![#2]\!]_{#3}^{#1}}
                   {[\![#2]\!]_{#3}^{#1}}}}

\DeclareMathOperator{\argmin}{argmin}
\DeclareMathOperator{\sph}{sph}
\DeclareMathOperator{\dev}{dev}
\DeclareMathOperator{\tr}{tr}
\DeclareMathOperator{\Div}{div}

\newcommand{\rB}{\rule{0pt}{20pt}}
\newcommand{\rE}{\rule[-20pt]{0pt}{1pt}}

\newcommand{\tc}{\text,}
\newcommand{\tb}{\text.}
\newcommand\et{\eta}
\newcommand\he[1][1]{^{\hpt[#1]\et}}

\newcommand{\PA}[2]{%
\put(-0.015,0.03){\makebox(0,0)[bl]{\includegraphics[scale=0.22,trim=190 40 170 20,clip]{#1}}}
\put(0.03,0.02){\makebox(0,0)[tl]{#2}}%
}
\newcommand{\PB}[2]{%
\put(0.180,0.03){\makebox(0,0)[bl]{\includegraphics[scale=0.22,trim=190 40 170 20,clip]{#1}}}
\put(0.225,0.02){\makebox(0,0)[tl]{#2}}%
}
\newcommand{\PC}[2]{%
\put(0.375,0.03){\makebox(0,0)[bl]{\includegraphics[scale=0.22,trim=190 40 170 20,clip]{#1}}}
\put(0.420,0.02){\makebox(0,0)[tl]{#2}}%
}
\newcommand{\PD}[2]{%
\put(0.570,0.03){\makebox(0,0)[bl]{\includegraphics[scale=0.22,trim=190 40 170 20,clip]{#1}}}
\put(0.615,0.02){\makebox(0,0)[tl]{#2}}%
}
\newcommand{\PE}[2]{%
\put(0.765,0.03){\makebox(0,0)[bl]{\includegraphics[scale=0.22,trim=190 40 170 20,clip]{#1}}}
\put(0.81,0.02){\makebox(0,0)[tl]{#2}}%
}

\newcommand{\PBF}[2]{%
\put(0.180,0.03){\makebox(0,0)[bl]{\includegraphics[scale=0.22,trim=170 40 100 20,clip]{#1}}}
\put(0.225,0.02){\makebox(0,0)[tl]{#2}}%
}

\newcommand{\PDF}[2]{%
\put(0.570,0.03){\makebox(0,0)[bl]{\includegraphics[scale=0.22,trim=170 40 100 20,clip]{#1}}}
\put(0.615,0.02){\makebox(0,0)[tl]{#2}}%
}

\newcounter{Pvar}
\newcommand{\Pvar}[3]{%
\setcounter{Pvar}{#3}
\put(0.\thePvar,0.03){\makebox(0,0)[bl]{\includegraphics[scale=0.22,trim=190 40 170 20,clip]{#1}}}
\addtocounter{Pvar}{45}
\put(0.\thePvar,0.02){\makebox(0,0)[tl]{#2}}%
}

\newcommand{\CB}[2][765]{%
\put(0.#1,0.03){\makebox(0,0)[bl]{\includegraphics[scale=0.22,trim=550 40 100 20,clip]{#2}}}%
}

\begin{document}
\maketitle
\begin{abstract}
The developed computational approach is capable of initiating and propagating cracks inside materials and along material interfaces of general multi-domain structures under quasi-static conditions.
Special attention is paid to particular situation of a  solid with inhomogeneities.
Description of the fracture processes are based on the theory of material damage.
It  introduces two independent damage parameters  to distinguish between  interface and internal cracks.
The parameter responsible for interface  cracks   is defined in a thin adhesive layer of the interface and renders relation between stress and strain quantities in fashion of cohesive zone models.
The second parameter is defined inside material domains and it is founded on the theory of phase-field fracture guaranteeing the material  damage  to occur in a thin material strip introducing  a regularised model of internal cracks.
Additional property of both interface and phase-field damage is its capability to distinguish between fracture modes which is useful if the structures is  subjected  to combined loading.
The solution methodology  is based on a variational approach which allows implementation of non-linear programming optimisation into standard  methods of FEM discretisation and time stepping method.
Computational implementation is prepared in MATLAB whose numerical data validate developed formulation for analysis of problems of fracture in multi-domain elements of structures.
\end{abstract}

%%%%%%%%%%%%%%%%%
\section{Introduction}\label{Sec_Intro}
%%%%%%%%%%%%%%%%%%

Nowadays, materials in engineering structures are made of multiple components. 
They usually contain in\-homogene\-iti\-es like grains or fibres distributed inside a mother material. 
An applied increasing force inevitably causes significant changes in material components which may cause failure of the structure.
The presence of interfaces due to inhomogeneities increases  the risk of such failure which is accompanied  by initiation and propagation of cracks inside the materials or along material interfaces.
Any forms of cracks may appear simultaneously.
The computational simulation  of  both forms of cracks is therefore a challenging problem.
The solution of fracture problems in engineering was  initiated by Griffith~\cite{griffith21A1} and a criterion for crack propagation.
Since then lots of improvements were discovered to predict various issues like nucleation, propagation, branching etc.\ of cracks.

The computational approaches for modelling fracture followed two basic directions of treating with the  cracks: discrete or continuous.
The former considers the crack as a geometrical discontinuity, while in the latter the discontinuity is diffused over the material and it is considered as a change in physical characteristics of the material.
Such changes are called damage and computationally they are represented by internal parameters~\cite{fremond85A1,maugin15A1}.
More recently such a damage concept led to phase-field models, where the zone of damaged material reduces to narrow material strips which represent regularised internal cracks.
Anyhow, the damage is expressed by the internal parameters and the computational efforts meet problems with appropriate nucleation and propagation of cracks along generally unknown paths.
Either along interfaces, where the crack path is known, the concept of internal variables may be advantageously used to determine exact crack extension.

One of the crucial modifications, definitely appropriate for implementation with contemporary  computational power, was proposed in~\cite{francfort98A1,bourdin00A1}, where a computational model was  introduced which solved the problems of fracture variationally through minimisation of global energy in a solid. 
The formulation was then modified to provide  a concept of regularised cracks  and their relation to discrete ones by mathematical tools of  $\Gamma$ convergence, see~\cite{dalmaso12B1}, and it induced the introduction of the phase-field approach as documented in~\cite{bourdin08A1}.
The development continued by definition of a robust computational  Phase-Field Model (PFM)  with a rigorous thermodynamical background in~\cite{miehe10A1,delpiero14A1,molnar17A1}.
Since then, a lot of enhancements of the phase-field algorithms appeared which tried to specify particular problems of the approach related to the characteristics of the computational model: scale parameter related to the width of the  regularised crack, degradation function characterising the damage in the material \cite{kuhn15A1,sargado18A1}, its effect on cracking process in various materials  \cite{fang22A1,freddi22A1,raj20A1,xu22A1}, crack nucleation conditions and subsequent processes \cite{tanne18A1,wu17A1,yin19A1,wang20A1} etc.
The variety of  degradation functions  is frequently explained by particular material behaviour, properties of the computational approach and many times they are supported by empirical results.
Additionally, the ability of such approaches was revealed to distinguish between cracking mechanisms in distinct modes of cracks, advantageously related to different amount of energy necessary to initiate and propagate  cracks in various crack modes  \cite{feng22A1} within variationally based approaches.
Hence, the phase-field model appeared  under combination of loading in tension, compression, or shear \cite{cao22A1,li23A1,luo22A1,yue22A1}.
Anyhow, all such enhancements helped  to overcome limitation of the original Griffith theory.

Simultaneously,  the problems of fracture may appear due to presence of material interfaces or directly along them.
 Though the problems of crack path is avoided along a given interface, issues of describing particular stress-strain relations on them may still appear.
The cohesive zone approach was developed~\cite{barenblatt62A1}  as a modification of the original Griffith concept with stress singularities at crack tips to  mere stress concentrations by modifying the energy release depending on the  displacement jump caused by a crack as shown  by \cite{lenci01A1,charlotte06A1}.
 Thus,  Cohesive Zone Models (CZM)  were introduced \cite{bazant97A1}, and as a result of modified assumptions considered stress-strain relations are bounded in terms of stress as seen e.g. in~\cite{bankssills00A1,ortiz99A1,park11A1}.
Though, the original cohesive-zone concept does not take into account any internal interface variable, CZMs can be used to be combined with phase-field approaches for  internal cracks  according to numerous works~\cite{paggi17A1,chen22A1,marulli22A1,wei22A1,zambrano22A1}, and it is interesting to unify both concepts of dealing with cracks.
It requires to introduce a new degradation function which controls changing of elastic properties  on the interface, considered as a thin layer of adhesive, related to the current level of damage.
Such treatment is based on the model for delamination \cite{mielke04A1,roubicek13A2,roubicek15A2} or adhesive contact~\cite{raous99A1,delpiero10A1} which  introduced necessary internal variable for interface damage.
The stress-strain response along the interface similar to that declared by classical CZMs was studied by the author in \cite{vodicka16A1,vodicka17A2,vodicka21A1} where also particular  interface degradation functions were proposed.
That was the reason for developing the present approach in terms of damage-like variables independently defined both in the bulk and along the interface.

In this study, a computational quasi-static approach is proposed, implemented and tested in a scheme which utilises PFM for  solids  simultaneously with an independent  application of CZM with an internal variable for interface  damage.
The main contributions of this work are: (i) a unified form of integrating for PFM damage and interface damage into an energy based formulation for fracture; (ii) a new term controlling dissipation of energy  depending on the mode of fracture in the quasi-static evolutionary process; (iii) a variationally based computational approach  implemented in  an in-house MATLAB Finite Element Method (FEM) computer code  for assessing particular mechanical problems independent of used models of material and interface  damage. 
The algorithms are verified under tensional, compressive and mixed type loading of structural elements made of elastic materials.
 The tested solids represent typical academic examples  of multi-domains containing inhomogeneities, though simulating real  experimental observations obtained by other authors.

The rest of the paper is organized as follows. In Section~\ref{Sec_Model},  PFM combined with interface damage model is  formulated so that the dependence of the cracking process on the mixed mode character of fracture is stressed. 
Some details of the computational implementation of the model are described in Section~\ref{Sec_Implementation}, focused in  the describing of the staggered approach for the evolution process and  schematic computational  details of  FEM approximation and application of methods of mathematical programming. 
Numerical examples are shown in Section~\ref{Sec_Examples}.
The calculations validate the the discussed aspects of the fracture model for combination of  interface and internal  cracks,  dependence of the cracking phenomena on a general stressed state originated in the structure under load.
Finally, concluding remarks are provided in Section~\ref{Sec_Concl}.

%%%%%%%%%%%%%%%%%%%%%%%%%%
\section{Description of the model}\label{Sec_Model}
%%%%%%%%%%%%%%%%%%%%%%%%

Finding the relations which control  crack formation process in structural elements will be described in this section. 
Starting from Griffith \cite{griffith21A1}, the modern analysis of  fracture in brittle materials was initiated. 
His findings led to  observations that a sufficient  energy release is required for an existing crack to grow. 
The critical value of such  energy release is now called fracture energy, \Gc.
Nevertheless, to overcome problems with crack nucleation, a fracture variational formulation was introduced by \cite{francfort98A1} based on energy minimisation ansatz. 
Therefore in what follows, the explanations also focus on such an energy formulation for the intended problem to be solved.

\begin{figure}[t]
\centering
\begin{picture}(80,25)
\put(40,25){\makebox(0,0)[tc]{\includegraphics[scale=1]{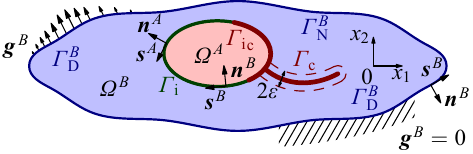}}}
\end{picture}
\caption{ Description of a solid with two subdomains, one of them is an inhomogeneity. 
It also describes a concept of  interface and internal  cracks, then boundary conditions and constraints.}\label{Fig_Model}
\end{figure}

 The solid subjected to load is presented by a bounded domain \OM{}.
It is assumed that the domain may contain several subdomains.
As an example, a two-domain case  is shown in Fig.~\ref{Fig_Model}, the subdomains are denoted  $\OM^A$ and $\OM^B$ here.
The respective subdomain boundaries are denoted $\GM^A$ and $\GM^B$.  
Interfaces, the common parts of the subdomain boundaries, are denoted \GM\I, they introduce also a contact zone between the subdomains. 

The domain is  under  an increasing displacement loading in such a way that a kind of  damage  arises in the domain material or along interfaces.
 It finally leads to a crack. 
The  internal crack  is distinguished by subscript '$c$': $\GM\C$.
Similarly, if the crack arises along $\GM\I$, it is denoted by ${\GM\I}_{\text{c}}$.
Let the inertial effects be neglected so that the state of the structure  evolves in time $t$ under quasi-static conditions. 
The prescribed hard-device loading introduces boundary conditions and also constraints for the displacement field \Vector{u}. The load itself is represented by a time dependent function $\Vector{u}(t)=\Vector{g}(t)$ on a part of the domain boundary.
For the two subdomain case  they are marked as $\GM^A\D$ and $\GM^B\D$. 
The rest of the outer boundary is supposed to be free of mechanical loading, it means that zero traction vector \Vector{p} is given here as only displacement loading is considered.
 Corresponding parts of the outer boundary are denoted $\GM^A\N$ and $\GM^B\N$ .

A deformed state of the structure at  particular time instant $t$ should be given by physical quantities.
In what follows, a description in terms of energies is used.
First of all, the variables of state include deformation quantities: the displacement field \Vector{u} in the interior of the domains including the  jump of displacements  \JUMP{\Vector{u}}{}{} characterising the deformed state along the interfaces $\GM\I$. 
The formulation of the stored energy, as developed  in a pioneer work of \cite{francfort98A1}, includes elastic stored energy and energy accumulated in cracks 
\begin{equation}\label{Eq_FunGriffith}
\Functional{E}(t;\Vector{u},\GM\C,{\GM\I}_{\text{c}})=
\sum_{\eta=A,B}\int_{\Omega^\eta\setminus\GM\C}\!\!
\frac12 \Tensor{e}(\Vector{u}^\eta):\Tensor{C}^\eta:\Tensor{e}(\Vector{u}^\eta)\,\dd\Omega
+\int_{\GM\C}\Gc[I]\,d\GM+\int_{{\GM\I}_{\text{c}}}{\Gc[iI]}\,d\GM\tb
\end{equation}
It is available for any admissible state satisfying the conditions
\begin{equation}\label{Eq_BCGriffith}
\Vector{u}^\eta|_{\GM\D^\eta}^{}=\Vector{g}^\eta\text,\quad \JUMP{\Vector{u}}{}=\Vector{u}^A{-}\Vector{u}^B\text, \quad \JUMP{\Vector{u}}{\sfn}\geq0\text{ on }\GM\I\text,\quad \JUMP{\Vector{u}}{}=0\text{ on }\GM\I\setminus{\GM\I}_{\text{c}}\text,
\end{equation}
where the subscript $\sfn$ refers to the outward normal vector \Vector{n} of  the matching domain  (distinguished by a superscript, if necessary) and $\JUMP{\Vector{u}}{\sfn}$ is meant in the sense: $\JUMP{\Vector{u}}{\sfn}=\JUMP{\Vector{u}}{}\cdot\Vector{n}^B$.
The constraints containing the displacement  jump  $\JUMP{\Vector{u}}{}$ introduce contact conditions at the interfaces.
The strain energy is expressed in terms of (small) strain tensor \Tensor{e} and material characteristics -- the tensor of elastic constants \Tensor{C}.
The energy accumulated in cracks has two resources,  internal cracks  and cracks along interfaces, therefore it is also distinguished between  required  fracture energy for  internal cracks  $\Gc[I]$, and  fracture energy for the interface cracks $\Gc[iI]$. 

The problem of searching for a minimiser \Vector{u} features also discontinuity of the  displacement field \Vector{u} across $\GM\C$ whose location is not known before the calculation is performed. 
Therefore, the functional was reformulated, e.g.\ by \cite{bourdin08A1}, to make the displacement response continuous, e.g.\ to be suitable for computational models based on finite elements. 
To this end, it was necessary to replace the term  $\int_{\GM\C}\Gc[I]\,d\GM$ by a regularised integral over whole domain $\OM^\eta$.
For such a purpose, a functional of Ambrosio-Tortorelli \cite{ambrosio90A1} was introduced defining an internal variable $\alpha$, which allows to continuously connect displacements across a crack, with a length parameter $\epsilon$ controlling amount of regularisation.
The introduced variable $\alpha$ is call phase-field damage variable.
This leads to  a form of a smeared crack  as due to the regularisation the crack is diffused to exhibit a finite width determined by $\epsilon$ as also indicated in Fig.~\ref{Fig_Model}.
Simultaneously, as presented e.g.\ by \cite{tanne18A1,sargado18A1}, this parameter can be used  to control a stress criterion  for initiating damaging due to evolution of the phase-field variable.
Nevertheless, such assumptions are then far from the classical Griffith crack interpretation, though it was also shown that  tending  $\epsilon$ to zero  reverts to the formulation with \eqref{Eq_FunGriffith} in the sense of $\Gamma$-convergence, see~\cite{bourdin08A1}.
This regularisation leads to the following functional of the stored energy 
\begin{equation}\label{Eq_FunRegul}
\Functional{E}(t;\Vector{u},\alpha,{\GM\I}_{\text{c}})=
\sum_{\eta=A,B}\int_{\Omega^\eta}
\frac12 \Tensor{e}(\Vector{u}^\eta):\left(\Fun{\Phi}(\alpha^\eta)\Tensor{C}^\eta\right):\Tensor{e}(\Vector{u}^\eta)
+\frac38\Gc[I]\left(\frac1{\epsilon}(1-\alpha^\eta)+{\epsilon}\left(\nabla\alpha^\eta\right)^2\right)\,\dd\Omega\\
+\int_{{\GM\I}_{\text{c}}}{\Gc[iI]}\,d\GM\text.
\end{equation}
The internal phase-field variable $\alpha\in[0;1]$ is here defined in a manner that $\alpha=1$ pertains to the intact material and $\alpha=0$ reflects the actual crack.
In that sense it is a damage-like parameter so that $\Fun{\Phi}$ is a degradation function having the properties $\Fun{\Phi}(1)=1$ (the intact state means full stiffness), $\Fun{\Phi}(0)=\delta$ ($\delta$ being a small positive number to guarantee positiveness of the bulk energy also in the crack formation process when $\alpha\rightarrow0$), $\Fun{\Phi}'(0)=0$ (for computational purposes below it is also assumed $\Fun{\Phi}''(x)>0$ for all $x\in[0;1]$).
It should be noted that more frequently the extreme values are interchanged.
Here, it is intended to complement the model with treating the interface in a way of an adhesive contact model as developed previously by the author in~\cite{vodicka16A1}, where  the present choice seems to be a natural notation for a variable related to adhesion (1 for full adhesion, 0 for no adhesion).

The reasoning for the regularisation functional may be naively explained within the solution process of an energy functional minimisation: the term $\frac1{\epsilon}(1-\alpha^\eta)$ controls in the minimisation process distribution of $\alpha$ so that for really small $\epsilon$ the phase-field variable tends to be mostly equal to 1 inside the domains in order the integral not to be very large.
It provides the domains where $\alpha\neq1$ in narrow strips of small width controlled by $\epsilon$.
 The term with gradient   ${\epsilon}\left(\nabla\alpha^\eta\right)^2$ then guarantees continuous distribution of the phase-field variable $\alpha$ in those regions though possibly with high gradients due to the same small value of $\epsilon$. 
The factor $\frac38$ appears due to particular form of the regularisation term for the  corresponding crack energy  to reach the same value $\Gc[I]$ as in Eq.~\eqref{Eq_FunGriffith}, see~\cite{tanne18A1}.

Similar expressions can be used for the interface term $\int_{{\GM\I}_{\text{c}}}{\Gc[iI]}\,d\GM$.
Though the location of the interface crack is obvious, there is still a need to determine position of the crack along the interface at each instant.  
Additionally, properties of the interface deterioration, like adhesiveness, can be captured by such formulation.
It can then also be seen  within the theory of adhesive contact presented in~\cite{raous99A1} and further developed in~\cite{vodicka16A1,vodicka17A2}, too, where it was used to treat the interface stress-strain relations like for cohesive zone models.
To maintain the same structure of the functional, an interface damage parameter $\zeta$ is defined so that the aforementioned interface integral is converted to $\int_{{\GM\I}}{\Gc[iI]}(1-\zeta)\,d\GM$ (here, the integral is over the whole interface, the unknown is $\zeta$), where $\zeta\in[0;1]$ is defined so that $\zeta=1$ (with full adhesion) pertains to the intact interface and $\zeta=0$  (no adhesion) reflects the actual crack.
Additionally, the interface is considered as an infinitesimally thin adhesive layer with its own stiffness \kk{}, which is degraded introducing a degradation function $\Fun{\phi}$ having the properties similar to the bulk degradation function $\Fun{\Phi}$: $\Fun{\phi}(1)=1$, $\Fun{\phi}(0)=0$ (no need to keep it positive for full damage), $\Fun{\phi}'(0)\geq0$, $\Fun{\phi}''(x)>0$ for all $x\in[0;1]$.
This requires to add a term corresponding to the elastic energy in such a definition of the adhesive $\int_{\GM\I}\frac12\left(\Fun{\phi}(\zeta)\kk\JUMP{\Vector{u}}{}\right){\cdot}\JUMP{\Vector{u}}{}\,\dd\Gamma$.

The energy functional containing dependence of both defined internal parameters is finally represented in the following form
\begin{multline}\label{Eq_StoredE}
\Functional{E}\left(t;\Vector{u},\Fun{\alpha},\Fun{\zeta}\right)=
\sum_{\eta=A,B}\int_{\OM^\eta}\!\Fun{\Phi}(\Fun{\alpha}^\eta)\left(\Kp\left|\sph^{+}\! \Tensor{e}(\Vector{u}^\eta)\right|^2+\mu\left|\dev \Tensor{e}(\Vector{u}^\eta)\right|^2\right)+\Kp\left|\sph^{-}\! \Tensor{e}(\Vector{u}^\eta)\right|^2\\
+\frac38 \Gc[I]\left(\frac1{\epsilon}\left(1-\alpha^\eta\right)+{\epsilon}\left(\nabla{\alpha}^\eta\right)^2\right)\,\dd\Omega\\
\displaystyle+\int_{\GM\I}\!\frac12\left(\kappa\Fun{\phi}(\zeta)\JUMP{\Vector{u}}{}\right){\cdot}\JUMP{\Vector{u}}{}+\frac12\kG\left(\JUMP[-]{\Vector{u}}{\sfn}\right)^2+\Gc[iI]\left(\left(1-\zeta\right)+\left(\epsilon\I\nabla_\sfs\zeta\right)^2\right)\,\dd\Gamma\text,
\end{multline}
where $v^\pm=\pm\max(0,\pm v)$.
The value of the functional is valid for any admissible state
\begin{equation}\label{Eq_Constr}
\Vector{u}^\eta|_{\GM\D^\eta}^{}=\Vector{g}^\eta(t)\text,\quad \JUMP{\Vector{u}}{}=\Vector{u}^A{-}\Vector{u}^B\text, \quad 0\leq\alpha^\eta\leq1\text,\quad 0\leq\zeta\leq1\text.
\end{equation}
It can be considered for computational purposes that the non-admissible states have infinite energy \Functional{E}. 
The term containing the new parameter \kG\ was added here to prohibit interpenetration along the interfaces instead of contact conditions, including $\JUMP{\Vector{u}}{\sfn}\geq0$ in Eq.~\eqref{Eq_BCGriffith}, \kG\ introduces normal compression stiffness as a large number.
Though, such a penalisation term enables small interpenetration (negative $\JUMP[-]{\Vector{u}}{\sfn}$), it can be practically explained by roughness or unevenness of the contacting surfaces.  
Further, $\epsilon\I$ is a small number accounting for nonlocal character of $\zeta$ and allowing high gradients (in the tangential space) $\nabla_\sfs$ in $\zeta$ distribution (as in gradient damage theory, see \cite{fremond02B1,vodicka14A1}).
The bulk elastic energy term was written here in a more specific form for an isotropic material defined by two parameters:  the (plain strain) bulk modulus $\Kp$ and the shear modulus $\mu$, see~\cite{kruzik19B1}.
The formulation includes  the additive orthogonal split of the (small) strain tensor $\Tensor{e}= \sph{\Tensor{e}}+\dev{\Tensor{e}}$ into spherical $\sph{\Tensor{e}}$ and deviatoric $\dev{\Tensor{e}}$ parts to demonstrate a possibility of different material  damage  related to volumetric or shear strain (though not considered in the formula), and a split of the spherical part into tensile (+) and compressive (-) parts due to a common assumption that there is no  damage  in pure compression ( the compressive part  does not include the degradation function).

It is also important to recall that damage and crack propagation in present situation is a unidirectional process.
It means that both introduced internal parameters may only decrease by satisfying the conditions $\dot{\zeta}\leq 0$ on \GM\I, and $\dot{{\alpha}}\leq 0$ in $\OM$, where 'dot' means time derivative.
This unidirectionality can be introduced into present energy formulation in a form of dissipation.
It is also known that other nonlinear processes which dissipate the energy exist, like those related to plastic deformations.
They may affect mainly stress states when  shear is significant. 
Even in the case of microscopic influence of those processes, the fracture energy may be different if the crack tends to propagate in a shearing mode. 
That was the reason for the superscript 'I' for \Gc above,  e.g.\ in Eq.~\eqref{Eq_StoredE}  -- to stress that  it is related to the crack mode I, opening.
For shearing, the crack mode II, the  internal and interface fracture energies  are respectively distinguished as $\Gc[II]$ and $\Gc[iII]$, where $\Gc[II]\geq\Gc[I]$ and $\Gc[iII]\geq\Gc[iI]$.
Mentioned processes  can be simulated by introducing  a mode dependent fracture energy as e.g. done by~\cite{hutchinson91A1,benzeggagh96A1} at interfaces.
The influence of the described phenomena can be expressed by a dissipation (pseudo)potential 
\begin{equation}\label{Eq_DissE}
\Functional{R}(\Vector{u};\dot{\alpha},\dot{\zeta})=-\sum_{\eta=A,B}\int_{\OM^\eta}\!\frac3{8\epsilon} \Fun{D}^\eta(\Vector{u})\dot\alpha^\eta\dd\Omega-\int_{\GM\I}\Fun{D}^{\text{i}}(\JUMP{\Vector{u}}{})\dot\zeta\,\dd\Gamma\text,
\quad\dot{\zeta}\leq 0 \text{ on } \GM\I\text,\quad \dot{{\alpha}}\leq 0 \text{ in } \OM\text,
\end{equation}
otherwise it is set to infinity. 

According to \cite{hutchinson91A1,vodicka17A2},  the function $\Fun{D}^{\text{i}}(\JUMP{\Vector{u}}{})$ pertaining to  the interface   can be introduced by the following relation:
\begin{equation}\label{Eq_Difun}
\Fun{D}^{\text{i}}(\Vector{w})=\Gc[iI]\tan^2\left(\frac2\pi\arctan\sqrt{\frac{\Gc[iII]}{\Gc[iI]}-1}\cdot\arctan\frac{w_\sfs}{w_\sfn^+}\right)\text.
\end{equation}
It is clearly seen that in pure opening $\JUMP{\Vector{u}}{\sfs}=0$ and $\Fun{D}^{\text{i}}(\JUMP{\Vector{u}}{})=0$ so that the  corresponding dissipation term  vanishes and no additional energy is dissipated in the crack mode I.
On the other hand, in pure shear  $\JUMP{\Vector{u}}{\sfn}=0$ and the same terms provides the required additional dissipation, as  now  $\Fun{D}^{\text{i}}(\JUMP{\Vector{u}}{})=\Gc[iII]-\Gc[iI]$.
For  internal cracks  the mode mixity character can be introduced by the aforementioned orthogonal split of the strain tensor used for the strain energy based on the formulae of~\cite{benzeggagh96A1,vodicka18A2}.
 From the asymptotic series analysis of the crack tip strain relation, see e.g.\ \cite{aliabadi91B1}, follows that in front of the crack in mode I holds: $\dev\Tensor{e}=0$, and similarly in front of the crack in mode II the relation $\sph\Tensor{e}=0$ is valid.
Thus, the function $\Fun{D}^\eta(\Vector{u})$ defined inside the solids  can be considered in the form
\begin{equation}\label{Eq_Dfun}
\Fun{D}^{\eta}(\Vector{w})=\frac{\Kp^\eta\left|\sph^{+}\! \Tensor{e}(\Vector{w})\right|^2+\mu^\eta\left|\dev \Tensor{e}(\Vector{w})\right|^2}{\frac{\Kp^\eta\left|\sph^{+}\! \Tensor{e}(\Vector{w})\right|^2}{\Gc[I]}+\frac{\mu^\eta\left|\dev \Tensor{e}(\Vector{w})\right|^2}{\Gc[II]}}-\Gc[I]\text,
\end{equation}
and  in the pure crack modes holds:  $\Fun{D}^\eta(\Vector{u})=0$ in the mode I, and $\Fun{D}^\eta(\Vector{u})=\Gc[II]-\Gc[I]$ in the mode II.

Finally, if external forces were taken into account,  their energy should be added  to the system, too. 
Here, such a case is not considered.

The relations which govern the quasi-static evolution in a deformable structure with cracks modelled by internal variables can be generally written in a form of nonlinear variational inclusions, see~\cite{kruzik19B1}.
The particular form here   reads as 
\begin{equation}\label{Eq_Incl}
\begin{alignedat}{2}
&\uppartial_{\Vector{u}}\Functional{E}(t;\Vector{u},\Fun{\alpha},\Fun{\zeta})&&\ni 0\text,\\
\uppartial_{\dot{\Fun{\alpha}}}\Functional{R}(\Vector{u};\dot{\Fun{\alpha}},\dot{\Fun{\zeta}}) +&\uppartial_{\Fun{\alpha}}\Functional{E}(t;\Vector{u},\Fun{\alpha},\Fun{\zeta})&&\ni 0\text,\\
\uppartial_{\dot{\Fun{\zeta}}}\Functional{R}(\Vector{u};\dot{\Fun{\alpha}},\dot{\Fun{\zeta}}) +&\uppartial_{\Fun{\zeta}}\Functional{E}(t;\Vector{u},\Fun{\alpha},\Fun{\zeta})&&\ni 0\text,
\end{alignedat}
\end{equation}
where \uppartial{} denotes a partial subdifferential as the functionals does not have to be smooth, e.\ g.\ \Functional{R} jumps from zero to infinity at zeros of the rate arguments. 
For smooth functionals, the subdifferentials can be replaced by the (Gateaux) differentials and the inclusions by equations.
The first relation in fact determines the deformation state, the other two are flow rules for propagation of internal parameters of phase-field damage and interface damage.
Along with the described relations, initial conditions  for the state variables have to be taken into account:
\begin{equation}\label{Eq_IC}
\rB\Vector{u}^\eta(0,\cdot)=\Vector{u}_{0}^\eta,\quad \Fun{\alpha}^\eta(0,\cdot)=\alpha_{0}^\eta=1\quad\text{ in }\Omega^\eta\text,\qquad\Fun{\zeta}(0,\cdot)=\zeta_0=1\quad\text{ on }\GM\I\text,\rE
\end{equation}
they correspond to an intact state.

The meaning of the relations in Eq.~\eqref{Eq_Incl}  may be read also after the differentiation.
When the first relation is evaluated, the application of divergence theorem and previously introduced boundary conditions (in the text above Eq.~\eqref{Eq_FunGriffith}, or in Eq.~\eqref{Eq_Constr}) render  following set of equations 
\begin{equation}\label{Eq_BVP}
\begin{aligned}
&\Div \sigma\he=0\tc\quad \sigma\he= \Fun{\Phi}(\alpha\he)\Tensor{C}\he\Tensor{e}(\Vector{u}\he)= \Fun{\Phi}(\alpha\he)\left(2\Kp\sph^{+}\! \Tensor{e}(\Vector{u}^\eta)+2\mu\dev \Tensor{e}(\Vector{u}^\eta)\right)+2\Kp\sph^{-}\! \Tensor{e}(\Vector{u}^\eta)\tc & \text{ in } &\OM\he\tc\\
&\Vector{u}\he = \Vector{g}\he(t)\tc & \text{ on } &\GM\he\D\tc\\
&\Vector{p}\he = 0\tc \text{ with } \Vector{p}\he=\sigma\he\Vector{n}\he\tc& \text{ on } &\GM\he\N\tc\\
&\Vector{p}^{A}+\Vector{p}^{B}=0\tc\ {p}_\sfn =\left(\kappa\Fun{\phi}(\zeta)\JUMP{\Vector{u}}{}\right)_\sfn+\kG\JUMP[-]{\Vector{u}}{n}\tc\
{p}_\sfs = \left(\kappa\Fun{\phi}(\zeta)\JUMP{\Vector{u}}{}\right)_\sfs\tc\text{ with }  {p}_\sfn=\Vector{p}^{B}\cdot\Vector{n}^{B}\tc\ {p}_\sfs=\Vector{p}^{B}\cdot\Vector{s}^{B} &\text{ on } &\GM\I\tc
\end{aligned}
\end{equation}
which represent stress equilibrium conditions  with definitions of stress variables  in the damaged domain $\OM\he$  and along its whole boundary  $\GM\he$ (including the interface) at the time instant $t$.

Differentiation in the second relation in Eq.~\eqref{Eq_Incl} requires calculation of  the subdifferential of the non-smooth functional \Functional{R}. 
It provides the relations in form  of complementarity inequalities in the domain $\OM\he$ as follows 
\begin{equation}\label{Eq_BVP_A}
\begin{aligned}
0&\geq\dot\alpha\he\tc\quad 0\leq\alpha\he\tc\quad 0\geq\omega\he\tc\quad 0=\omega\he\alpha\he\tc\\
0&\stackrel{(\ast)}{\geq}\Fun{\Phi}'(\alpha\he)\cdot\left(\Kp\left|\sph^{+}\! \Tensor{e}(\Vector{u}^\eta)\right|^2+\mu\left|\dev \Tensor{e}(\Vector{u}^\eta)\right|^2\right) - \omega\he -\frac3{8\epsilon}\left(\Gc[I]+\Gc[I]\epsilon^2\Delta\alpha\he + \Fun{D}\left(\Vector{u}\right)\right)\tc\\ 
0&=\dot\alpha\he\left(\Fun{\Phi}'(\alpha\he)\cdot\left(\Kp\left|\sph^{+}\! \Tensor{e}(\Vector{u}^\eta)\right|^2+\mu\left|\dev \Tensor{e}(\Vector{u}^\eta)\right|^2\right) - \omega\he -\frac3{8\epsilon}\left(\Gc[I]+\Gc[I]\epsilon^2\Delta\alpha\he + \Fun{D}\left(\Vector{u}\right)\right)\right)\tc\\
& \frac{\uppartial \alpha\he}{\uppartial\Vector{n}}=0\tc \text{ on } \GM\he\tc
\end{aligned}
\end{equation}
where also the divergence theorem was used  (converting the gradient term $\left(\nabla{\alpha}^\eta\right)^2$ of Eq.~\eqref{Eq_StoredE} to the Laplace operator term $-\Delta\alpha\he$) , and the boundary conditions for the phase-field parameter $\alpha\he$ were set in the last equation.
It should be also noted that a Lagrange multiplier  $\omega\he$ was introduced in order to guarantee the constraints from Eq.~\eqref{Eq_Constr}.
The upper bound does not have to be cared about due to  the constraint on the rate variable and the initial condition from Eq.~\eqref{Eq_IC} which pertains to the undamaged state.
These relations  define the flow rule for the evolution of the phase-field variable.

A stress criterion for the phase-field damage triggering and crack nucleation is revealed, if the relation \eqref{Eq_Dfun} is substituted into the $(\ast)$-inequality of Eq.~\eqref{Eq_BVP_A}. 
 At the instant of damage initiation $\alpha\he=1$ (with $\Delta\alpha\he=0$, $\omega\he =0$), $\sph\sigma\he= 2\Kp\he\sph \Tensor{e}(\Vector{u}^\eta)$, $\dev\sigma\he = 2\mu\he\dev \Tensor{e}(\Vector{u}^\eta)$, and the condition for a related critical stress $\sigma\he_{\text{crit}}$ reads
\begin{equation}\label{Eq_StressCrit_A}
\frac{\left|\sph^{+}\! \sigma\he_{\text{crit}}\right|^2}{4\Kp\he\Gc[I]}+\frac{\left|\dev \sigma\he_{\text{crit}}\right|^2}{4\mu\he\Gc[II]}=\frac3{8\epsilon\Fun{\Phi}'(1)}\tb
\end{equation}

One another similar differentiation in the last inclusion of Eq.~\eqref{Eq_Incl} renders relations in a complementarity form valid for the interface  $\GM\I$.
Those relations read 
\begin{equation}\label{Eq_BVP_Z}
\begin{aligned}
0&\geq\dot\zeta\tc\quad 0\leq\zeta\tc\quad 0\geq\omega^{\text{i}}\tc\quad 0=\omega^{\text{i}}\zeta\tc\\
0&\stackrel{(\ast)}{\geq}\Fun{\phi}'(\zeta)\cdot\frac12\left(\kappa\JUMP{\Vector{u}}{}\right)\cdot\JUMP{\Vector{u}}{} - \omega^{\text{i}} -\Gc[iI]-\epsilon\I^2\Delta_\sfs\zeta - \Fun{D}^{\text{i}}\left(\JUMP{\Vector{u}}{}\right)\tc\\ 
0&=\dot\zeta\left(\Fun{\phi}'(\zeta)\cdot\frac12\left(\kappa\JUMP{\Vector{u}}{}\right)\cdot\JUMP{\Vector{u}}{} - \omega^{\text{i}} -\Gc[iI]-\epsilon\I^2\Delta_\sfs\zeta - \Fun{D}^{\text{i}}\left(\JUMP{\Vector{u}}{}\right)\right)\tc\\
& \frac{\uppartial \zeta}{\uppartial\Vector{s}}=0\tc \text{ on } \uppartial\GM\I\tc 
\end{aligned}
\end{equation}
where the by-parts integration was used in the tangential space (providing the  tangential Laplace operator  $\Delta_\sfs$),  and the boundary conditions (at the end points of the interface) for the interface damage parameter $\zeta$ were set in the last equation.
Nevertheless, if the interface is closed (for  inhomogeneities), this condition is omitted.
The interface Lagrange multiplier  $\omega^{\text{i}}$ was introduced in order to guarantee the constraints from Eq.~\eqref{Eq_Constr}, though the upper bound is not active due to  the constraint on the rate variable and the initial condition from Eq.~\eqref{Eq_IC}.
These relations  define the flow rule for the evolution of the interface damage variable.

Analogously to the stress criterion for the phase-field damage, the condition for critical interface stress is obtained, if the relation \eqref{Eq_Difun} is substituted into the $(\ast)$-inequality of Eq.~\eqref{Eq_BVP_Z} considering the value of  intact damage variable  $\zeta=1$ (with $\Delta_\sfs\zeta=0$, $\omega^{\text{i}} =0$).
If the interface stiffness $\kappa$ is represented by a diagonal matrix $\kappa=\left(\begin{smallmatrix} \kappa_\sfn & 0\\ 0 & \kappa_\sfs\end{smallmatrix}\right)$ and $\JUMP{\Vector{u}}{\sfn}\geq0$, the relation for critical interface stresses $p_      {\sfn\,\text{crit}}$, $p_{\sfs\,\text{crit}}$ is expressed as
\begin{equation}\label{Eq_StressCrit_Z}
\frac{p_{\sfn\,\text{crit}}^2}{2\kappa_\sfn}+\frac{p_{\sfs\,\text{crit}}^2}{2\kappa_\sfs}=\frac{\Gc[iI]}{\phi'(1)}\left(1+\tan^2\left(\frac2\pi\arctan\sqrt{\frac{\Gc[iII]}{\Gc[iI]}-1}\cdot\arctan\frac{p_{\sfs\,\text{crit}}\,\kappa_\sfn}{p_{\sfn\,\text{crit}}\,\kappa_\sfs}\right)\right)\tc
\end{equation}
where the formulae for the interface stress from Eq.~\eqref{Eq_BVP} were used.

The solution of the quasi-static evolution problem was based on an energy formulation.
Thus, a form of energy balance should be  available for the system~\eqref{Eq_Incl}.
Anyhow, it is necessary to make a shift of the solution in order to separate the time dependence from the dependence on the displacement \Vector{u}, which is presently bound by the boundary condition in Eq.~\eqref{Eq_Constr}. 
The shift is $\Vector{u}\he=\widetilde{\Vector{u}}\he+\widetilde{\Vector{g}}(t)\he$, where $\widetilde{\Vector{g}}(t)\he$ is a suitably smooth function in $\OM\he$ satisfying the boundary condition on $\GM\D\he$, see also Eq.~\eqref{Eq_BVP}.
The new displacement $\widetilde{\Vector{u}}\he$ then satisfies a vanishing boundary condition which is  time independent.
The energy functional $\Functional{E}$ can be rearranged according to this separation as $\Functional{E}(t;\Vector{u},\Fun{\alpha},\Fun{\zeta})=\widetilde{\Functional{E}}(\widetilde{\Vector{g}}(t);\widetilde{\Vector{u}},\Fun{\alpha},\Fun{\zeta})$.
For  inhomogeneities  or under an assumption that $\GM\D\he$ is far from $\GM\I$ (distance between them is positive), it is legitimate supposing $\JUMP{\widetilde{\Vector{g}}(t)}{}=0$ at the interface.
The interface energy terms, in Eq.~\eqref{Eq_StoredE}, then do not change the form.

Let Eq.~\eqref{Eq_Incl} be written weakly using duality pairings $\langle\cdot,\cdot\rangle$ expressed by some integrals   (though not fully mathematically correctly due to discontinuities in some functions,  e.g.\ in \Functional{R}).
Multiplication of  the relations in Eq.~\eqref{Eq_Incl} in the respective order by  $\dot{\widetilde{\Vector{u}}}$, $\dot\alpha$, $\dot\zeta$, integration over the space domain, and summing them up  renders
\begin{multline}\label{Eq_TestingIncl}
\left\langle\uppartial_{\widetilde{\Vector{u}}}\widetilde{\Functional{E}}(\widetilde{\Vector{g}}(t);\widetilde{\Vector{u}},\Fun{\alpha},\Fun{\zeta}),\dot{\widetilde{\Vector{u}}}\right\rangle+
\left\langle\uppartial_{\dot\alpha}\widetilde{\Functional{E}}(\widetilde{\Vector{g}}(t);\widetilde{\Vector{u}},\Fun{\alpha},\Fun{\zeta}),\dot\alpha\right\rangle+
\left\langle\uppartial_{\dot\zeta}\widetilde{\Functional{E}}(\widetilde{\Vector{g}}(t);\widetilde{\Vector{u}},\Fun{\alpha},\Fun{\zeta}),\dot\zeta\right\rangle\\ +
\left\langle\uppartial_{\dot{\Fun{\alpha}}}\Functional{R}(\Vector{u};\dot{\Fun{\alpha}},\dot{\Fun{\zeta}}),\dot\alpha\right\rangle +\left\langle\uppartial_{\dot{\Fun{\zeta}}}\Functional{R}(\Vector{u};\dot{\Fun{\alpha}},\dot{\Fun{\zeta}}),\dot\zeta\right\rangle =0\tb
\end{multline}
The sum of the terms containing $\widetilde{\Functional{E}}$ can be seen as as a derivative $\frac{\dd \widetilde{\Functional{E}}}{\dd t}-\frac{\uppartial \widetilde{\Functional{E}}}{\uppartial \widetilde{\Vector{g}}}\dot{\widetilde{\Vector{g}}}(t)$ and the sum of the terms containing $\Functional{R}$ represents in view of Eq.~\eqref{Eq_DissE} the value of $\Functional{R}$.
Integrating it over a time span $[0;T]$ (starting from the initial conditions~\eqref{Eq_IC}) provides
\begin{equation}\label{Eq_Ebal_tilde}
\widetilde{\Functional{E}}(\widetilde{\Vector{g}}(T);\widetilde{\Vector{u}}(T),\Fun{\alpha}(T),\Fun{\zeta}(T))-\widetilde{\Functional{E}}(\widetilde{\Vector{g}}(0);\widetilde{\Vector{u}}(0),\Fun{\alpha}_0,\Fun{\zeta}_0)
-\int_0^T\frac{\uppartial \widetilde{\Functional{E}}}{\uppartial \widetilde{\Vector{g}}}(\widetilde{\Vector{g}}(t);\widetilde{\Vector{u}}(t),\Fun{\alpha},\Fun{\zeta})\dot{\widetilde{\Vector{g}}}(t)+\Functional{R}(\Vector{u};\dot\alpha,\dot\zeta)\dd t=0\tb
\end{equation}
Returning back to the original displacements \Vector{u}, the following relation is obtained:
\begin{equation}\label{Eq_Ebal}
\Functional{E}(T;\Vector{u}(T),\Fun{\alpha}(T),\Fun{\zeta}(T))
+\int_0^T\Functional{R}(\Vector{u};\dot\alpha,\dot\zeta)\dd t=\Functional{E}(0;\Vector{u}_0,\Fun{\alpha}_0,\Fun{\zeta}_0)+
\int_0^T\frac{\uppartial \widetilde{\Functional{E}}}{\uppartial \widetilde{\Vector{g}}}(\widetilde{\Vector{g}}(t);\Vector{u}(t)-\widetilde{\Vector{g}}(t),\Fun{\alpha},\Fun{\zeta})\dot{\widetilde{\Vector{g}}}(t)\dd t\tc
\end{equation}
which represents the energy balance: the energy of the system at time $T$ plus energy dissipated during the time interval $[0,T]$ is equal to the the energy of the system at time $0$ plus energy due to the work of external forces at the same time range  represented here by the hard-device loading prescribed by function $\Vector{g}(t)$ at the boundary $\GM\D\he$.
Satisfying an energy balance relation confirms an idea of physical accuracy of the relations governing the quasi-static evolution expressed by Eq.~\eqref{Eq_Incl}. 
Seeing the relation in a thermodynamical view, the dissipation appearing in the energy balance equation~\eqref{Eq_Ebal} is controlled to be positive as Eq.~\eqref{Eq_DissE} states. 
This guarantees non-decreasing entropy and satisfaction of the second law of thermodynamics.

%%%%%%%%%%%%%%%%%%%%%%%%%
\section{Numerical solution and computer implementation }\label{Sec_Implementation}
%%%%%%%%%%%%%%%%%%%%%%%%%

The evolution problem expressed by Eq.~\eqref{Eq_Incl} is to be solved numerically.
Such solution requires various numerical procedures.
They include discretisations  with respect to the time variable  and those describing the deformation and  damage  state of the structural domain.
Appropriate algorithms should be used to both of them. 
Some properties of the algorithms are described below.

%%%%%%%%%%%%%%%%%%%%%%%%%
\subsection{Discretisation of the evolution process}\label{SecImplTime}
%%%%%%%%%%%%%%%%%%%%%%%%%

The model is considered in a quasi-static way with time dependent evolution according to the prescribed load.
The time dependence is approximated by  making a time stepping algorithm.
The time steps are considered to have a fixed size $\tau$, though generally the time step can be variable in the present quasi-static model.
In each time instant $t^k$, for $k=0,1\text,\ \dots\text,\ \frac{T}{\tau}$ of a fixed time range $[0,T]$,  $t^k{=}k\tau$, the state of the system is expressed in terms of the displacement variable $\Vector{u}^k$ and the damage variables $\zeta^k$ and $\alpha^k$.
Therefore, the governing relations from Eq.~\eqref{Eq_Incl} have to be written for separate time instants after discretisation.
At each instant the rates of the state variables are approximated by the finite difference, which means replacements $\dot{\Fun{\zeta}}\approx\frac{\Fun{\zeta}^k{-}\Fun{\zeta}^{k{-}1}}{\tau}$, $\dot{\Fun{\alpha}}\approx\frac{\Fun{\alpha}^k{-}\Fun{\alpha}^{k{-}1}}{\tau}$, where $\zeta^k$, $\alpha^k$ approximate the damage solution at the instant   $t^k$.
It also means that the differentiations with respect to the rates are  substituted by differentiation with respect to  the matching state variable  at the instant $t^k$.

According to the assumptions made  in description of the model, the energy functional \Functional{E}  is separately convex in variables for displacements and in variables for damage  (the degradation functions $\Fun{\Phi}$, $\Fun{\phi}$ are considered with positive second derivative in Eq.~\eqref{Eq_StoredE} and the text above), though such convexity does not hold with respect to all state variables unseparated,  the dissipation functional \Functional{R} is positively homogeneous of degree one in the rate variables ($\Functional{R}(\Vector{u};p\dot{\alpha},q\dot{\zeta})=pq\Functional{R}(\Vector{u};\dot{\alpha},\dot{\zeta})$ for $p>0$, $q>0$).
Therefore, for simplification of solution processes, the time-stepping algorithm is split to follow a staggered solution scheme, in which the solution within each time step is first found for deformation quantities with fixed damage variables taken from the previous step, and afterwards the damage solution (phase-field and interface damage together) is calculated with fixed deformation variable of the current time step.
These assumptions render Eq.~\eqref{Eq_Incl} in the following form:
\begin{equation}\label{Eq_InclDiscr}
\begin{alignedat}{2}
&\uppartial_{\Vector{u}^k}\Functional{E}(t^k;\Vector{u}^k,\Fun{\alpha}^{k-1},\Fun{\zeta}^{k-1})&&\ni 0\text,\\
\tau\,\uppartial_{\Fun{\alpha}^k}\Functional{R}\left(\Vector{u}^k;\frac{\Fun{\alpha}^k{-}\Fun{\alpha}^{k{-}1}}{\tau},\frac{\Fun{\zeta}^k{-}\Fun{\zeta}^{k{-}1}}{\tau}\right) +&\uppartial_{\Fun{\alpha}^k}\Functional{E}(t^k;\Vector{u}^k,\Fun{\alpha}^k,\Fun{\zeta}^k)&&\ni 0\text,\\
\tau\,\uppartial_{\Fun{\zeta}^k}\Functional{R}\left(\Vector{u}^k;\frac{\Fun{\alpha}^k{-}\Fun{\alpha}^{k{-}1}}{\tau},\frac{\Fun{\zeta}^k{-}\Fun{\zeta}^{k{-}1}}{\tau}\right) +&\uppartial_{\Fun{\zeta}^k}\Functional{E}(t^k;\Vector{u}^k,\Fun{\alpha}^k,\Fun{\zeta}^k)&&\ni 0\text,
\end{alignedat}
\end{equation}
and the initial conditions \eqref{Eq_IC} read
\begin{equation}\label{Eq_ICDiscr}
\Vector{u}^0=\Vector{u}_{0}\text,\quad \Fun{\alpha}^0=\Fun{\alpha}_{0}\quad\text{ in }\OM^\eta\text,\qquad\Fun{\zeta}^0=\Fun{\zeta}_0\quad\text{ on }\GM\I\text.
\end{equation}
Having considered the separation of the variables, a variational character of the discretised relations is revealed.
Namely, each time step provides two minimisations which have to be solved:
the first relation in Eq.~\eqref{Eq_InclDiscr}, is in fact first order minimisation condition of the (convex) functional
\begin{eqnarray}\label{Eq_FunctionalsMinU}
\Functional{H}_{\Vector{u}}^k(\hat{\Vector{u}})= \Functional{E}(t^k;\hat{\Vector{u}},\Fun{\alpha}^{k-1},\Fun{\zeta}^{k-1}) 
\end{eqnarray}
and therefore it provides a unique minimiser $\Vector{u}^k=\argmin\Functional{H}_{\Vector{u}}^k(\hat{\Vector{u}})$.
The other two relations in Eq.~\eqref{Eq_InclDiscr}, represent the first order minimisation conditions of the another (convex) functional given by the relation (utilising homogeneity of the funcional \Functional{R})
\begin{eqnarray}\label{Eq_FunctionalsMinZ}
\Functional{H}_{\text{d}}^k(\hat{\Fun{\alpha}},\hat{\Fun{\zeta}})= \Functional{E}(t^k;\Vector{u}^k,\hat{\Fun{\alpha}},\hat{\Fun{\zeta}})+\Functional{R}\left(\Vector{u}^k;\hat{\Fun{\alpha}}-\Fun{\alpha}^{k{-}1},\hat{\Fun{\zeta}}-\Fun{\zeta}^{k{-}1}\right)\text.
\end{eqnarray}
It should be noted that due to unidirectional character of the damage process ($\dot{\Fun{\alpha}}\leq 0$, $\dot{\Fun{\zeta}}\leq 0$),  and the constraints given in Eq.~\eqref{Eq_Constr}, the overall constraints for the minimisation in terms of discretised quantities can be written by the inequalities $0\leq\hat{\Fun{\alpha}}\leq\Fun{\alpha}^{k-1}$, $0\leq\hat{\Fun{\zeta}}\leq\Fun{\zeta}^{k-1}$.

The minimisation here provides the minimiser $(\alpha^k$, $\zeta^k)=\argmin\Functional{H}_{\text{d}}^k(\hat{\Fun{\alpha}},\hat{\Fun{\zeta}})$.

%%%%%%%%%%%%%%%%%%%%%%%%%
\subsection{Finite element approximation and implementation to quadratic programming algorithms}\label{Sec_FEM}
%%%%%%%%%%%%%%%%%%%%%%%%%

Analysing the structure of the functionals in Eqs.~\eqref{Eq_StoredE} and~\eqref{Eq_DissE}, it can be seen that, making an appropriate spatial discretisation, the first minimisation presents a quadratic functional $\Functional{H}_{\Vector{u}}^k$, which  leads to an implementation of minimisation by a  Quadratic Programming (QP)  algorithm, as formulated in~\cite{dostal09B1}. 
In the present model, the normal contact condition provides additional constraints to the minimisation, e.\ g.\  due to the term $\JUMP[-]{\Vector{u}}{}$ in Eq.~\eqref{Eq_StoredE}.
This term can be implemented by bound constraints and used in a constrained conjugate gradient scheme as applied in previous author works~\cite{vodicka14A1,vodicka16A1}. 
The functional $\Functional{H}_{\text{d}}^k$, however, does not have to be quadratic, as the assumptions for the degradation functions \Fun{\Phi} and \Fun{\phi} in Eq.~\eqref{Eq_StoredE} only state that the functions are convex. 
The minimisation with respect to the damage variables thus needs to use  QP sequentially -- Sequential Quadratic Programming (SQP), see~\cite{boggs95A1,bjorkman95} and it is constrained by aforementioned box constraints (below Eq.~\eqref{Eq_FunctionalsMinZ}).
Some details of the computational procedures follow.

The functional~\eqref{Eq_StoredE} contains some terms which make it piecewise quadratic, like that of  $\JUMP[-]{\Vector{u}}{}$ or distinguishing between $\sph^{+}\! \Tensor{e}$ and $\sph^ {-}\! \Tensor{e}$ (expressed in terms of the trace of the strain tensor $\tr\Tensor{e}$), which can be reformulated introducing  new variables $\psi$ and $\omega$  satisfying additional constraints 
\begin{equation}\label{Eq_MoscoRestrictions}
\begin{alignedat}{3}
\psi\he&\geq0\tc &\quad \psi\he+\tr\Tensor{e}(\Vector{u}\he)&\geq 0 \tc&\quad &\text{ in }\OM\he\tc\\
\omega&\geq0\tc & \omega+\JUMP{\Vector{u}}{}&\geq 0\tc& &\text{ on }\GM\I\tb
\end{alignedat}
\end{equation}
This is a classical scheme, also referred to as a Mosco-type transformation~\cite{mosco67A1}.
It was implemented in~\cite{vodicka16A1,vodicka17A2}, too.

A spatial discretisation made by the generated finite element mesh, characterised by its typical mesh size $h$, is introduced by adequate implementation of FEM.
The formulae for the approximation of the state variables and the  newly defined auxiliary variables at the time instant $t^k$, can be written in the form
\begin{equation}\label{Eq_ABapproximation}
\begin{array}{c}\displaystyle
\Vector{u}^k_h(x)=\sum_{n}^{} \Vector{N}_n(x)\Vector{u}^k_{n}\tc\quad
\Fun{\alpha}^k_h(x)=\sum_{n}^{} {N}_n(x)\Fun{\alpha}^k_{n}\tc\quad
\Fun{\zeta}^k_h(x)=\sum_{n}^{} {M}_n(x)\Fun{\zeta}^k_{n}\tc\\[1ex]\displaystyle
\Fun{\psi}^k_h(x)=\sum_{n}^{} {P}_n(x)\Fun{\psi}^k_{n}\tc\quad
\Fun{\omega}^k_h(x)=\sum_{n}^{} {M}_n(x)\Fun{\omega}^k_{n}\tc
\end{array}
\end{equation}
introducing nodal variables  $\Vector{u}^k_n$, $\Fun{\alpha^k}_n$, $\Fun{\zeta}^k_n$, $\Fun{\psi}^k_n$, $\Fun{\omega}^k_n$ and appropriate nodal shape functions according to the type of approximation considered in the FEM mesh ${N}_n(x)$, ${P}_n(x)$, ${M}_n(x)$.
Matrices generated from them and those necessary for computations include the following ones:
\begin{equation}\label{Eq_Nmatrices}
\Vector{N}_n=\begin{pmatrix} {N}_n & 0 \\ 0 & {N}_n\end{pmatrix}\tc\quad
\Vector{B}_n=\begin{pmatrix} {N}_{n\,,1} & 0 \\ 0 & {N}_{n\,,2}\\ {N}_{n\,,2} & {N}_{n\,,1}\end{pmatrix}\tc\quad
\bar{\Vector{N}}_n=\begin{pmatrix} {N}_{n\,,1} & {N}_{n\,,2}\end{pmatrix}\tc\quad
\bar{M}_n= \nabla_\sfs {M}_{n}\tc
\end{equation}
where subscripts separated by comma refer to differentiation with respect to  the corresponding spatial variable  $x_1$, or $x_2$, cf.\ Fig.~\ref{Fig_Model}.
Additionally for the split in strains, two constant matrices are used
\begin{equation}\label{Eq_SDmatrices}
\Matrix{S} = \begin{pmatrix} 1 & 1 & 0 \\ 1 & 1 & 0 \\ 0 & 0 & 0 \end{pmatrix}\tc\quad
\Matrix{D} = \begin{pmatrix} 1 & -1 & 0 \\ -1 & 1 & 0 \\ 0 & 0 & 1 \end{pmatrix}\tb
\end{equation}
Now, by means of the functional~\eqref{Eq_StoredE}, the restricted functional ~\eqref{Eq_FunctionalsMinU} is written at the current time step after eliminating terms which contain only $\alpha^{k-1}$ or $\zeta^{k-1}$, as those do not affect the minimisation with respect to the displacement variables.
The modified functional $\widetilde{\Functional{H}}_{\Vector{u}}^k(\hat{\Vector{u}}_h,\hat{\psi}_h,\hat{\omega}_h)$ then  reads 
\begin{multline}\label{Eq_HuDiscr}
\widetilde{\Functional{H}}_{\Vector{u}}^k(\hat{\Vector{u}}_h,\hat{\psi}_h,\hat{\omega}_h)=
\sum_{\eta=A,B}\left(\sum_{m}^{}\sum_{n}^{}\hat{\Vector{u}}^{\top}_{m}\left(\int_{\OM^\eta}\!\Fun{\Phi}(\Fun{\alpha}^{\eta\,k-1}_h)\left(\Kp\Vector{B}^\top_m\Matrix{S}\Vector{B}_n+\mu\Vector{B}^\top_m\Matrix{D}\Vector{B}_n\right)\,\dd\Omega\right)\hat{\Vector{u}}^{}_{n}\right.\\
\left.+\sum_{m}^{}\sum_{n}^{}\hat{\psi}^{}_{m}\left(\int_{\OM^\eta}\left(1-\Fun{\Phi}(\Fun{\alpha}^{\eta\,k-1}_h)\right)\Kp {P}_m{P}_n\,\dd\Omega\right)\hat{\psi}^{}_{n}\right)\\
\displaystyle+\sum_{m}^{}\sum_{n}^{}\JUMP[\top]{\hat{\Vector{u}}^{}_{m}}{}\left(\int_{\GM\I}\!\frac12\kappa\Fun{\phi}(\zeta^{k-1}_h)\Vector{N}^\top_m\Vector{N}_n\,\dd\Gamma\right)\JUMP{\hat{\Vector{u}}^{}_{n}}{}
+\sum_{m}^{}\sum_{n}^{}\hat{\omega}^{}_{m}\left(\int_{\GM\I}\frac12\kG M_m M_n\,\dd\Gamma\right)\hat{\omega}^{}_{n}\tc
\end{multline}
with additional constraints on auxiliary variables provided according to the conditions~\eqref{Eq_MoscoRestrictions} as follows:
\begin{equation}\label{Eq_MoscoDiscr}
\begin{alignedat}{3}
\hat{\psi}_n&\geq0\tc &\quad \hat{\psi}_n+\bar{\Vector{N}}_n\hat{\Vector{u}}^{}_{n}&\geq 0&\quad &\text{ for }x_n\in\OM\he\tc\\
\hat{\omega}_n&\geq0\tc & \hat{\omega}_n+\JUMP{\hat{\Vector{u}}_n}{\sfn}&\geq 0& &\text{ for }x_n\in\GM\I\tb
\end{alignedat}
\end{equation} 
If the nodal values are gathered to column vectors $\hat{\Matrix{u}}_h$, $\hat{\Matrix{\psi}}_h$, $\hat{\Matrix{\omega}}_h$  with respect to the previous notation  and the integrals in Eq.~\eqref{Eq_HuDiscr} are used to determine matrices $\Matrix{K}_{\Vector{u}}$, $\Matrix{K}_{\psi}$, $\Matrix{K}_{\omega}$ the algebraic minimisation with constraints~\eqref{Eq_MoscoDiscr} is resolved  by the functional
\begin{equation}\label{Eq_DiscreteGeneralU}
\widetilde{\Functional{H}}^k_{\Vector{u},h}(\hat{\Matrix{u}}_h, \hat{\Matrix{\psi}}_h, \hat{\Matrix{\omega}}_h)
=\hat{\Matrix{u}}^{\top}_h\Matrix{K}_{\Vector{u}}(\Fun{\alpha}^{k-1}_h,\Fun{\zeta}^{k-1}_h)\hat{\Matrix{u}}_h +
\hat{\Matrix{\psi}}{\top}_h\Matrix{K}_{\psi}(\Fun{\alpha}^{k-1}_h)\hat{\Matrix{\psi}}_h+
\hat{\Matrix{\omega}}^{\top}_h\Matrix{K}_{\omega}\hat{\Matrix{\omega}}_h\tc
\end{equation}
in which the stiffness matrices $\Matrix{K}_{\Vector{u}}$, $\Matrix{K}_{\psi}$ depend on the actual state of the damage variables taken from the previous time step, as can be read in  Eq.~\eqref{Eq_HuDiscr}.
This dependence is stressed in the relation.
The minimiser $\Matrix{u}^k_h$ is obtained by a QP algorithm (implemented by a conjugate gradient based scheme with bound constraints \cite{dostal09B1}) and provides the nodal displacements at the $k$th time step, i.\ e.\ for $\hat{\Matrix{u}}_h=\Matrix{u}^k_h$, which determines the (approximated) solution $\Vector{u}^k_h$ of~Eq.~\eqref{Eq_ABapproximation}.

Similarly, the substitutions of approximations~\eqref{Eq_ABapproximation} into the functional~\eqref{Eq_FunctionalsMinZ} and using the definitions in Eqs.~\eqref{Eq_StoredE} and~\eqref{Eq_DissE} lead to another expression which  provides the solution in the damage variables.

Before writing the formula, recall that the degradation functions $\Phi$ and $\phi$ in Eq.~\eqref{Eq_StoredE} are considered to be convex.
If they are not  quadratic, the aforementioned QP algorithm should be applied sequentially. 
It means that the degradation functions are approximated by  the quadratic Taylor polynomial and the solution is  found iteratively.
Namely, at each time step $t^k$, an iteration starts with $r=1$ by putting $\Fun{\alpha}^{k,r-1}_{h}=\Fun{\alpha}_h^{k{-}1}$, $\Fun{\zeta}^{k,r-1}_{h}=\Fun{\zeta}_h^{k{-}1}$, and it finds the minimum of an iterated functional $\Functional{H}^{k,r}_{\text{d},h}(\hat{\Fun{\alpha}}_h,\hat{\Fun{\zeta}}_h)$  which differs from  $\Functional{H}_{\text{d}}^k(\hat{\Fun{\alpha}}_h,\hat{\Fun{\zeta}}_h)$ of Eq.~\eqref{Eq_FunctionalsMinZ} in a replacement of degradation functions by their quadratic approximations
\begin{equation}\label{Eq_Kapprox}
\begin{alignedat}{1}
\Fun{\Phi}(\hat{\Fun{\alpha}}_h)&\approx\Fun{\Phi}(\Fun{\alpha}^{k,r-1}_h)+\Fun{\Phi}'(\Fun{\alpha}^{k,r-1}_h)\left(\hat{\Fun{\alpha}}_h-\Fun{\alpha}^{k,r-1}_{h}\right)
+\frac12\Fun{\Phi}''(\Fun{\alpha}^{k,r-1}_h)\left(\hat{\Fun{\alpha}}_h-\Fun{\alpha}^{k,r-1}_{h}\right)^2\tc\\
\Fun{\phi}(\hat{\Fun{\zeta}}_h)&\approx\Fun{\phi}(\Fun{\zeta}^{k,r-1}_h)+\Fun{\phi}'(\Fun{\zeta}^{k,r-1}_h)\left(\hat{\Fun{\zeta}}_h-\Fun{\zeta}^{k,r-1}_{h}\right)
+\frac12\Fun{\phi}''(\Fun{\zeta}^{k,r-1}_h)\left(\hat{\Fun{\zeta}}_h-\Fun{\zeta}^{k,r-1}_{h}\right)^2\tb
\end{alignedat}
\end{equation} 
where $\Fun{\alpha}^{k,r-1}_{h}$, $\Fun{\zeta}^{k,r-1}_{h}$ are known from the previous iteration.
It should be stressed one again that the second derivatives are supposed to be positive to satisfy the required convexity of the functions $\Phi$ and  $\phi$.

Now, by means of the functionals~\eqref{Eq_StoredE} and~\eqref{Eq_DissE}, the restricted functional~\eqref{Eq_FunctionalsMinZ} at the current time step is written after eliminating terms which contain only displacement variables or constants (the functional is denoted $\widetilde{\Functional{H}}^{k,r}_{\text{d},h}$), they have no effect in minimisation with respect to $\alpha$ and $\zeta$.
It reads
\begin{multline}\label{Eq_HdDiscr}
\widetilde{\Functional{H}}^{k,r}_{\text{d},h}(\hat{\Fun{\alpha}}_h,\hat{\Fun{\zeta}}_h)=
\sum_{\eta=A,B}\left(\sum_{n}^{}\left(\int_{\OM^\eta}\!\left(\Fun{\Phi}'(\Fun{\alpha}^{k,r-1}_h)-\Fun{\alpha}^{k,r-1}_{h}\Fun{\Phi}''(\Fun{\alpha}^{k,r-1}_h)\right)\right.\right.\\
\shoveright{\left.\times\left(\Kp\left|\sph^{+}\! \Tensor{e}(\Vector{u}^{\eta\,k}_h)\right|^2+\mu\left|\dev \Tensor{e}(\Vector{u}^{\eta\,k}_h)\right|^2\right)N_n\,\dd\Omega\right)\hat{\alpha}_n}\\
+\sum_{m}^{}\sum_{n}^{}\hat{\alpha}_m\left(\int_{\OM^\eta}\!\frac12\Fun{\Phi}''(\Fun{\alpha}^{k,r-1}_h)\left(\Kp\left|\sph^{+}\! \Tensor{e}(\Vector{u}^{\eta\,k}_h)\right|^2+\mu\left|\dev \Tensor{e}(\Vector{u}^{\eta\,k}_h)\right|^2\right)N_mN_n\,\dd\Omega\right)\hat{\alpha}_n\\
\left.-\sum_{n}^{}\left(\int_{\OM^\eta}\frac3{8\epsilon}\left(\Gc[I]+\Fun{D}^\eta(\Vector{u}^{\eta\,k}_h)\right)N_n\,\dd\Omega\right)\hat{\alpha}_n+
\sum_{m}^{}\sum_{n}^{}\hat{\alpha}_m\left(\int_{\OM^\eta}\frac{3\epsilon}8\Gc[I]\bar{\Vector{N}}_m\bar{\Vector{N}}^\top_n\,\dd\Omega\right)\hat{\alpha}_n\right)\\
\displaystyle+\sum_{n}^{}\left(\int_{\GM\I}\!\frac12\left(\kappa\left(\Fun{\phi}'(\Fun{\zeta}^{k,r-1}_h)-\Fun{\zeta}^{k,r-1}_{h}\Fun{\phi}''(\Fun{\zeta}^{k,r-1}_h)\right)\JUMP{\Vector{u}^k_h}{}\right){\cdot}\JUMP{\Vector{u}^k_h}{}M_n\,\dd\Gamma\right)\hat{\zeta}_n\\
\displaystyle+\sum_{m}^{}\sum_{n}^{}\hat{\zeta}_m\left(\int_{\GM\I}\!\frac12\left(\kappa\frac12\Fun{\phi}''(\Fun{\zeta}^{k,r-1}_h)\JUMP{\Vector{u}^k_h}{}\right){\cdot}\JUMP{\Vector{u}^k_h}{}M_m M_n\,\dd\Gamma\right)\hat{\zeta}_n\\
-\sum_{n}^{}\left(\int_{\GM\I}\left(\Gc[iI]+\Fun{D}^{\text{i}}(\JUMP{\Vector{u}^k_h}{})\right)M_n\,\dd\Gamma\right)\hat{\zeta}_n
+\sum_{m}^{}\sum_{n}^{}\hat{\zeta}_m\left(\int_{\GM\I}\Gc[iI]\bar{M}_m\bar{M}_n\,\dd\Gamma\right)\hat{\zeta}_n\tb
\end{multline}
If also here the nodal values are gathered to column vectors $\hat{\Matrix{\alpha}}_h$, $\hat{\Matrix{\zeta}}_h$ in accordance with the previous notation and the integrals in Eq.~\eqref{Eq_HdDiscr} are used to introduce matrices $\Matrix{K}_{\alpha}$, $\Matrix{K}_{\zeta}$, $\Matrix{Q}_{\alpha}$, $\Matrix{Q}_{\zeta}$, the algebraic  minimisation accounts for the matrix functional
\begin{equation}\label{Eq_DiscreteGeneralD}
\Functional{H}^{k,r}_{\text{d},h}(\hat{\Matrix{\alpha}}_h,\hat{\Matrix{\zeta}}_h)
=\hat{\Matrix{\alpha}}^{\top}_h\Matrix{K}_{\alpha}(\Fun{\alpha}^{k,r-1}_h,\Vector{u}^{\eta,k}_h)\hat{\Matrix{\alpha}}_h +
\hat{\Matrix{\zeta}}^{\top}_h\Matrix{K}_{\zeta}(\Fun{\zeta}^{k,r-1}_h,\Vector{u}^{\eta,k}_h)\hat{\Matrix{\zeta}}_h+
\Matrix{Q}_{\alpha}(\Fun{\alpha}^{k,r-1}_h,\Vector{u}^{\eta,k}_h)\hat{\Matrix{\alpha}}_h+\Matrix{Q}_{\zeta}(\Fun{\zeta}^{k,r-1}_h,\Vector{u}^{\eta,k}_h)\hat{\Matrix{\zeta}}_h\tc
\end{equation}
and constraints as the nodal equivalence of those introduced below Eq.~\eqref{Eq_FunctionalsMinZ}
\begin{equation}\label{Eq_DConstrDiscr}
0\leq\hat{\Matrix{\alpha}}_h\leq\Matrix{\alpha}^{k-1}_h\tc\qquad 0\leq\hat{\Matrix{\zeta}}_h\leq\Matrix{\zeta}^{k-1}_h\tb
\end{equation}
All the  matrices in Eq.~\eqref{Eq_DiscreteGeneralD} depend on the actual displacement state calculated in the first minimisation within the current time step and the values of the damage variables taken form the previous iteration, as can be read in  Eq.~\eqref{Eq_HdDiscr}.
This dependences are stressed in the relation.

The minimum within each iteration is obtained by a constrained conjugate gradient scheme of a QP algorithm.
The constrained minimiser is denoted $(\Matrix{\alpha}^{k,r}_h,\Matrix{\zeta}^{k,r}_h)$ and determines nodal values (for $(\hat{\Matrix{\alpha}}_h,\hat{\Matrix{\zeta}}_h)=(\Matrix{\alpha}^{k,r}_h,\Matrix{\zeta}^{k,r}_h)$) of the approximation $(\Fun{\alpha}^{k,r}_{h}, \Fun{\zeta}^{k,r}_h)$ given by  corresponding approximation formula  from Eq.~\eqref{Eq_ABapproximation}.
Such a minimisation is repeated by stepping the counter $r$ until a convergence criterion 
$\max(\|\Fun{\alpha}^{k,r}_{h}-\Fun{\alpha}^{k,r-1}_{h}\|,\|\Fun{\zeta}^{k,r}_{h}-\Fun{\zeta}^{k,r-1}_{h}\|)<\varepsilon$, for a small $\varepsilon$,  is met for $r=\bar{r}$.
Then, it is put $\Fun{\alpha}^k_{h}=\Fun{\alpha}^{k,\bar{r}}_h$, $\Fun{\zeta}^k_{h}=\Fun{\zeta}^{k,\bar{r}}_h$, which stops the SQP iteration and completes the solution at the time step $t^k$.

The time-stepping process is applied recursively up to given time T.

%%%%%%%%%%%%%%%%%%%%%%%%%%%%
\section{ Numerical examples}\label{Sec_Examples}
%%%%%%%%%%%%%%%%%%%%%%%%%%%%

The computational procedures are tested  in solution of several problems which include combination of  damage  processes appearing along an interface between an  inhomogeneity  and matrix of a material and inside those materials.
It is intended to present situations where both such crack formation processes appear simultaneously.
The calculations utilise an own author  computer code written in Matlab, which covers implementation of a FEM algorithm originally based on~\cite{alberty02A1}, presently used to couple the elastic solutions with a phase-field discretisation and an interface damage discretisation as shown in the author's works~\cite{vodicka19A3,vodicka22A1}.
The general meshes introduced below for particular tests have been generated by the mesh generator GMSH~\cite{geuzaine09A1}.
The minimisation procedures for above described approached were particularly implemented  using a standard algorithm called Modified Proportioning with Reduced Gradient Projections based on the schemes of~\cite{dostal09B1}.

The solved problems follow two directions.
First,  mutual interaction of crack formation processes in bulk and along interface is tested in a structural domain including one  inhomogeneity or three inhomogeneities subjected  to a simple tensional load.
The solved domains are shown in Figs.~\ref{Fig_MaFi3} and~\ref{Fig_Incl}. 
Second, the effect of dependence of fracture energy on  a type of loading and state of the stress or strain in the material is discussed in examples which  cover the cases of compressive loading and a combined load of compression and tension which causes shear.
The domains are shown in Figs.~\ref{Fig_TP} and~\ref{Fig_NoMoB}.

%%%%%%%%%%%%%%%%%%%%%%%%%%%%
\subsection{A domain with several  inhomogeneities under  tension}\label{Sec_MaFi3}
%%%%%%%%%%%%%%%%%%%%%%%%%%%%

\begin{figure}[!ht]
\centering
\begin{subfigure}{0.49\textwidth}
\centering
\includegraphics[scale=1]{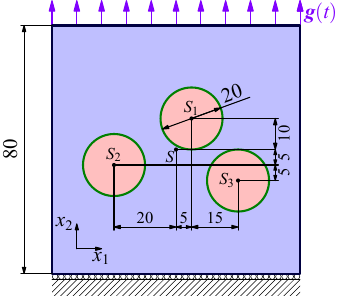}
\caption{The domain with with three  inhomogeneities. }\label{Fig_MaFi3_G}
\end{subfigure}
\begin{subfigure}{0.49\textwidth}
\centering
\includegraphics[scale=0.32,trim=170 30 150 20,clip]{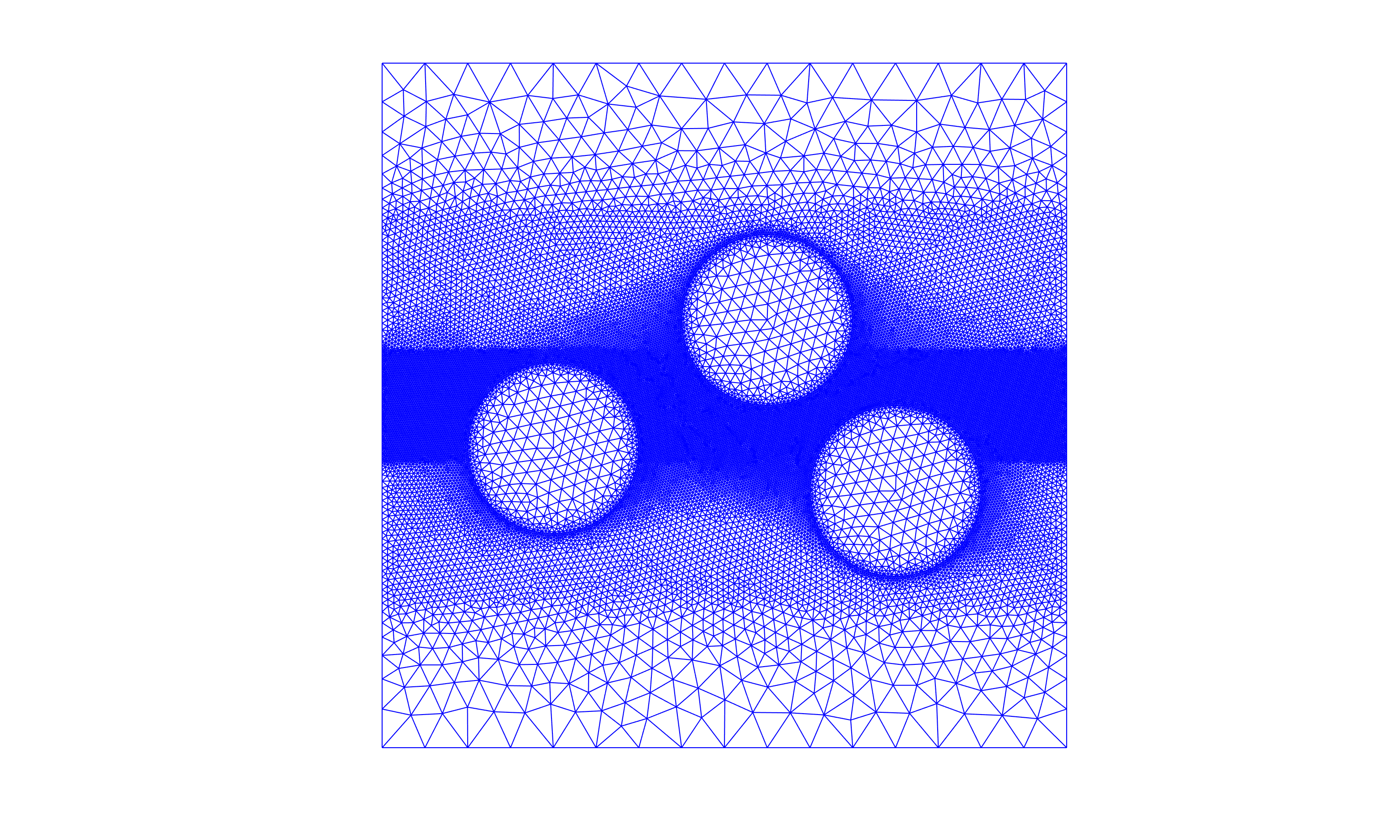}
\caption{Refinement of the mesh.}\label{Fig_MaFi3_M}
\end{subfigure}
\caption{ Description and discretisation of the solid with three inhomogeneities.}\label{Fig_MaFi3}
\end{figure}

The analysis is started with a structural element which contains three  inhomogeneities  of a material different from the matrix domain.
The scheme is shown in Fig.~\ref{Fig_MaFi3}.
The figure also presents a mesh which is used in the calculations.
It is refined in the zones where cracks are  expected, determined according to some preliminary calculations not shown here.
Here, the mesh is refined down to the smallest element size of $0.25$ mm.
 This refinement is required to achieve an appropriate approximation  of gradients in the distribution of  the phase-field variable $\alpha$ within the band width determined by the parameter $\epsilon$, cf.\ Eq.~\eqref{Eq_StoredE}, which is set to  the value $0.75$ mm.
Moreover, additional horizontal constraints are put at the midpoints of the horizontal faces of the matrix domain for the elastic solution to be unique even after total rupture.

The parameters needed for the computations include the material stiffness values which according to Eq.~\eqref{Eq_StoredE} mean the  bulk modulus $\Kp=51.8$ GPa, the shear modulus $\mu=29.0$ GPa for the  inhomogeneity , and $\Kp=3.1$ GPa, $\mu=1.0$ GPa for the matrix. 
The interface, considered as a thin adhesive layer, is characterised by $\kappa = \left(\begin{smallmatrix}1 & 0\\0 & 0.5\end{smallmatrix}\right)
\text{PPam}^{-1}$, $\kG=10\ \text{PPam}^{-1}$. 

The fracture energy in the matrix domain where the cracks are observed is $\Gc[I]=750\ \text{Jm}^{-2}$. 
Additionally, it is supposed that the shear mode fracture energy may have different values as introduced in the additional dissipation term in Eq.~\eqref{Eq_Dfun}.
The particular values $\Gc[II]\in\{\Gc[I],2\Gc[I],10\Gc[I]\}$ are considered to demonstrate the differences. 

 Damage  of the interface is controlled by the interface fracture energy $\Gc[iI]=1\ \text{Jm}^{-2}$. 
The mode II interface fracture energy is set to a high value here not to influence the appearing of the interface crack according to the stress condition~\eqref{Eq_StressCrit_Z}.
Its influence will be observed in the next examples below. 
It was also studied  in \cite{vodicka16A1}.
In particular,   with the interface degradation function $\Fun{\phi}(\zeta)=\frac{\exp\left(-\gamma^{-1}(\beta\zeta+\delta)\right)-\exp\left(-\gamma^{-1}(\delta)\right)}{\exp\left(-\gamma^{-1}(\beta+\delta)\right)-\exp\left(-\gamma^{-1}(\delta)\right)}$ with $\gamma(z)=\exp(-z)\left(1+z+\frac12z^2\right)$, the stress condition provides maximum of normal interface stress  by $\sigma_\text{crit}=\sqrt{\frac{\exp\left(\gamma^{-1}(\beta+\delta)-2\right)\kappa_\sfn\Gc[iI]}{\beta}}=13.85$ MPa, where the parameters were set as follows: $\beta=0.99$, $\delta=0.005$. 
It should be noted that with the present degradation function the stress maximum occurs  for initiated damage $\Fun{\zeta}<1$ (approx 0.9, see~\cite{vodicka16A1}).

Similar analysis for the domain fracture is done for the basic degradation function $\Fun{\Phi}(\alpha) = \alpha^2+10^{-6}$ used in the calculations and based on the stress condition~\eqref{Eq_StressCrit_A}, or better on its modification using plane stress trace $\tr\sigma$: $(\sph \sigma)^2=\frac{(\tr\sigma)^2}2$ (used also in graphical representation of results below)
\begin{equation}\label{Eq_StressCrit_Amod}
\left(1+\frac\mu\Kp\left(1-\frac{\Gc[I]}{\Gc[II]}\right)\right)\left(\tr^{+}\! \sigma_{\text{crit}}\right)^2+\frac{2\Kp}{\mu}\frac{\Gc[I]}{\Gc[II]}\left|\dev \sigma_{\text{crit}}\right|^2=\frac{3\Kp\Gc[I]}{\epsilon\Fun{\Phi}'(1)}\tb
\end{equation}
It means that for moderate values of shear stresses  contained in the deviatoric part of stress $|\dev\sigma|$, and for the cases where $\Gc[II]$ is substantially greater than $\Gc[I]$, the critical value of $\tr\sigma$ is given by an approximate relation $\tr\sigma_{\text{crit}}=\sqrt{\frac{3\Gc[I]}{\epsilon\Fun{\Phi}'(1)}\frac{\Kp^2}{\Kp+\mu}}$.
This gives the value $59.1$ MPa in the present case, with the present value of the length scale parameter $\epsilon$.

The structural element is loaded by the displacement loading $g(t)=v_0t$, at the velocity $v_0=1\ \text{mm\,s}^{-1}$. 
This load is applied incrementally in  time steps refined to $0.1$ ms.

\begin{figure}[!ht]
\centering
\includegraphics[scale=0.3,trim=5 5 40 50,clip]{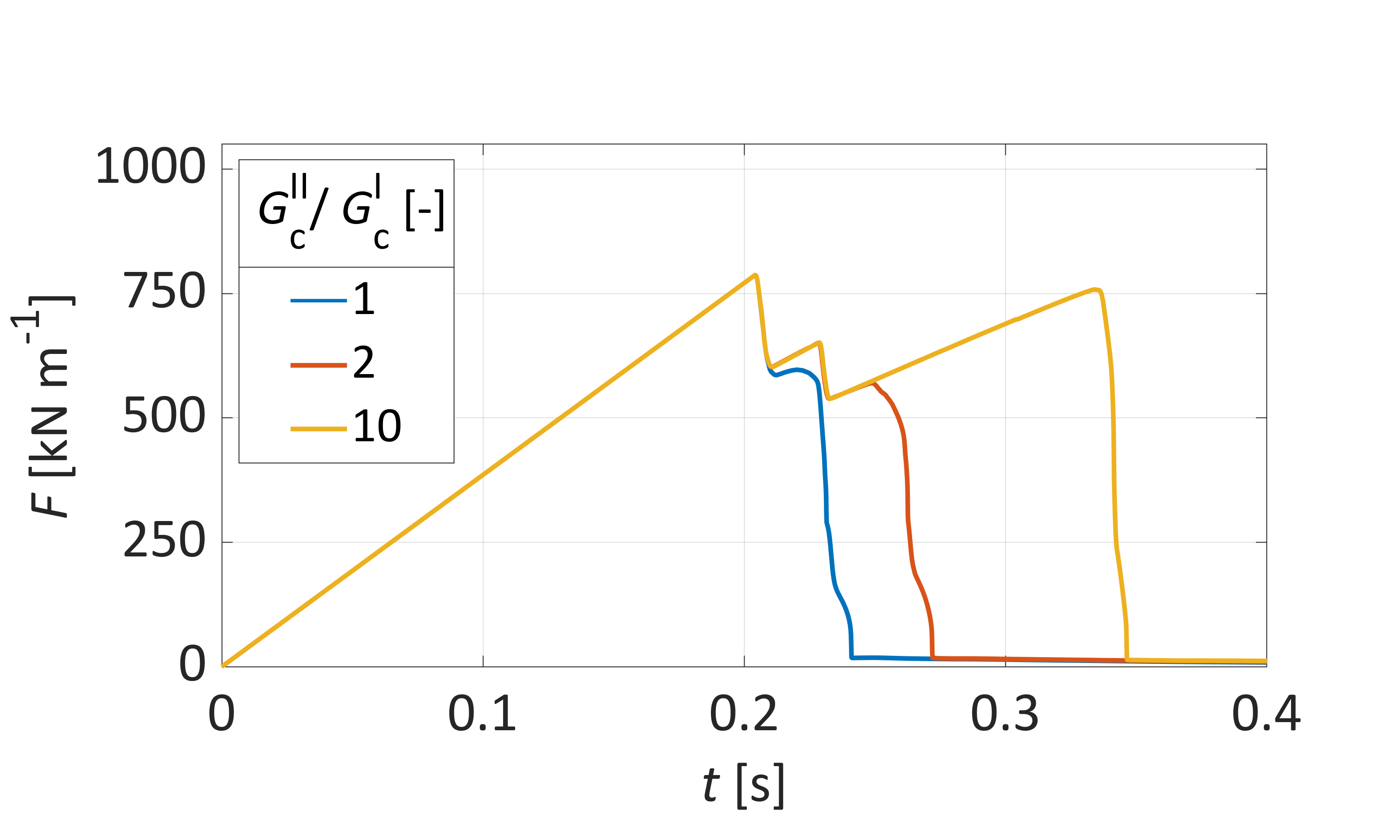}
\caption{Total force applied at the top face during loading for  the solid with three inhomogeneities subjected  to tension.}\label{Fig_MaFi3_F}
\end{figure}

The global force response of the structural element is shown in~Fig.~\ref{Fig_MaFi3_F}.
The evolution of the total vertical force $F$ applied at the top face of the matrix domain in time as a representation of the prescribed displacement $g$ is demonstrated.
The graphs include all three options of the shear fracture energy $\Gc[II]$.
As it will be seen bellow, the two jumps in the force are caused by appearing of interface cracks at various interfaces.
In the case of the smallest value of $\Gc[II]$ the second jump approximately corresponds to crack propagation in matrix domain which finally causes total rupture of the analysed domain.
As the stress state close to the  inhomogeneity  is general, the instant of crack initiation in the matrix domain depends also on the value of $\Gc[II]$: with its higher value, a higher value of loading parameter $g$ is needed.
Anyhow, in all cases the loading is terminated by an abrupt crack propagation across the whole domain.

\begin{figure}[!ht]
\centering
\includegraphics[scale=0.3,trim=5 5 40 50,clip]{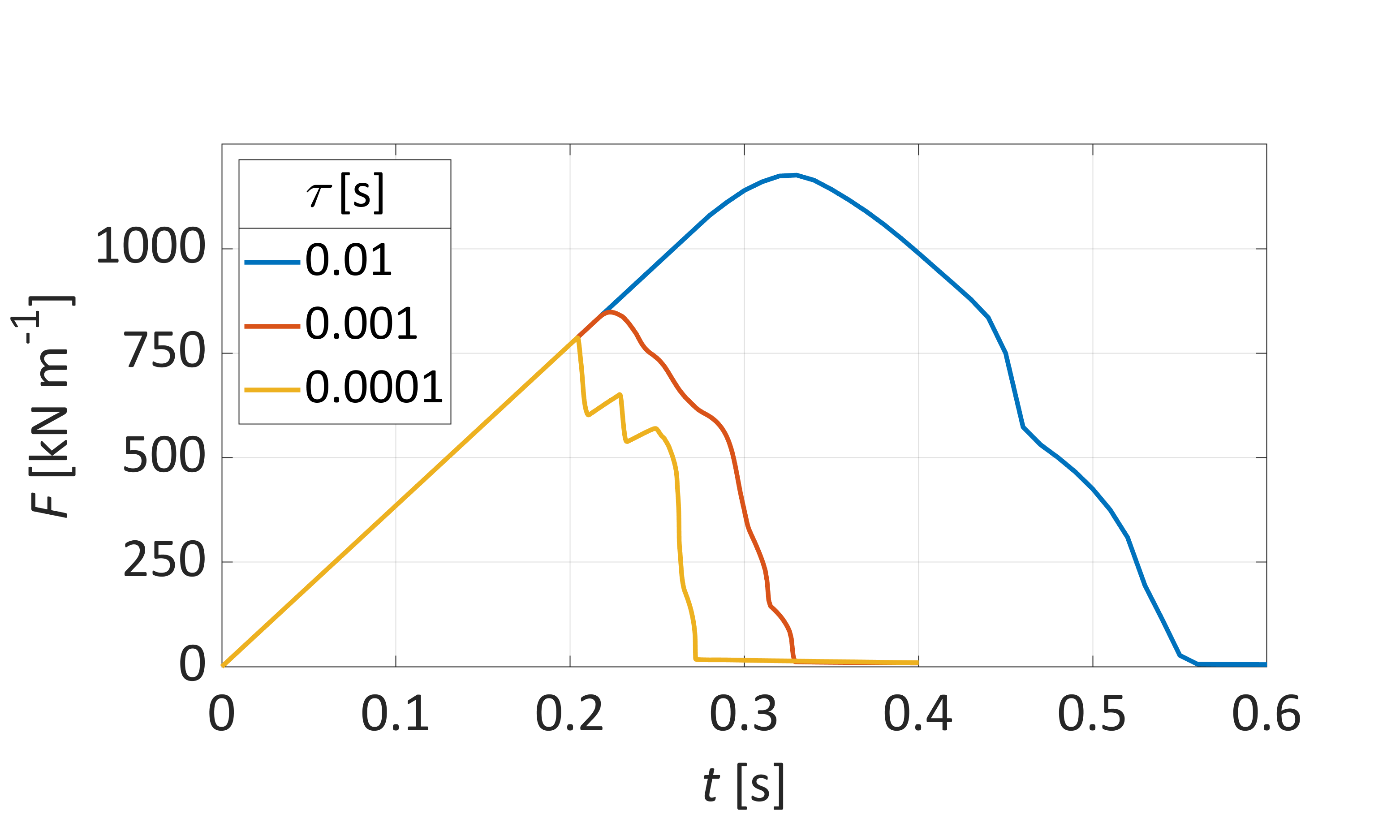}
\caption{Total force applied at the top face during loading, the case of  the  solid with three inhomogeneities , influence of the time step $\tau$ for the case with $\Gc[II]=2\Gc[I]$.}\label{Fig_MaFi3_F_conv}
\end{figure}

In the present staggered computational model the evolution of damage, either along the interface or in the solid domain, strongly depends on the value of the time step. 
A substantially refined time step is required to get the observed natural behaviour.
A comparison of the results for different time steps $\tau$  is shown in Fig.~\ref{Fig_MaFi3_F_conv}.
As it does not depend on the value of $\Gc[II]$ only one option is selected. 
The larger value of $\tau$ leads in staggered approach to slower crack propagation which does not have to reach its natural abrupt character.
Nevertheless, the previous proposed value of the time step parameter provides satisfactorily good approximation of possibly unstable crack propagation. 

\begin{figure}[!ht]
\centering
\begin{subfigure}{0.49\textwidth}
\centering
\includegraphics[scale=0.22,trim=5 5 45 45,clip]{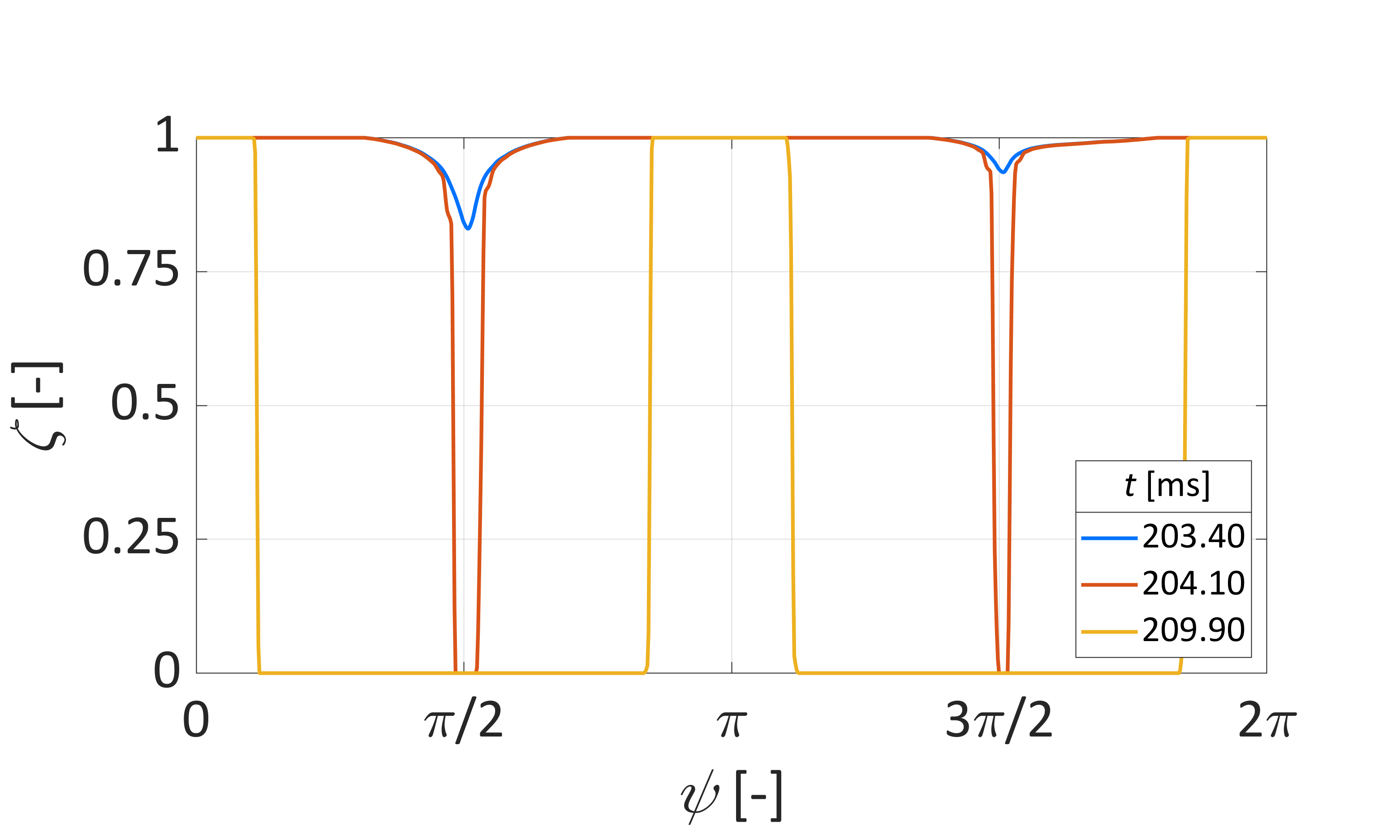}
\caption{Interface damage, the  inhomogeneity  at $S_1$}\label{Fig_MaFi3_ratG2_IF1_1}
\end{subfigure}
\begin{subfigure}{0.49\textwidth}
\centering
\includegraphics[scale=0.22,trim=5 5 45 45,clip]{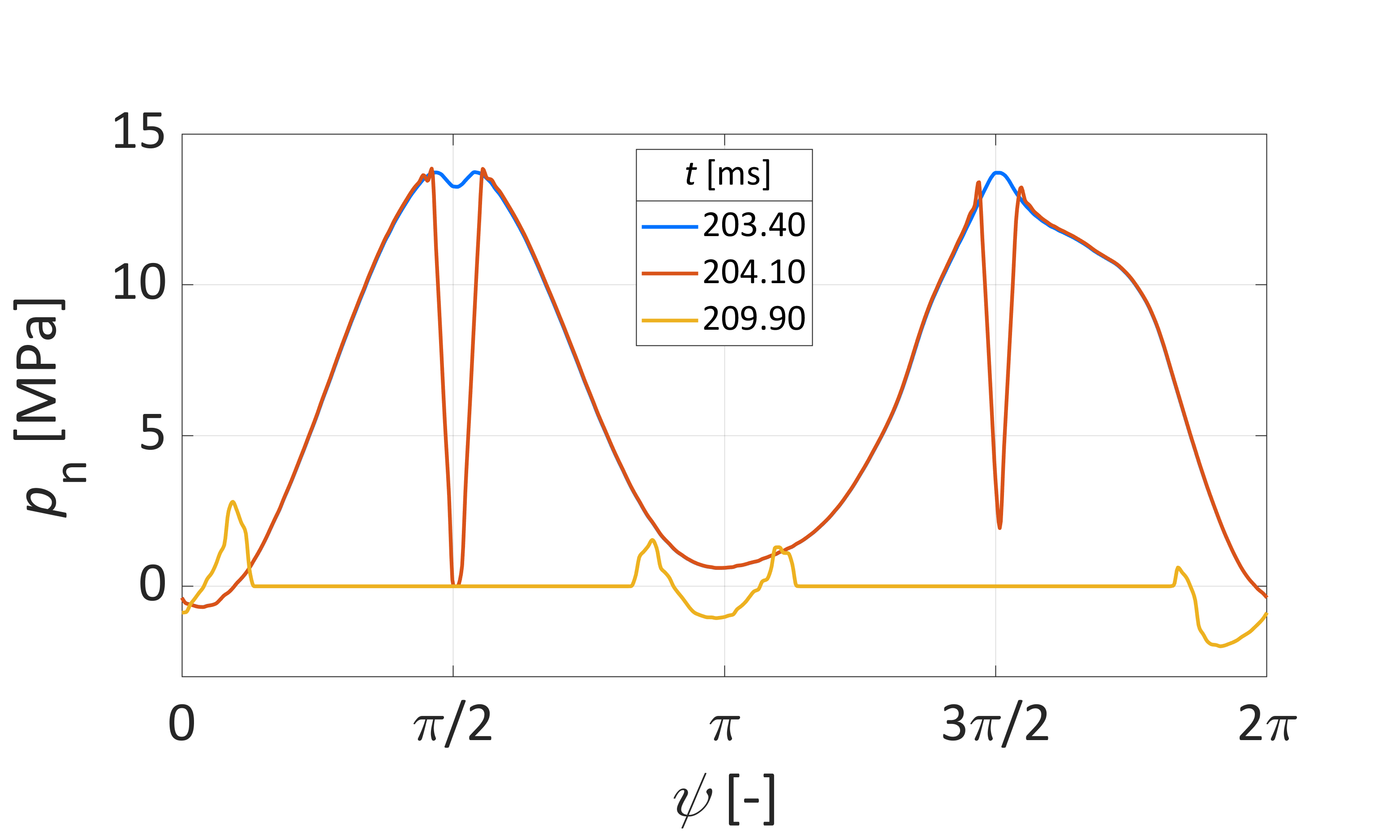}
\caption{Normal stress, the  inhomogeneity  at $S_1$}\label{Fig_MaFi3_ratG2_IF1_2}
\end{subfigure}
\begin{subfigure}{0.49\textwidth}
\centering
\includegraphics[scale=0.22,trim=5 5 45 45,clip]{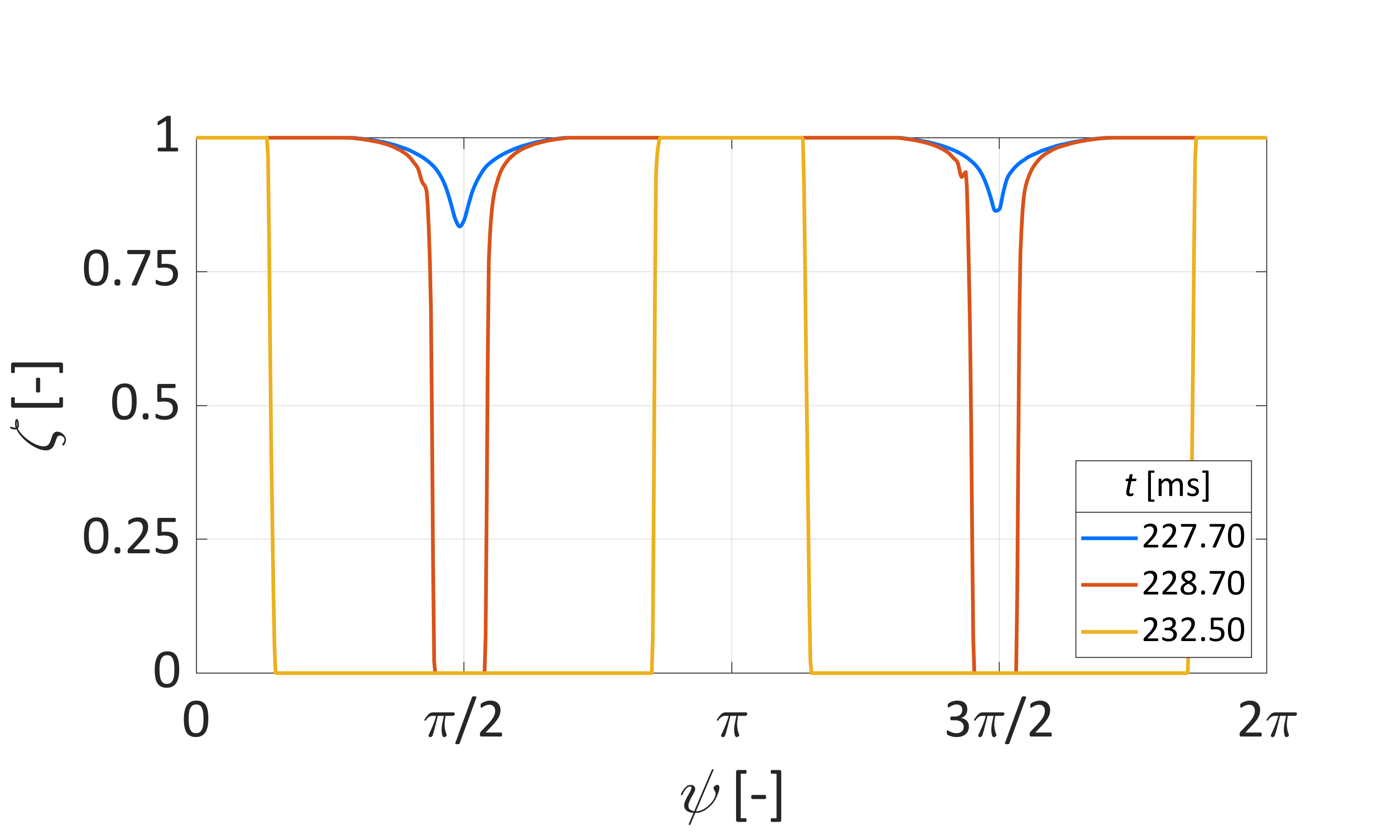}
\caption{Interface damage, the  inhomogeneity  at $S_2$}\label{Fig_MaFi3_ratG2_IF2_1}
\end{subfigure}
\begin{subfigure}{0.49\textwidth}
\centering
\includegraphics[scale=0.22,trim=5 5 45 45,clip]{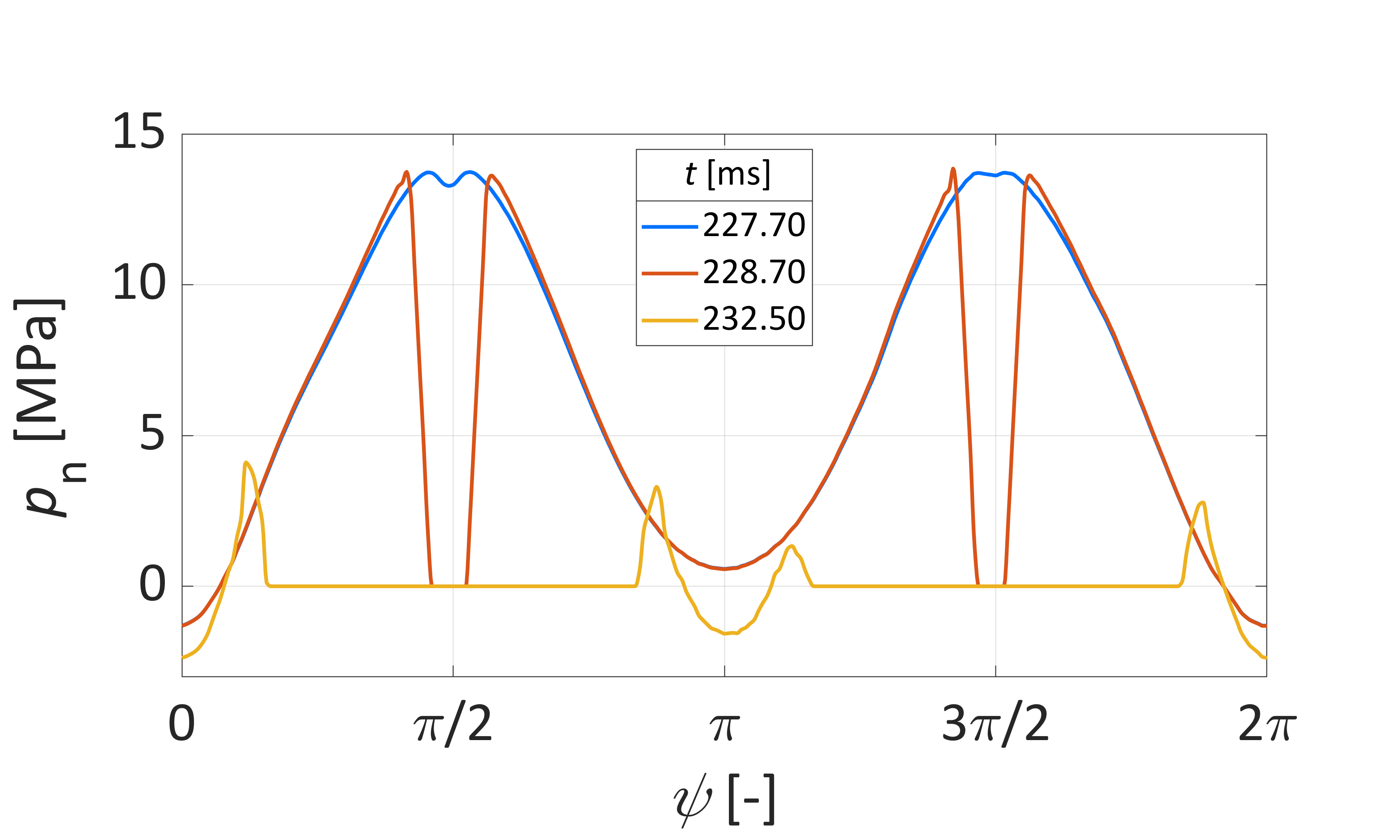}
\caption{Normal stress, the  inhomogeneity  at $S_2$}\label{Fig_MaFi3_ratG2_IF2_2}
\end{subfigure}
\caption{Distribution of the interface quantities during the debonding phase at selected time instants for the case with three  inhomogeneities , $\Gc[II]=2\Gc[I]$.}\label{Fig_MaFi3_ratG2_IF}
\end{figure}

The interface debonding appears for all three options of $\Gc[II]$ at the same time instants therefore only one case is shown.
Graphs in Fig.~\ref{Fig_MaFi3_ratG2_IF} show the state of the interface crack in terms of the interface damage parameter and distribution of the normal contact stress.
The instants pertain to situations where the maximum stress value was reached (it corresponds to the above calculated critical stress value for  activated damage variable $\zeta<1$), when the first crack was detected (at least at one interface point $\zeta=0$), and when  final extent of the interface crack is reached.
Naturally, at the cracks the normal stresses vanish.

The selected times verify that the jumps appearing in  Fig.~\ref{Fig_MaFi3_F} actually correspond to crack propagation along the interfaces.
The first crack appeared at the interface  of the inhomogeneity  with the centre at the point $S_1$, see Fig.~\ref{Fig_MaFi3}, this is the first jump (simultaneously with that a crack along the  inhomogeneity  around $S_3$ has been formed).
After that, a crack appeared at the interface related to the point $S_2$ as selected time instants document in agreement with the other force jump in Fig.~\ref{Fig_MaFi3_F}.

\begin{figure}[!ht]
\centering
\setlength{\unitlength}{\textwidth}
\begin{subfigure}{0.99\textwidth}
\begin{picture}(0.97,0.22)
\PB{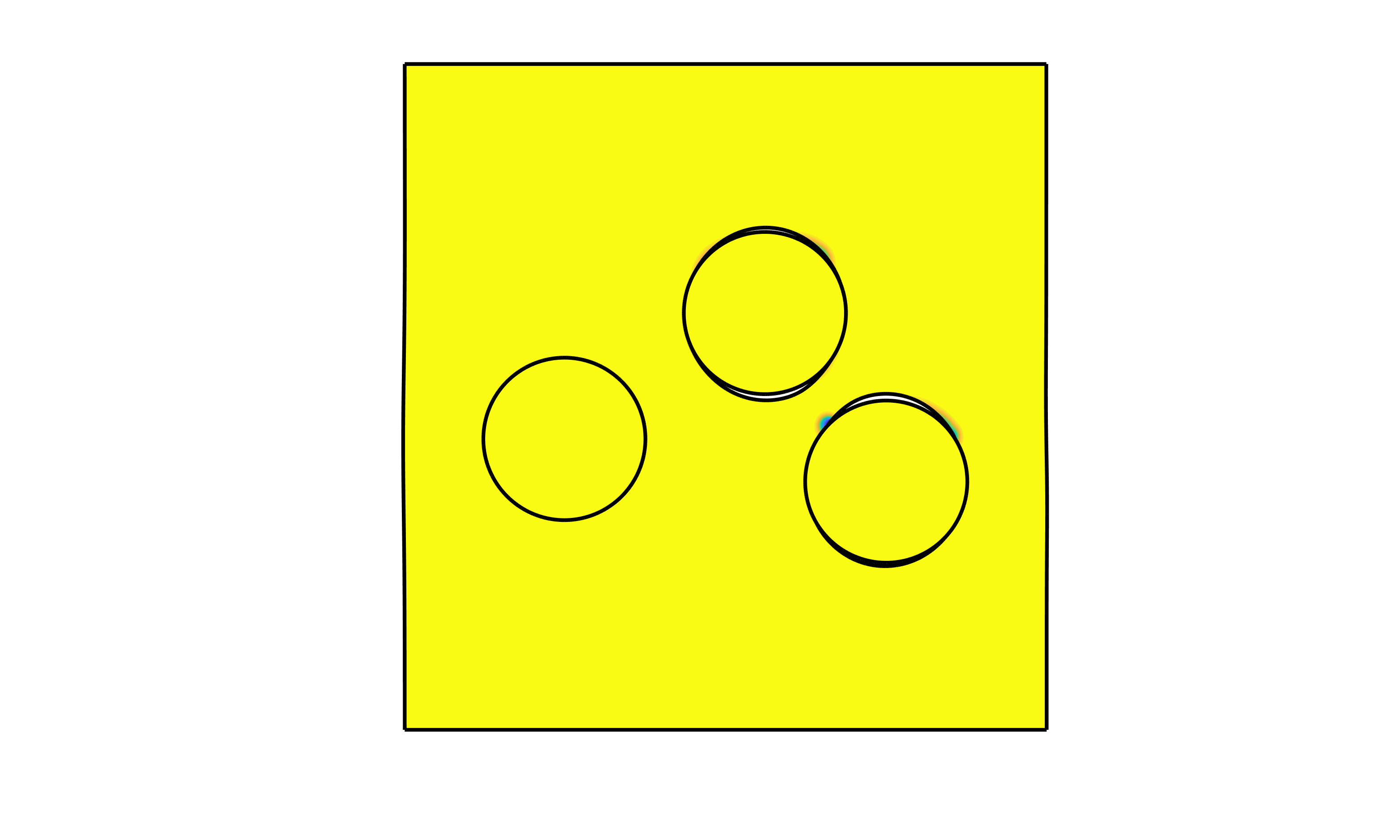}{$t=210$ms}
\PC{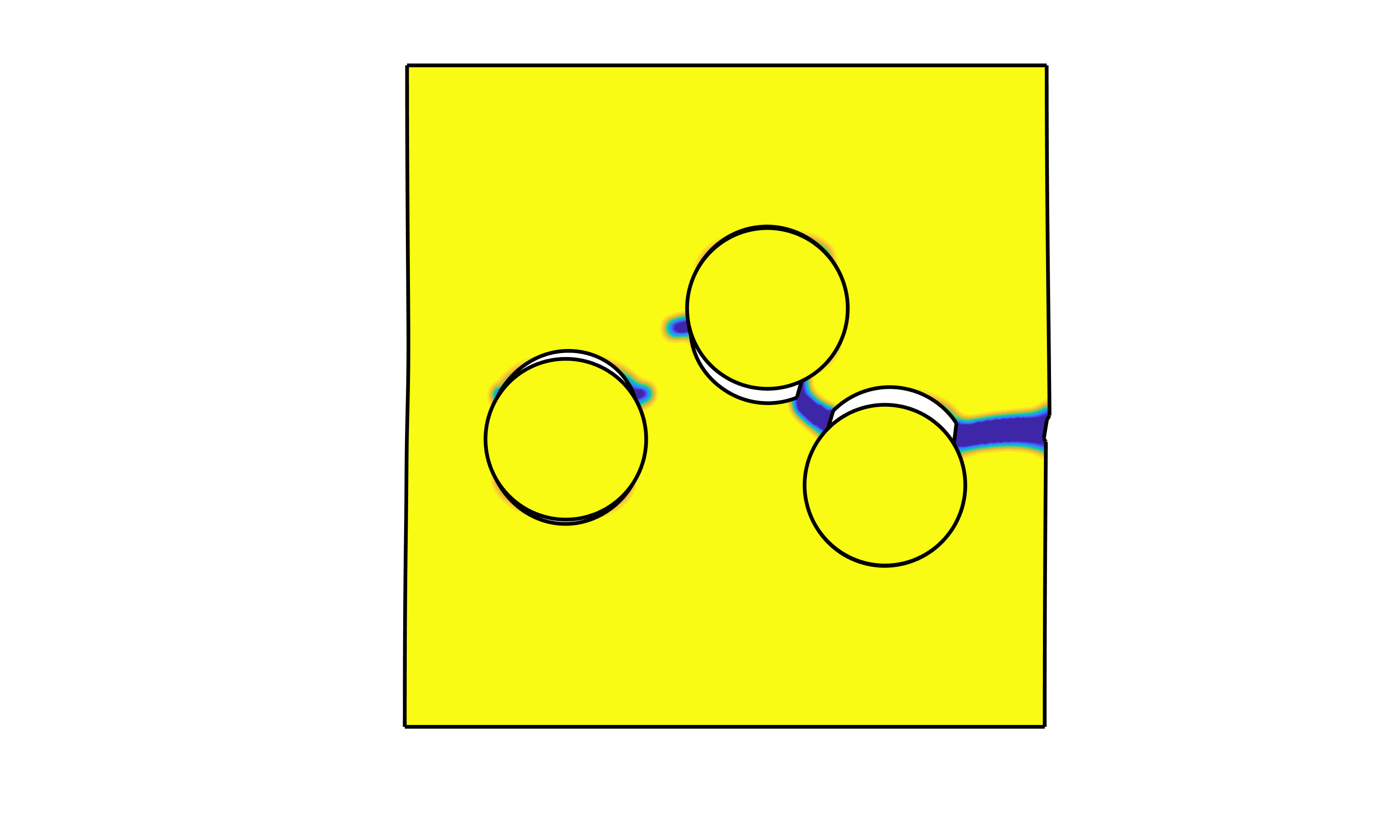}{$t=233$ms}
\PDF{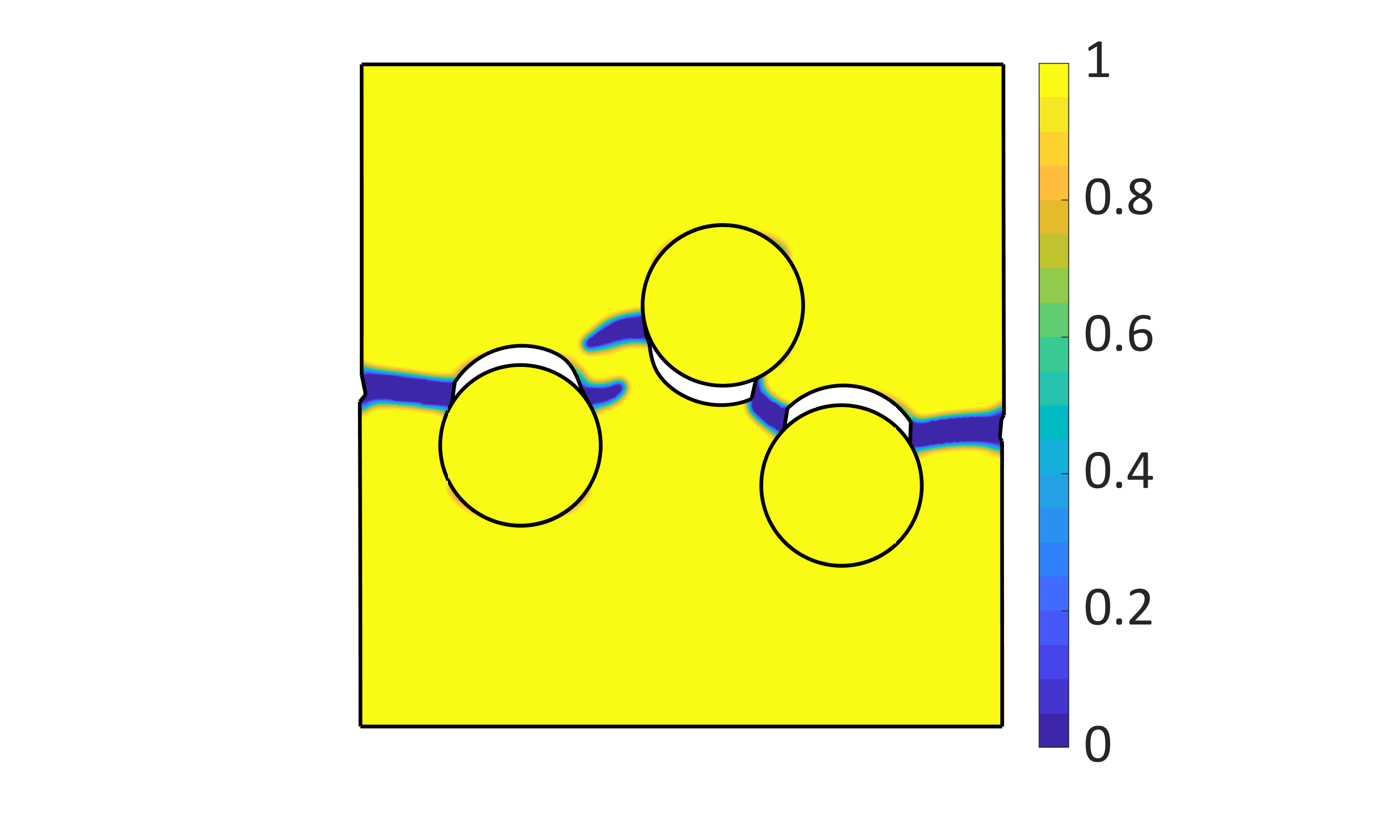}{$t=244$ms}
\end{picture}
\caption{$\Gc[II]=\Gc[I]$.}\label{Fig_MaFi3_ratG1_DAM}
\end{subfigure}
\begin{subfigure}{0.99\textwidth}
\begin{picture}(0.97,0.22)
\PB{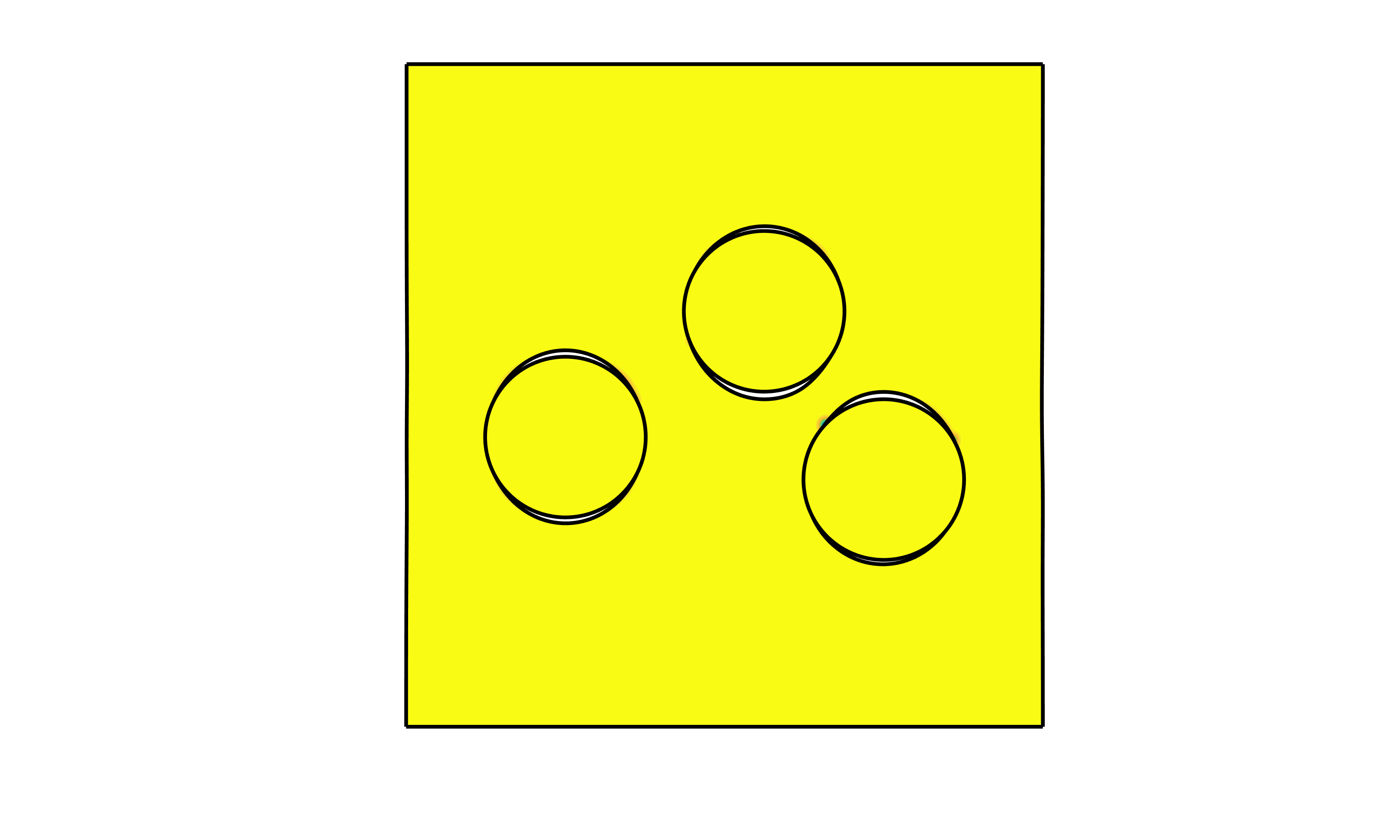}{$t=244$ms}
\PC{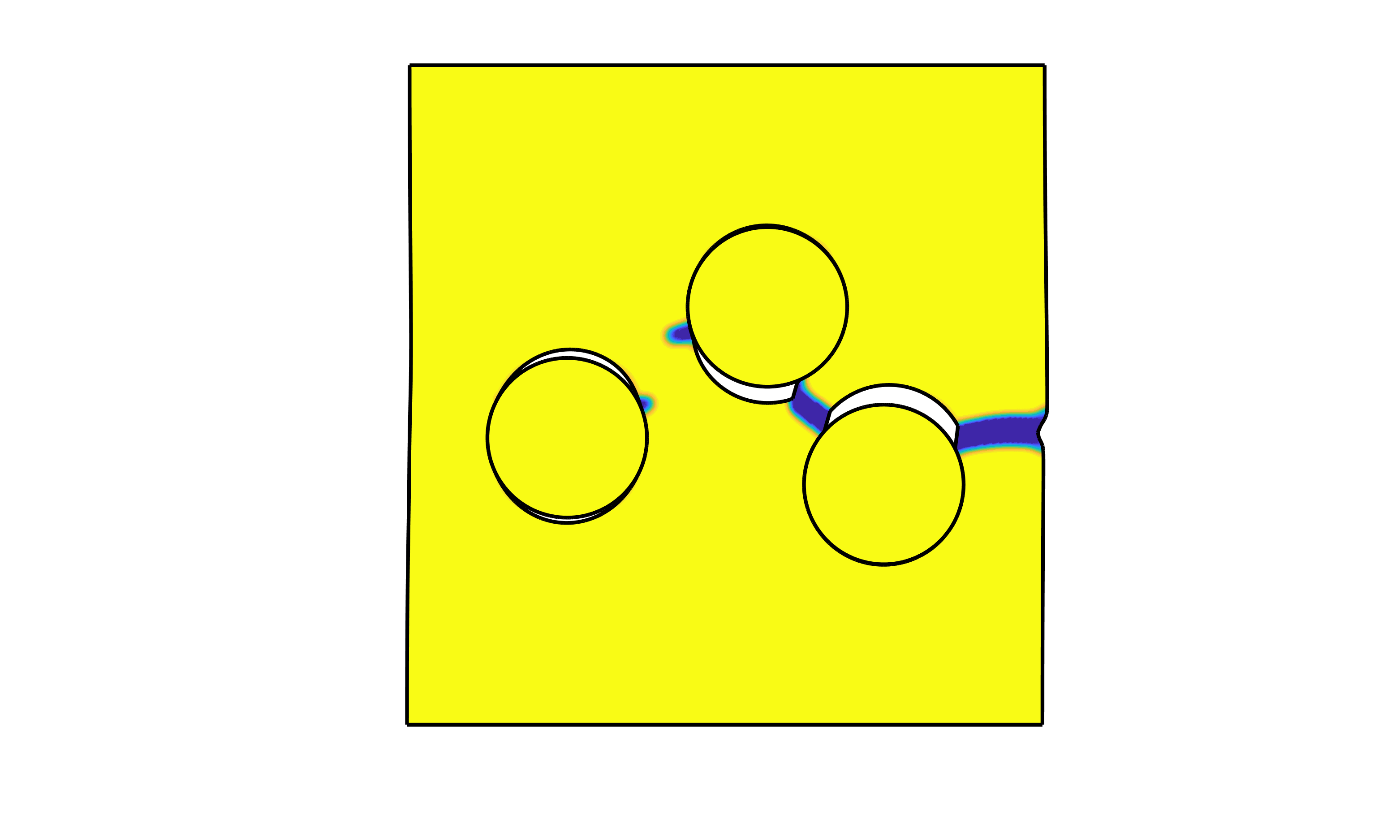}{$t=263$ms}
\PDF{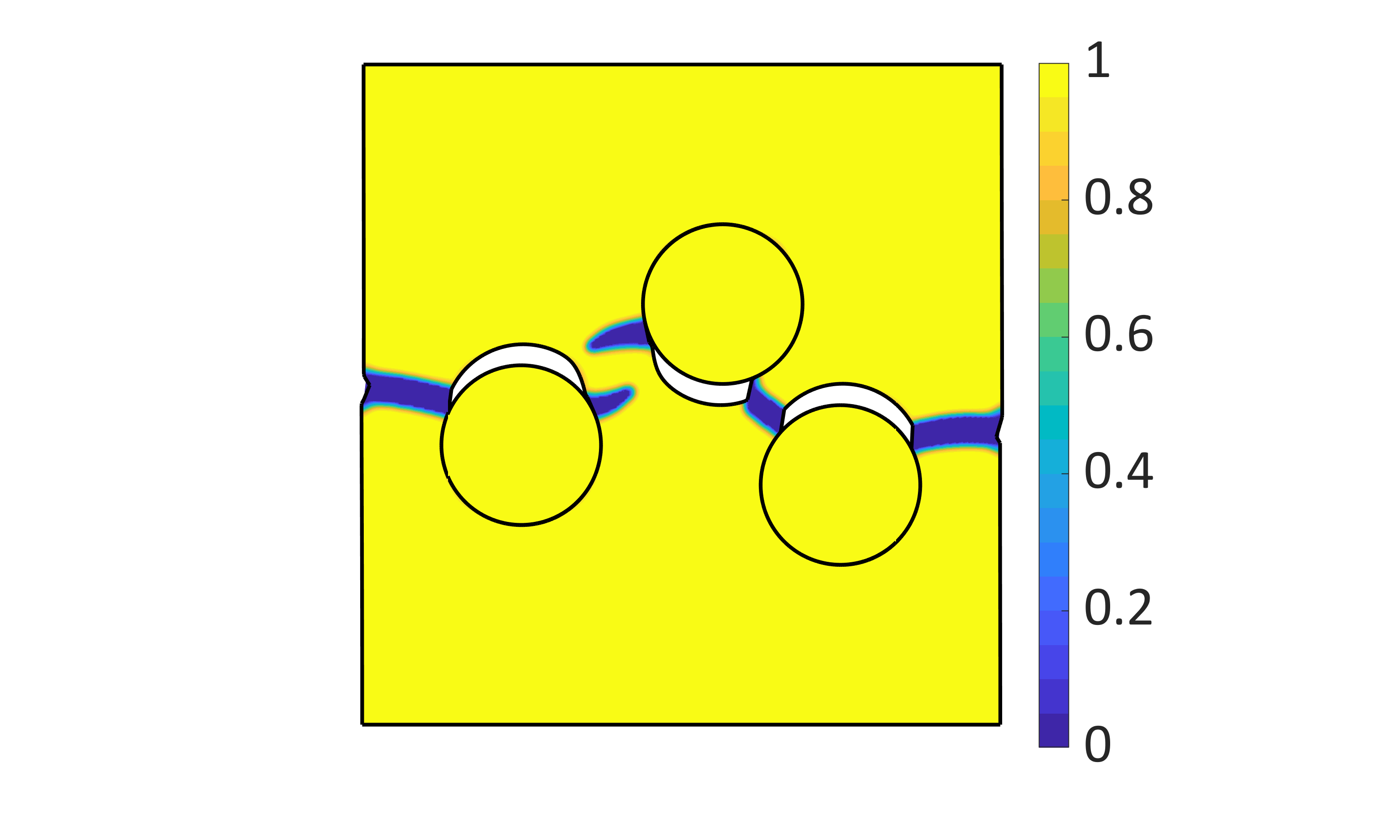}{$t=272$ms}
\end{picture}
\caption{$\Gc[II]=2\Gc[I]$.}\label{Fig_MaFi3_ratG2_DAM}
\end{subfigure}
\begin{subfigure}{0.99\textwidth}
\begin{picture}(0.97,0.22)
\PB{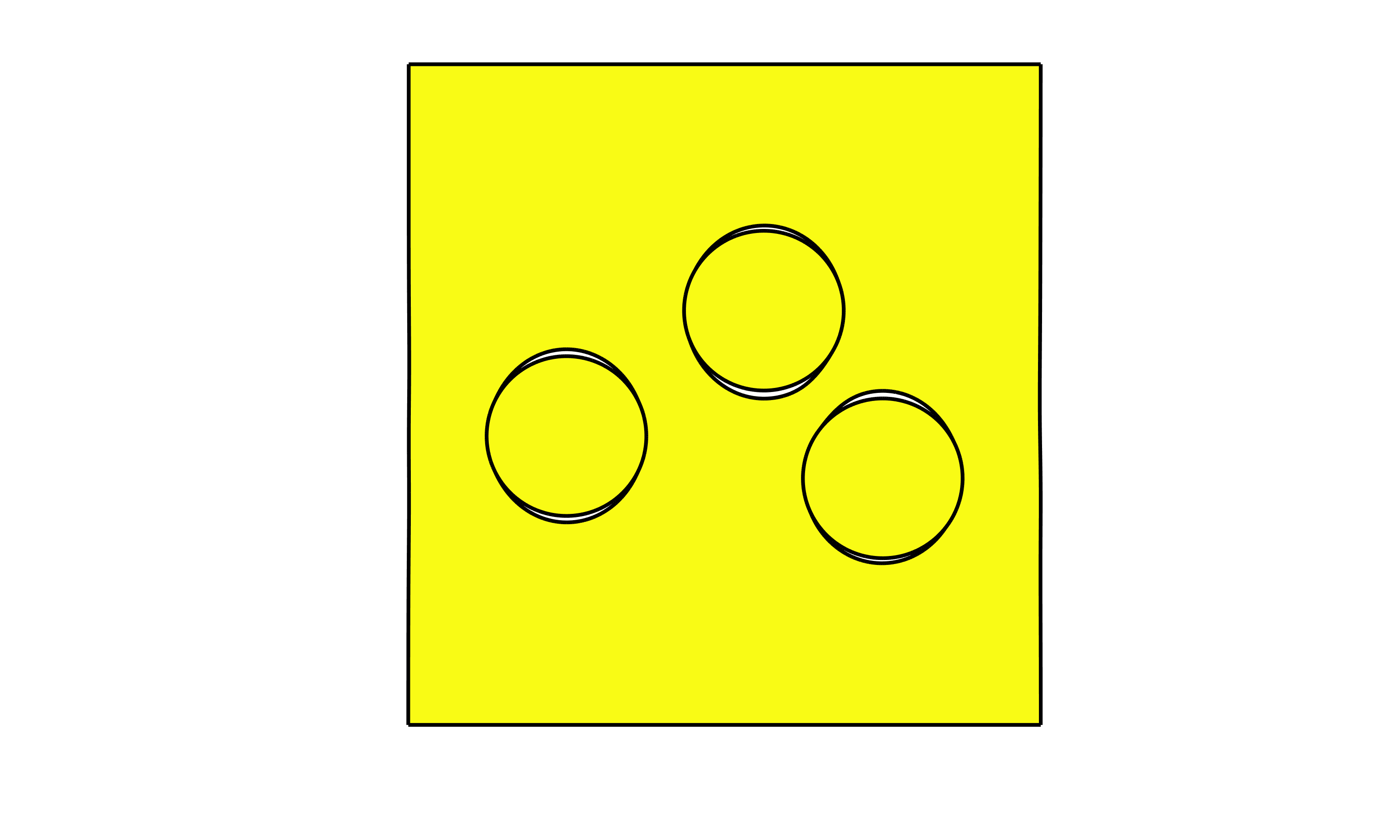}{$t=272$ms}
\PC{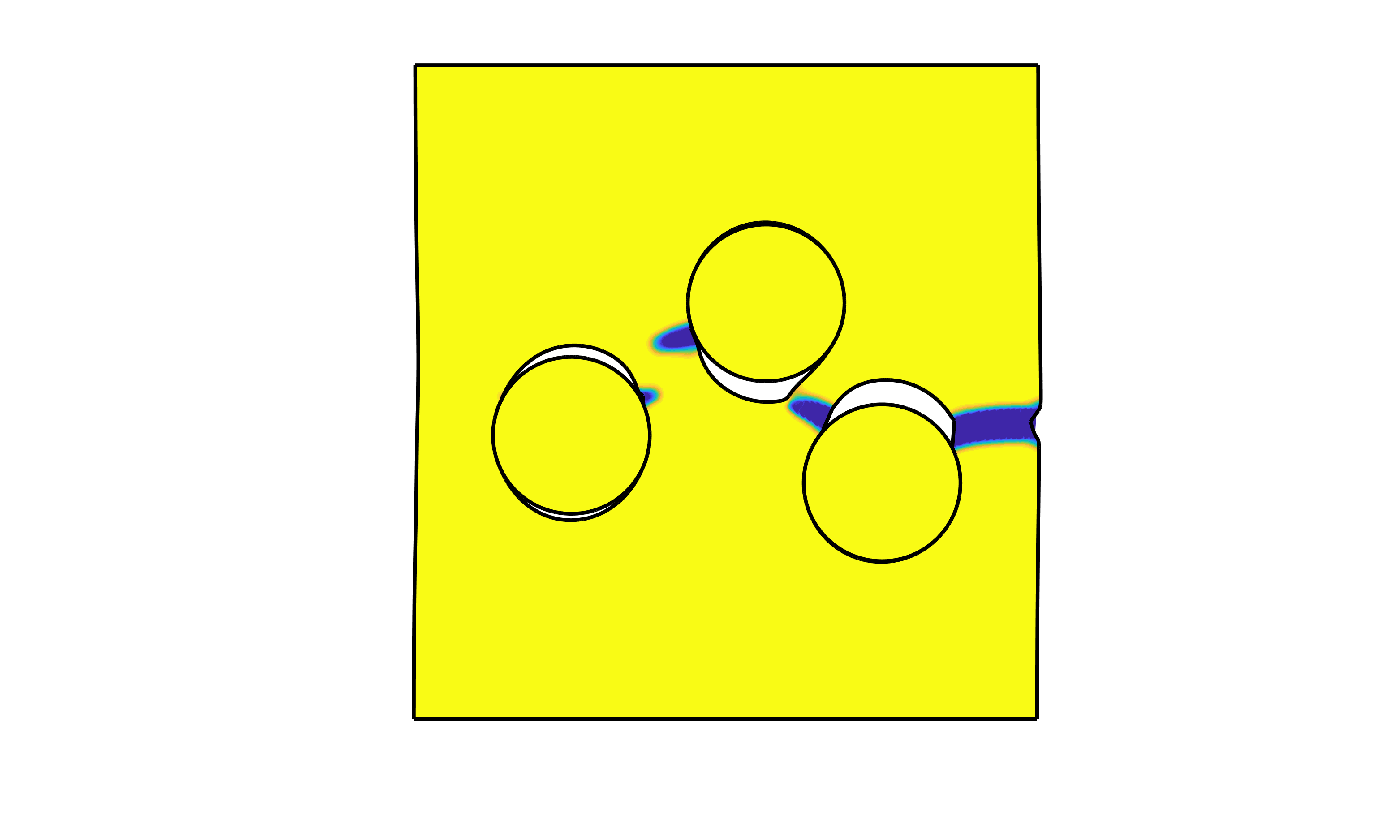}{$t=342$ms}
\PDF{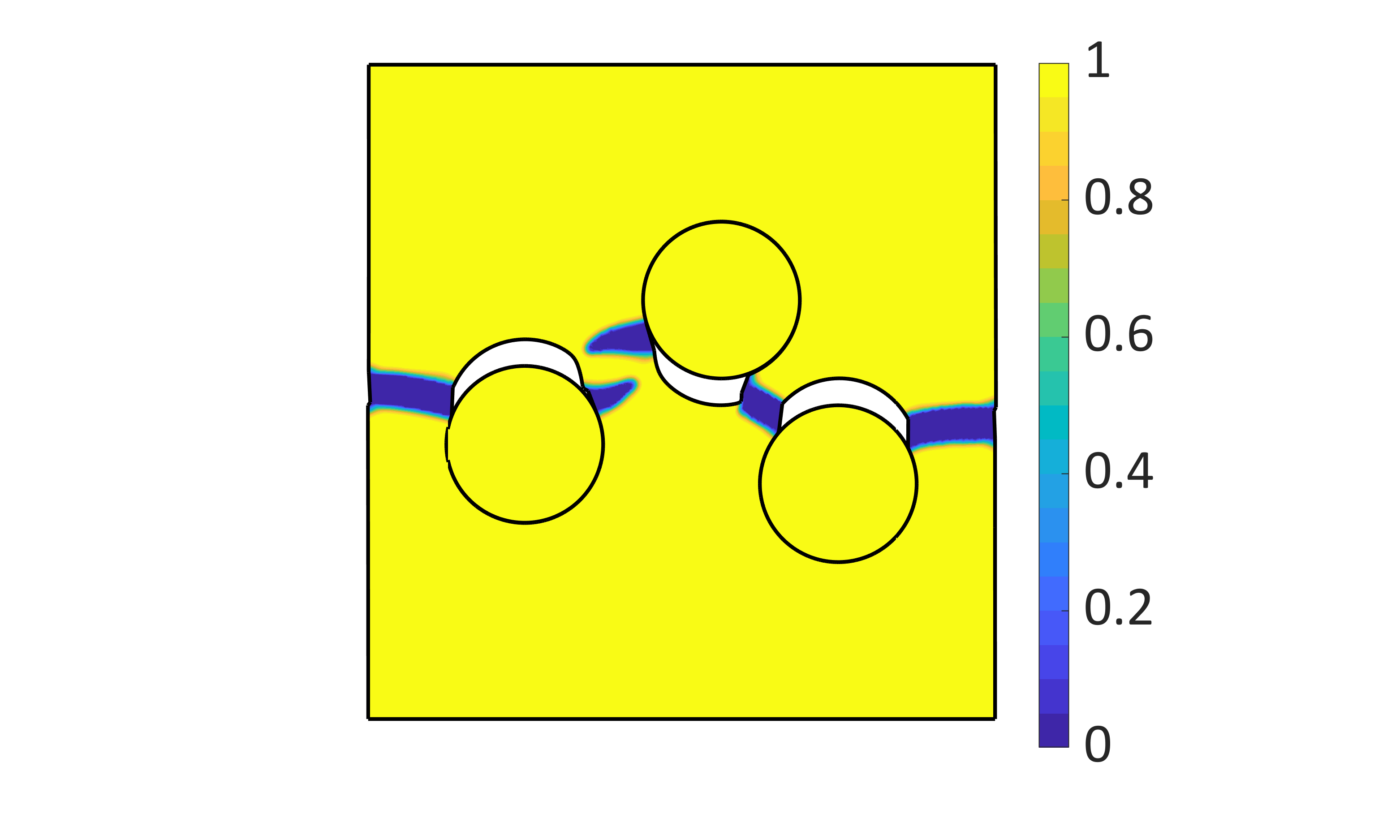}{$t=346$ms}
\end{picture}
\caption{$\Gc[II]=10\Gc[I]$.}\label{Fig_MaFi3_ratG10_DAM}
\end{subfigure}
\caption{Distribution of the phase-field variable  documenting crack propagation at selected time instants for the case with three  inhomogeneities.
Deformation is $10\times$ magnified. }\label{Fig_MaFi3_DAM}
\end{figure}

The varying shear fracture energy affects the initiation of  damage  process inside the material.
Therefore, the crack is initiated for higher load if this parameter is greater relatively to the opening mode fracture energy.
To show that, the distribution of the phase-field damage parameter is plotted.
The drawings in Fig.~\ref{Fig_MaFi3_DAM} then document the differences in crack initiation.
The data also include deformation of the structure which is magnified for the interface crack would be clearly observable.
While the form and location of the crack are the same for all three cases of $\Gc[II]$, there are different instants at which the crack is initiated and subsequently grown.
The used instants can also be viewed at the evolution in Fig.~\ref{Fig_MaFi3_F}, where the abrupt crack propagation pertains to vertical segments in  corresponding graphs.
This again corresponds to unstable crack propagation.

\begin{figure}[!ht]
\centering
\setlength{\unitlength}{\textwidth}
\begin{subfigure}{0.99\textwidth}
\begin{picture}(0.97,0.22)
\PB{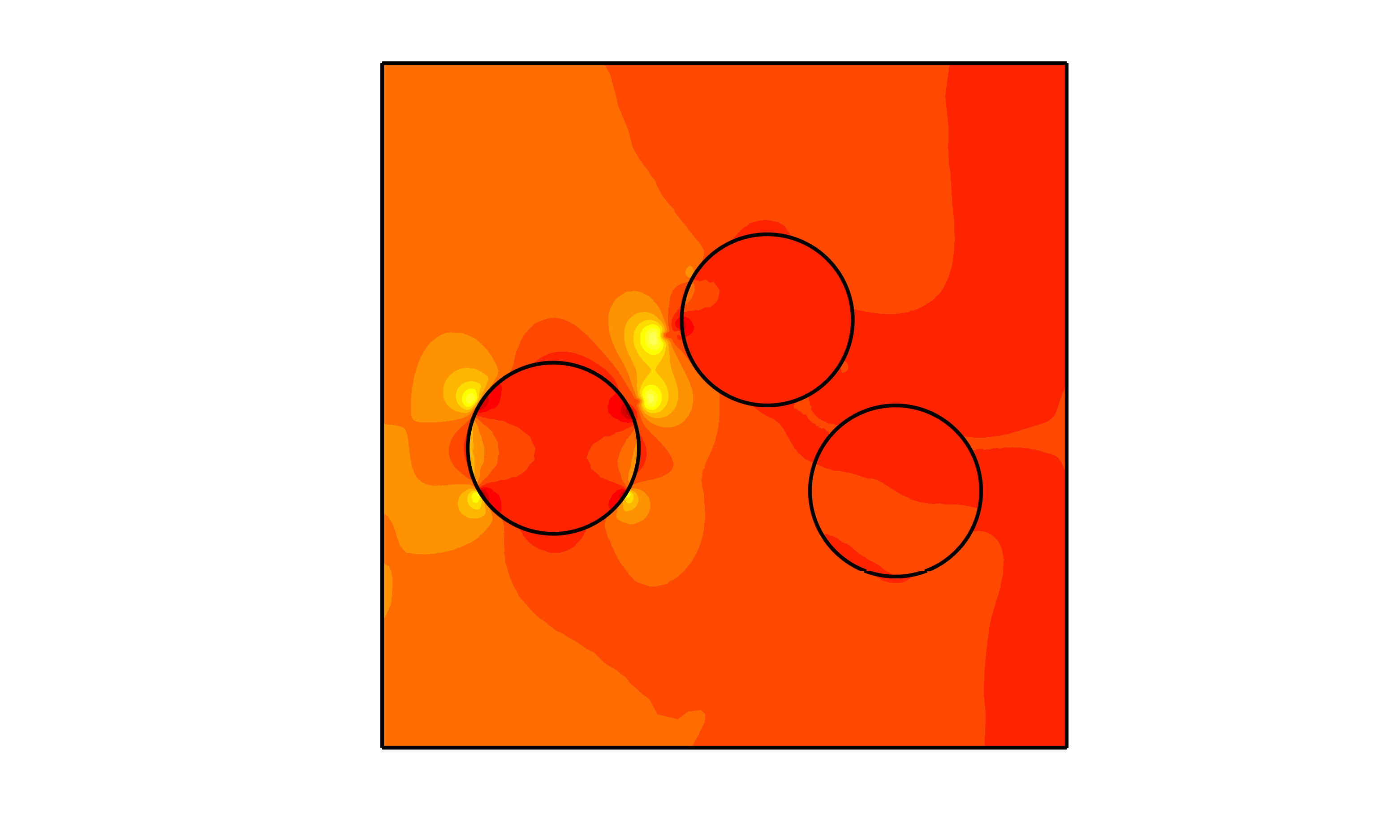}{$\Gc[II]=\Gc[I]$}
\PC{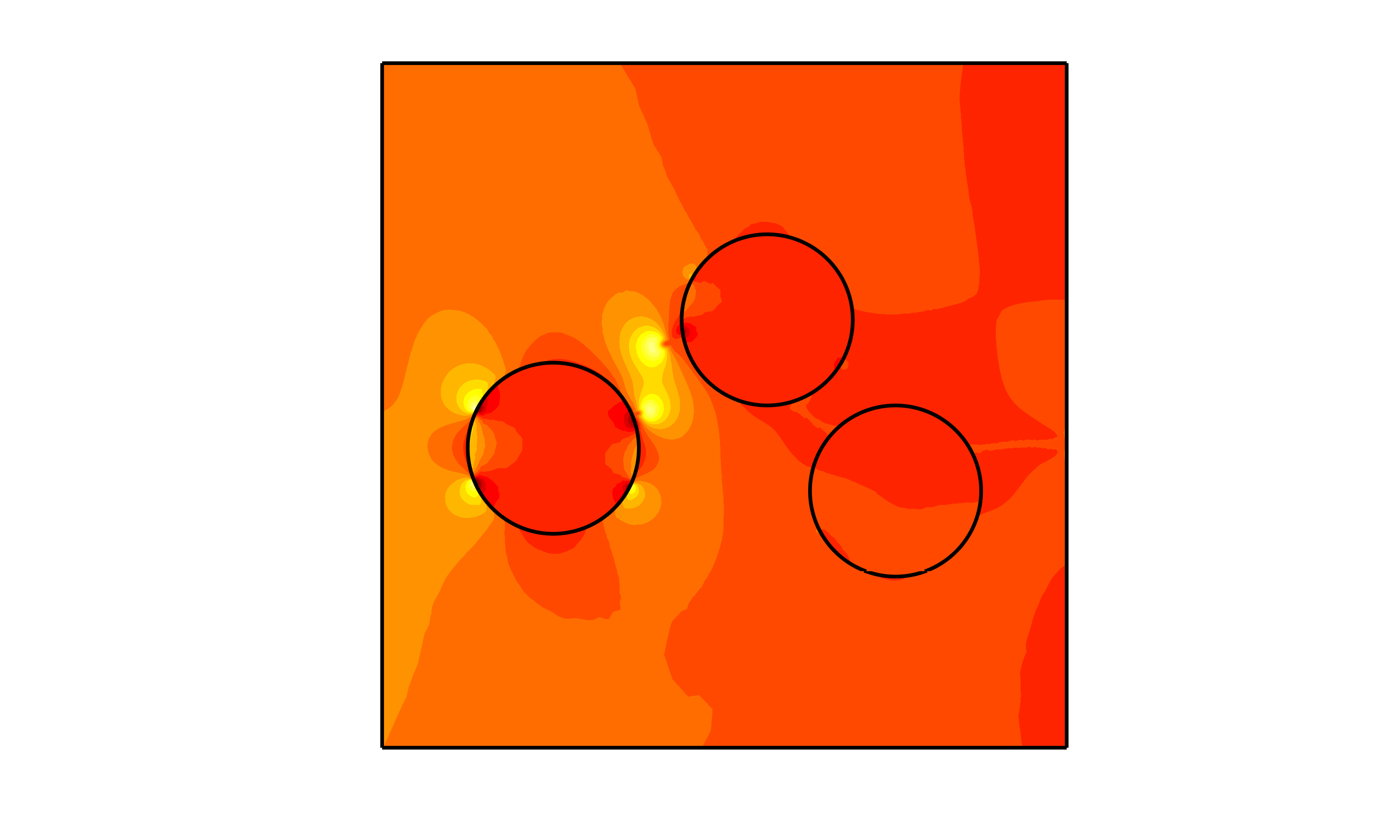}{$\Gc[II]=2\Gc[I]$}
\PD{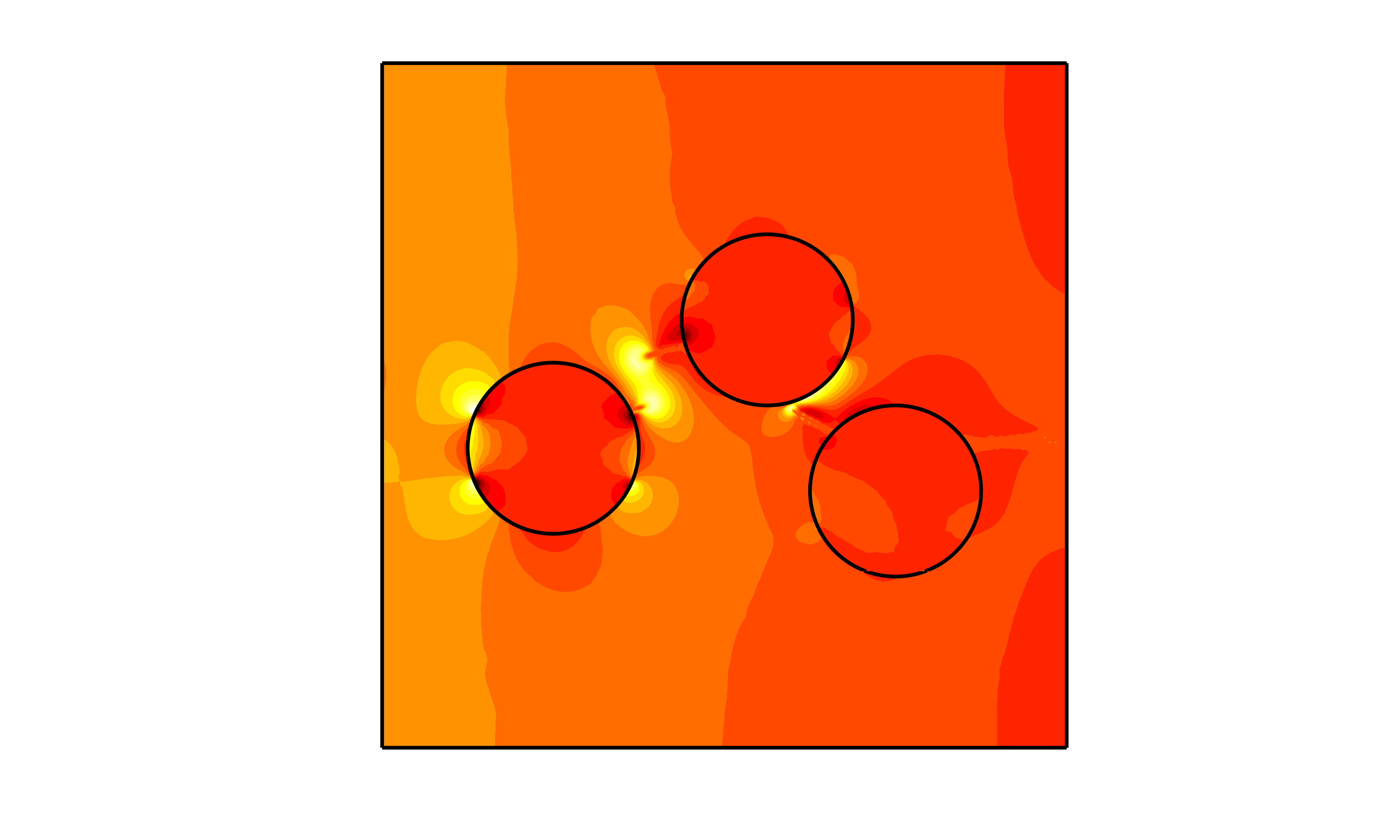}{$\Gc[II]=10\Gc[I]$}
\CB{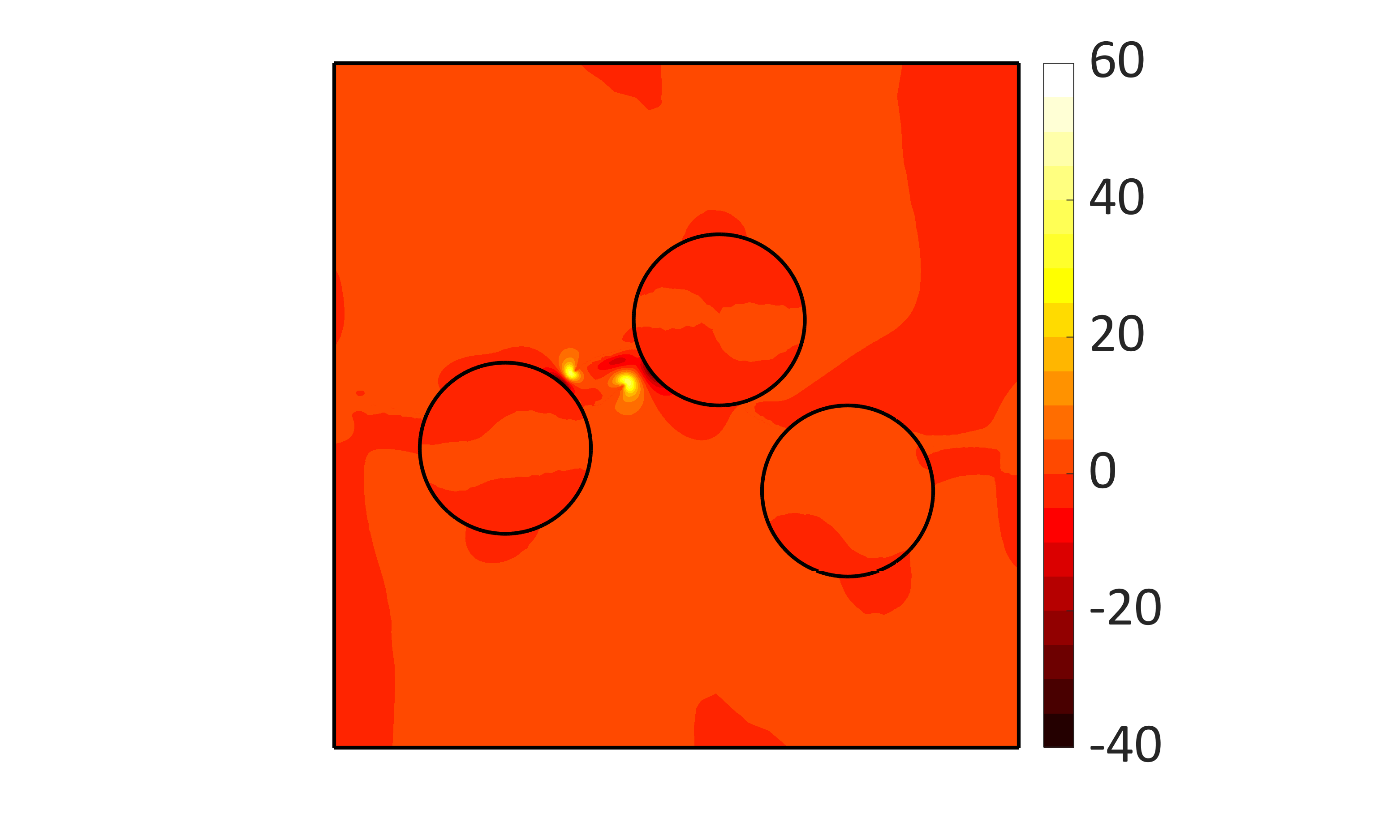}
\end{picture}
\caption{Stress trace.}\label{Fig_MaFi3_TRS}
\end{subfigure}
\begin{subfigure}{0.99\textwidth}
\begin{picture}(0.97,0.22)
\PB{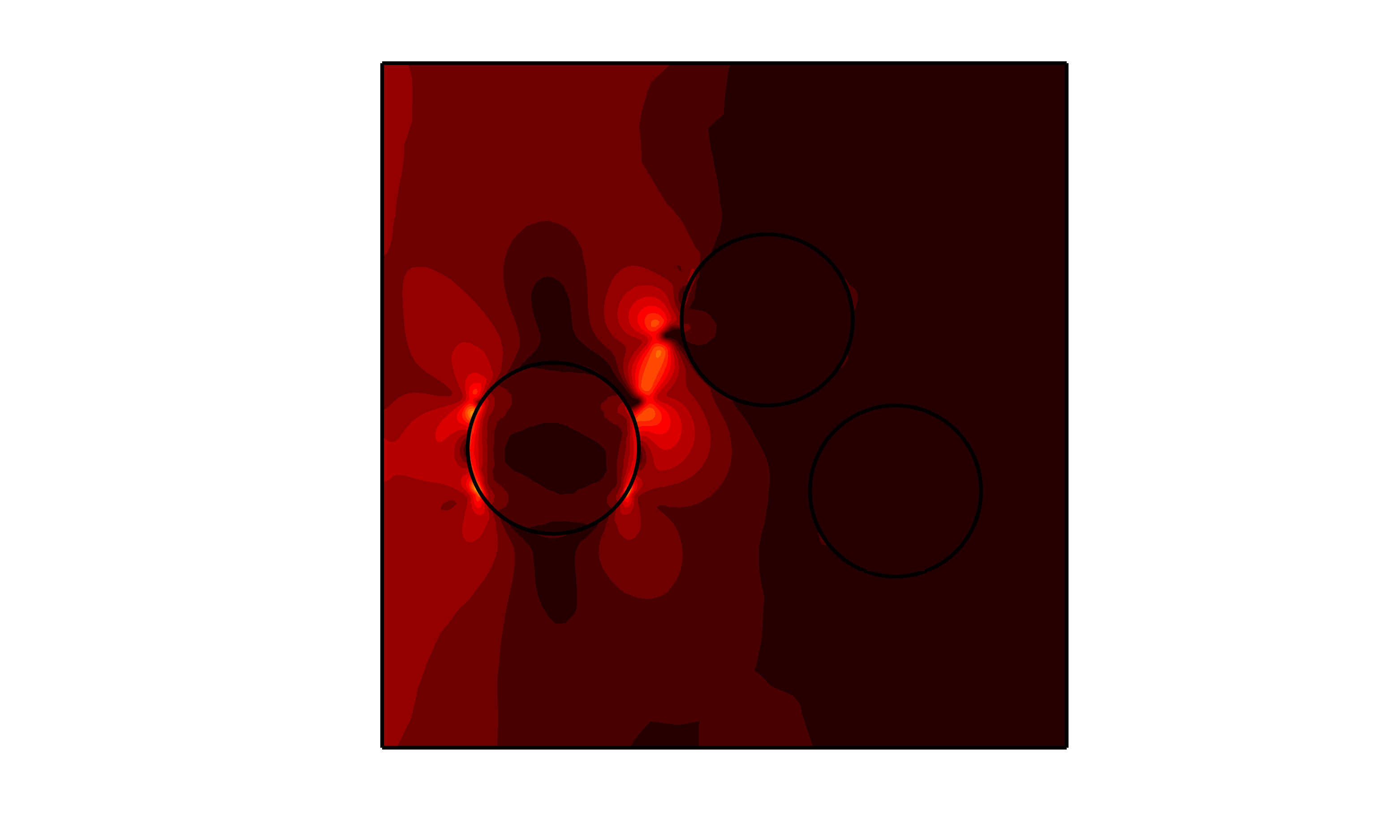}{$\Gc[II]=\Gc[I]$}
\PC{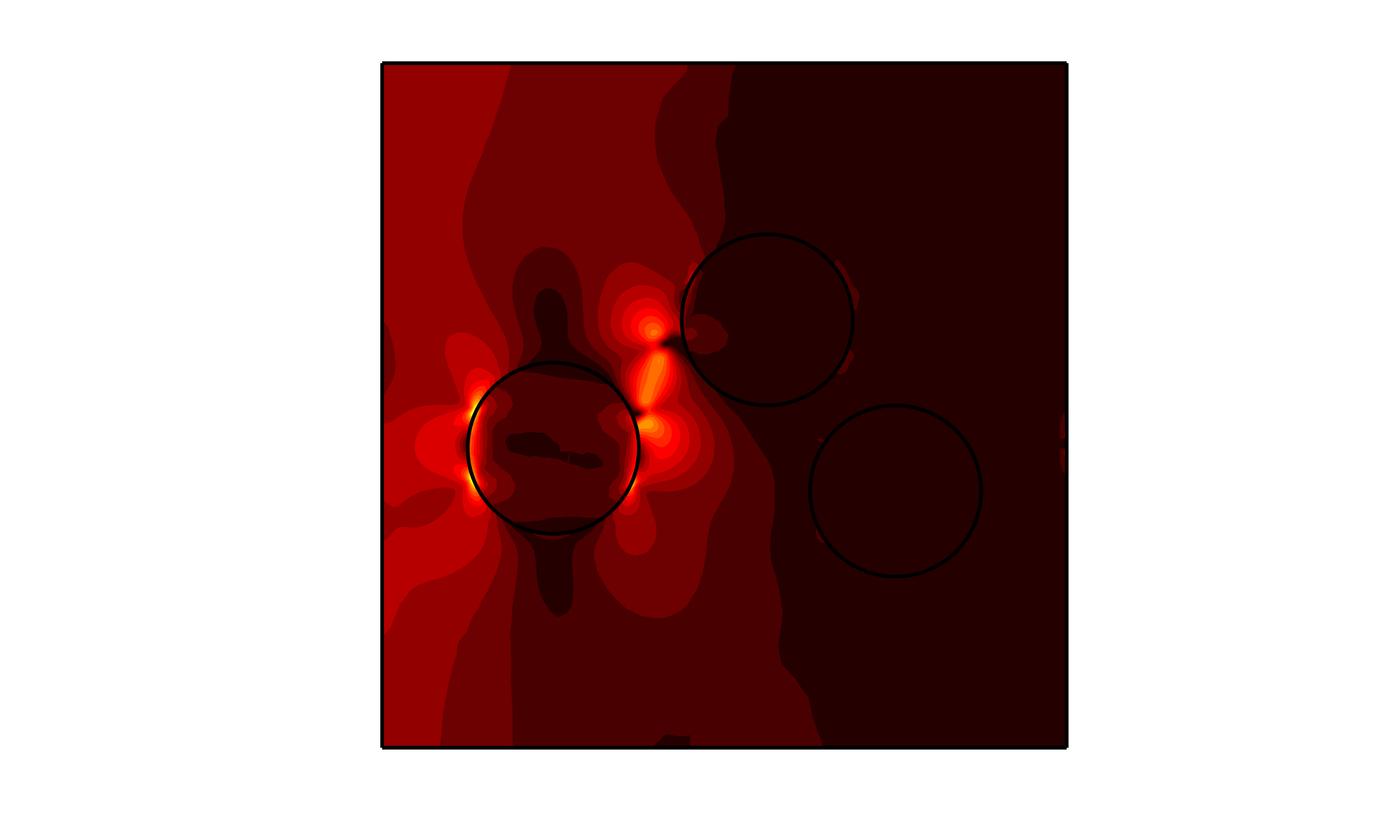}{$\Gc[II]=2\Gc[I]$}
\PD{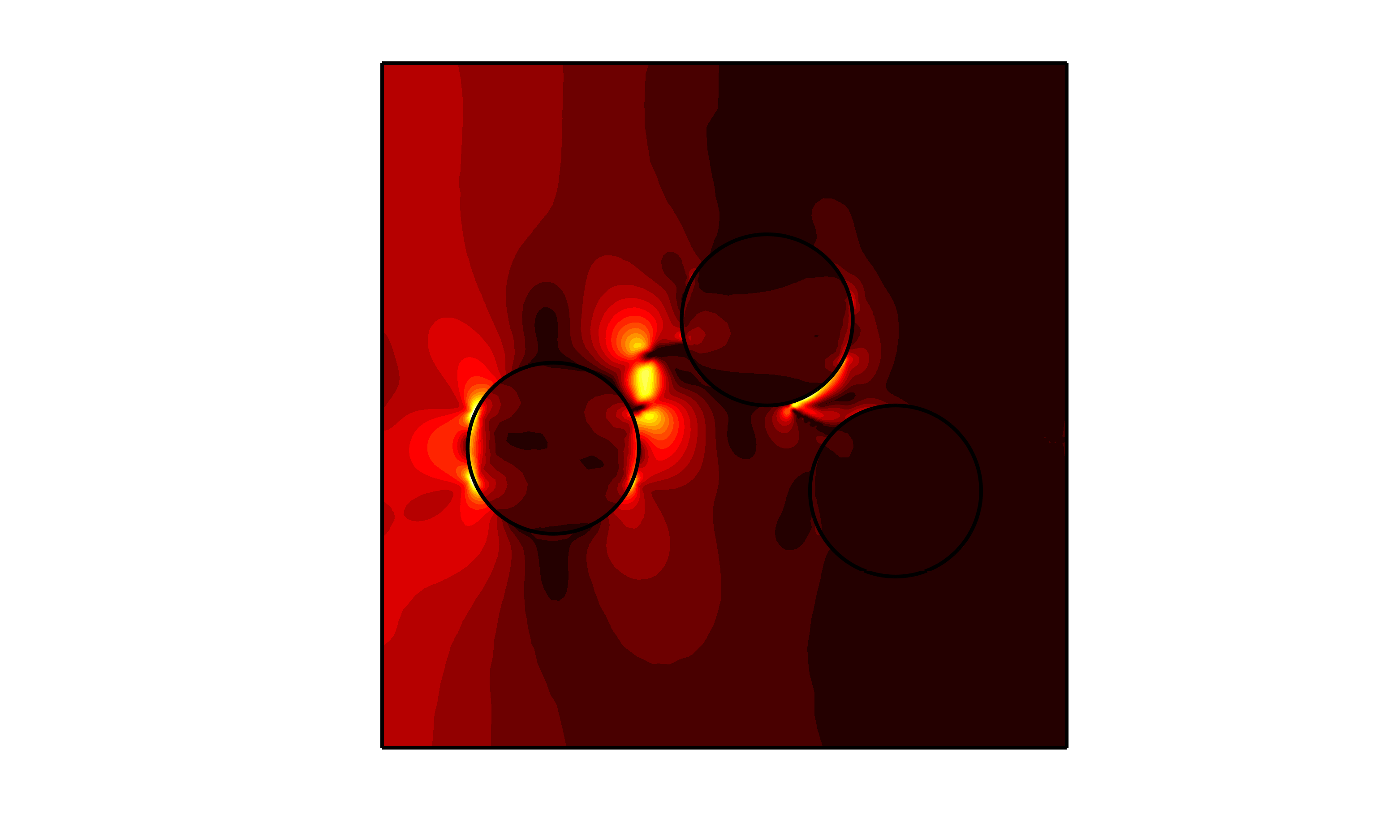}{$\Gc[II]=10\Gc[I]$}
\CB{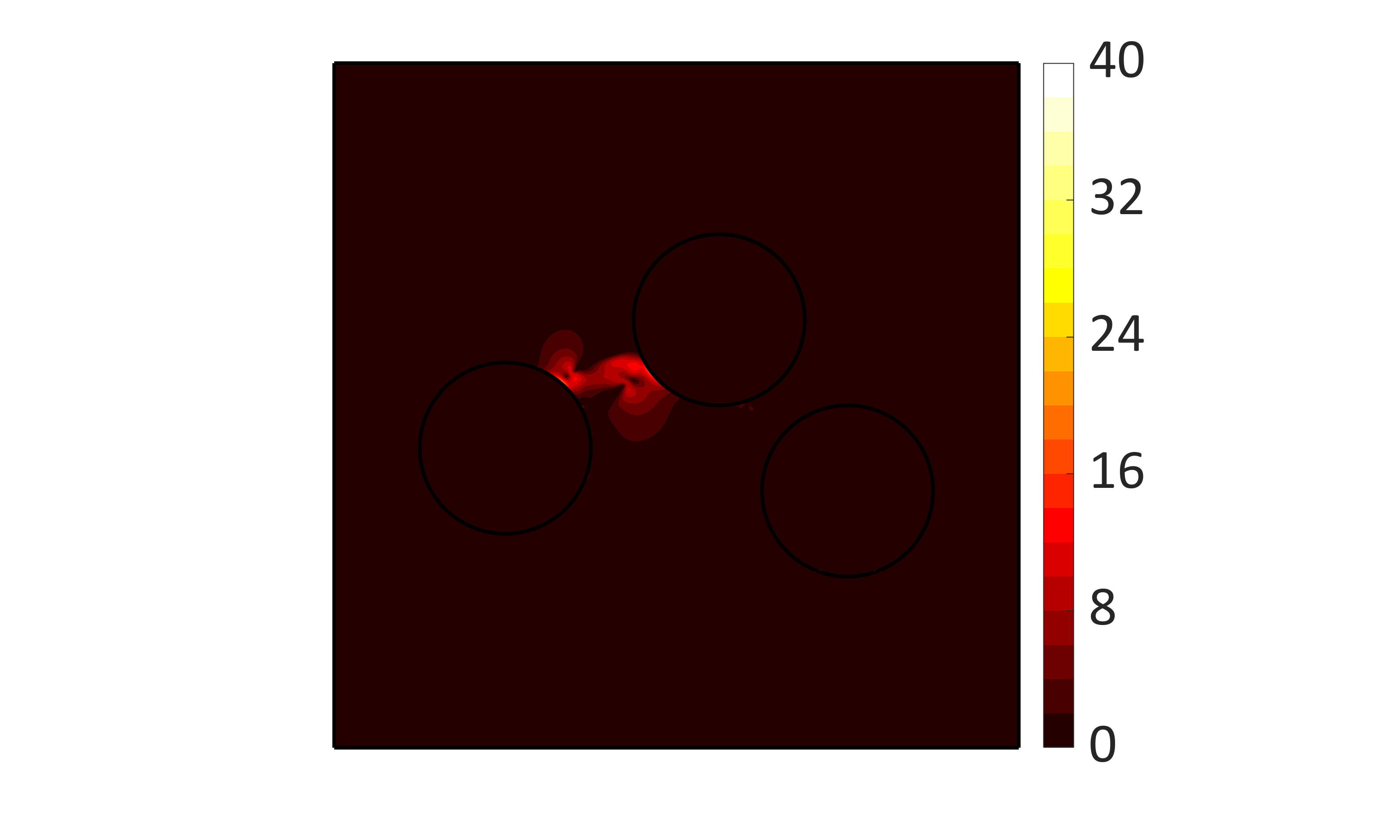}
\end{picture}
\caption{Norm of deviatoric stress.}\label{Fig_MaFi3_NDS}
\end{subfigure}
\caption{Distribution of the stresses  documenting crack propagation at second time instants used in Fig.~\ref{Fig_MaFi3_DAM} for the case with three  inhomogeneities.}\label{Fig_MaFi3_STRESS}
\end{figure}

\begin{figure}[!ht]
\centering
\includegraphics[scale=0.3,trim=5 5 40 50,clip]{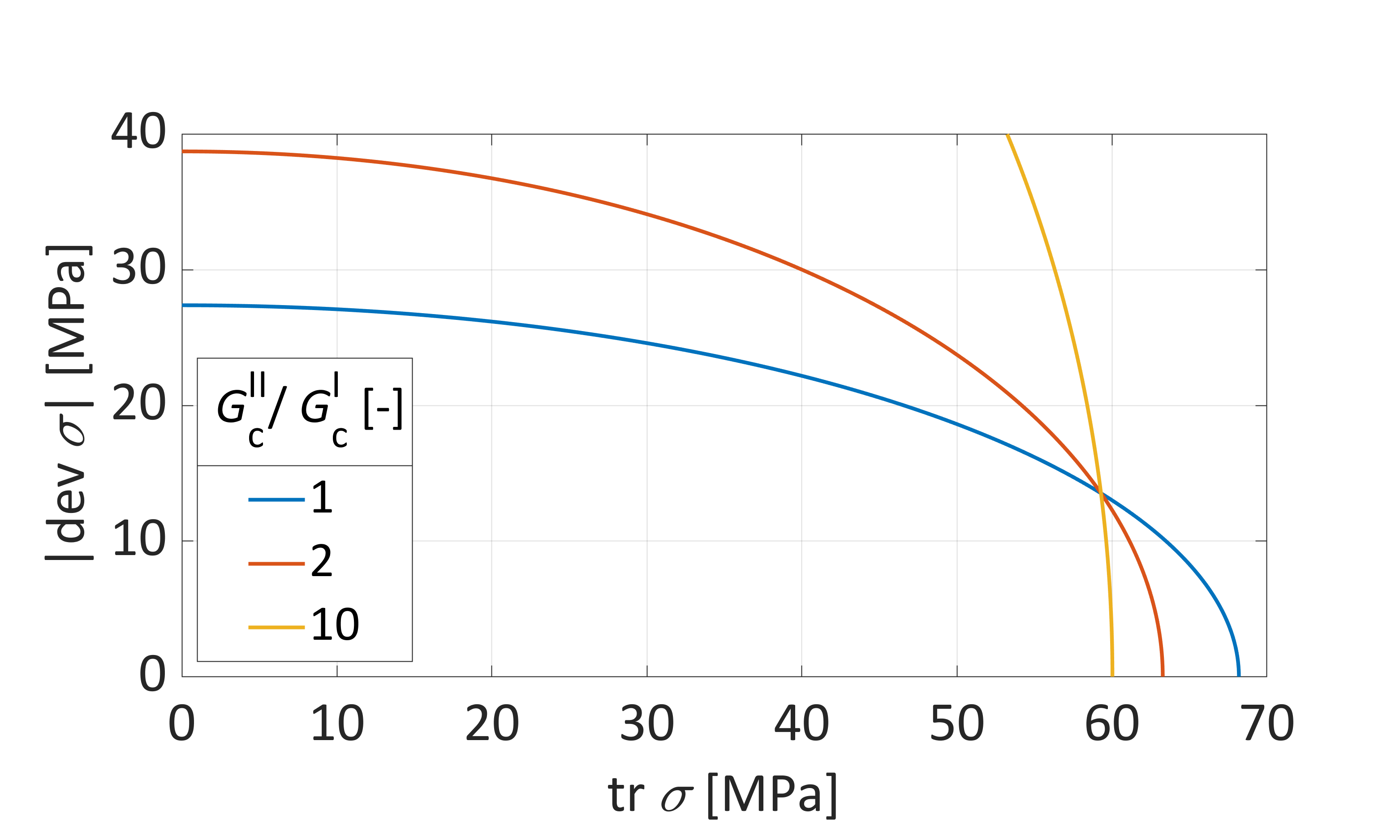}
\caption{ Relation  between  the norm of deviatoric stress $|\dev\sigma|$ (as a measure of shear stress) and stress trace $\tr \sigma$ (as a characteristic of opening  stress) corresponding to triggering of the phase-field evolution producing an internal crack  for the data corresponding to the case with three  inhomogeneities.}\label{Fig_StressCrit}
\end{figure}

It is also worth to see the stress distribution at some instants for a comparison of stresses for various ratios $\Gc[II]/\Gc[I]$.
The second instants of those in Fig.~\ref{Fig_MaFi3_DAM} were used for displaying the stresses in Fig.~\ref{Fig_MaFi3_STRESS}.
The interesting location is between the  inhomogeneities  at $S_1$ and $S_2$, where the crack is propagated at those selected moments.
As the crack propagation is related to the stress state by the relation~\eqref{Eq_StressCrit_Amod}, the curves corresponding to that relation are shown in Fig.~\ref{Fig_StressCrit} for the present material parameters.
In any case,  the crack appears due to tensile loading. Anyhow, the complicated stress state related to the presence of several  inhomogeneities  causes that also shear stresses play a role in  crack formation process.
 For the case $\Gc[II]=\Gc[I]$ the crack appears while the stresses are lower, because combined stress state satisfies requirements of total fracture energy consisting of the two ingredient $\Gc[I]$, $\Gc[II]$.
 Switching  subsequently to the other options, increasing value of $\Gc[II]$ suppresses influence of the shear state and to reach the damage triggering point higher load is needed. 
Therefore, the adjusting of the fracture energy values can be used in the proposed phase-field approach to avoid shear effect to the fracture propagation where it is not physically reasonable.

%%%%%%%%%%%%%%%%%%%%%%%%%%%%
\subsection{A domain with an  inhomogeneity under  tension}\label{Sec_MaFi}
%%%%%%%%%%%%%%%%%%%%%%%%%%%%

\begin{figure}[!ht]
\centering
\begin{subfigure}{0.49\textwidth}
\centering
\includegraphics[scale=1]{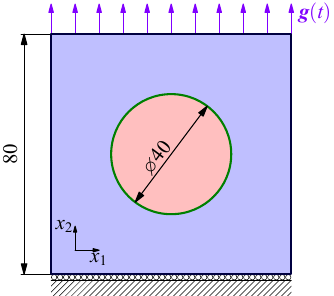}
\caption{The domain with with three inclusions}\label{Fig_Incl_G}
\end{subfigure}
\begin{subfigure}{0.49\textwidth}
\centering
\includegraphics[scale=0.32,trim=170 30 150 20,clip]{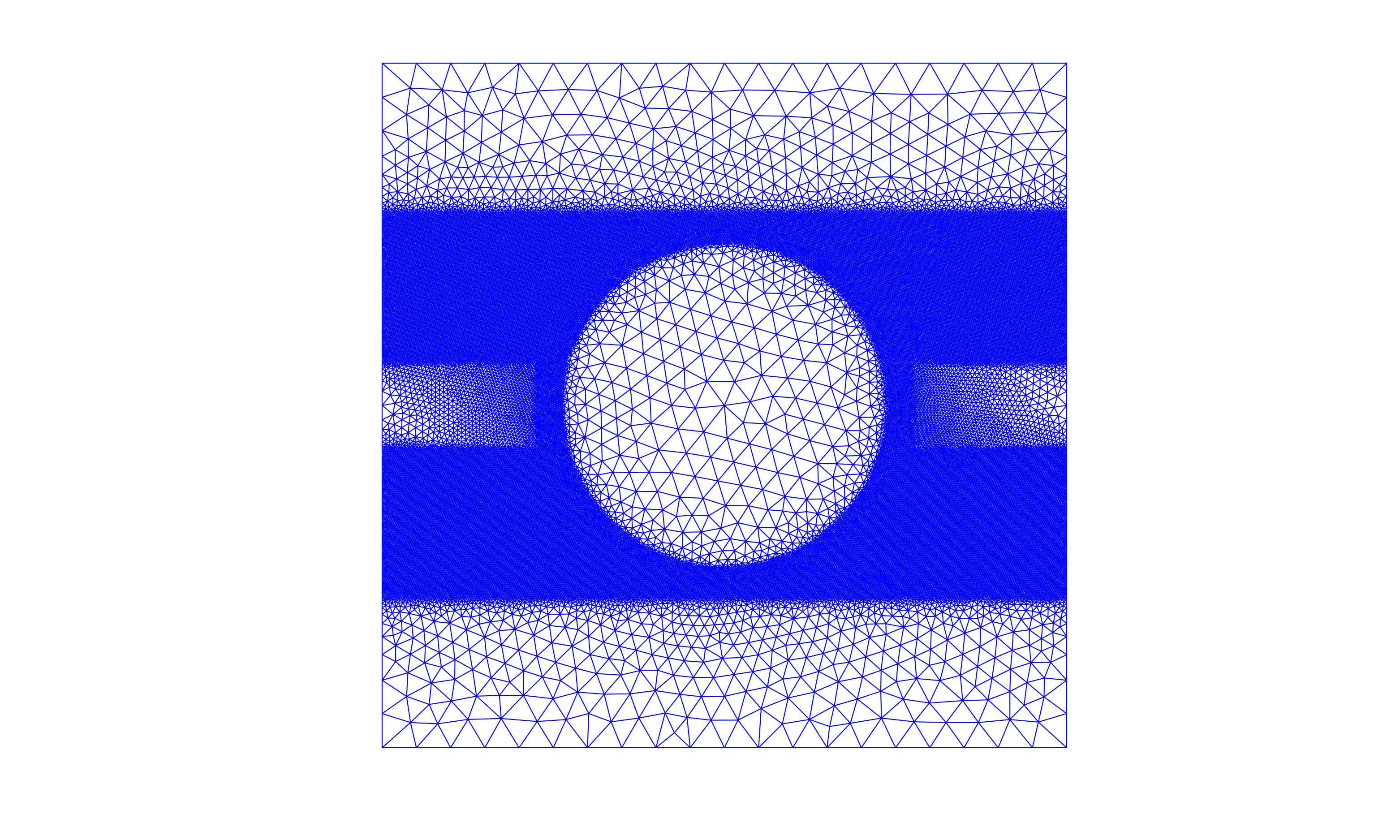}
\caption{Refinement of the mesh}\label{Fig_Incl_M}
\end{subfigure}
\caption{ Description and discretisation of the domain with an inhomogeneity.}\label{Fig_Incl}
\end{figure}

An interface can be considered as an initialiser of a  damage  process, especially if the  inhomogeneity  is made of different material than the matrix.
Based on changing parameters of the interface, various scenarios of cracking processes can be observed  at the vicinity of the interface.
The analysis in this example compares two such scenarios.
 The problem scheme is shown in Fig.~\ref{Fig_Incl}, which also presents a mesh which was used in the calculations.
For computational purposes, additional horizontal constraints are put at the midpoints of the horizontal faces of the block.
The refinements in the band regions are made because the crack is expected to appear in the domain, depending on the material characteristics of the two-component domain and on characteristics of the fracture.
The mesh refinements contain element of the minimal size equal to $0.2$ mm.

The parameters needed for the computations include the material stiffness values: the  bulk modulus $\Kp=192.3$ GPa, the shear modulus $\mu=76.9$ GPa for the  inhomogeneity, and $\Kp=22.0$ GPa, $\mu=12.3$ GPa for the matrix. 
The interface, considered as a thin adhesive layer is characterised by $\kappa = 6.88\left(\begin{smallmatrix}1 & 0\\0 & 1\end{smallmatrix}\right)\text{ TPam}^{-1}$, $\kG=1\ \text{PPam}^{-1}$. 

The fracture energy in the matrix domain is $\Gc[I]=0.25\ \text{Jm}^{-2}$, for the shear mode a different value is considered: $\Gc[II]=100\Gc[I]$ to reduce influence of shear on the cracking process in the matrix.
The length scale parameter of the phase-field model is  $\epsilon= 0.5$ mm.

 Damage  of the interface is controlled by the interface fracture energy, which takes various values to obtain alternative behaviour in crack formation process along the interface:  $\Gc[iI]=\{0.032,0.8,20\}\ \text{Jm}^{-2}$. 
The mode II interface fracture energy is again set to a high value for the  stress condition~\eqref{Eq_StressCrit_Z} to be reduced to normal component $p_\sfn$.
In particular,   with the interface degradation function $\Fun{\phi}(\zeta)=\frac{\beta\zeta}{1+\beta-\zeta}$, the provided maximum of normal interface stress  is $p_{\sfn\,\text{crit}}=\sqrt{\frac{2\kappa_\sfn\Gc[iI]\beta}{\beta+1}}=\{0.2,1,5\}$ MPa (respectively to $\Gc[iI]$), where the parameter was set to $\beta=0.1$. 
Here, the stress maximum occurs  at damage triggering for $\Fun{\zeta}=1$.

Similar analysis for the domain fracture is done for the degradation function $\Fun{\Phi}(\alpha) = \frac{\alpha^2}{\alpha^2+\beta(1-\alpha)}+10^{-6}$ (see~\cite{wu17A1}), with $\beta=3$, used in the calculations based on the stress condition~\eqref{Eq_StressCrit_Amod}.
It means that when the shear stresses can be omitted by considering $\Gc[II]$ substantially greater than $\Gc[I]$, the critical value of  $\tr\sigma$ is given by the same approximate relation as in the previous example and the actual values of the parameters provide the value $\tr\sigma_{\text{crit}}=2.7$ MPa.

The structural element is loaded by displacement loading $g(t)=v_0t$, at the velocity $v_0=1\ \text{mm\,s}^{-1}$. 
This load is applied incrementally in  time steps refined to $0.01$ ms.

\begin{figure}[!ht]
\centering
\includegraphics[scale=0.3,trim=5 5 40 50,clip]{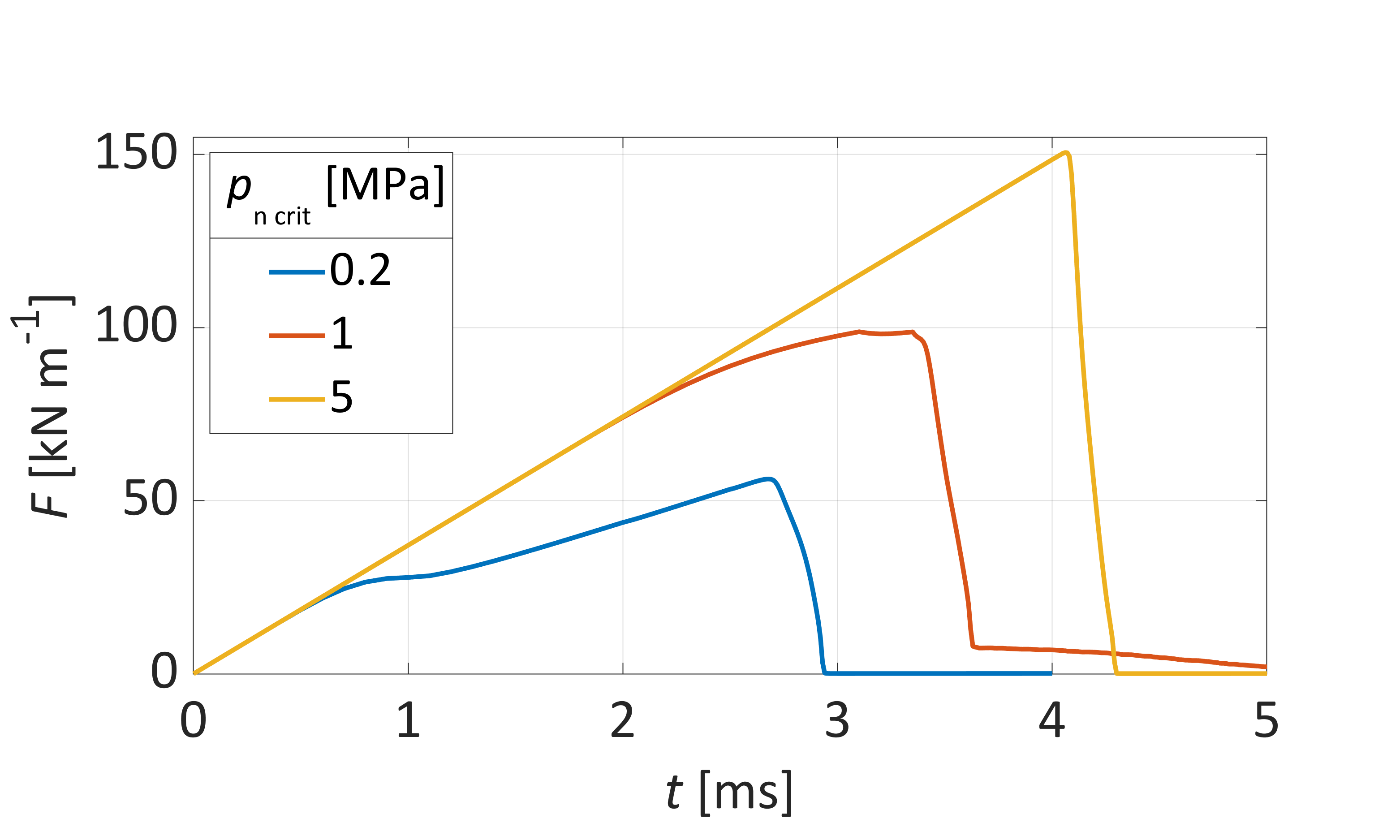}
\caption{Total force applied at the top face during loading for the  solid with an inhomogeneity subjected  to tension.}\label{Fig_MaFi_F}
\end{figure}

The force response of the  structural element  in terms of  the total vertical force $F$ applied at the top face of the matrix domain is shown in~Fig.~\ref{Fig_MaFi_F}.
The dependence is related to the time $t$ as the principal factor of the prescribed displacement $g$.
The graphs include all three options of the interface fracture energy $\Gc[iI]$ distinguished by the critical value of the normal stress $p_{\sfn\,\text{crit}}$.
For smaller values of $p_{\sfn\,\text{crit}}$ the deviation from the linear relation is caused by appearing of an interface crack.
For the smallest value, the crack is propagated very early so that there is a stabilised increase in the applied force with a fixed interface crack.
The last case does not show such a decrease of overall stiffnesses which can be read such that no interface crack was developed.
It will also be confirmed below in the direct plot containing all developed  cracks.
In all the cases,  total rupture of the analysed domain can be guessed in the final jump of the force and fast  crack propagation across the whole domain.

\begin{figure}[!ht]
\centering
\begin{subfigure}{0.325\textwidth}
\centering
\includegraphics[scale=0.21,trim=5 5 45 45,clip]{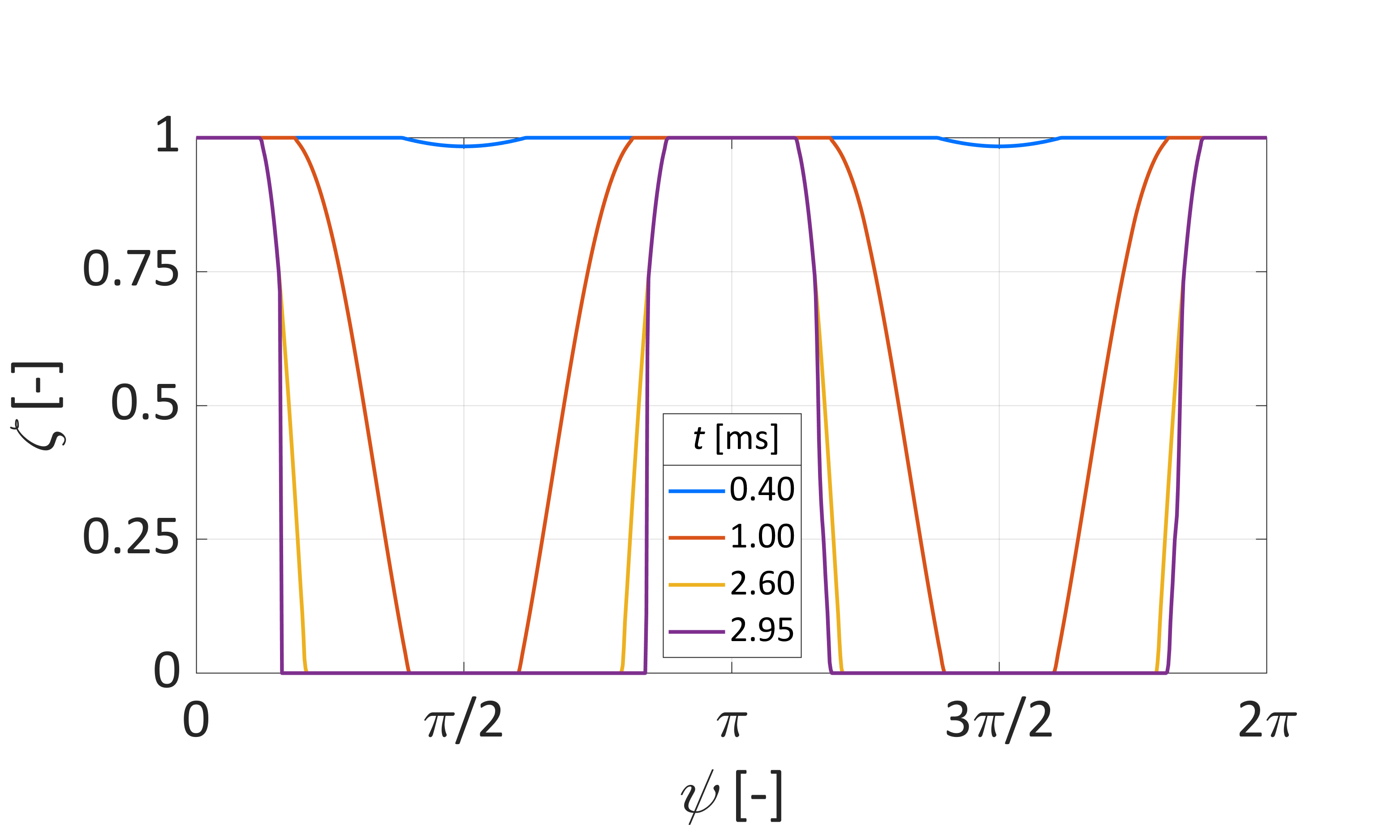}
\caption{Interface damage, $p_{\sfn\,\text{crit}}{=}0.2$ MPa}\label{Fig_MaFi_SNI02_IF_1}
\end{subfigure}
\begin{subfigure}{0.325\textwidth}
\centering
\includegraphics[scale=0.21,trim=5 5 45 45,clip]{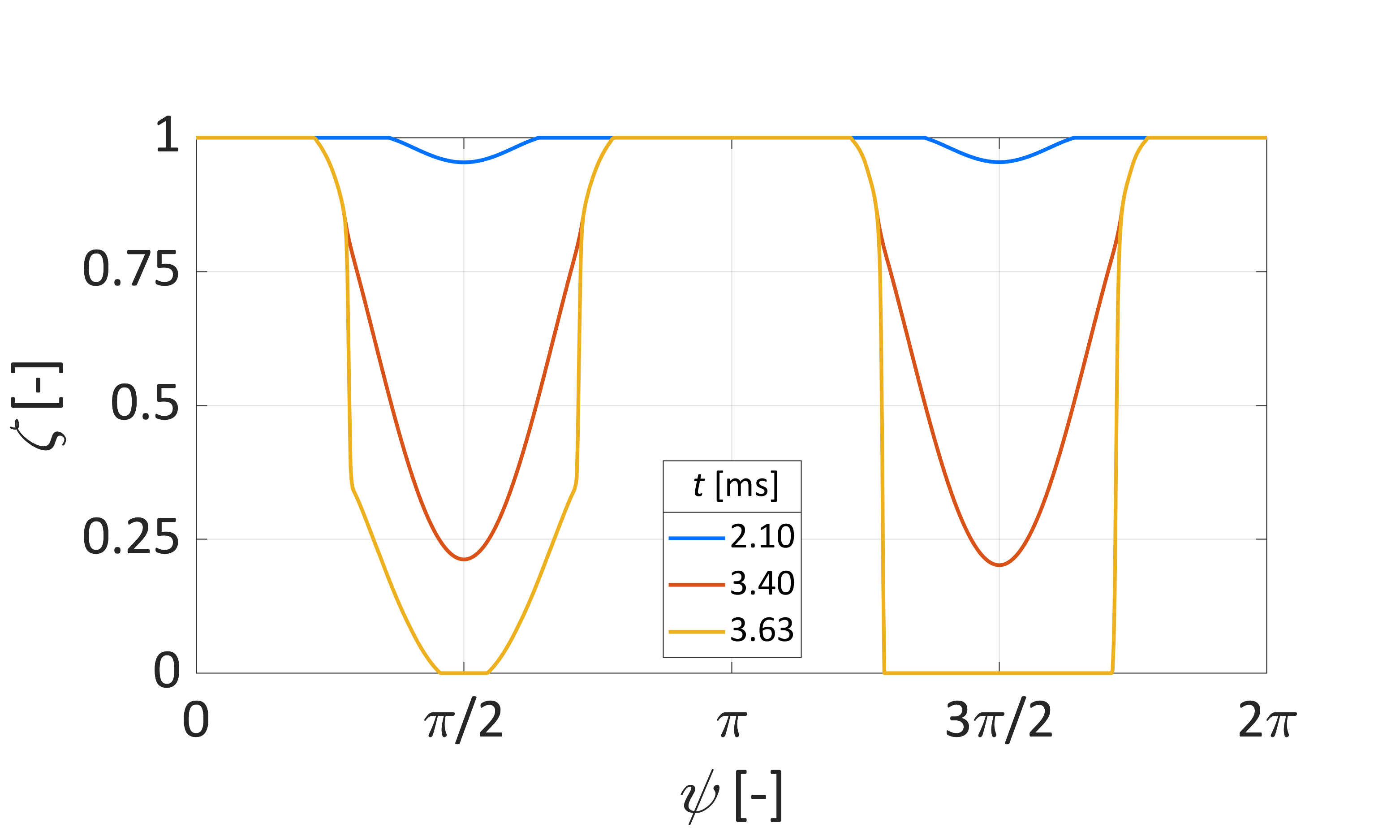}
\caption{Interface damage, $p_{\sfn\,\text{crit}}=1$ MPa}\label{Fig_MaFi_SNI1_IF_1}
\end{subfigure}
\begin{subfigure}{0.325\textwidth}
\centering
\includegraphics[scale=0.21,trim=5 5 45 45,clip]{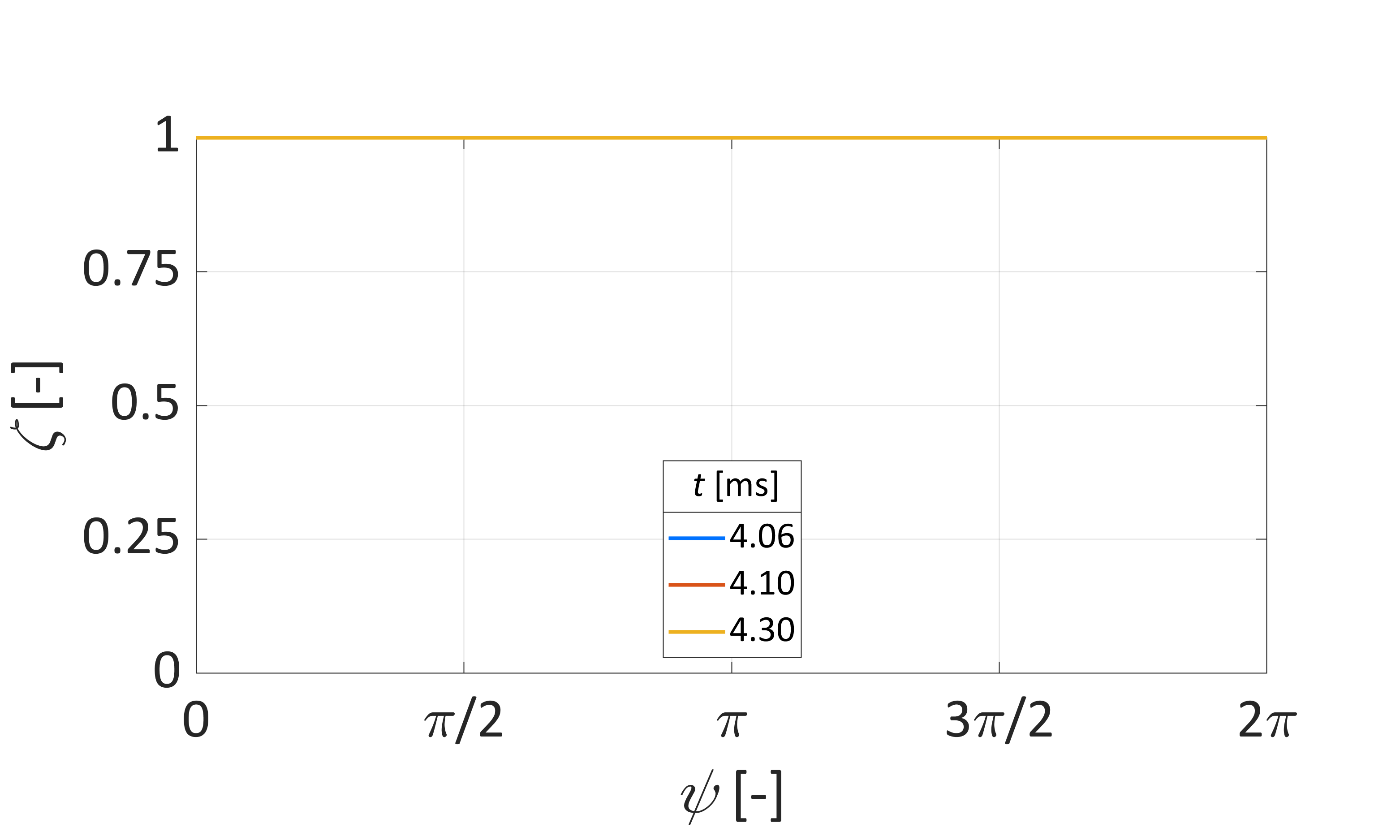}
\caption{Interface damage, $p_{\sfn\,\text{crit}}=5$ MPa}\label{Fig_MaFi_SNI5_IF_1}
\end{subfigure}
\begin{subfigure}{0.325\textwidth}
\centering
\includegraphics[scale=0.21,trim=5 5 45 45,clip]{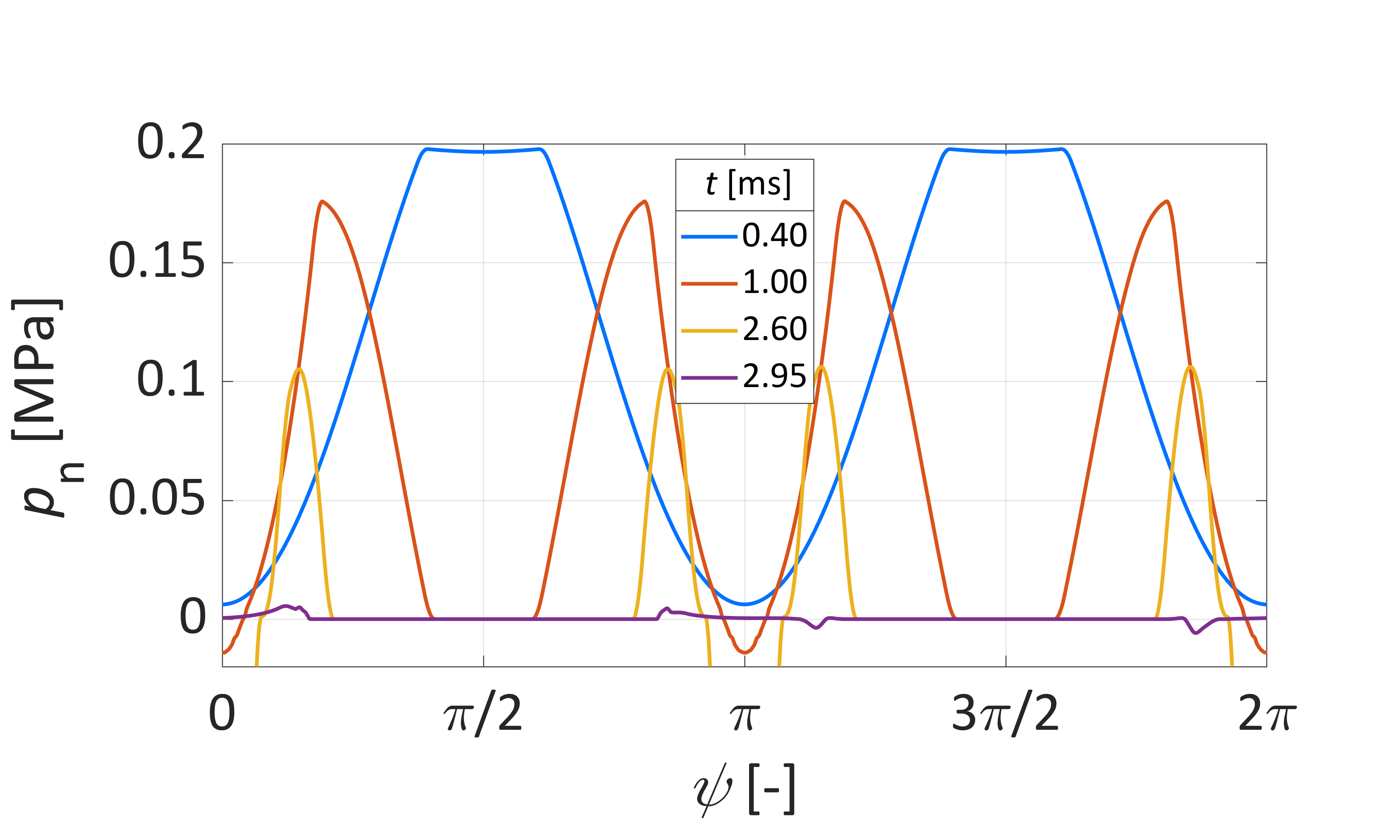}
\caption{Normal stress, $p_{\sfn\,\text{crit}}=0.2$ MPa}\label{Fig_MaFi_SNI02_IF_2}
\end{subfigure}
\begin{subfigure}{0.325\textwidth}
\centering
\includegraphics[scale=0.21,trim=5 5 45 45,clip]{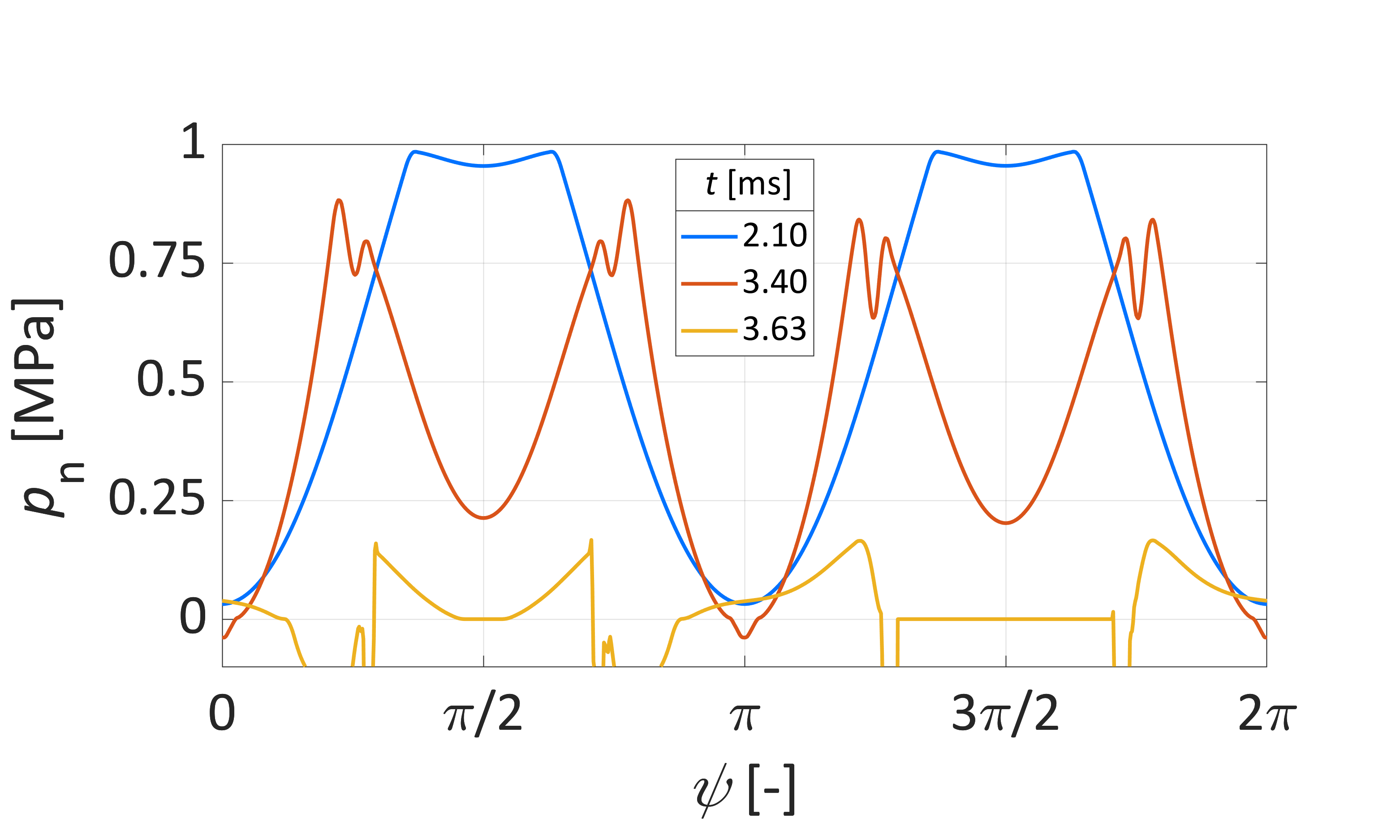}
\caption{Normal stress, $p_{\sfn\,\text{crit}}=1$ MPa}\label{Fig_MaFi_SNI1_IF_2}
\end{subfigure}
\begin{subfigure}{0.325\textwidth}
\centering
\includegraphics[scale=0.21,trim=5 5 45 45,clip]{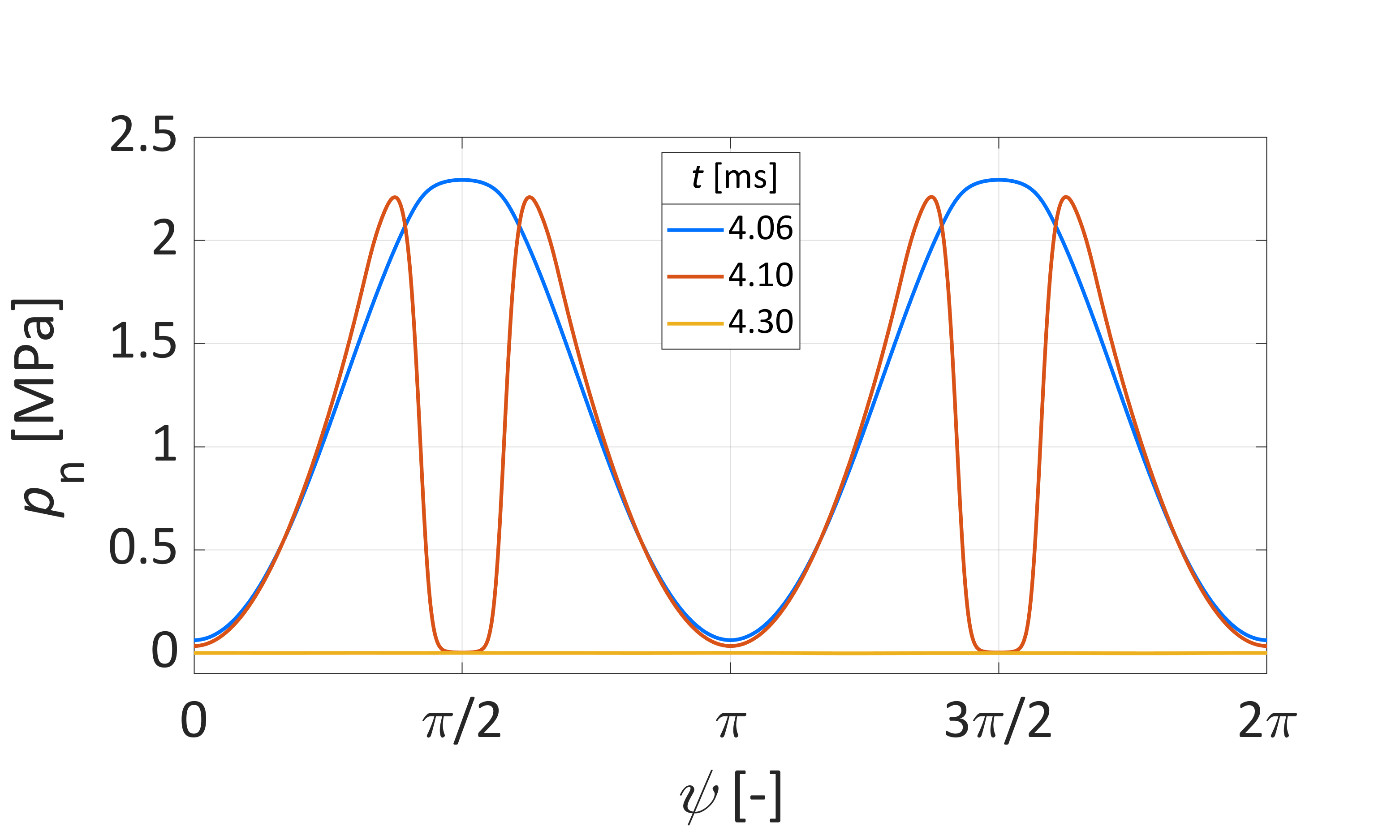}
\caption{Normal stress, $p_{\sfn\,\text{crit}}=5$ MPa}\label{Fig_MaFi_SNI52_IF_2}
\end{subfigure}
\caption{Distribution of the interface quantities during the debonding phase at selected time instants for the case with an  inhomogeneity.}\label{Fig_MaFi_SNIx_IF}
\end{figure}

Distributions of the interface damage parameter and the normal contact stress are displayed in Fig.~\ref{Fig_MaFi_SNIx_IF} to show that the interface debonding actually appears for the largest value of $p_{\sfn}$ set to  $p_{\sfn\,\text{crit}}$, and to demonstrate propagation of the interface crack.
The first selected instants belong to situations when the maximum stress value was reached (and can be seen in the graphs), which corresponds to initiating interface damage, beside the last case where the reached stress corresponds to the stress at the matrix domain at the moment of phase-field damage initiation.
The next instants  document damage growth and the interface crack propagation. 
The graphs in Figs.~\ref{Fig_MaFi_SNI1_IF_1} and~\ref{Fig_MaFi_SNI1_IF_2} also show that for this case the  damage  in the matrix starts  during the interface  damage evolution as the debonding angle is visibly smaller than observed in Fig.~\ref{Fig_MaFi_SNI02_IF_1}. 
With the largest value of the critical interface stress, the  stress  goes down to zero due to  a crack appearing in the matrix in the vicinity of the interface, cf.\ also Fig.~\ref{Fig_MaFi_SNI} below, rather than  due to a cracking process in the interface itself which remains at an intact state $\zeta=1$.
Anyhow, for developed global cracks the normal stresses vanish.

The parameters for phase-field fracture were set the same for all three option, nevertheless varying properties of the interface affect cracking in the matrix.
Drawings in Fig.~\ref{Fig_MaFi_SNI} present differences between  any two options.
\begin{figure}[!ht]
\centering
\setlength{\unitlength}{\textwidth}
\begin{subfigure}{0.99\textwidth}
\begin{picture}(0.97,0.22)
\PA{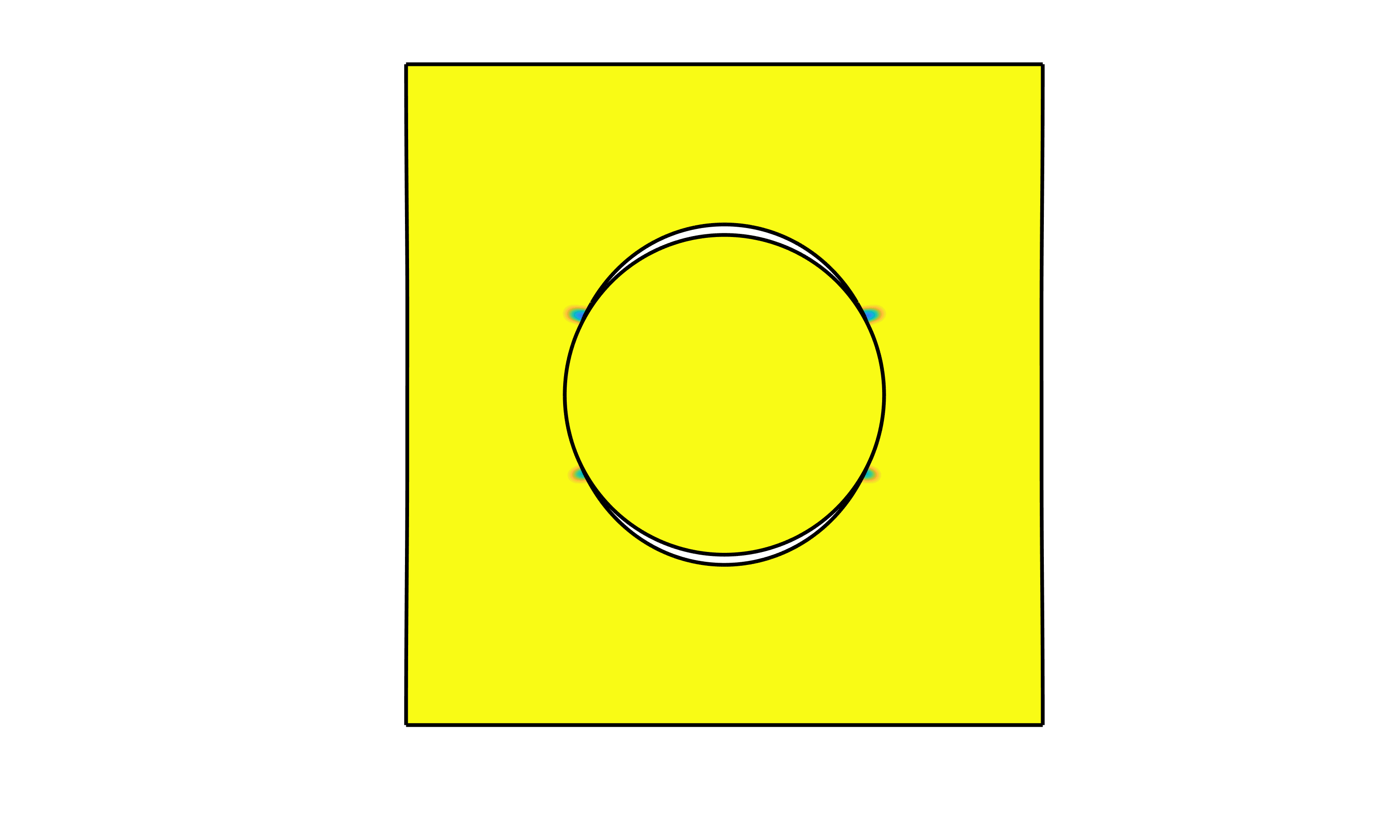}{$t=2.7$ms}
\PBF{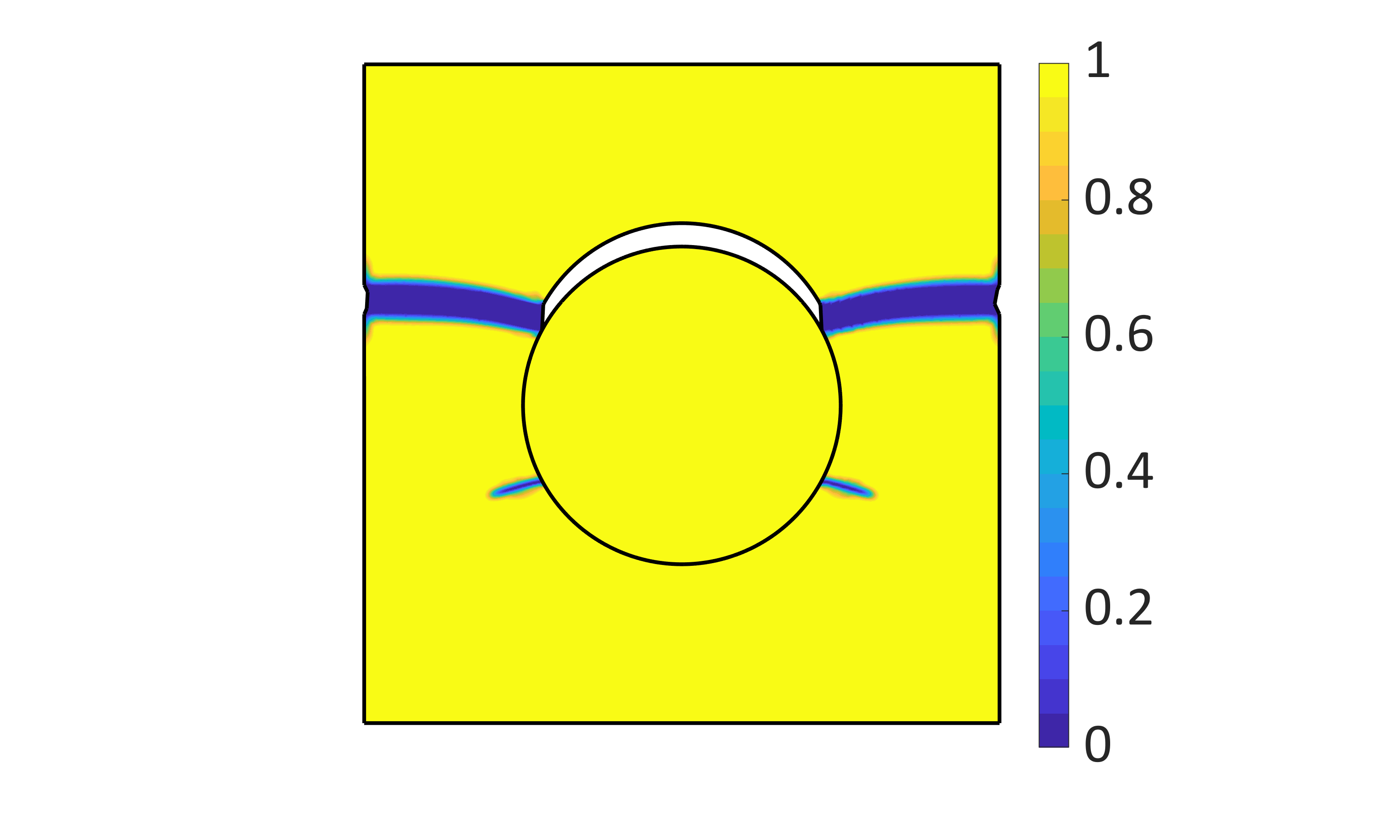}{$t=2.95$ms}
\PD{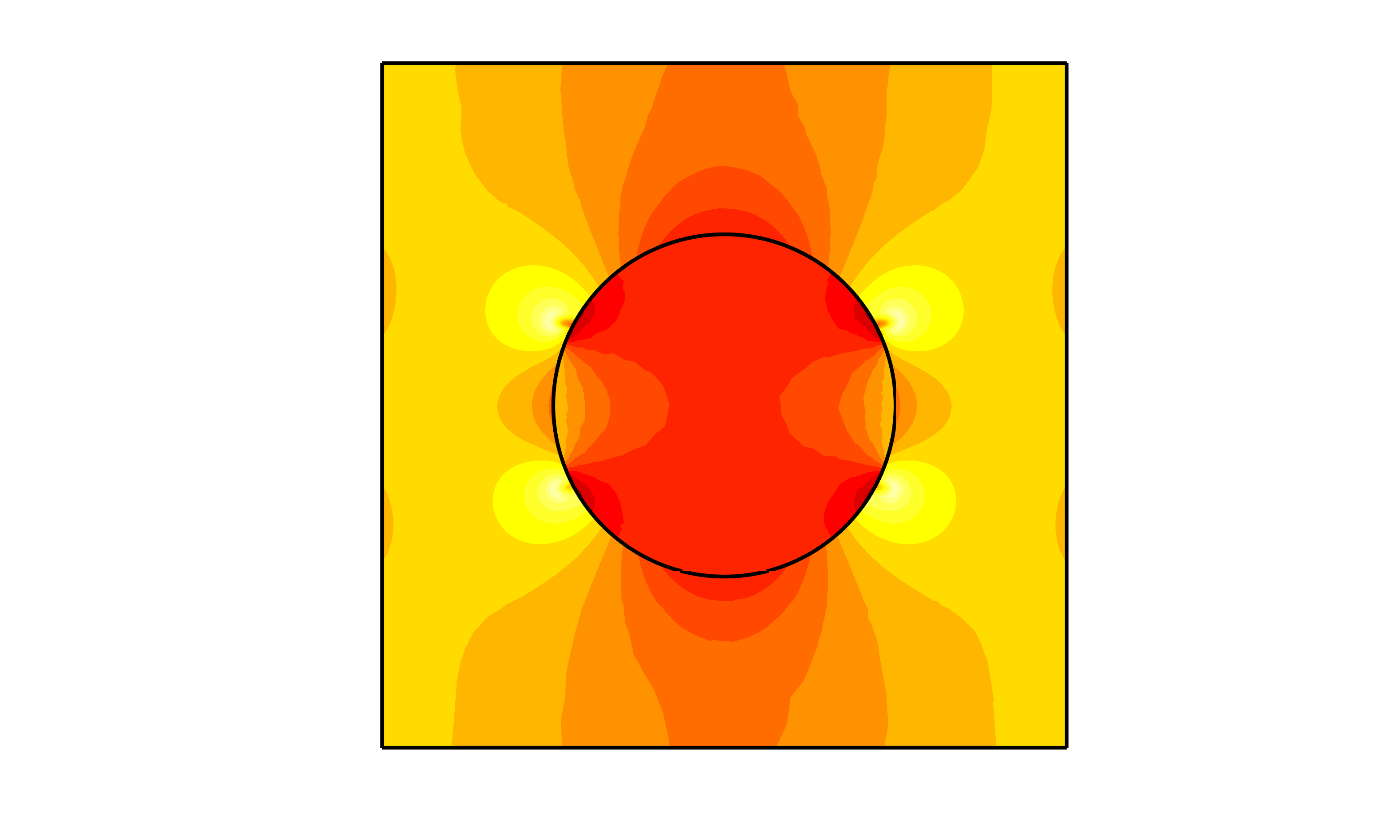}{Stress trace}
\PE{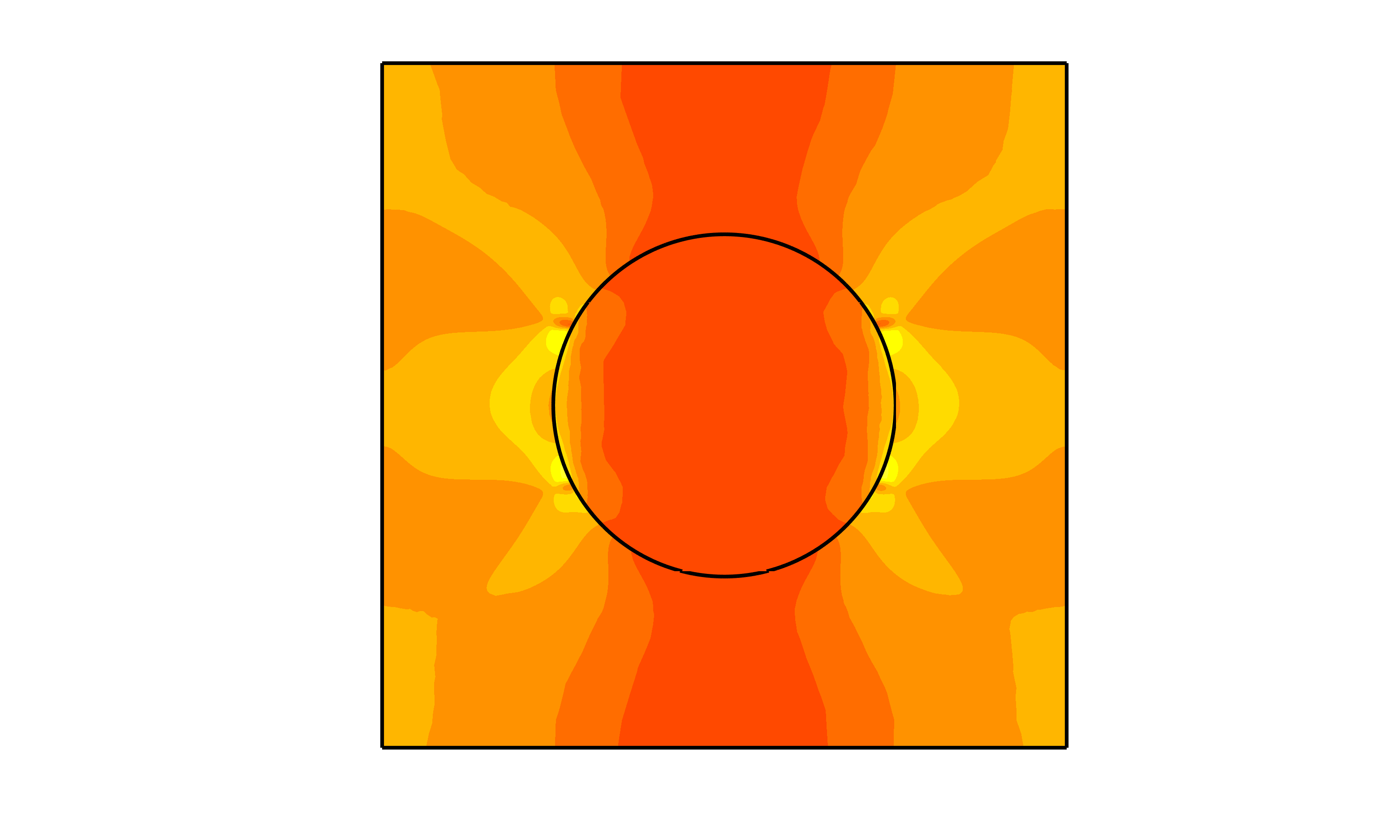}{Deviatoric stress}
\CB[96]{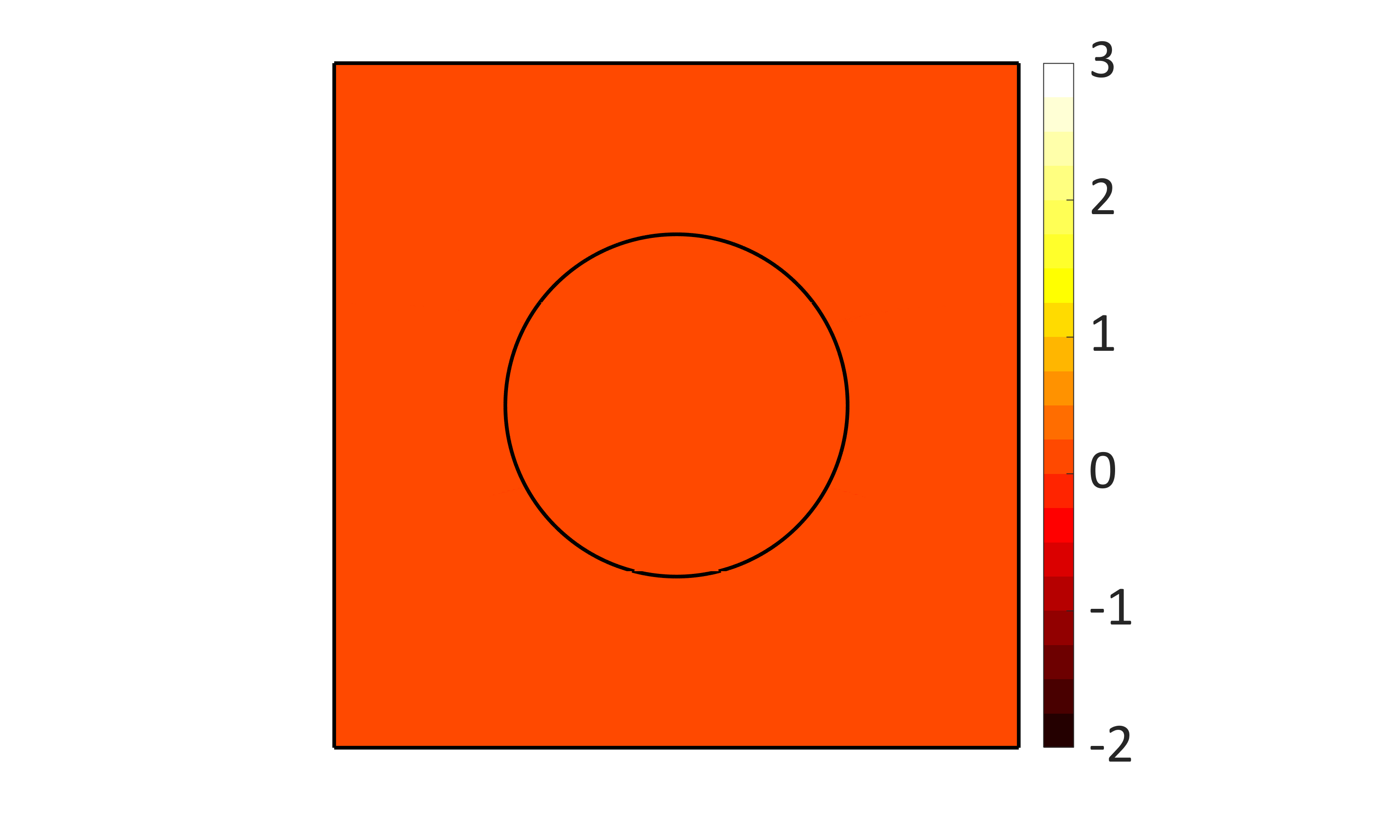}
\end{picture}
\caption{$\pcri=0.20$ MPa}\label{Fig_MaFi_SNI02}
\end{subfigure}
\begin{subfigure}{0.99\textwidth}
\begin{picture}(0.97,0.22)
\PA{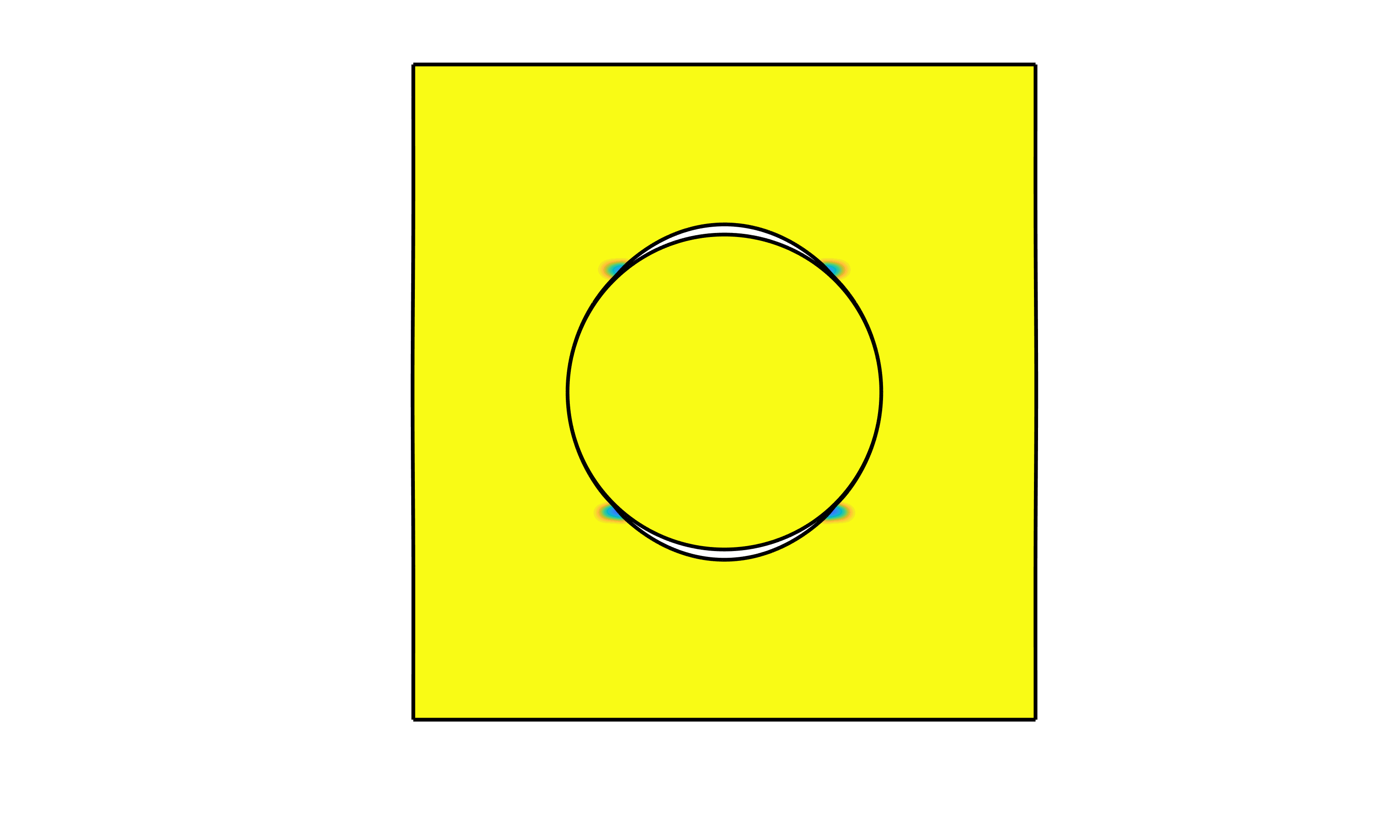}{$t=3.4$ms}
\PBF{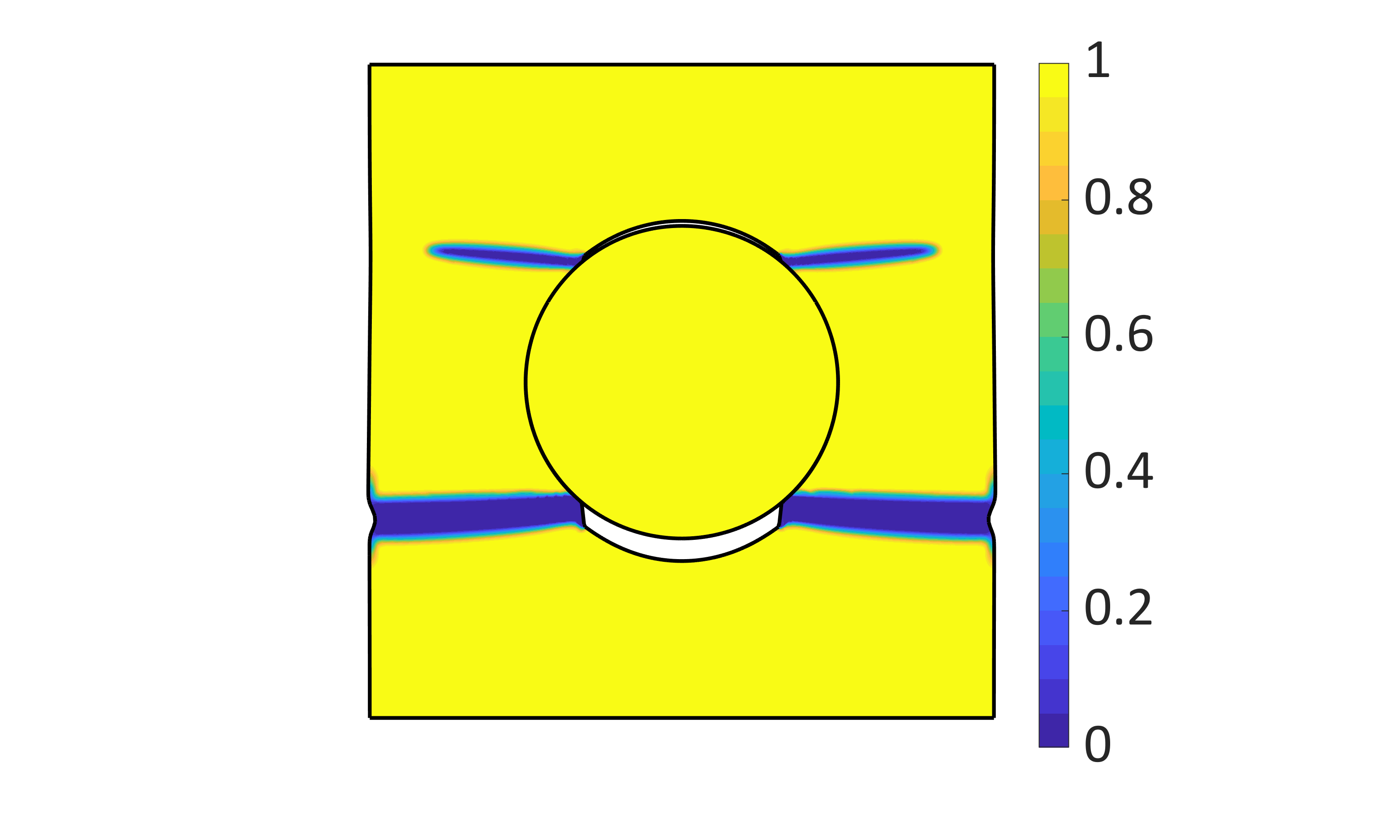}{$t=3.63$ms}
\PD{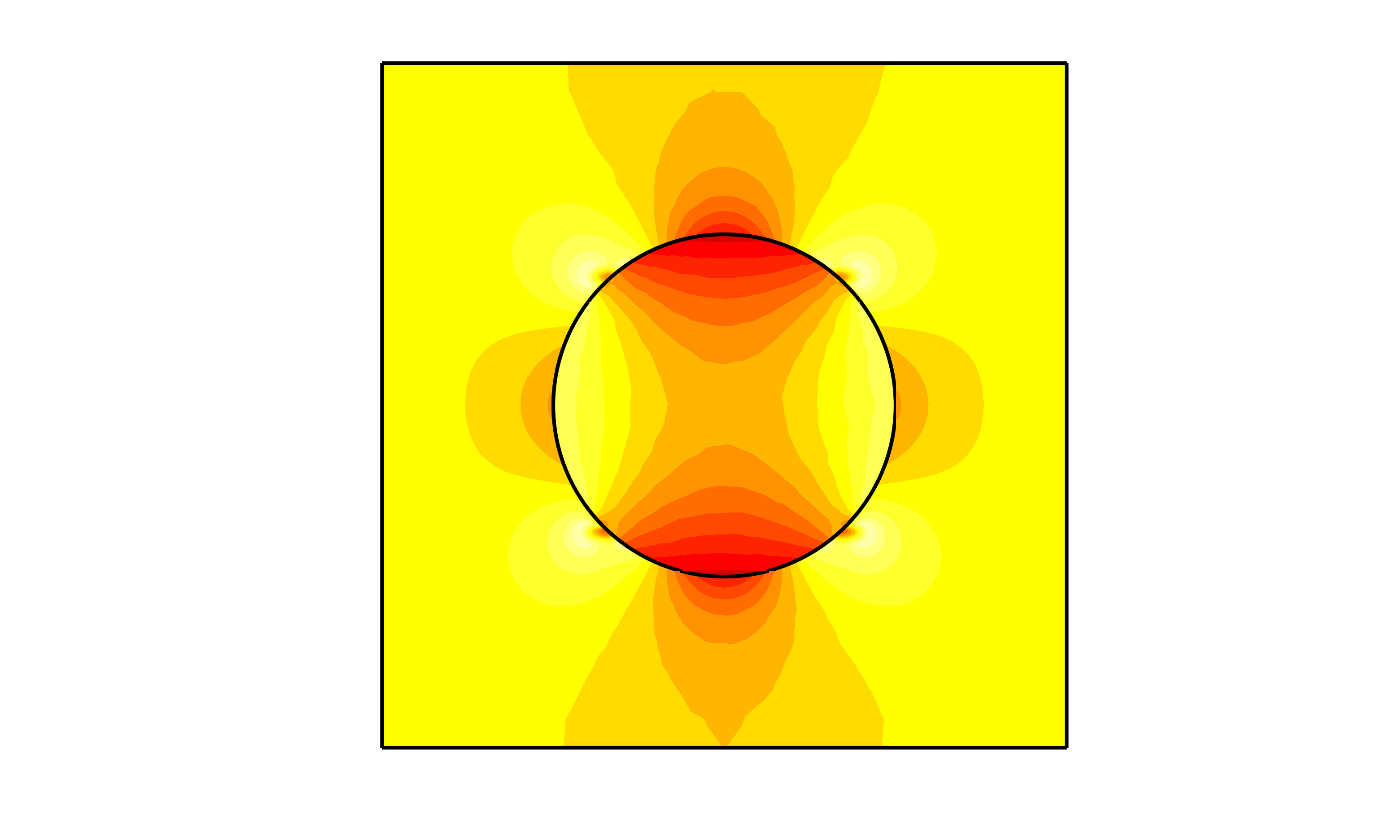}{Stress trace}
\PE{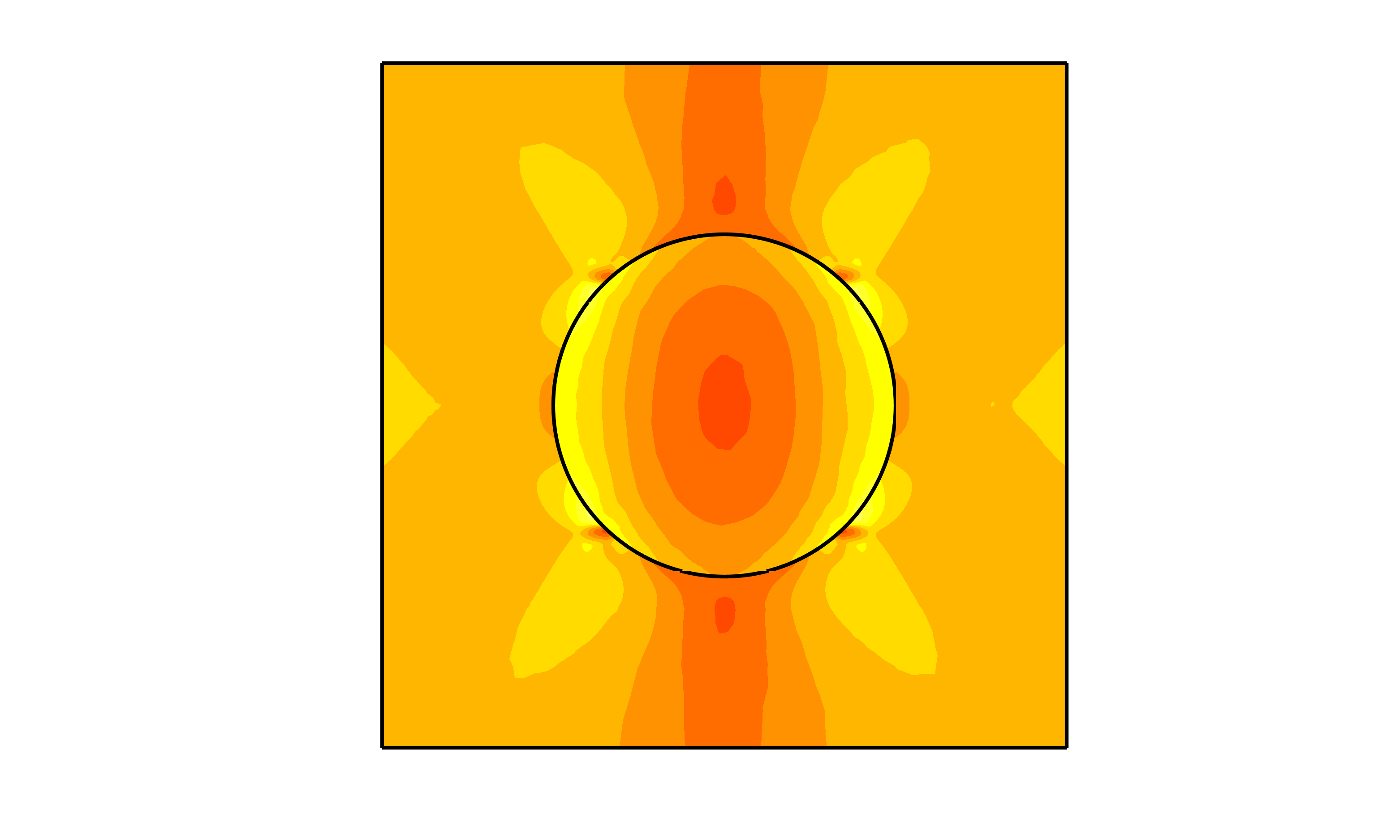}{Deviatoric stress}
\CB[96]{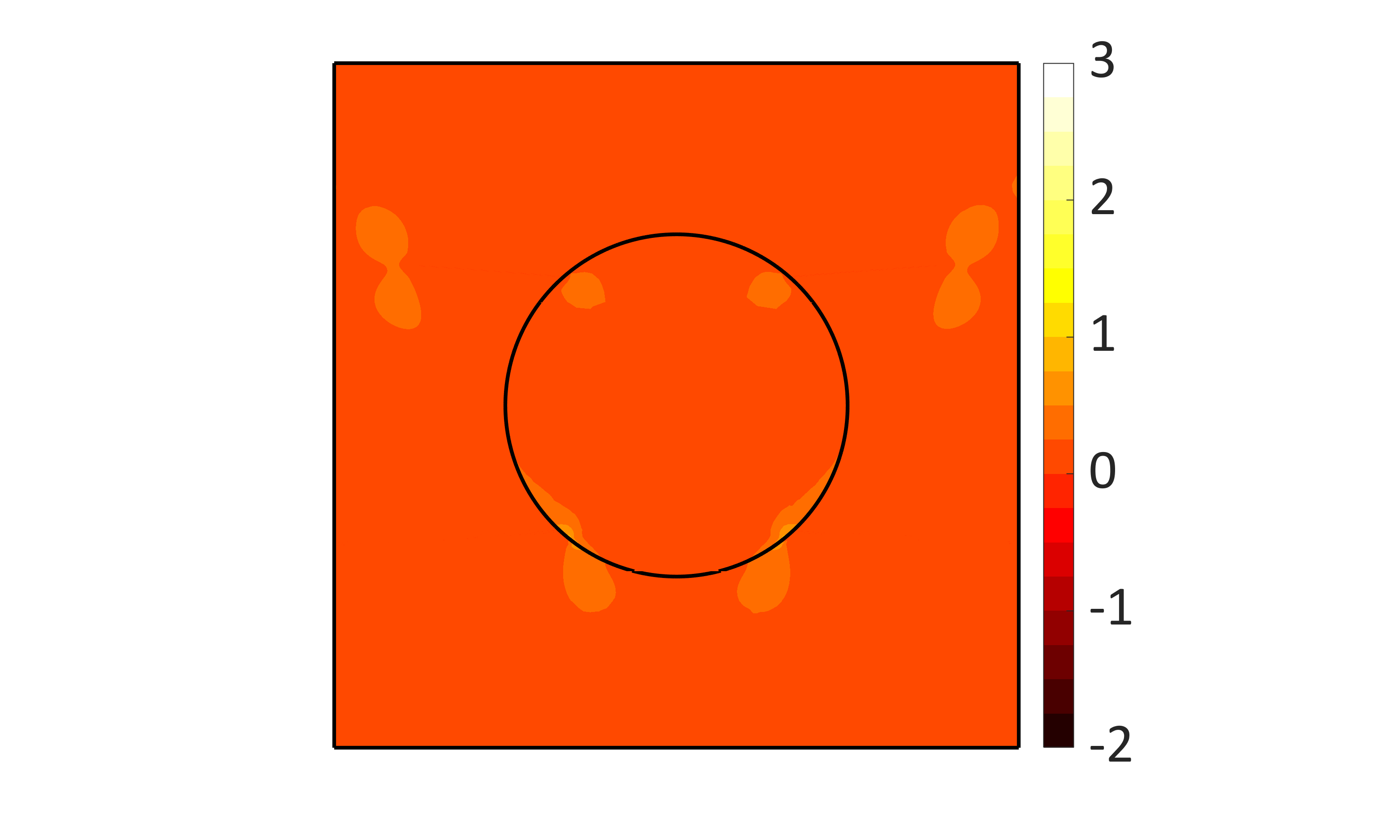}
\end{picture}
\caption{$\pcri=1$ MPa}\label{Fig_MaFi_SNI1}
\end{subfigure}
\begin{subfigure}{0.99\textwidth}
\begin{picture}(0.97,0.22)
\PA{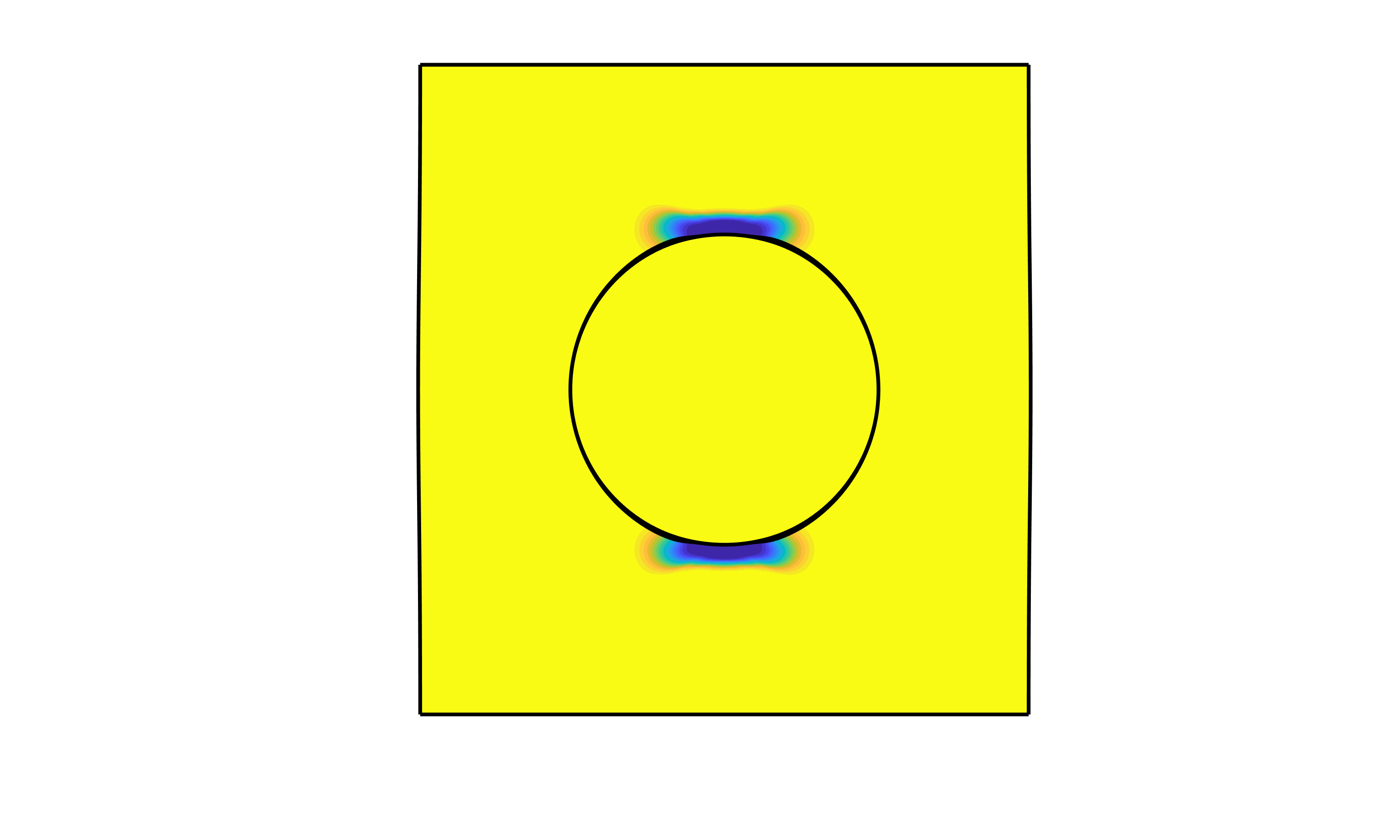}{$t=4.1$ms}
\PBF{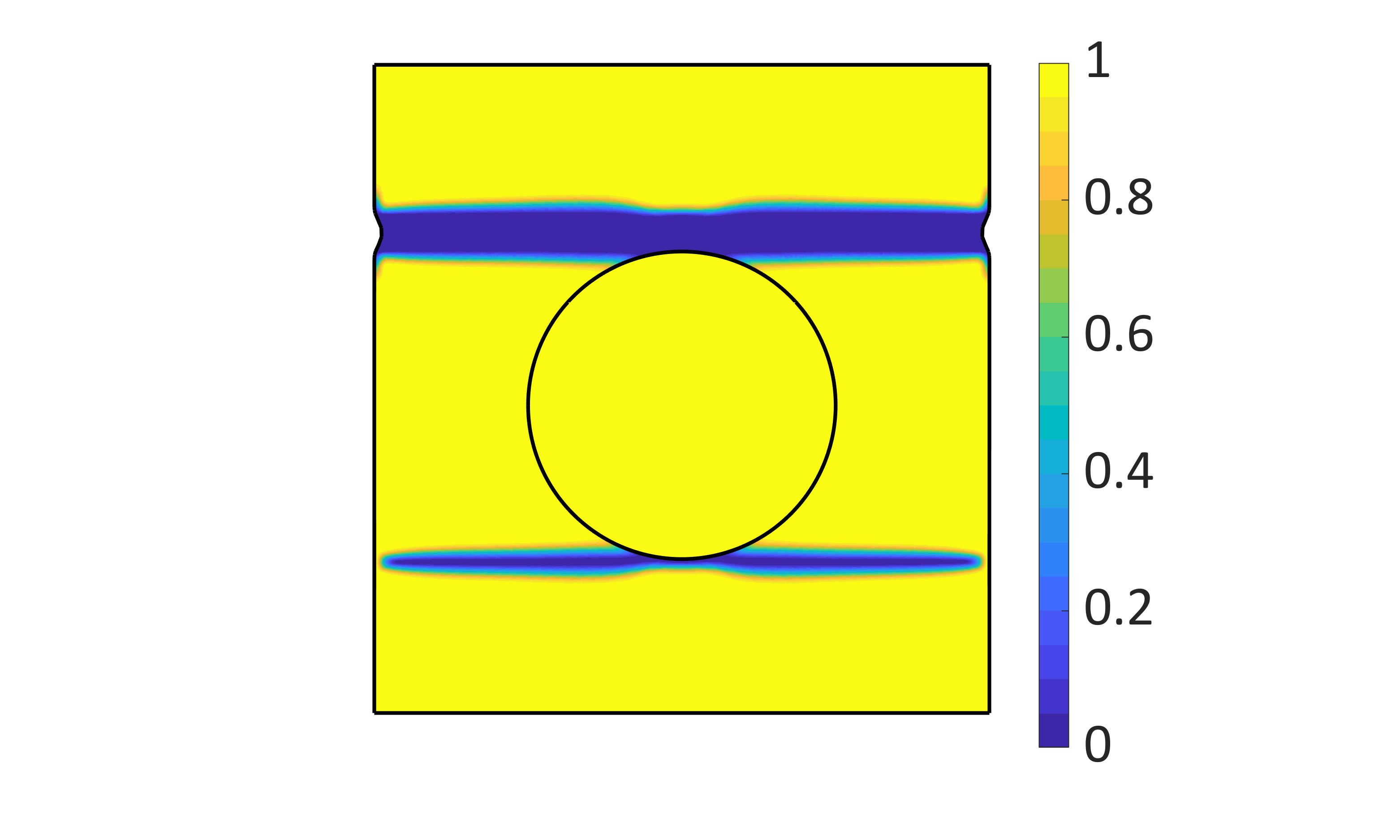}{$t=4.3$ms}
\PD{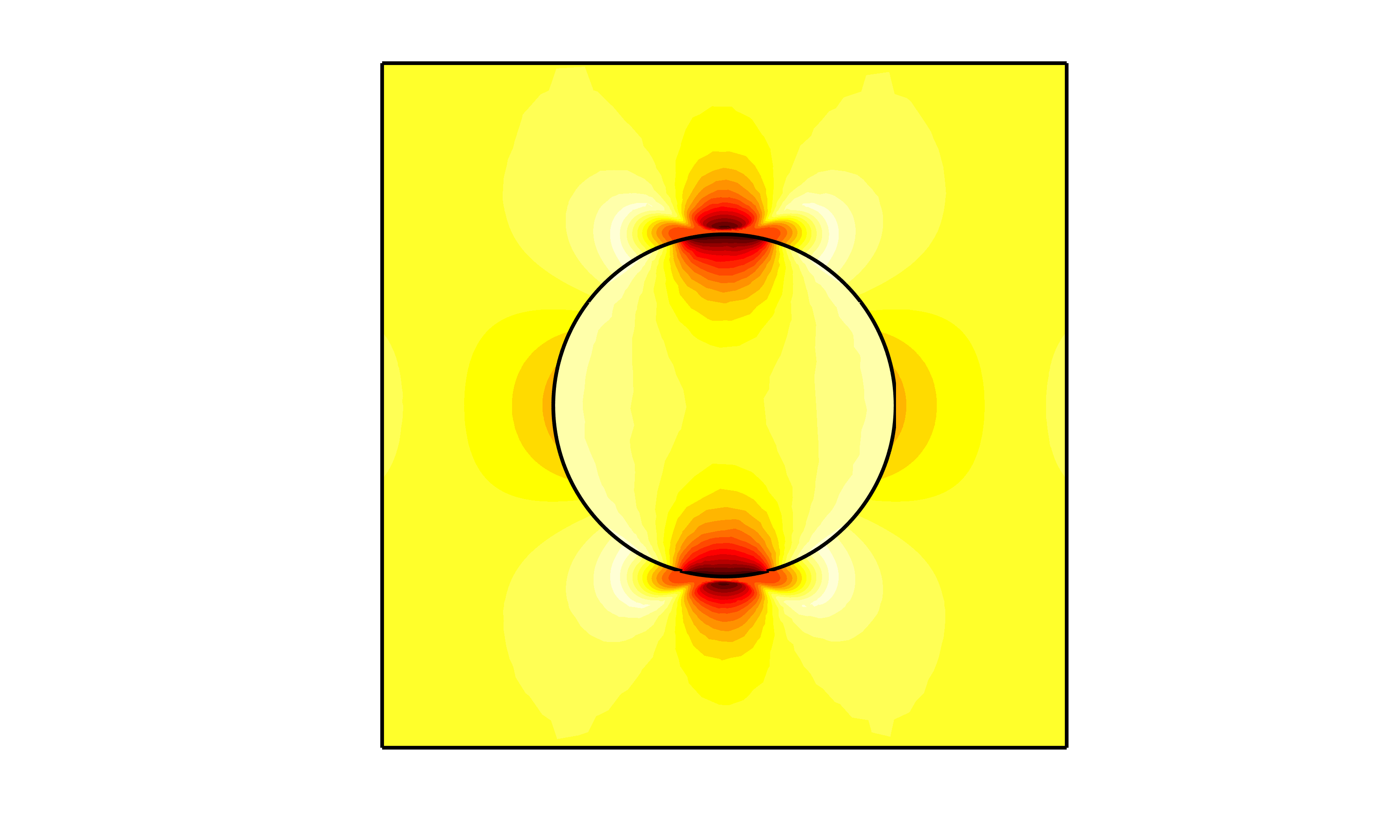}{Stress trace}
\PE{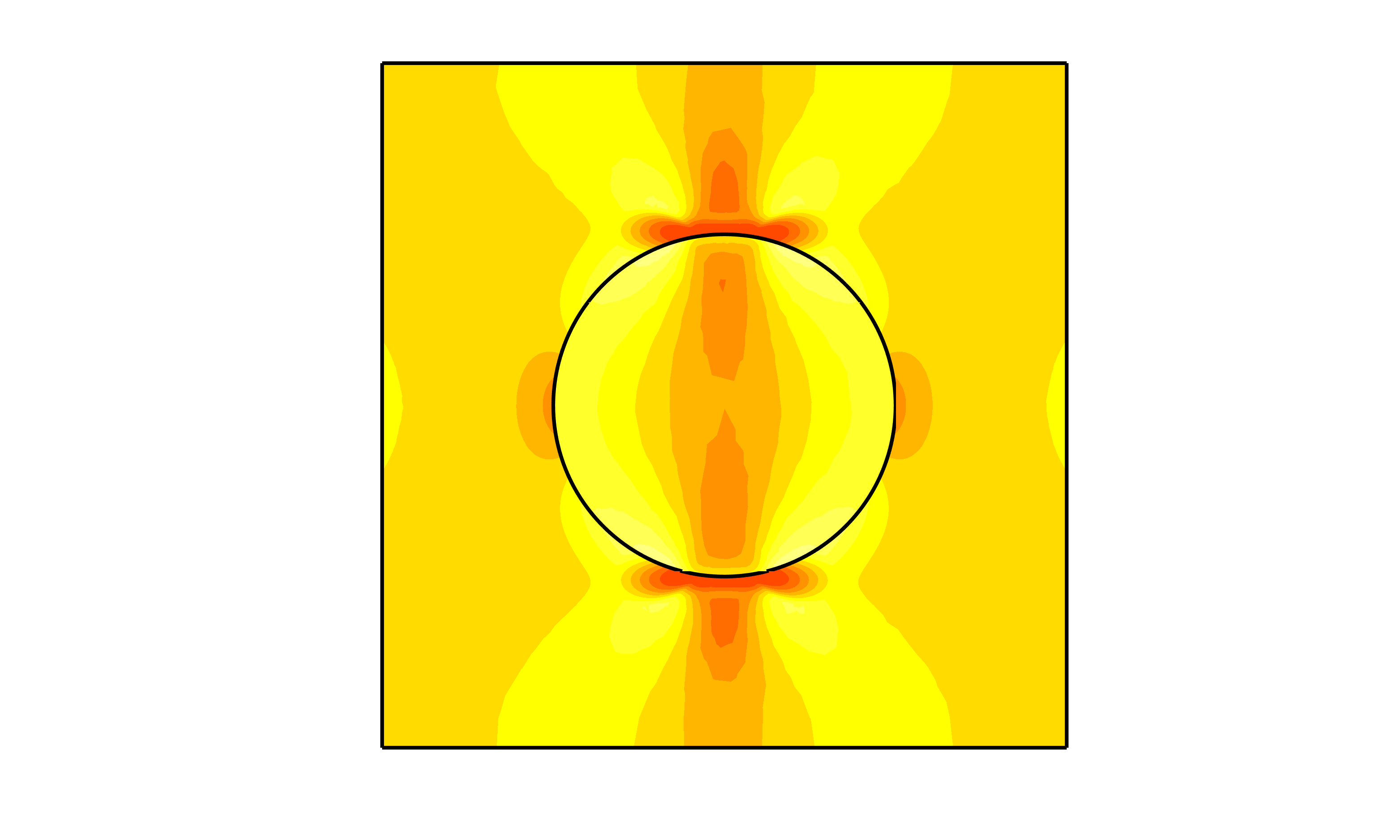}{Deviatoric stress}
\CB[96]{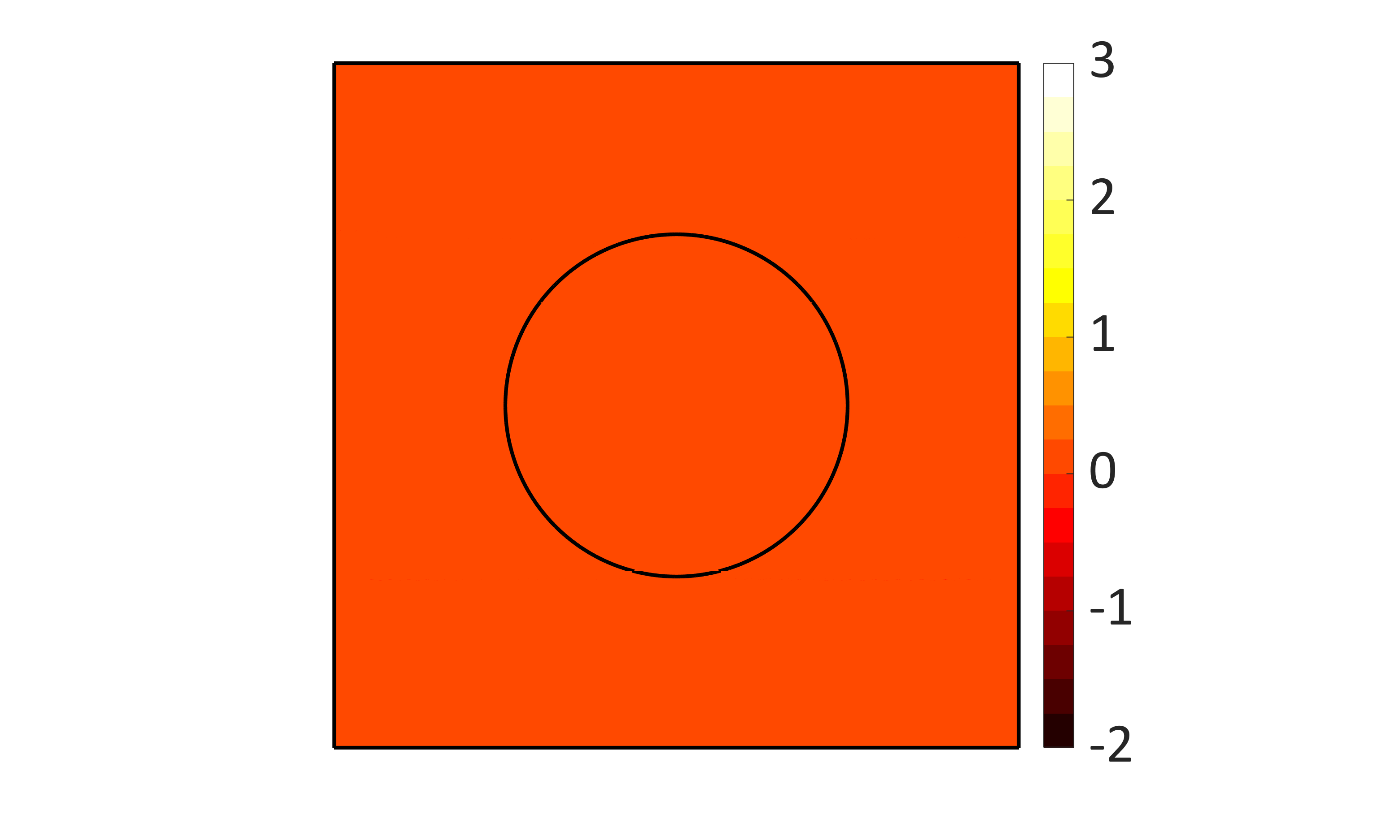}
\end{picture}
\caption{$\pcri=5$ MPa}\label{Fig_MaFi_SNI5}
\end{subfigure}
\caption{Distribution of the phase-field variable  documenting crack propagation (left) and stress values [MPa] (right) at selected time instants for the case with one  inhomogeneity. 
Stresses belong to the first used instant  and the deformation in the crack plot is $1000\times$ magnified.}\label{Fig_MaFi_SNI}
\end{figure}
The difference between the first two reveal different debonding angle as mentioned above.
The fact that  total rupture in matrix appears at other location is caused just by the numerical computational algorithm.
The third option shows a crack only in matrix, but caused by the different elastic properties of the  inhomogeneity and the matrix.
In fact, the crack is initiated at a layer adjacent to the interface, though no interface  damage  is obtained as it is also seen in Fig.~\ref{Fig_MaFi_SNI5_IF_1}.
The stress plots show  that the stress distribution in terms of stress trace responsible for opening of the crack approaches the critical value   $\tr\sigma_{\text{crit}}=2.7$ MPa calculated above. 
The distributions are displayed for those instants when the crack modeled by the phase-field approach starts to propagate.
 The magnitude of the shear stress  (expressed by the deviatoric stress norm)  shows where this stress reaches its  maximal values, nevertheless it does not influence the process of crack formation in a large amount.

%%%%%%%%%%%%%%%%%%%%%%%%%%%%
\subsection{Compression in a domain with an initial  crack}\label{Sec_TP}
%%%%%%%%%%%%%%%%%%%%%%%%%%%%

\begin{figure}[!ht]
\centering
\begin{subfigure}{0.49\textwidth}
\centering
\includegraphics[scale=1]{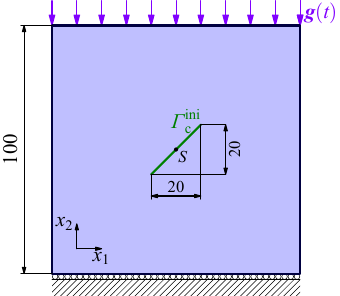}
\caption{A domain with a crack  for compressive loading}\label{Fig_Press}
\end{subfigure}
\begin{subfigure}{0.49\textwidth}
\centering
\includegraphics[scale=0.32,trim=170 30 150 20,clip]{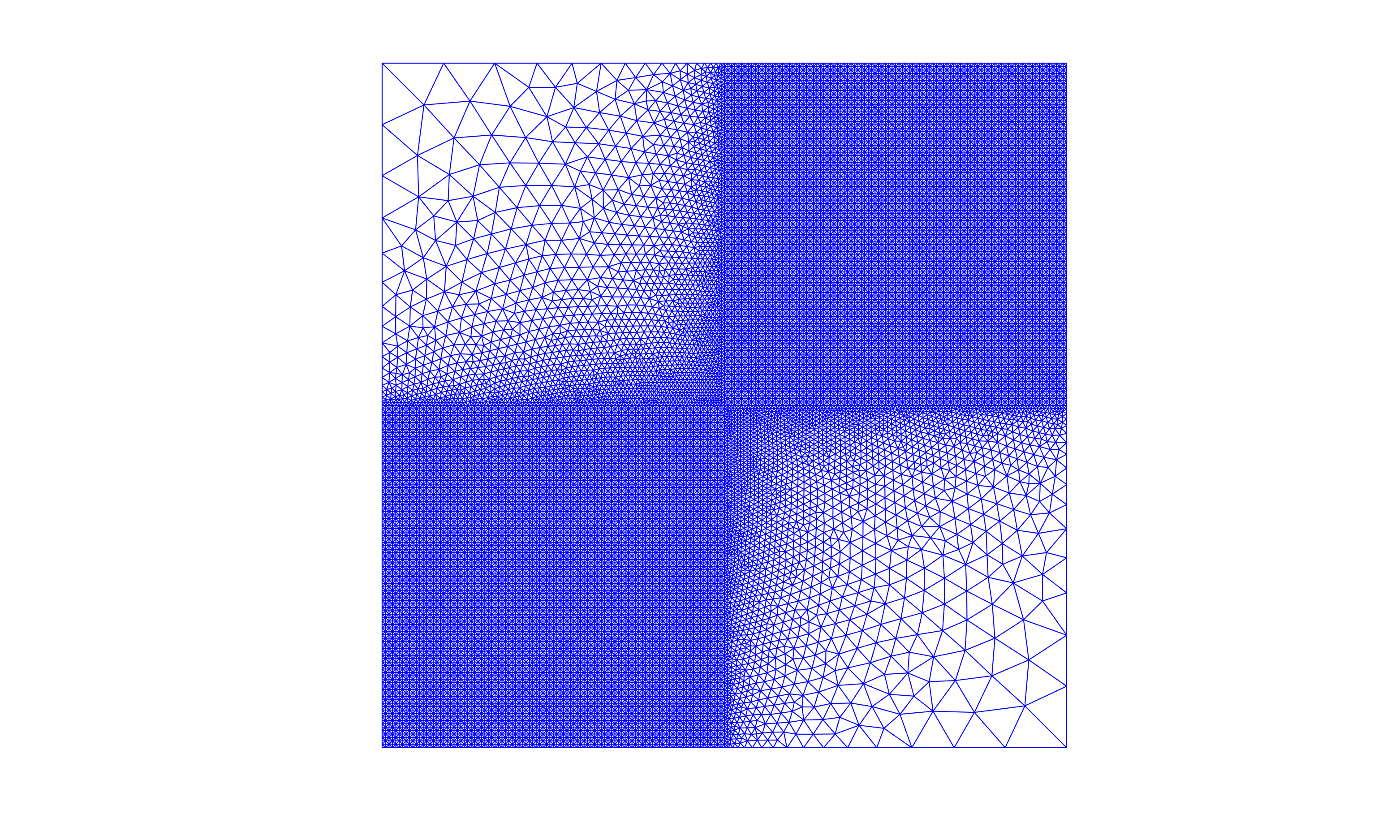}
\caption{Refinement of the mesh}\label{Fig_Press_M}
\end{subfigure}
\caption{ Description and discretisation of the domain subjected  to compression.}\label{Fig_TP}
\end{figure}

Operation of a loaded structures related to arising  of cracks may be highly affected  in situations where shear becomes dominant.
Then it is questionable, whether it may cause cracks inside the material. 
Such situation appears  for example  in compressed  structural elements.
Therefore, properties of the proposed approach are tested under such conditions in the present computational example. 
The scheme is shown in Fig.~\ref{Fig_TP}.
No  inhomogeneity is considered here, only an initial  crack  is modelled by prescribing the values of phase-field damage parameter to zero at the slit.
The figure also contains a mesh used in calculations with refinements at expected crack domains set according to a preliminary calculation with a coarser mesh.
The mesh refinements here contain elements of the minimal size equal to $0.35$ mm.

The parameters of the material stiffness are: $\Kp=22.0$ GPa, $\mu=12.3$ GPa. 
The fracture energy in the domain is $\Gc[I]=0.1\ \text{Jm}^{-2}$, for the shear mode three different values are considered: $\Gc[II]=\{1,25,100\}\cdot\Gc[I]$, to compare the influence of shear on the cracking process in the compressed domain.
The length scale parameter of the phase-field model is  $\epsilon= 1$ mm.

The structural element is loaded by displacement loading $g(t)=v_0t$, at the velocity $v_0=10\ \text{mm\,s}^{-1}$. 
This load is applied incrementally in  time steps refined to $0.01$ ms.

The analysis of the phase-field fracture is performed with the degradation function as before $\Fun{\Phi}(\alpha) = \frac{\alpha^2}{\alpha^2+\beta(1-\alpha)}+10^{-6}$, though using the value of the parameter $\beta=10$. 
Satisfaction of the stress condition~\eqref{Eq_StressCrit_Amod} can be checked in particular by substituting the values of the parameters which provide in the present case the relations depending on the ratio  $\Gc[II]/\Gc[I]$ as follows:
\begin{equation}\label{Eq_StressCritVal}
\begin{aligned}
\Gc[II]/\Gc[I]&=1 & &:& \left(\tr^{+}\! \scri\right)^2+3.33\left|\dev \scri\right|^2&=0.624\text{ MPa}^2\tc \\
\Gc[II]/\Gc[I]&=25 & &:& \left(\tr^{+}\! \scri\right)^2+0.084\left|\dev \scri\right|^2&=0.396\text{ MPa}^2\tc \\
\Gc[II]/\Gc[I]&=100 & &:& \left(\tr^{+}\! \scri\right)^2+0.021\left|\dev \scri\right|^2&=0.391\text{ MPa}^2\tb \\
\end{aligned}
\end{equation}
It e.g.\ means that in the last option,  if a region appears  with positive opening stress, the critical value of  $\tr\sigma$ is given by the value $\tr\scri=0.625$ MPa under small load condition which eliminates influence of shear.
On the other hand, the first option leads under compression to  damage  in shear mode with  $|\dev\scri|=0.433$ MPa, the second one to $|\dev\scri|=2.17$ MPa, and the last one to $|\dev\scri|=4.32 $ MPa.

\begin{figure}[!ht]
\centering
\includegraphics[scale=0.3,trim=5 5 40 50,clip]{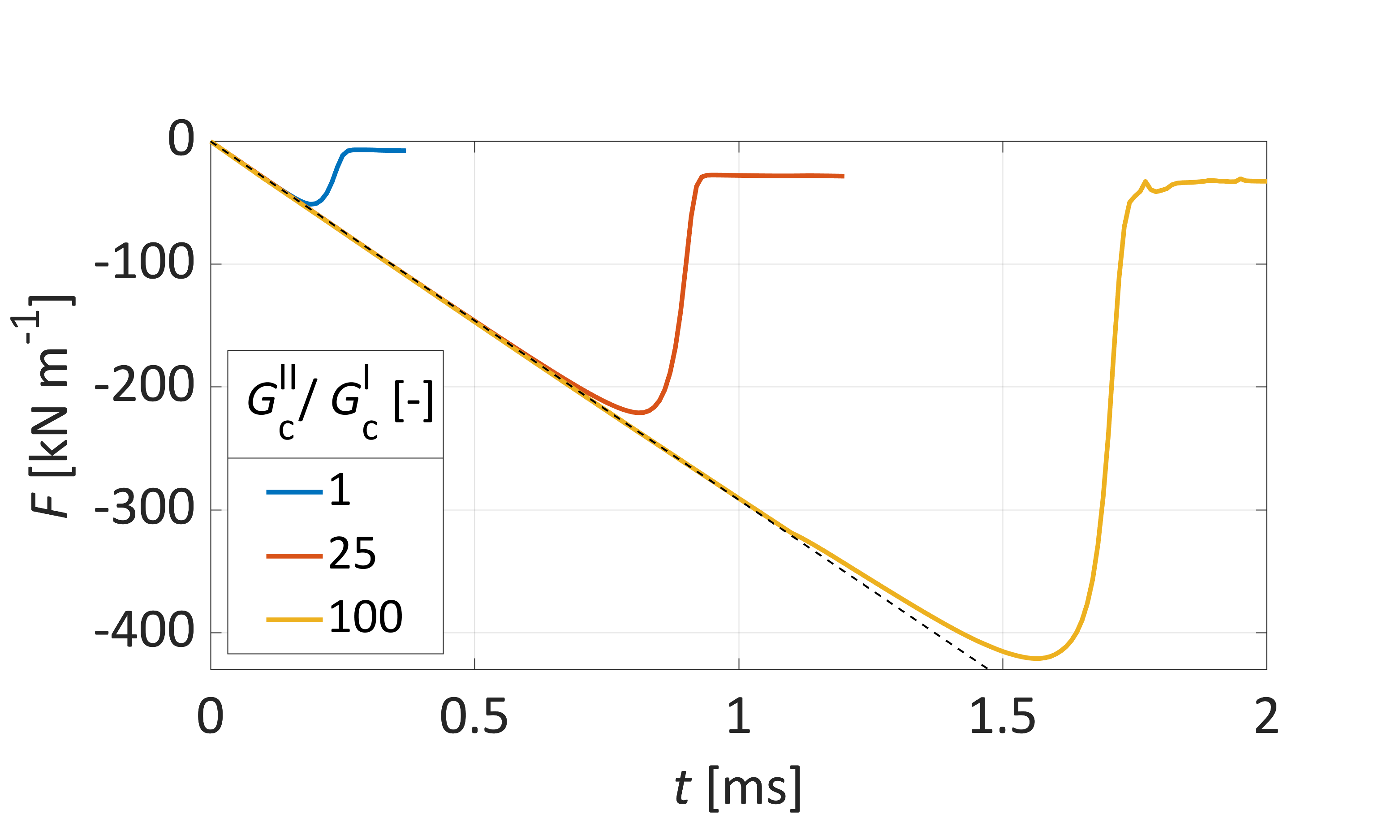}
\caption{Total force applied at the top face during loading for the compressed domain.
Dashed line pertains to  the elastic response}\label{Fig_TP_F}
\end{figure}

The force response of the  structural element  in terms of  the vertical compressive force $F$ applied at the top face of the matrix domain is shown in~Fig.~\ref{Fig_TP_F}.
The dependence is related to the time $t$ and the graphs include all three options of the fracture energy $\Gc[II]$.
The curves show that the total force increases linearly until the first changes in the material appear.
For identifying such an instant, theoretical elastic response line is included, it reveals the deviation of the actual force evolution from the straight line .
The parameters are set so that for the smallest value of $\Gc[II]$ the crack propagates very early and as will be seen below preferably in the shear mode.
It will also be confirmed below in direct plot containing all developed  cracks.
In all the cases,  total rupture of the compressed domain is observed in the final jump of the force. 
Principally, it corresponds to  a shear  crack propagation across the whole domain.

\begin{figure}[!ht]
\centering
\setlength{\unitlength}{\textwidth}
\begin{subfigure}{0.99\textwidth}
\begin{picture}(0.97,0.22)
\PA{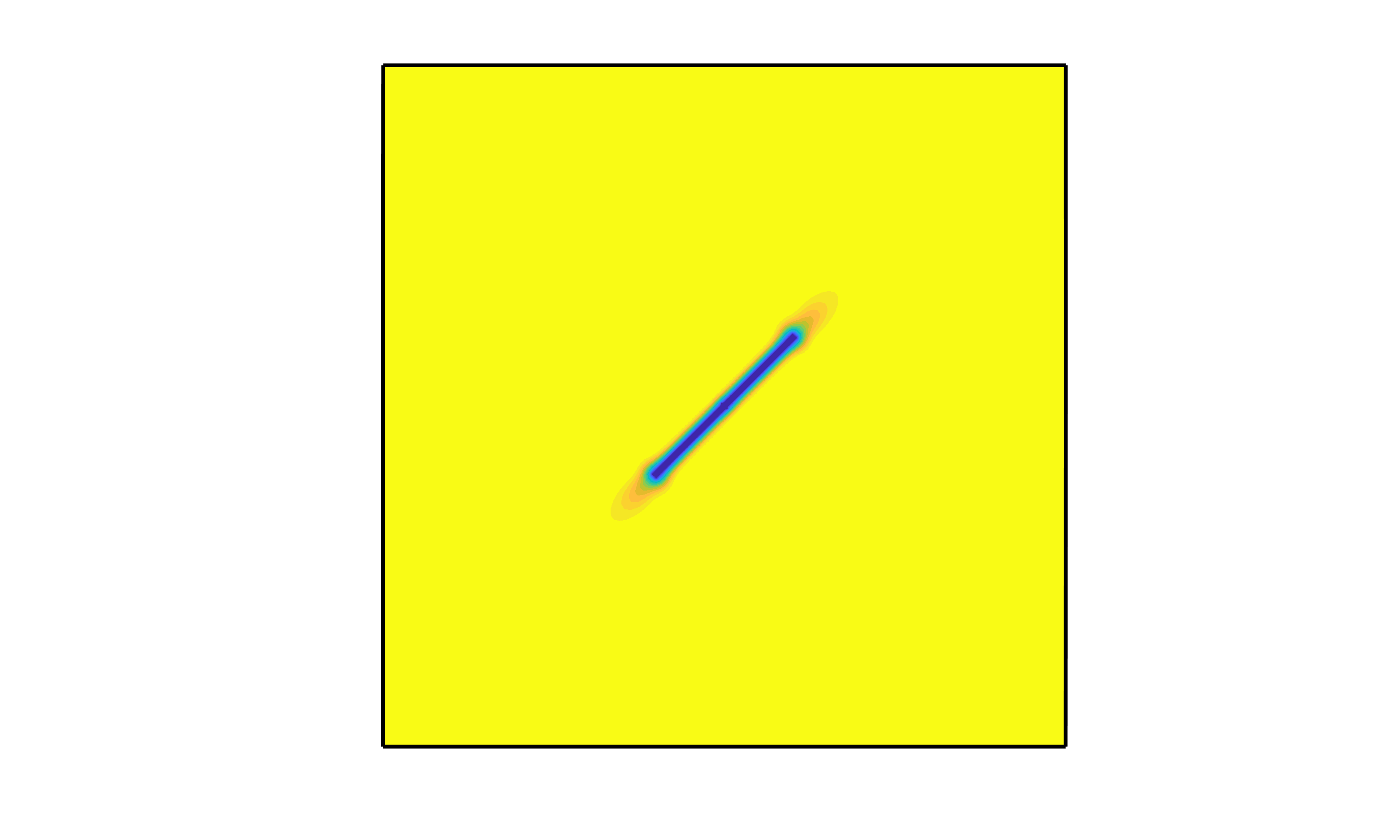}{$t=0.15$ms}
\PBF{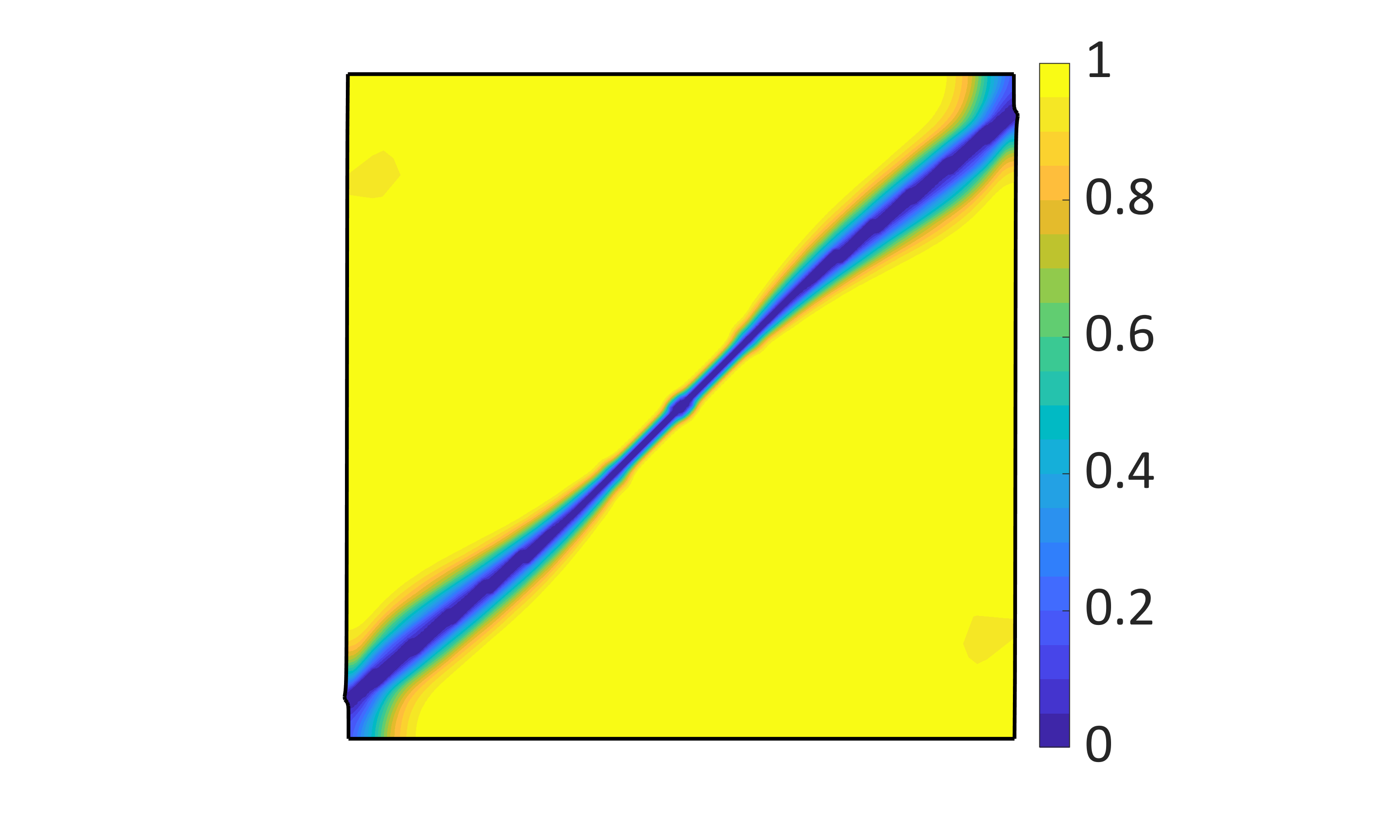}{$t=0.3$ms}
\Pvar{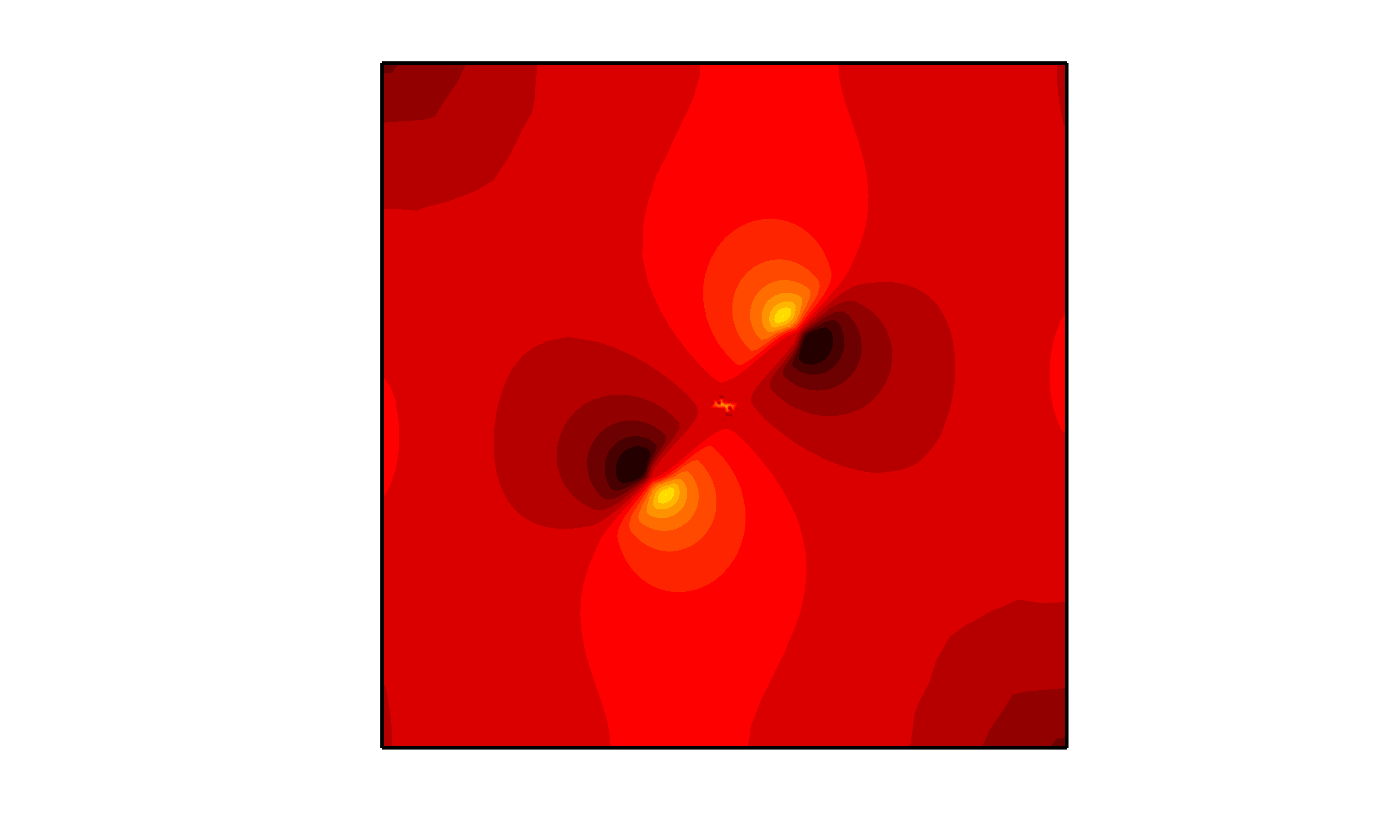}{Stress trace}{530}
\CB[720]{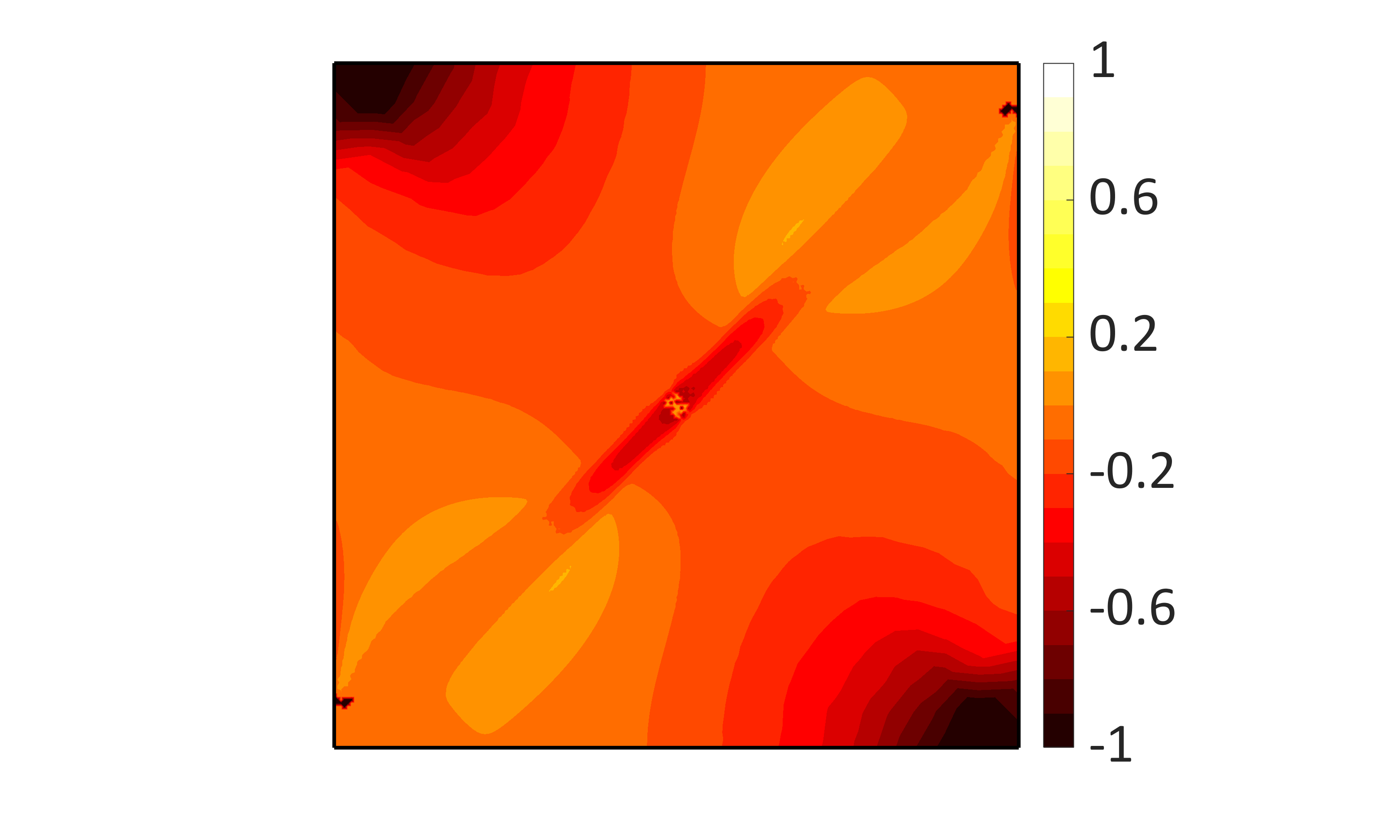}
\PE{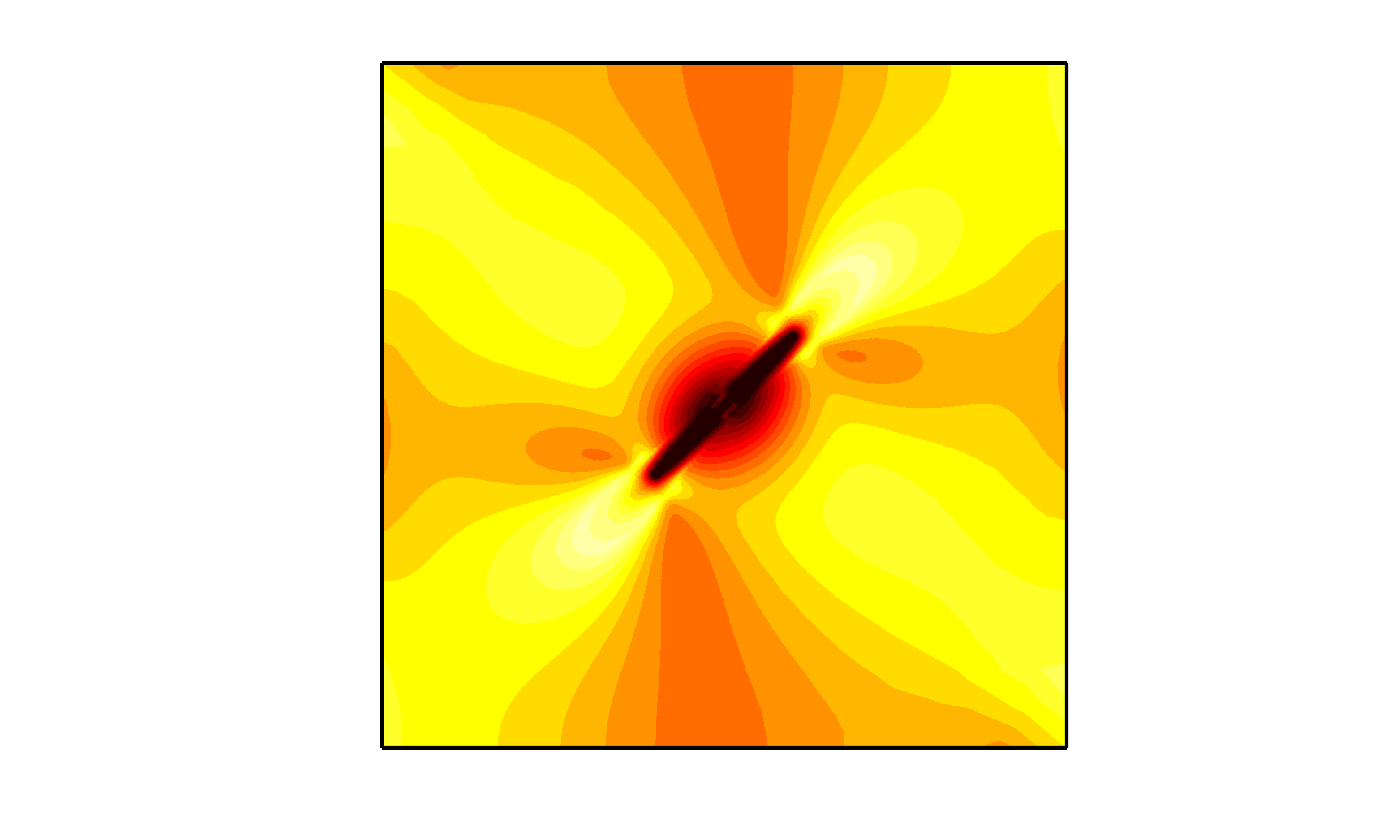}{Deviatoric stress}
\CB[955]{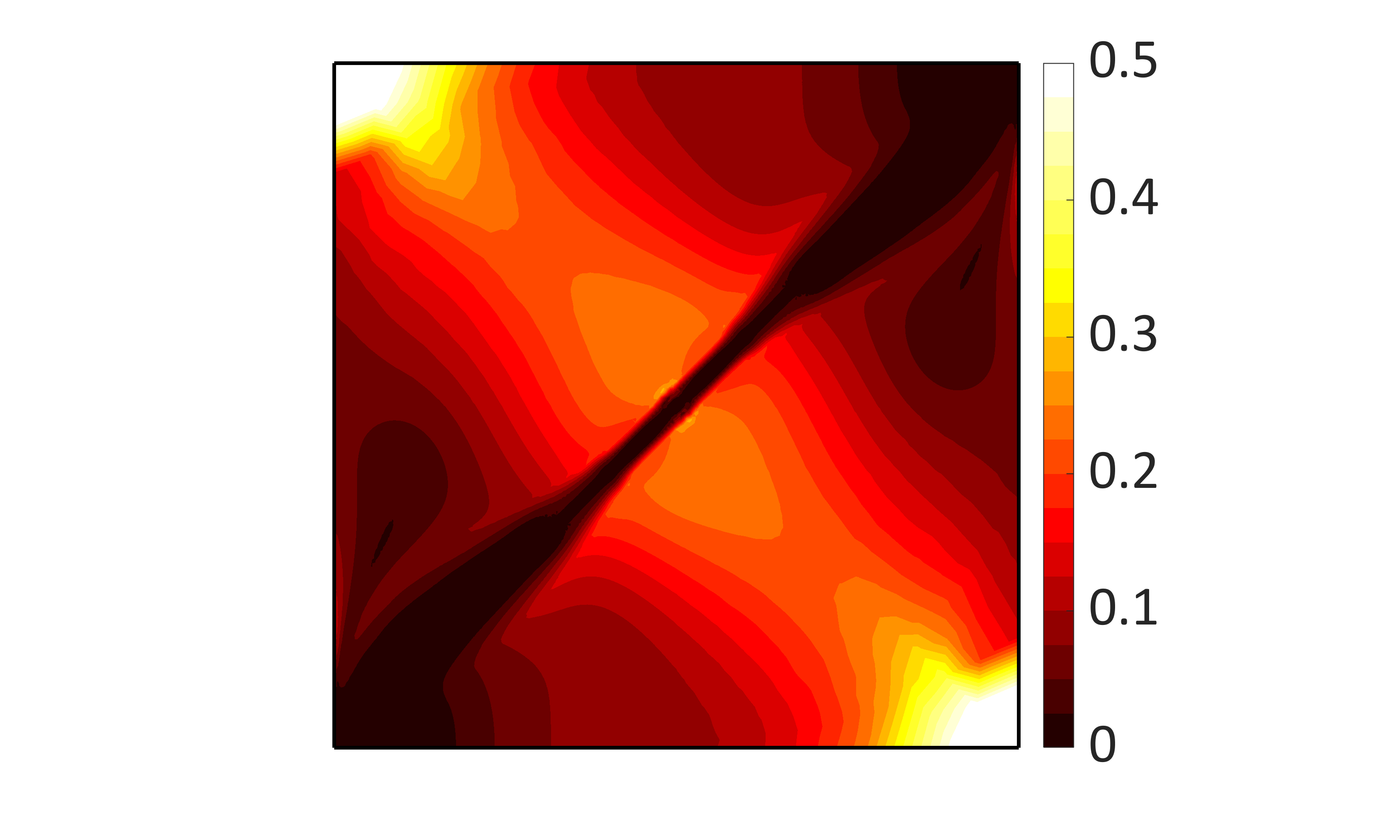}
\end{picture}
\caption{$\Gc[II]/\Gc[I]=1$}\label{Fig_TP_1_Grat1}
\end{subfigure}
\begin{subfigure}{0.99\textwidth}
\begin{picture}(0.97,0.22)
\PA{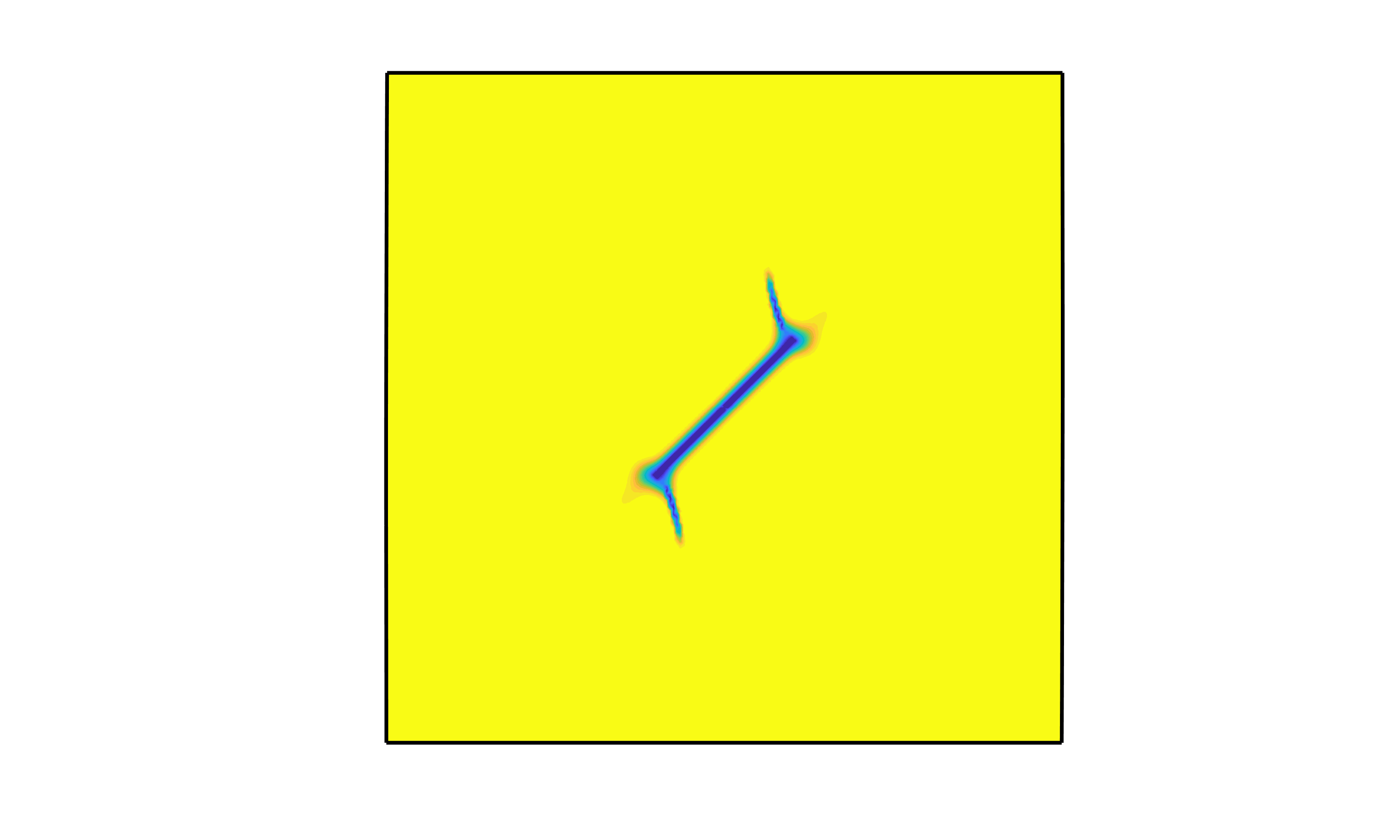}{$t=0.7$ms}
\PBF{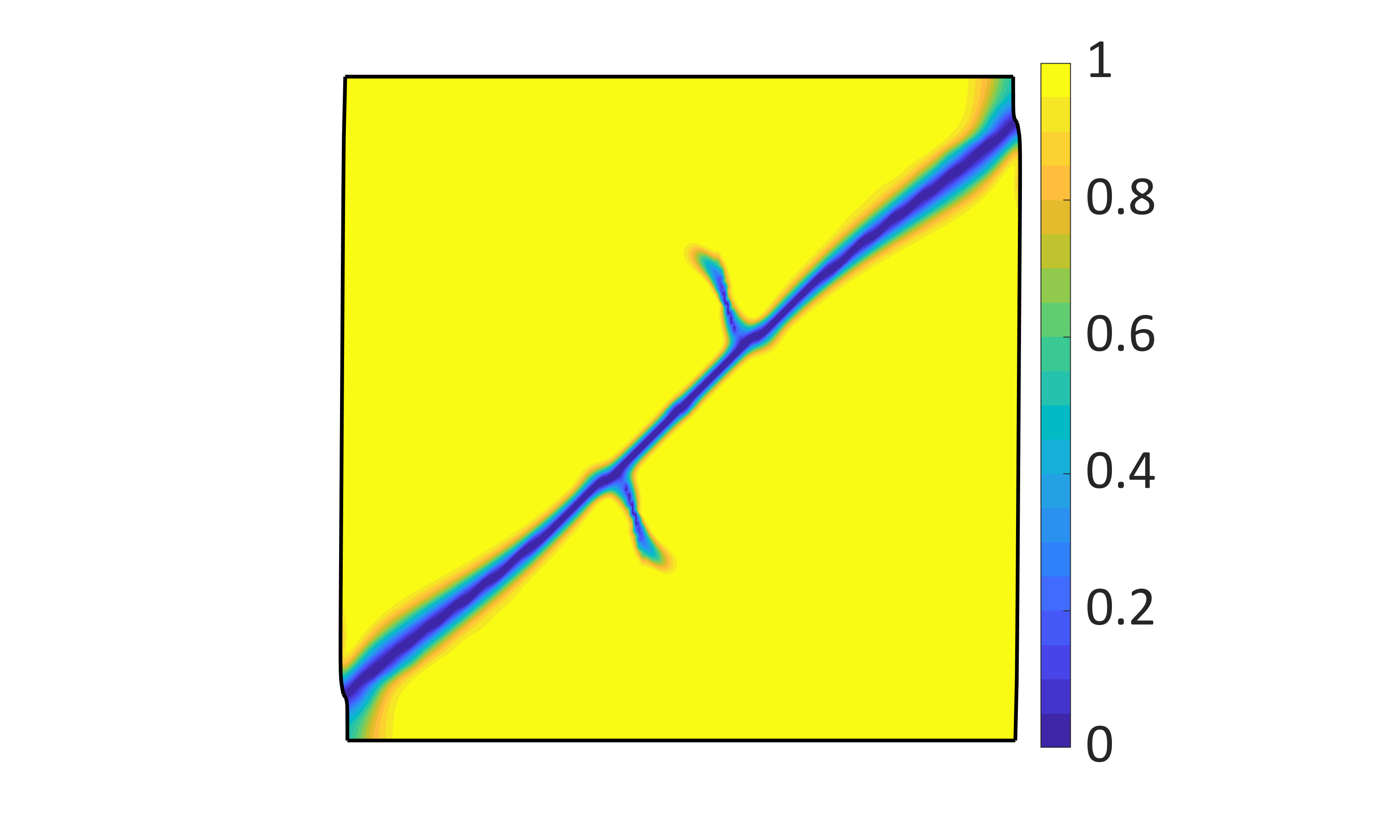}{$t=0.93$ms}
\Pvar{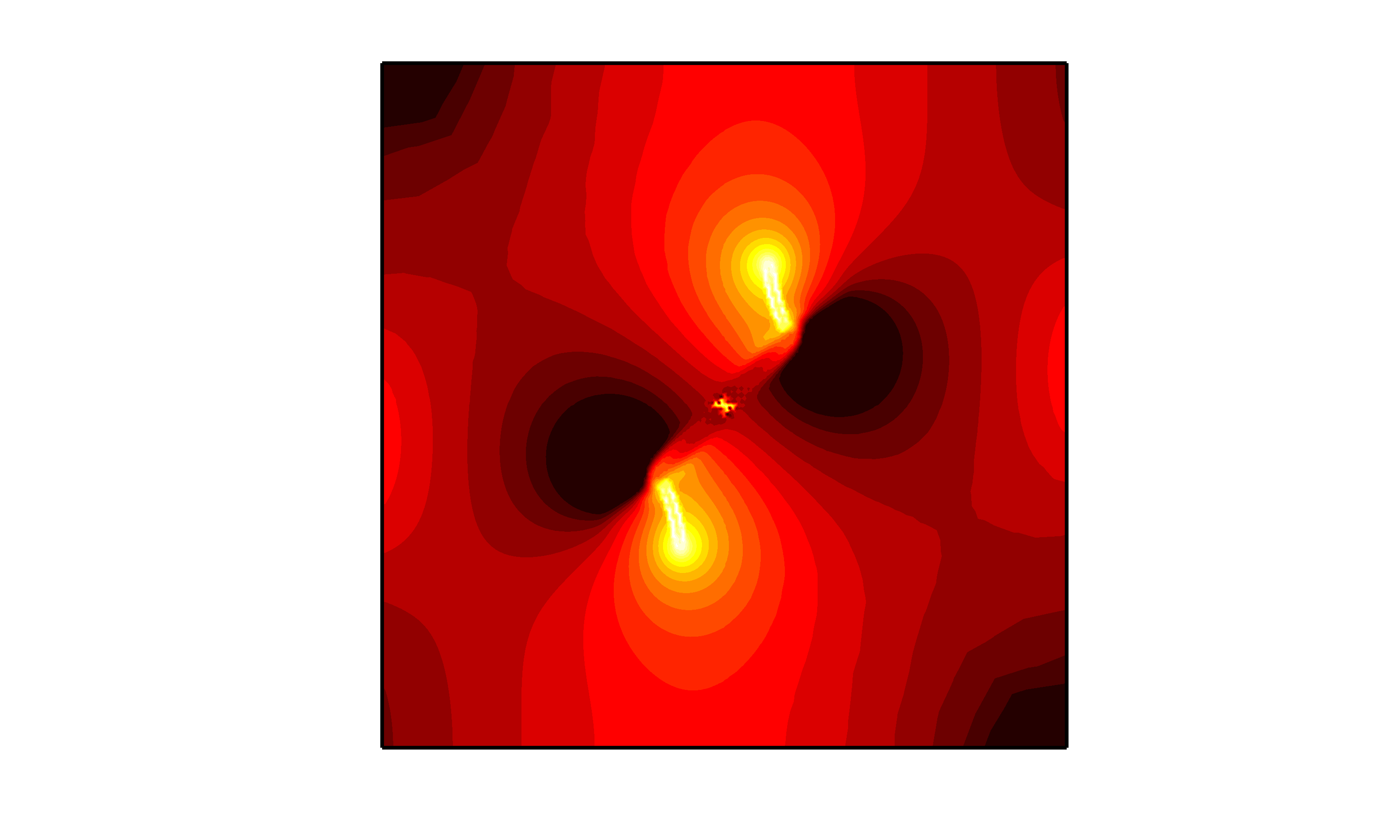}{Stress trace}{530}
\CB[720]{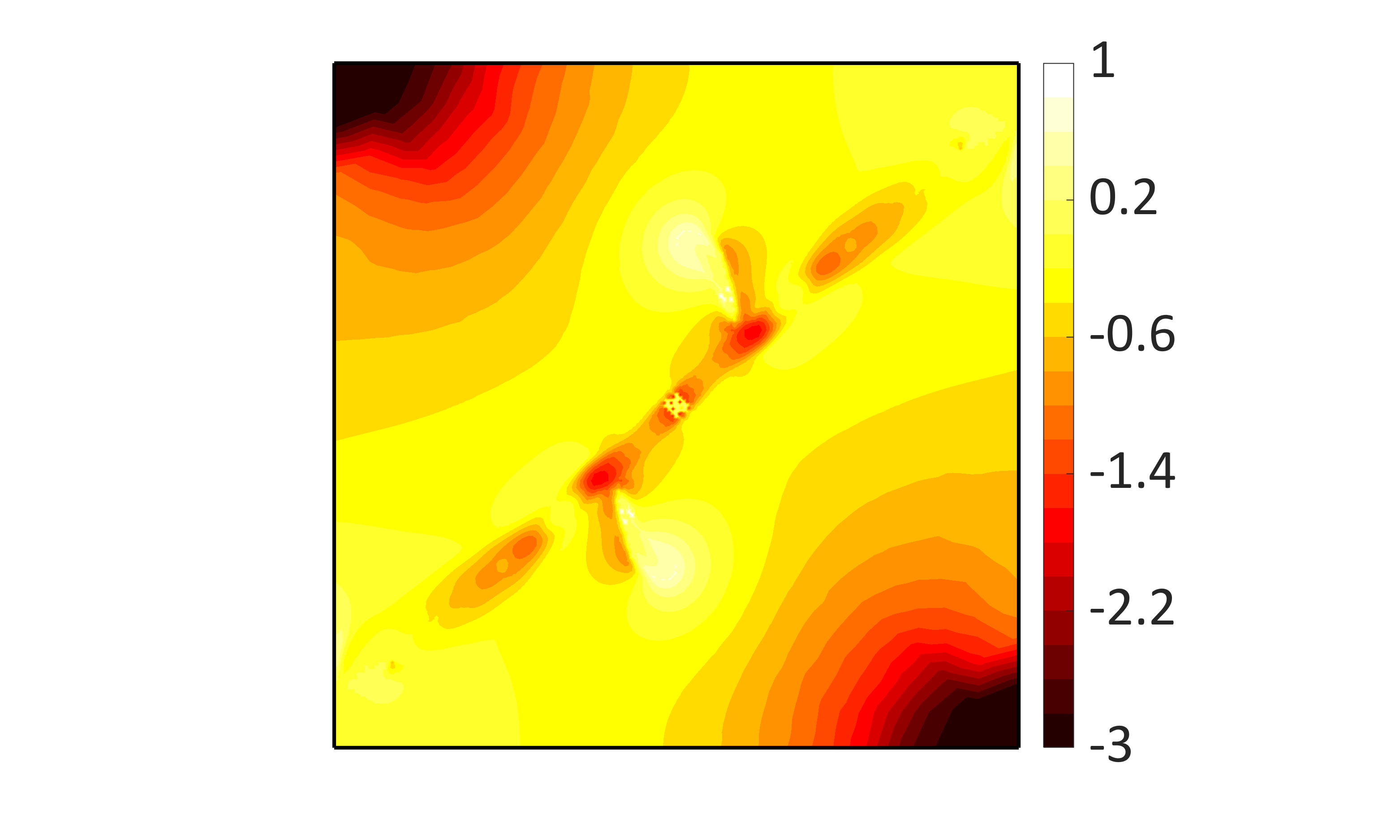}
\PE{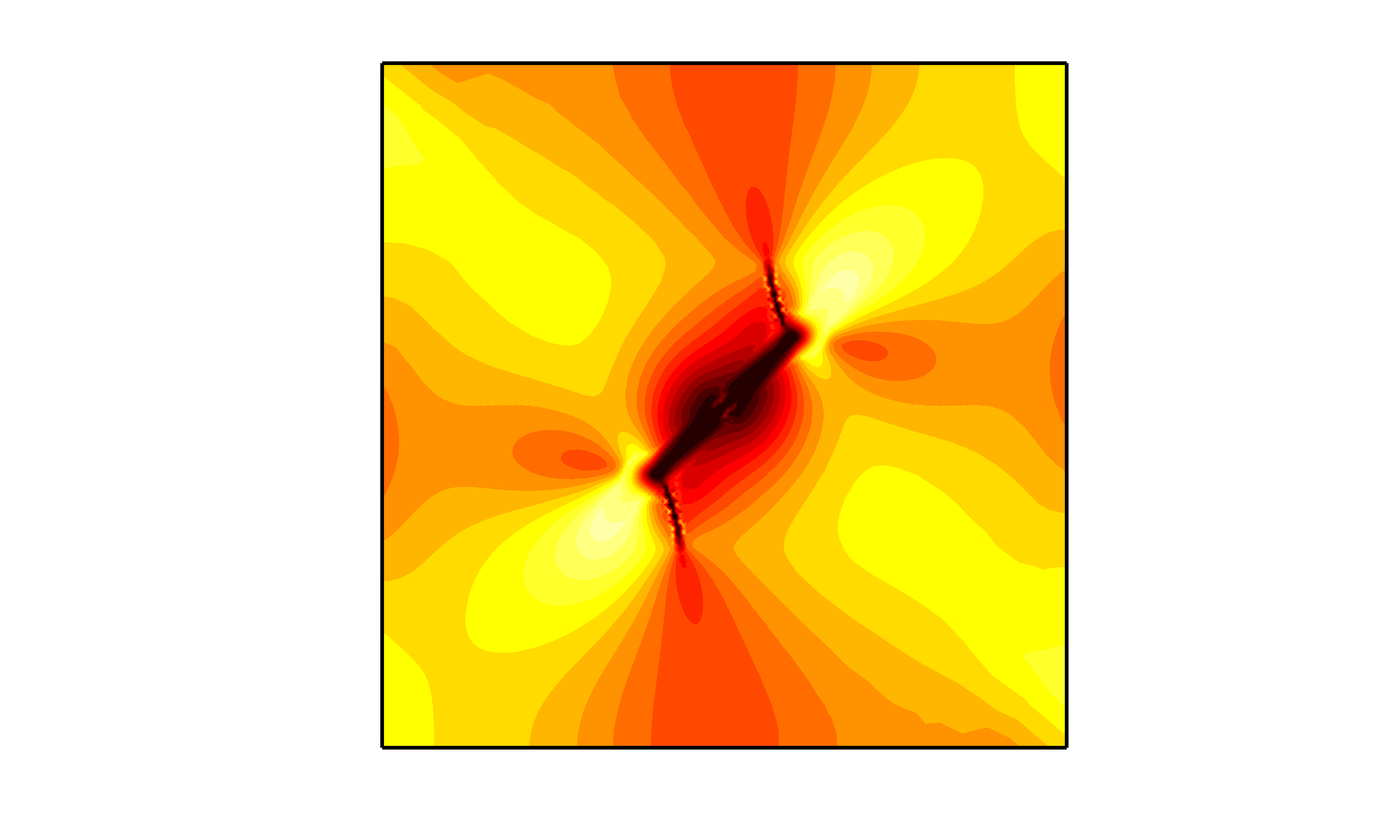}{Deviatoric stress}
\CB[955]{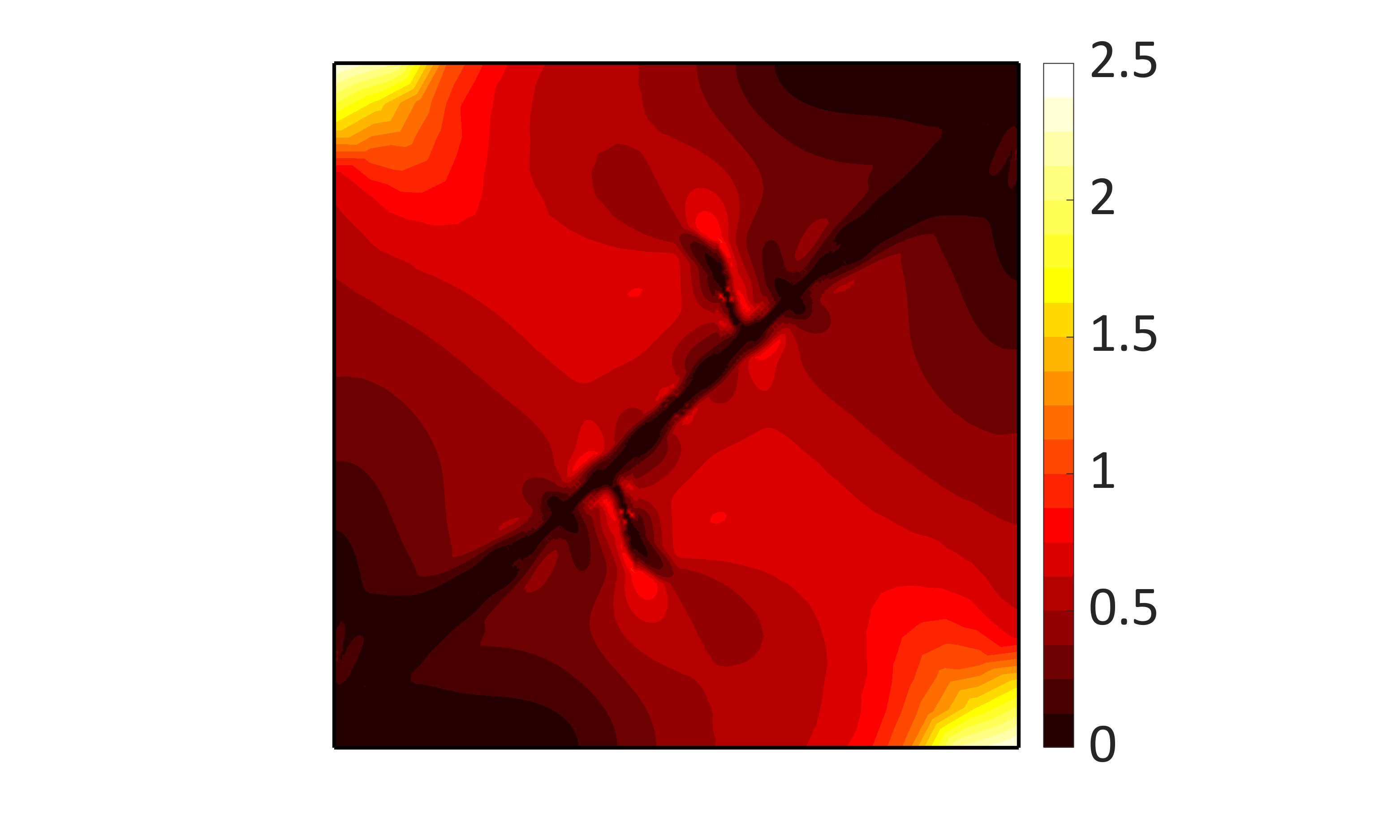}
\end{picture}
\caption{$\Gc[II]/\Gc[I]=25$}\label{Fig_TP_1_Grat25}
\end{subfigure}
\begin{subfigure}{0.99\textwidth}
\begin{picture}(0.97,0.22)
\PA{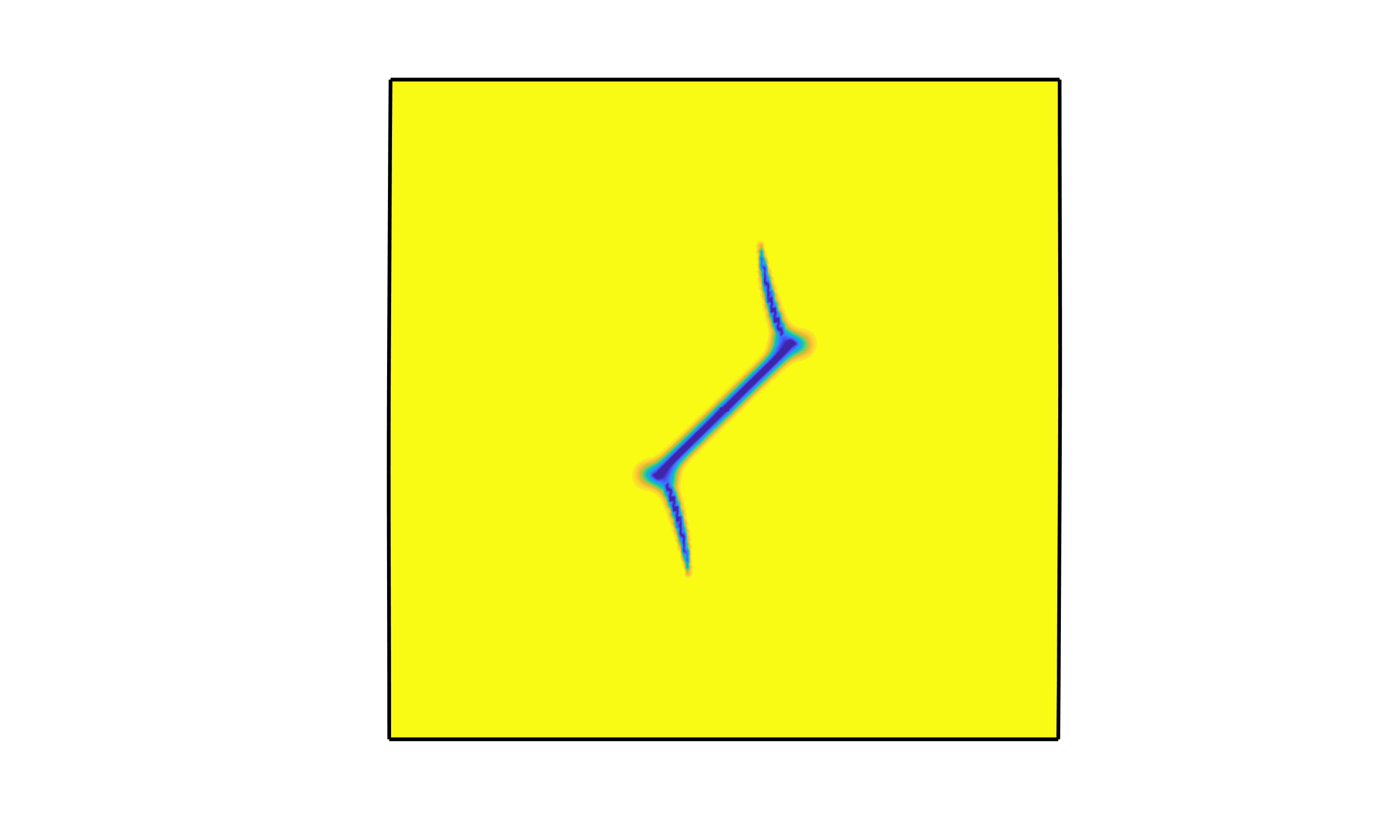}{$t=1.2$ms}
\PBF{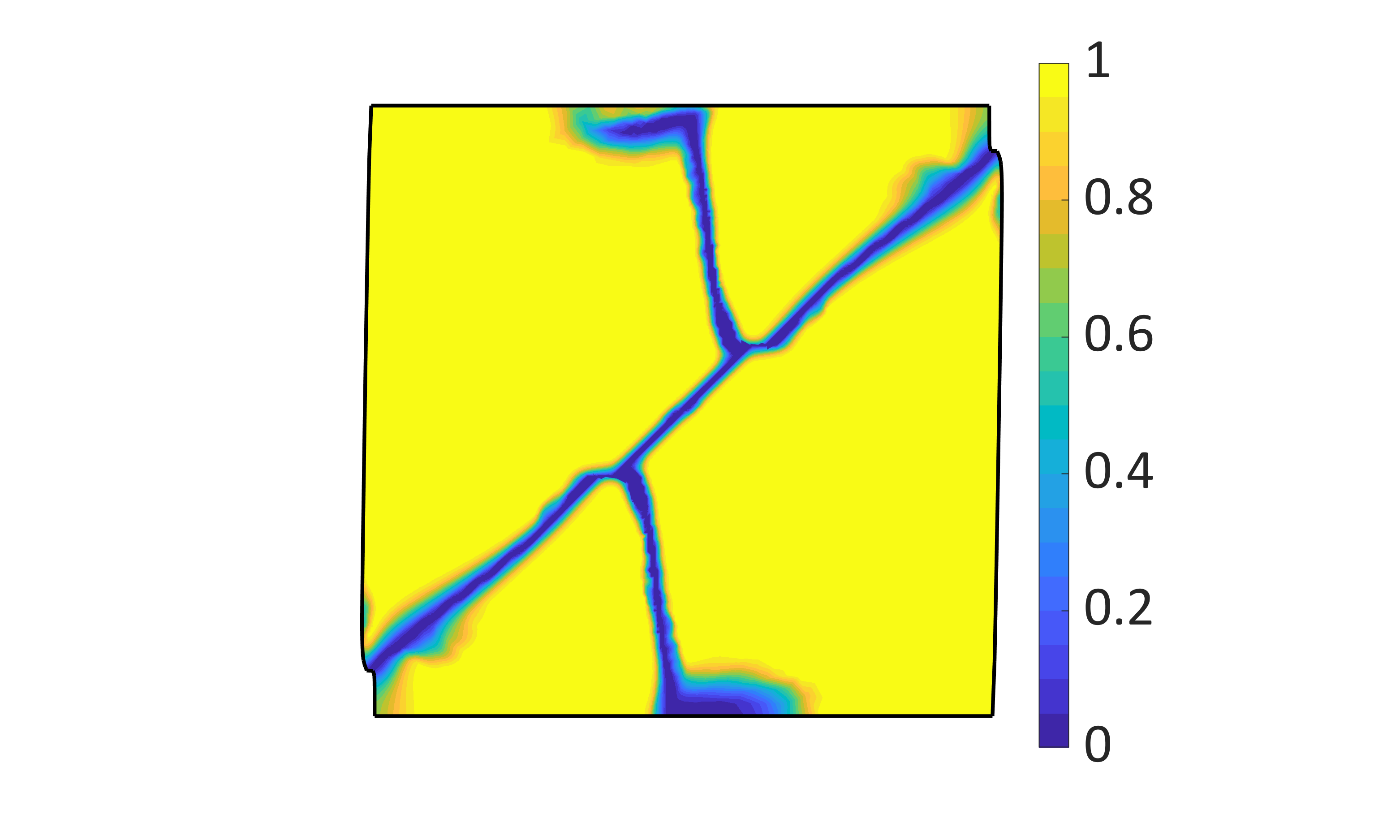}{$t=1.74$ms}
\Pvar{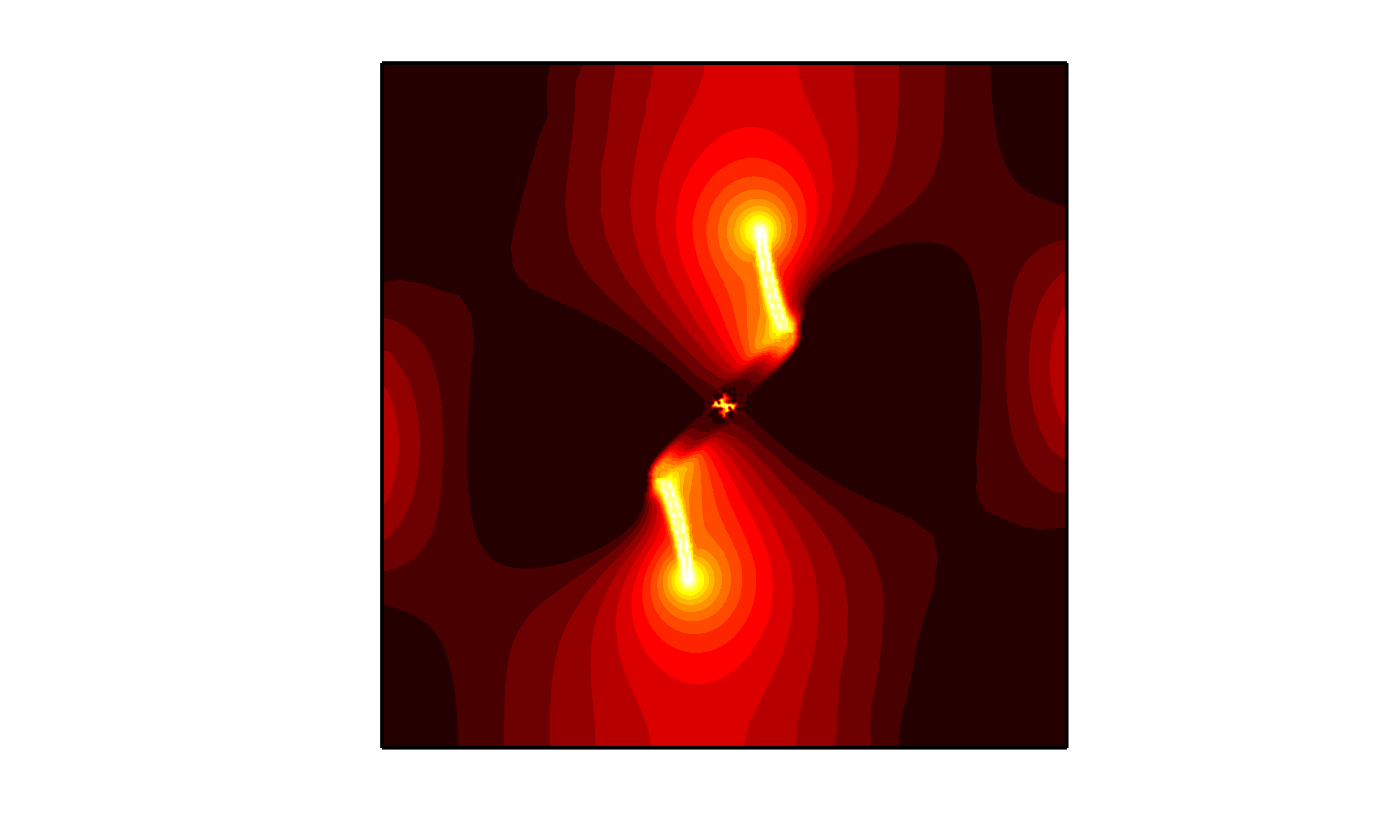}{Stress trace}{530}
\CB[720]{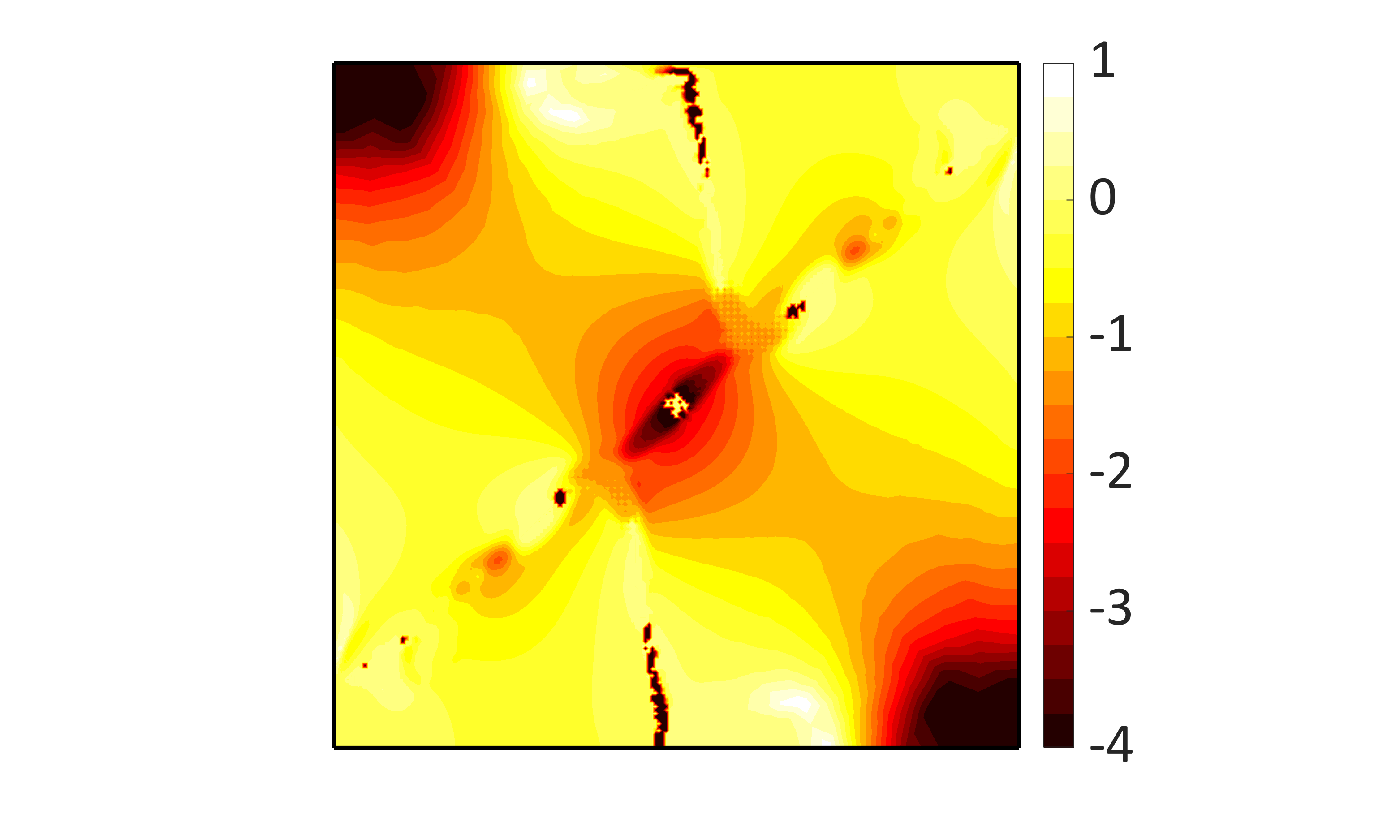}
\PE{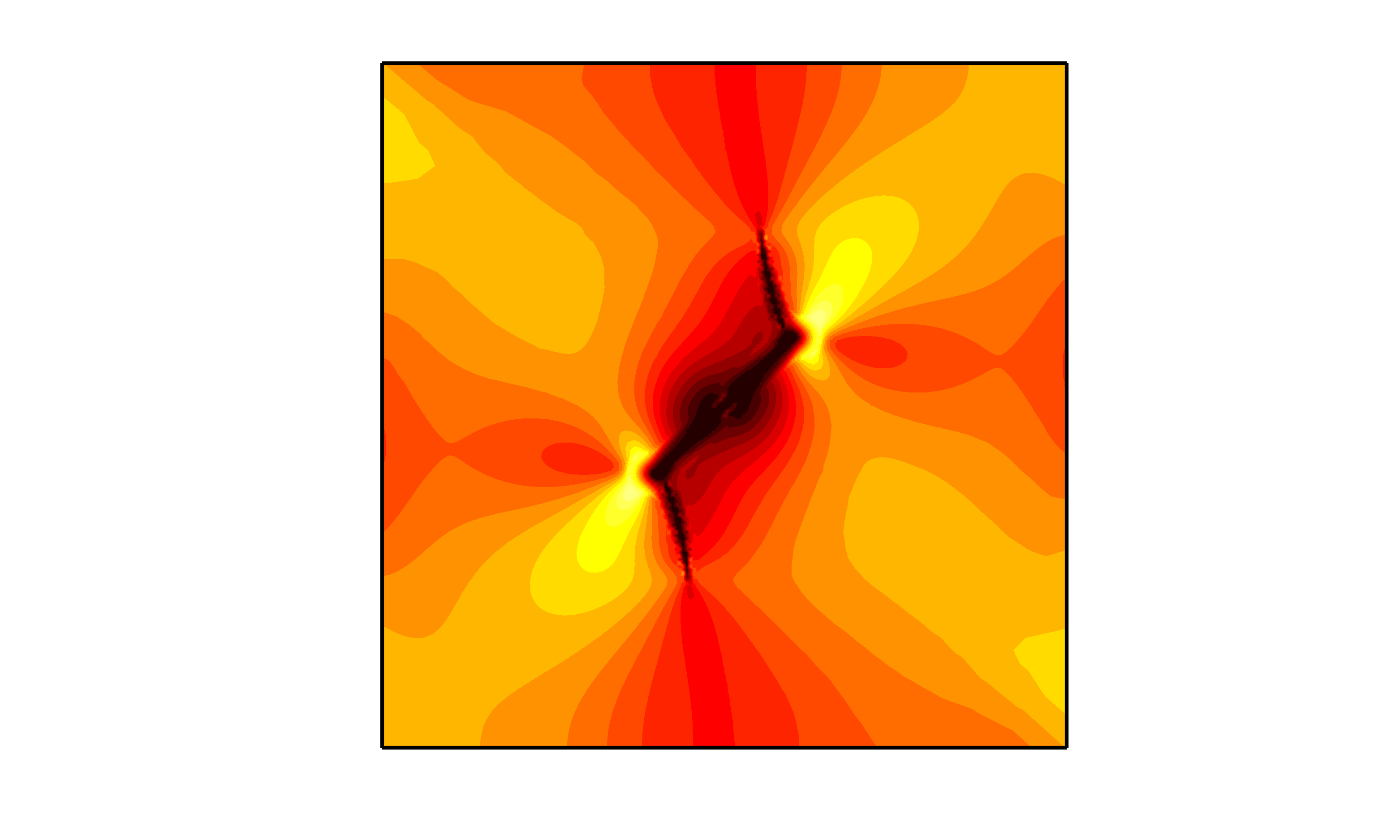}{Deviatoric stress}
\CB[955]{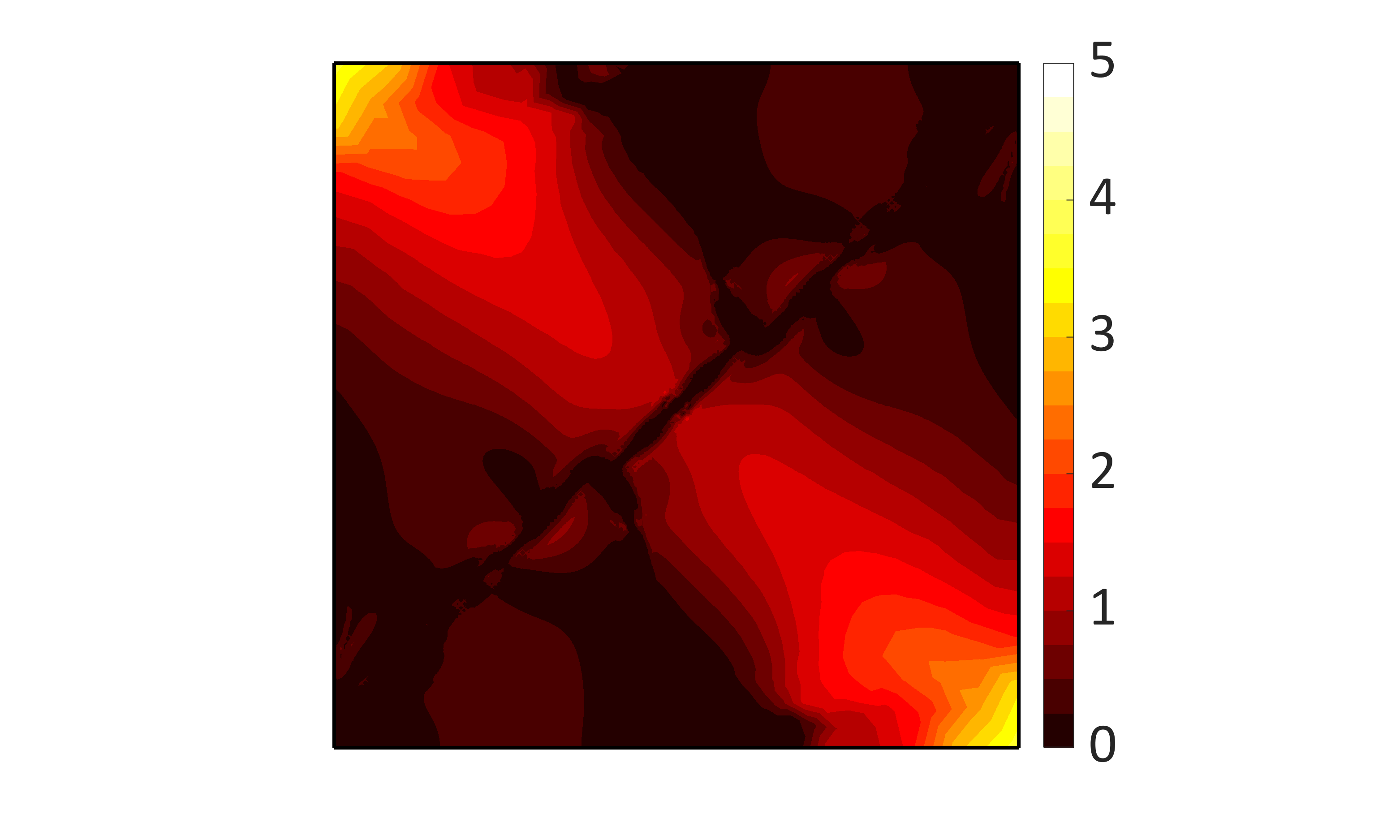}
\end{picture}
\caption{$\Gc[II]/\Gc[I]=100$}\label{Fig_TP_1_Grat100}
\end{subfigure}
\caption{Distribution of the phase-field variable  documenting crack propagation (left) and stress values [MPa] (right) at selected time instants for the compressed block. 
Stresses belong to the first used instant  and the deformation in the crack plot is $100\times$ magnified.}\label{Fig_TP_DS}
\end{figure}

The detailed study of the phase-field fracture evolution shows some additional details to the previous explanations, namely, the form of the  cracks  which appear during increasing loading  and the stress distribution near  the crack tips.
Drawings in Fig.~\ref{Fig_TP_DS} present those details.
As it was considered in this model, the rupture in compression, if a sufficiently large load is applied, is terminated by rupture in shear.
Nevertheless, appropriate adjustment of the ratio $\Gc[II]/\Gc[I]$  may postpone such a response in the structure and take into the consideration  tensile forces which appear  close to the given  crack tip.
This phenomenon has been actually captured by the model  for the two larger ratios.
As can be seen in damage plots, wing cracks appear at the initial crack tips where the opening stress reaches the aforementioned magnitude observable in  stress trace diagrams.
Anyhow, the effect of the shear stress appears in the shear domain located along the diagonal of the block and when it reaches the critical value expressed in terms of deviatoric stress, there is sufficient energy to propagate the crack in the shear mode.
It was observed in all three cases.
In the case with $\Gc[II]=\Gc[I]$, however,   it is the only  mode of crack developed as the region in tensile state does not bear sufficient energy related to small stress trace values in the zone close to the crack tip.

Finally, it is worth mentioning that for computational purposes with elastic solution, additional horizontal constraints were put at two corners as can be guessed form the stress plots and from the final deviation of the wing cracks.

%%%%%%%%%%%%%%%%%%%%%%%%%%%%
\subsection{Combined loading imposed on a domain with grooves}\label{Sec_NoMoB}
%%%%%%%%%%%%%%%%%%%%%%%%%%%%

The capability of the proposed model for mixed-mode fracture can be verified by a structural element with an  inhomogeneity  and two initial grooves which was motivated by the model analysed in~\cite{feng22A1}. 
Nevertheless, within the present model it includes also an  inhomogeneity.
The calculations are focused on influence of  the  inhomogeneity  and its material and interface properties and complement results obtained by the author in~\cite{vodicka22A1}.

\begin{figure}[!ht]
\centering
\begin{subfigure}{0.37\textwidth}
\centering
\includegraphics[scale=1]{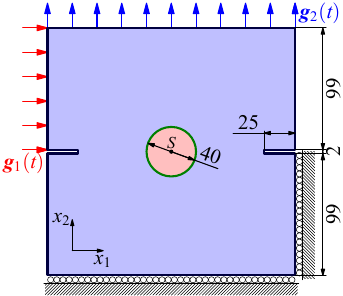}
\caption{A domain subjected to combined loading}\label{Fig_NoMoB_D}
\end{subfigure}
\begin{subfigure}{0.24\textwidth}
\centering
\includegraphics[scale=0.85]{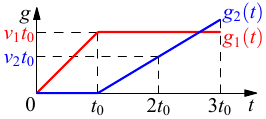}
\caption{Imposed loading}\label{Fig_NoMoB_Load}
\end{subfigure}
\begin{subfigure}{0.37\textwidth}
\centering
\includegraphics[scale=0.32,trim=170 30 150 20,clip]{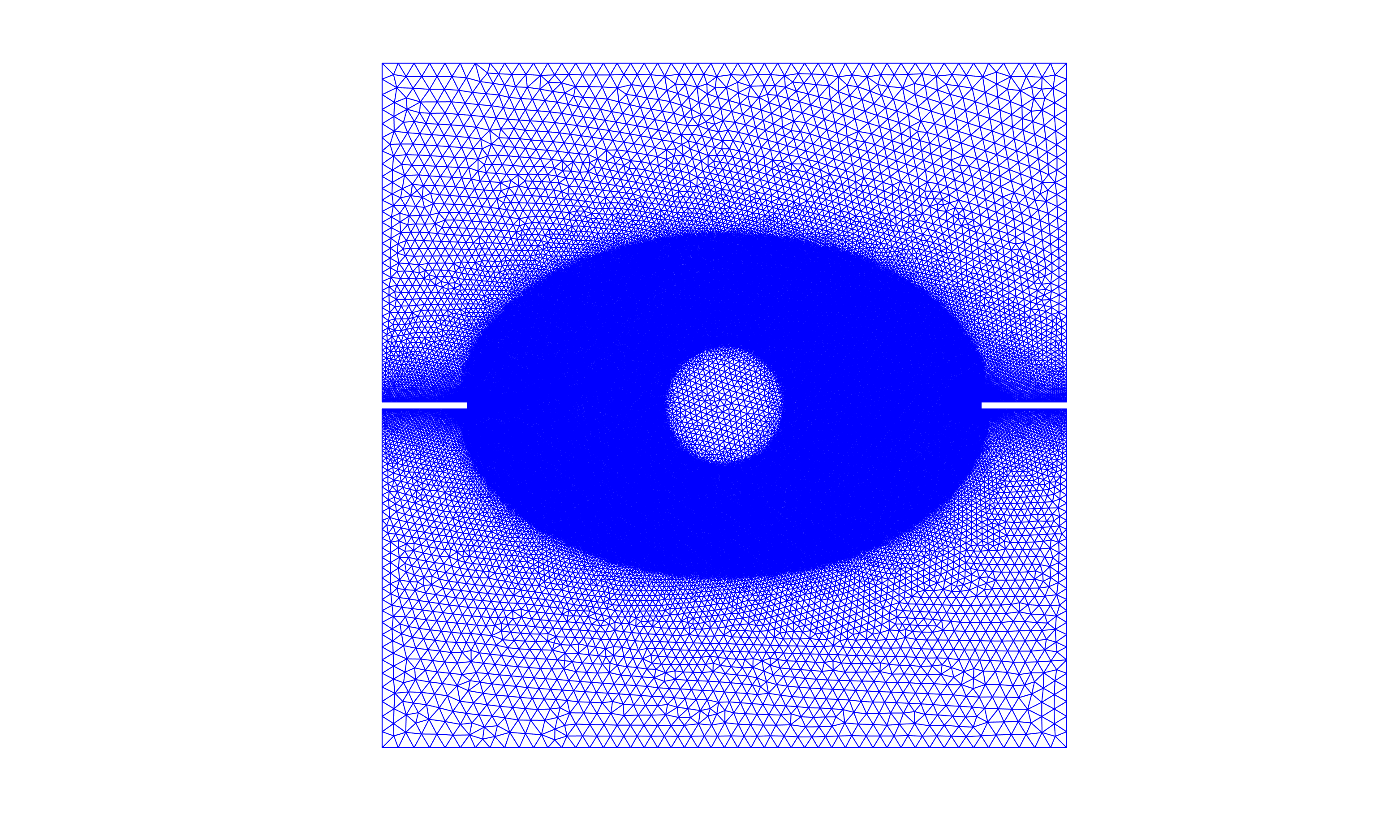}
\caption{Refinement of the mesh}\label{Fig_NoMoB_MSH}
\end{subfigure}
\caption{ Description and discretisation of the domain under  a combined loading.}\label{Fig_NoMoB}
\end{figure}

The scheme of the problem is shown in Fig.~\ref{Fig_NoMoB}.
The initial  internal cracks  are represented by two grooves of finite width which prevent contact of their faces not considered in the proposed computational model.
The figure also contains a mesh used in calculations with refinements in regions where cracks are expected.
The mesh refinements contain elements of the minimal size equal to $0.1$ mm close to the groove tips.

The domain is loaded by displacement loading in a way shown in Fig.~\ref{Fig_NoMoB_Load}: first, the horizontal load $g_1$ is applied which causes shear  between the grooves, and afterwards, with a constant horizontal load, the vertical one $g_2$ is gradually increased to define a kind of mixed-load.
The loading parameters are:  $v_1=v_2=1\ \text{mm\,s}^{-1}$, $t_0=0.1$ s. 
This load is applied incrementally in  time steps refined to $0.5$ ms.

The elastic parameters of the materials are: the  bulk modulus $\Kp=51.8$ GPa, the shear modulus $\mu=29.0$ GPa for the  inhomogeneity, and $\Kp=3.1$ GPa, $\mu=1.0$ GPa for the matrix. 
Alternatively, the  inhomogeneity  is considered of the same material as the matrix.
The interface is characterised by $\kappa = 10\left(\begin{smallmatrix}1 & 0\\0 & 0.5\end{smallmatrix}\right)\text{ PPam}^{-1}$, $\kG=100\ \text{PPam}^{-1}$. 

The fracture energy in the matrix domain is $\Gc[I]=10\ \text{Jm}^{-2}$, for the shear mode the  value $\Gc[II]=10\Gc[I]$ is considered to suppress effect of shear.
The length scale parameter of the phase-field model is  $\epsilon= 1$ mm.
The degradation function $\Phi$ and its parameter are the same as in previous example, Sec.~\ref{Sec_TP}.
Different material parameters make the stress condition~\eqref{Eq_StressCrit_Amod} to be expressed by the relation $\left(\tr^{+}\! \scri\right)^2+0.45\left|\dev \scri\right|^2=7.05\text{ MPa}^2$ ,
which  means that the  value $\tr\sigma_{\text{crit}}=2.66$ MPa obtained for vanishing shear stress presents an upper bound for opening stress distribution.

The interface degradation function is taken from Sec.~\ref{Sec_MaFi3}. 
For the present value of interface fracture energy $\Gc[iI] = \{0.01,100\}\ \text{Jm}^{-2}$, two options are obtained which provide (the shear mode interface fracture energy is again set to a high value) critical normal stress $\pcri = \{0.138,13.8\}$ MPa, respectively. 
Comparing to the stress condition  inside the solid, the first option should provide initiation of the interface crack, while the other one should not, and if  damage occurs near the inhomogeneity then only inside the matrix.

\begin{figure}[!ht]
\centering
\begin{subfigure}{0.49\textwidth}
\centering
\includegraphics[scale=0.3,trim=5 5 40 50,clip]{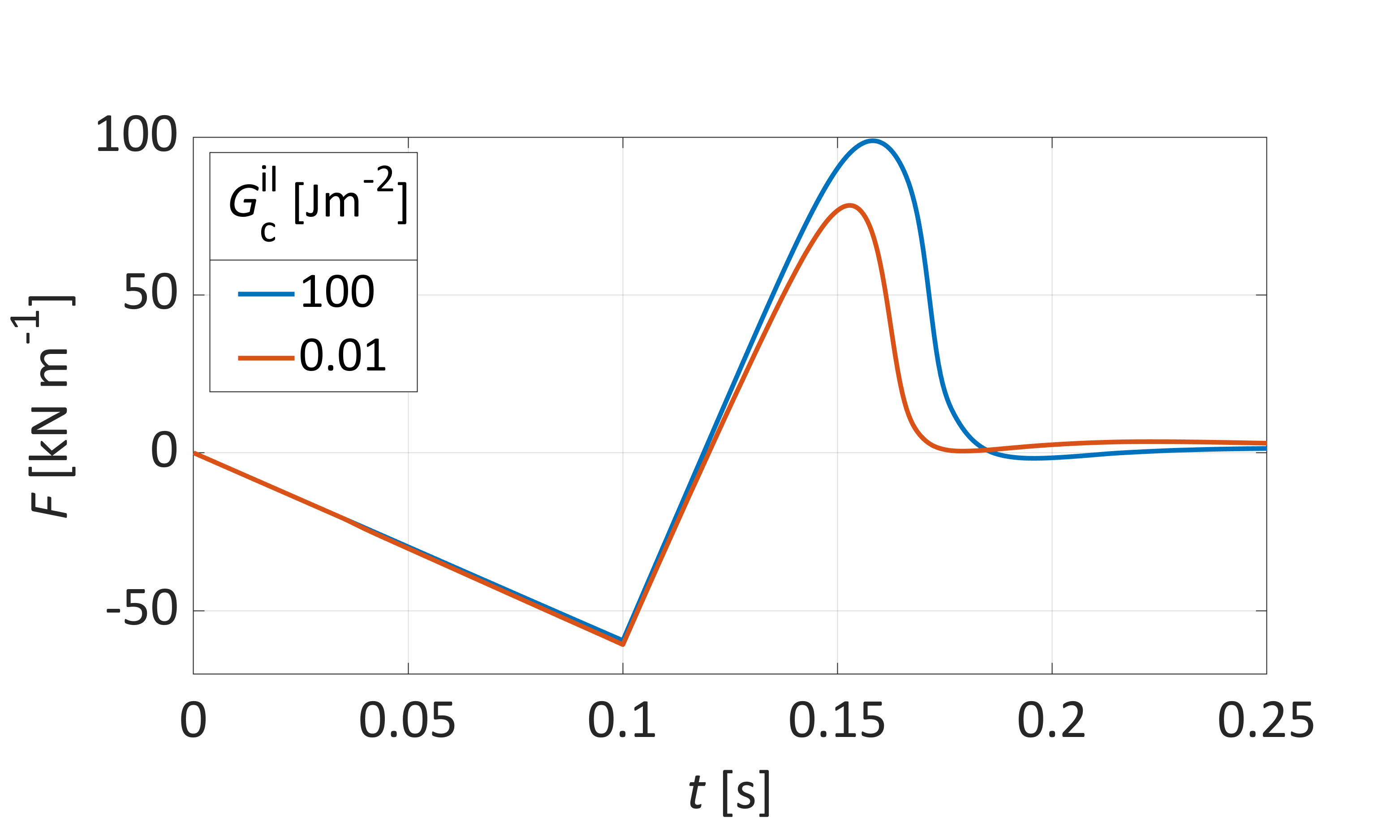}
\caption{${\Kp}_{\text{incl}} = 51.8$ GPa}\label{Fig_NoMoB_EF70_F}
\end{subfigure}
\begin{subfigure}{0.49\textwidth}
\centering
\includegraphics[scale=0.3,trim=5 5 40 50,clip]{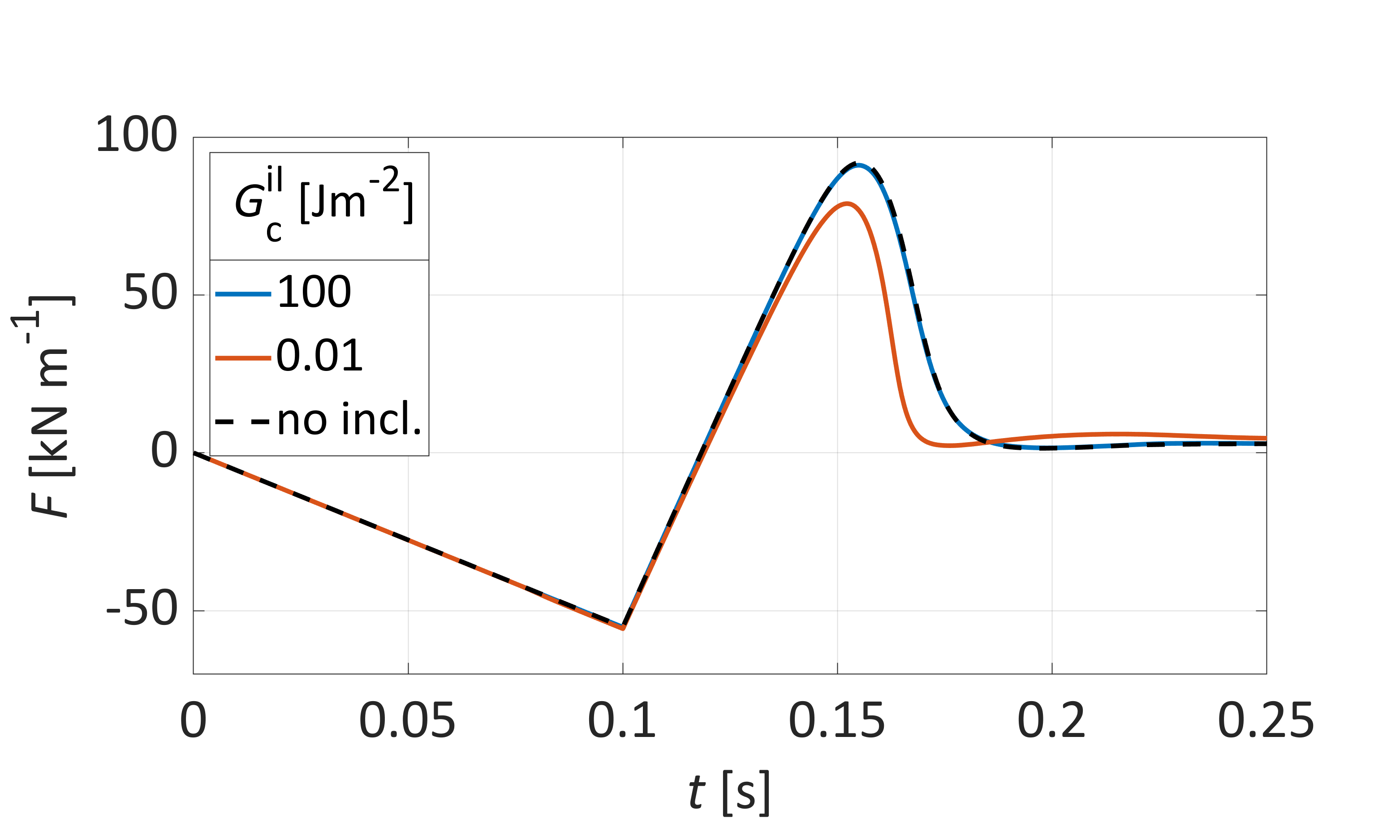}
\caption{${\Kp}_{\text{incl}} = 3.1$ GPa}\label{Fig_NoMoB_EF2_F}
\end{subfigure}
\caption{Total force applied at the top face for the combined loading case.}\label{Fig_NoMoB_F}
\end{figure}

The force response of the loaded block in terms of  the vertical force $F$ applied at the top face of the matrix domain is shown in~Fig.~\ref{Fig_NoMoB_F}.
The dependence is related to the time $t$.
As the first phase of loading includes lateral pressure, also this vertical response is compressive.
The graphs show differences caused by change of interface fracture energy and material stiffness .
The character of the global response is similar for all considered cases. 
For smaller value if interface fracture energy, the reaction reaches smaller value because an interface crack appears and the structural element becomes less stiff independently of the  inhomogeneity  stiffness.
In the  case of higher fracture energy,  no explicit interface crack exists.
Additionally, if the material of the  inhomogeneity  is the same as that of the matrix, the compound domain naturally behaves as there were no  inhomogeneity.
This is demonstrated by a comparison with  an homogeneous  case (dashed line).
The force finally falls to (almost) zero values though no crack  developed across the whole domain  because the lateral compression prevents the crack to hit the outer contour of the domain, as it was expected in accordance with experimental observations in~\cite{feng22A1}.

\begin{figure}[!ht]
\centering
\begin{subfigure}{0.49\textwidth}
\centering
\includegraphics[scale=0.22,trim=5 5 45 45,clip]{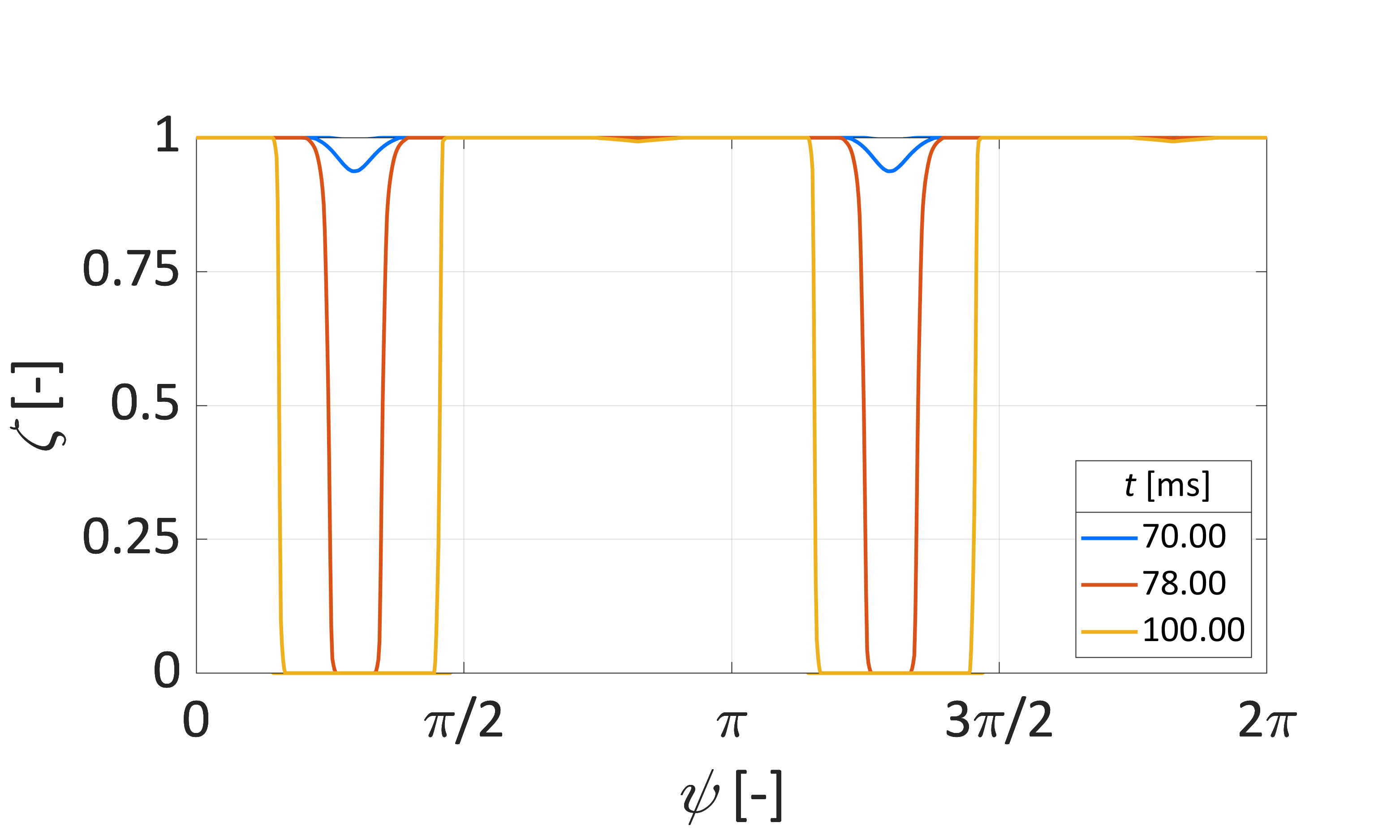}
\caption{Interface damage, ${\Kp}_\text{incl}=3.1$ GPa, $\Gc[iI]=0.01$ Jm$^{-2}$}\label{Fig_NoMoB_EF2_Gm5_IF_1}
\end{subfigure}
\begin{subfigure}{0.49\textwidth}
\centering
\includegraphics[scale=0.22,trim=5 5 45 45,clip]{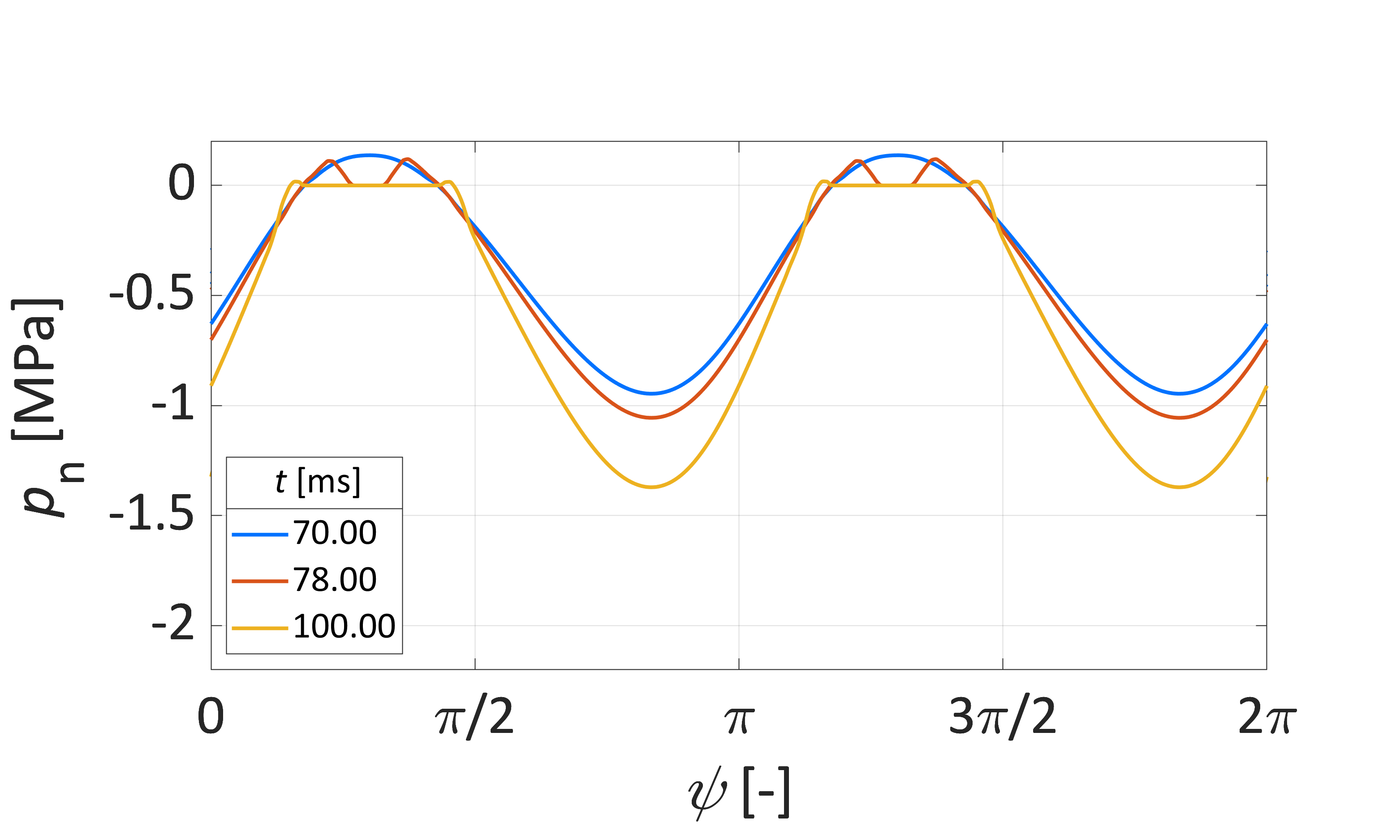}
\caption{Normal stress, ${\Kp}_\text{incl}=3.1$ GPa, $\Gc[iI]=0.01$ Jm$^{-2}$}\label{Fig_NoMoB_EF2_Gm5_IF_2}
\end{subfigure}
\begin{subfigure}{0.49\textwidth}
\centering
\includegraphics[scale=0.22,trim=5 5 45 45,clip]{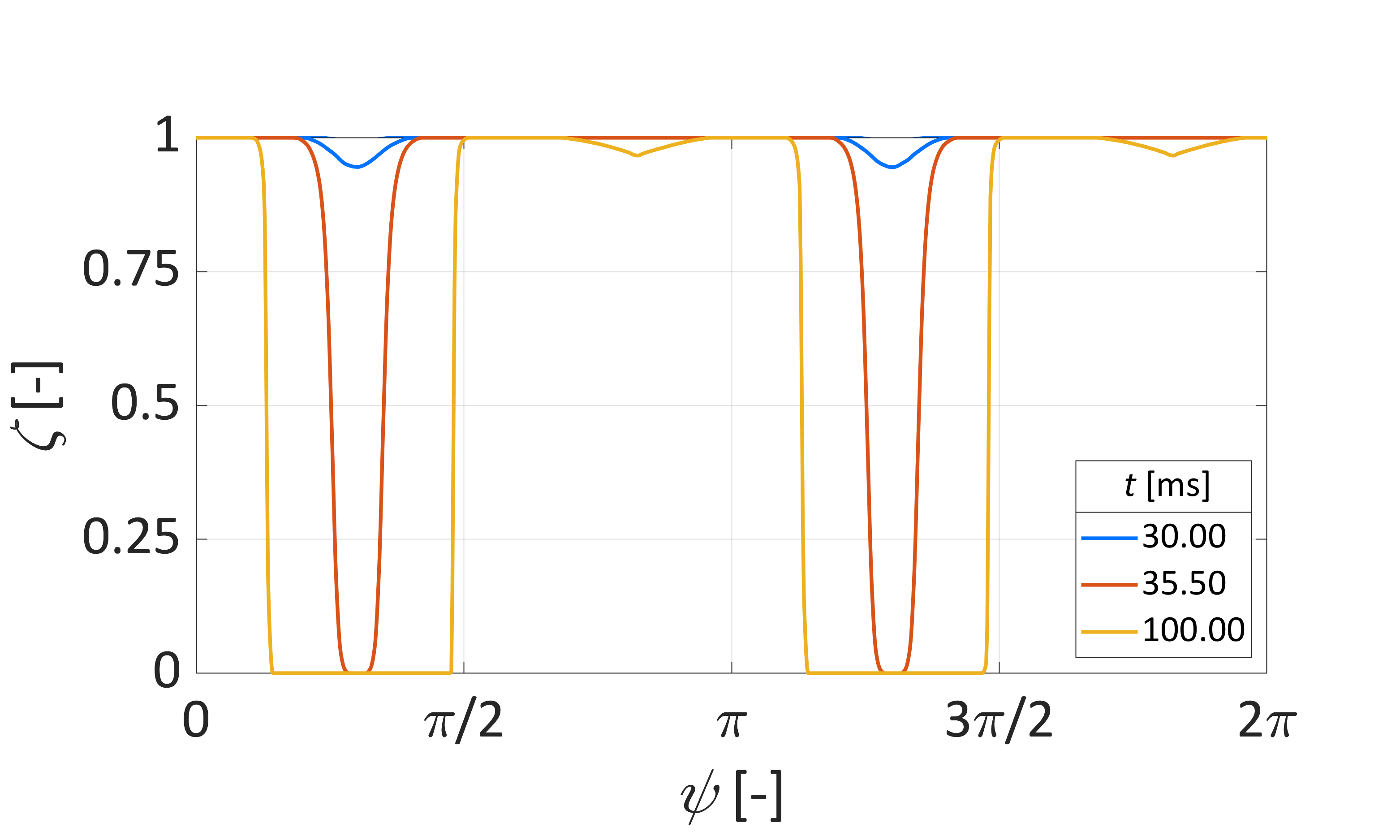}
\caption{Interface damage, ${\Kp}_\text{incl}=51.8$ GPa, $\Gc[iI]=0.01$ Jm$^{-2}$}\label{Fig_NoMoB_EF70_Gm5_IF_1}
\end{subfigure}
\begin{subfigure}{0.49\textwidth}
\centering
\includegraphics[scale=0.22,trim=5 5 45 45,clip]{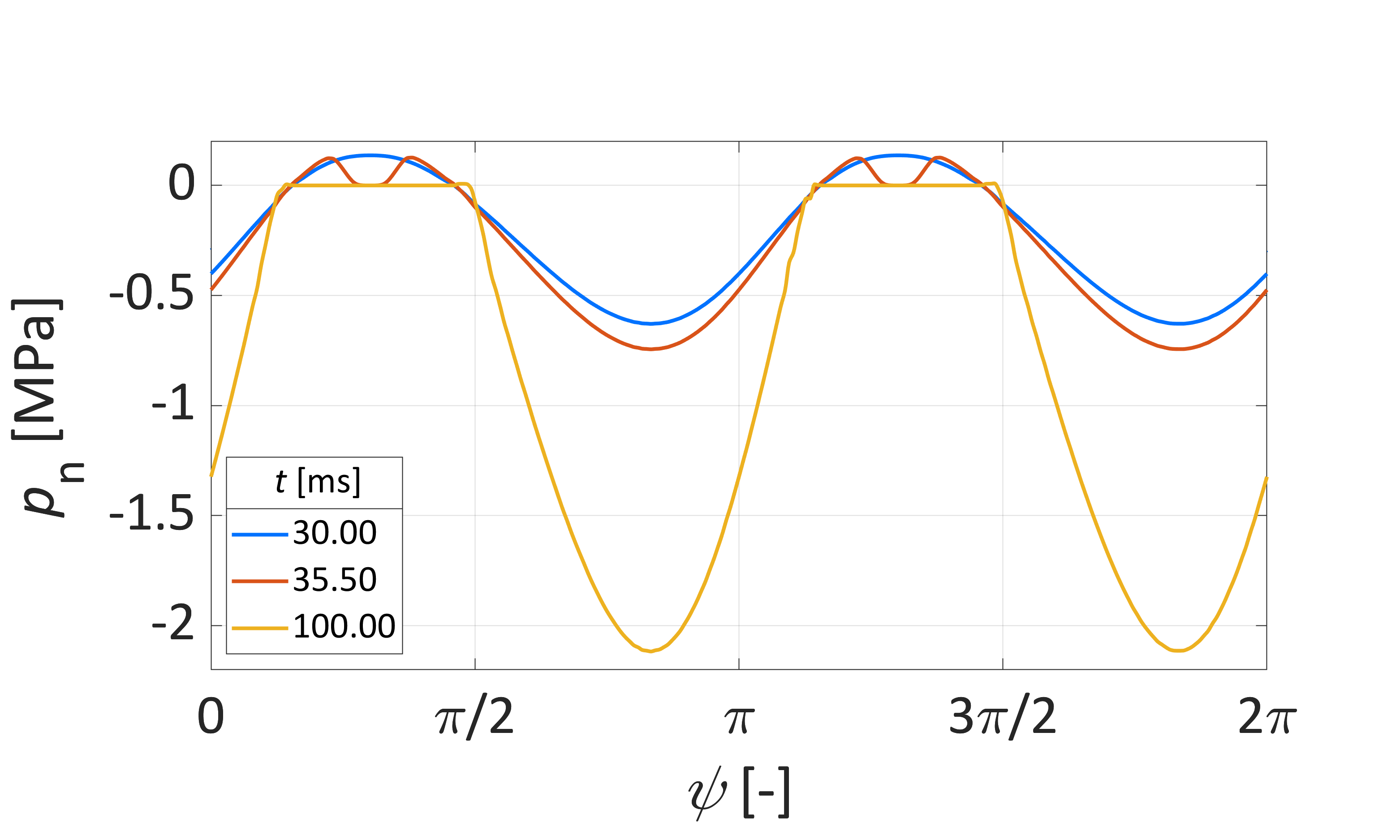}
\caption{Normal stress, ${\Kp}_\text{incl}=51.8$ GPa, $\Gc[iI]=0.01$ Jm$^{-2}$}\label{Fig_NoMoB_EF70_Gm5_IF_2}
\end{subfigure}
\begin{subfigure}{0.49\textwidth}
\centering
\includegraphics[scale=0.22,trim=5 5 45 45,clip]{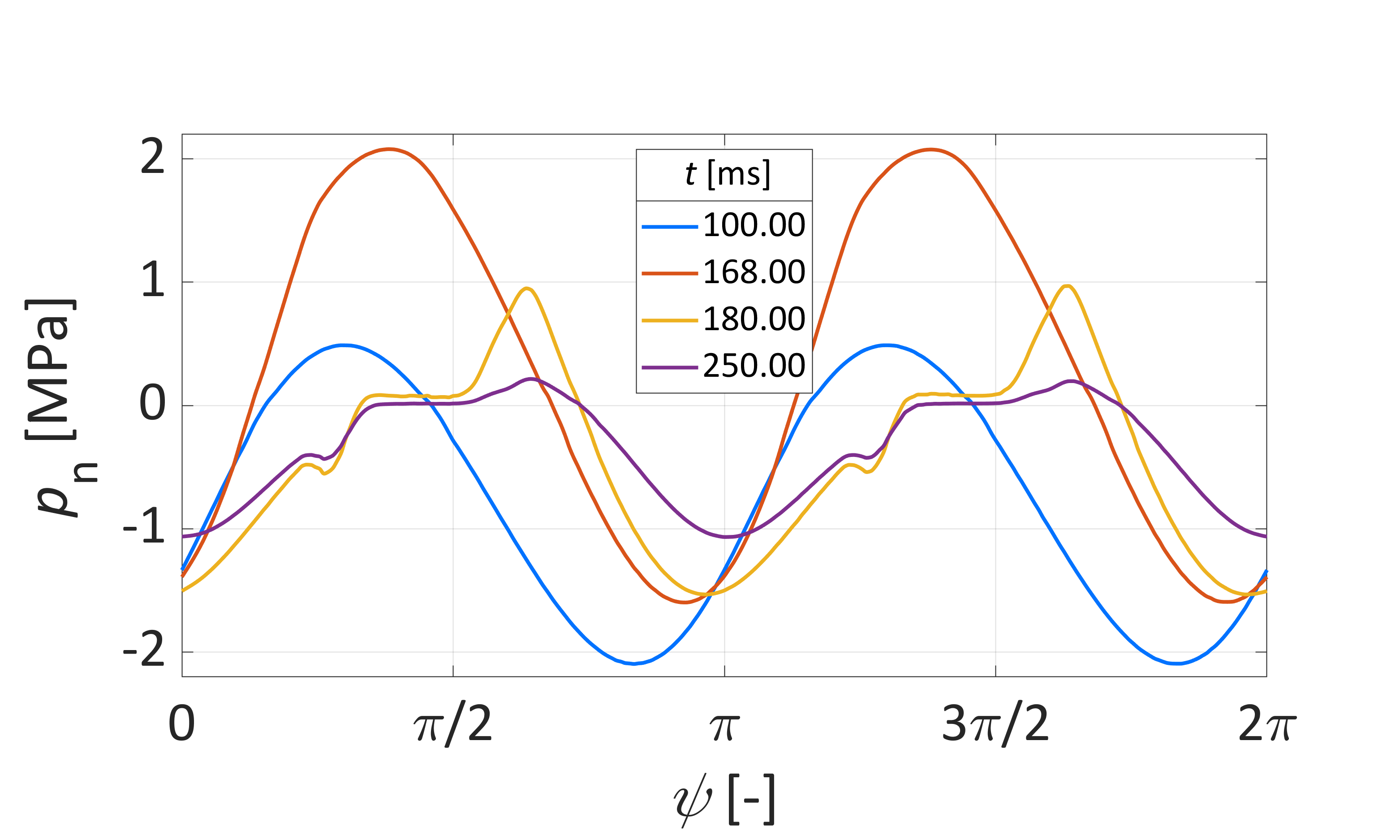}
\caption{Normal stress, ${\Kp}_\text{incl}=51.8$ GPa, $\Gc[iI]=100$ Jm$^{-2}$}\label{Fig_NoMoB_EF70_Gm1_IF_2}
\end{subfigure}
\caption{Distribution of the interface normal stress during the debonding phase at selected time instants for the combined loading case.
Interface damage is not initiated in the last case.}\label{Fig_NoMoB_IF}
\end{figure}

Direct observations of the interface variables provide another aspect of the results.
They are shown in Fig.~\ref{Fig_NoMoB_IF}.
The cases with higher value of  $\Gc[iI]$ contain a strong interface which is not damaged during load process, therefore the distribution of the interface damage is presented only for the other cases.
Nevertheless, if the materials of the  inhomogeneity  and matrix are different,   a damaged zone appears  near the interface due to mutual affecting as it can also be seen below in Fig.~\ref{Fig_NoMoB_DS}. 
Here, at least the normal stress shown in Fig.~\ref{Fig_NoMoB_EF70_Gm1_IF_2} document that material  damage  appears close to the interface.
It is also seen that the maximal stress does not approach its maximal possible value responsible for initiation of the interface damage (13.8 MPa).
Anyhow, the graph enables to read the extent of the damaged zone which projects to the interface in a form of  vanishing stress.

Unlike the previously described situation, the cases with the small value of $\Gc[iI]$ indicate the interface damage which finally developed into an interface crack.
The maximal achievable normal stress is detected at appropriate time instants shown in the graphs.
With the current damage function, the maximal stress is reached for $\zeta<1$.
The distribution of the interface  quantities at selected instants is similar for both material cases, but they are made at various moments, to document that the stress states  are different depending on the elastic material characteristics of the domains.

Initiation and propagation of cracks is influenced by three factors.
First, stress singularity of the elastic solution at the tips of the slits is a natural origin for initiation of a crack in the matrix domain. 
Second, the quality of the interface holding the two domains together affects initiation of the interface crack as observed in Fig.~\ref{Fig_NoMoB_IF}.
Third, various elasticity parameters of the two material component induce another stress concentration close to the interface.
The combinations of these effects can be observed in diagrams of Fig.~\ref{Fig_NoMoB_DS}.
\begin{figure}[!ht]
\centering
\setlength{\unitlength}{\textwidth}
\begin{subfigure}{0.99\textwidth}
\begin{picture}(0.97,0.22)
\PA{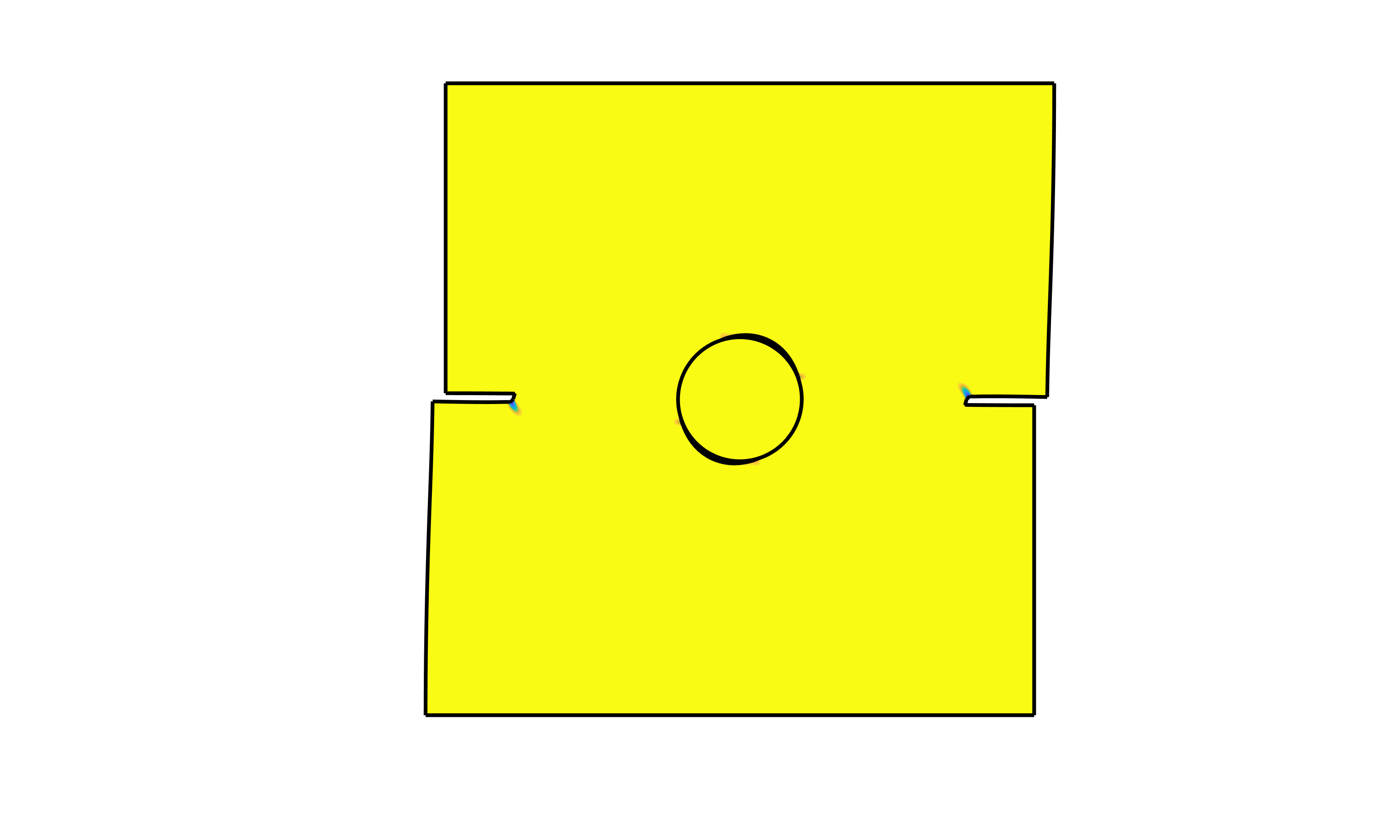}{$t=0.14$ s}
\PBF{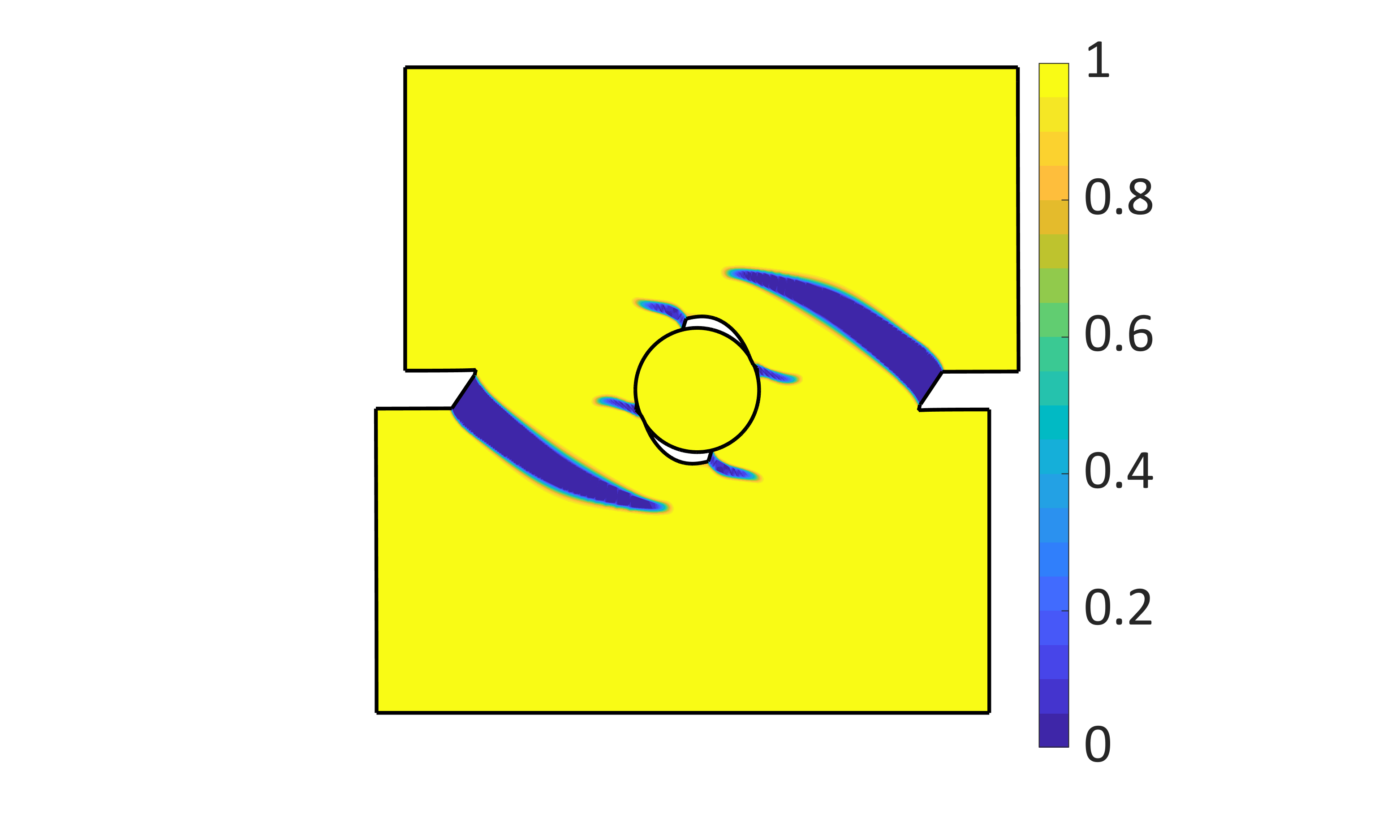}{$t=0.2$ s}
\Pvar{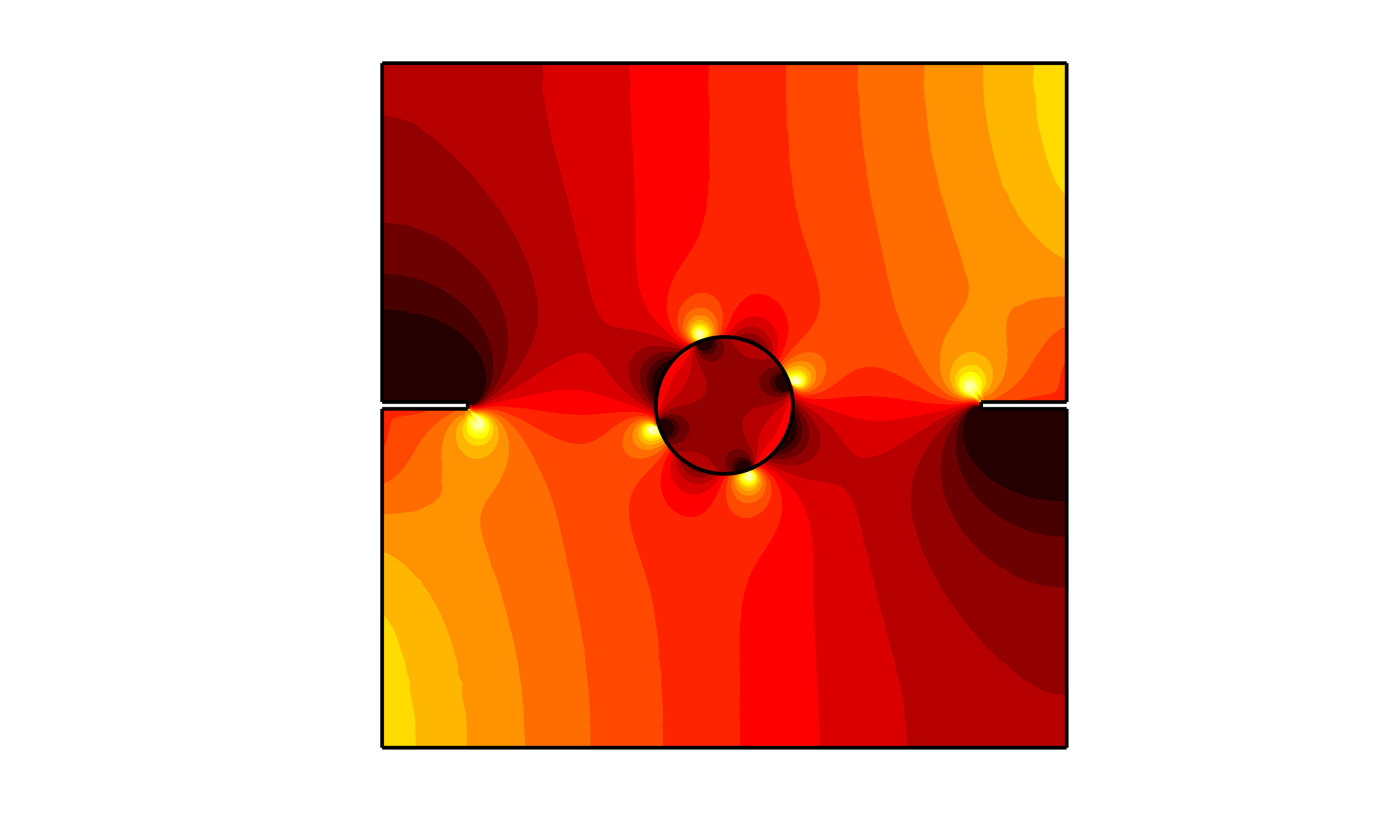}{Stress trace}{530}
\CB[720]{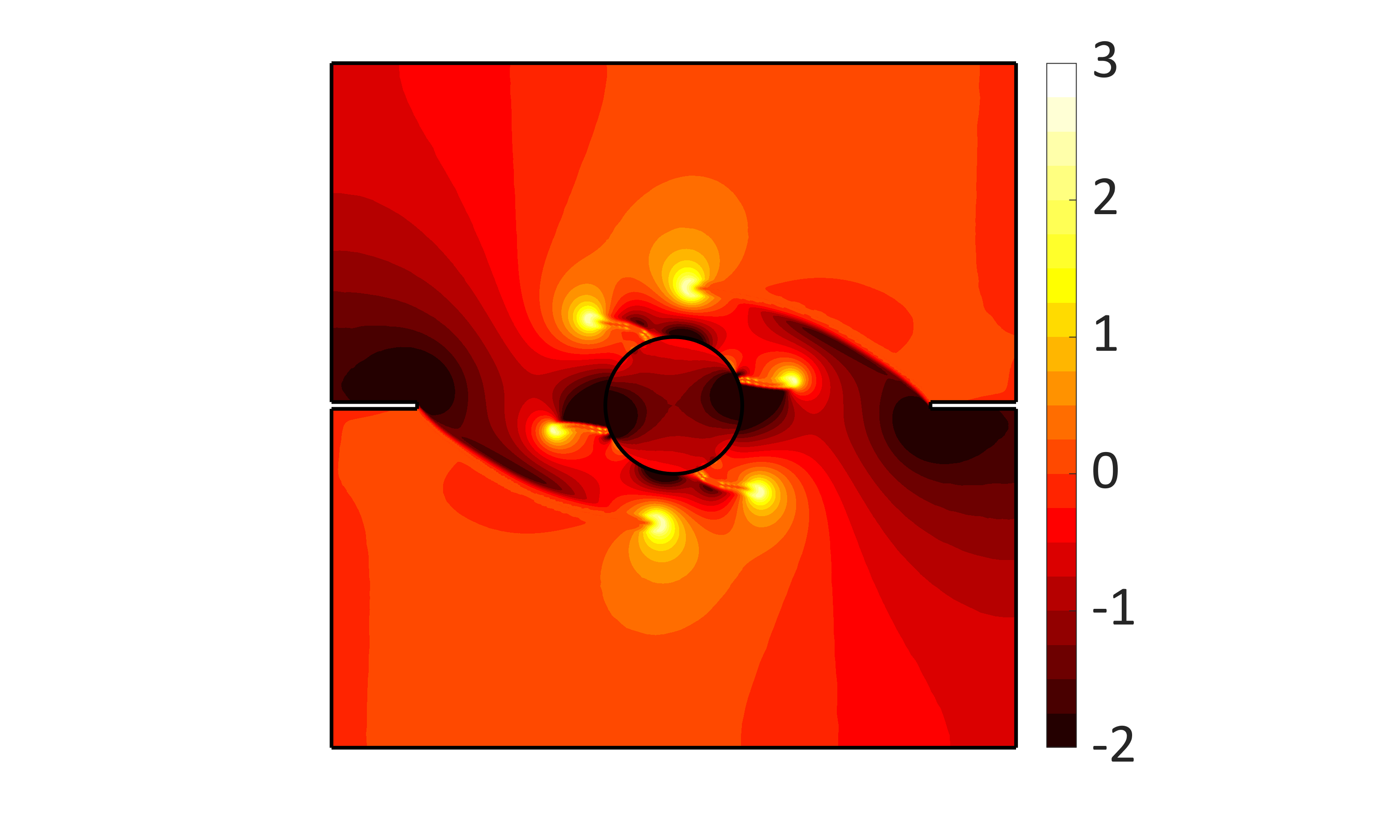}
\PE{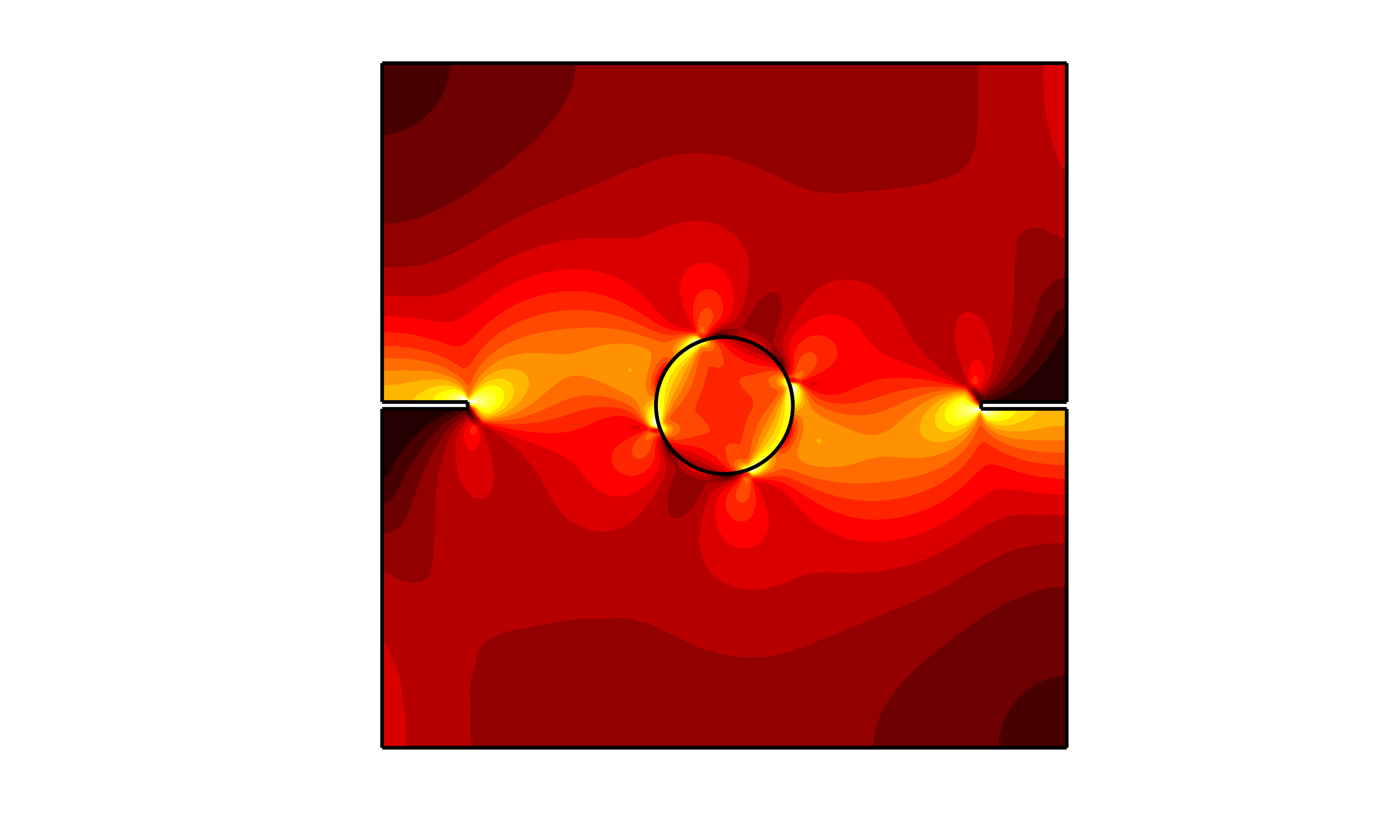}{Deviatoric stress}
\CB[955]{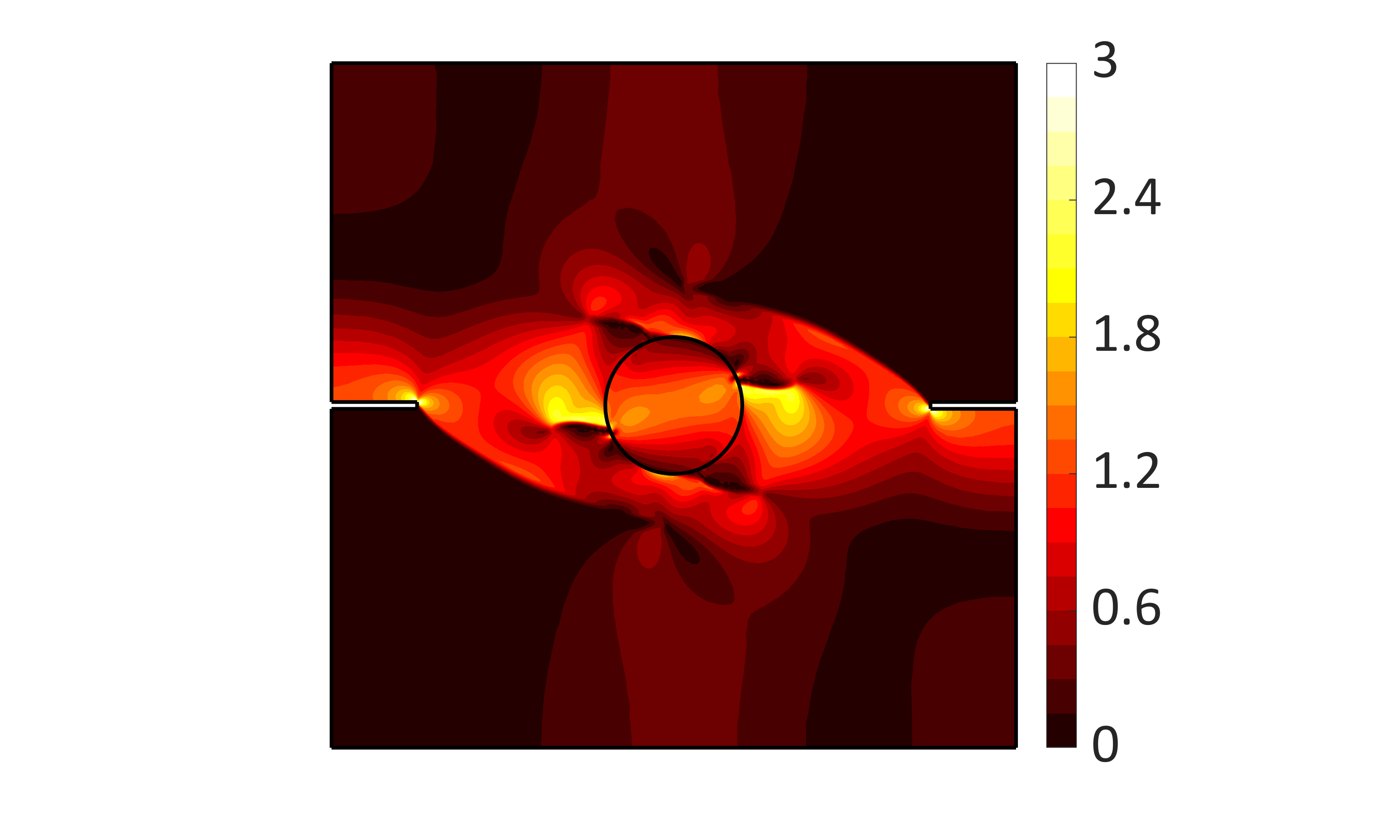}
\end{picture}
\caption{${\Kp}_\text{incl}=51.8$ GPa, $\Gc[iI]=0.01$ Jm$^{-2}$}\label{Fig_NoMoB_EF70_Gm5}
\end{subfigure}
\begin{subfigure}{0.99\textwidth}
\begin{picture}(0.97,0.22)
\PA{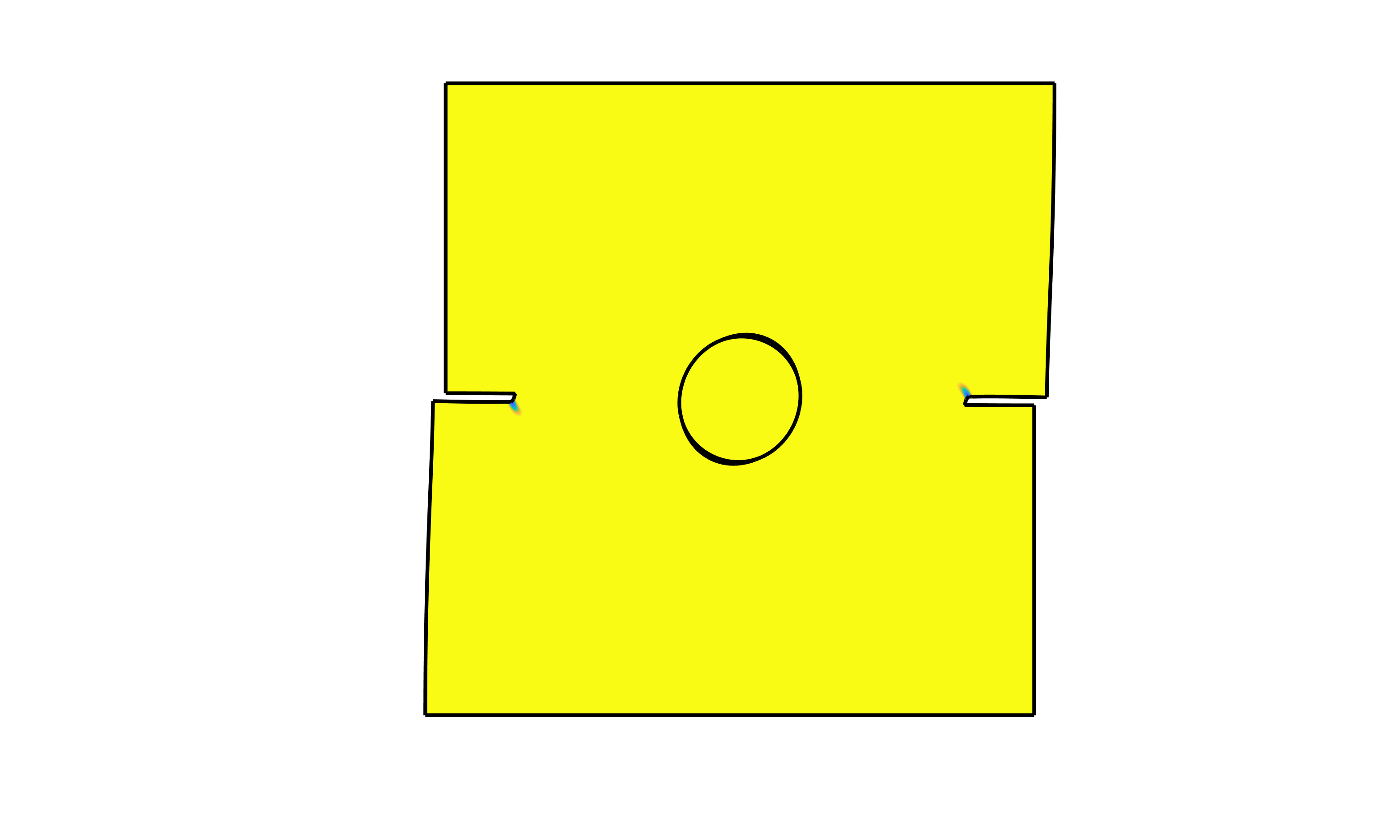}{$t=0.14$ s}
\PBF{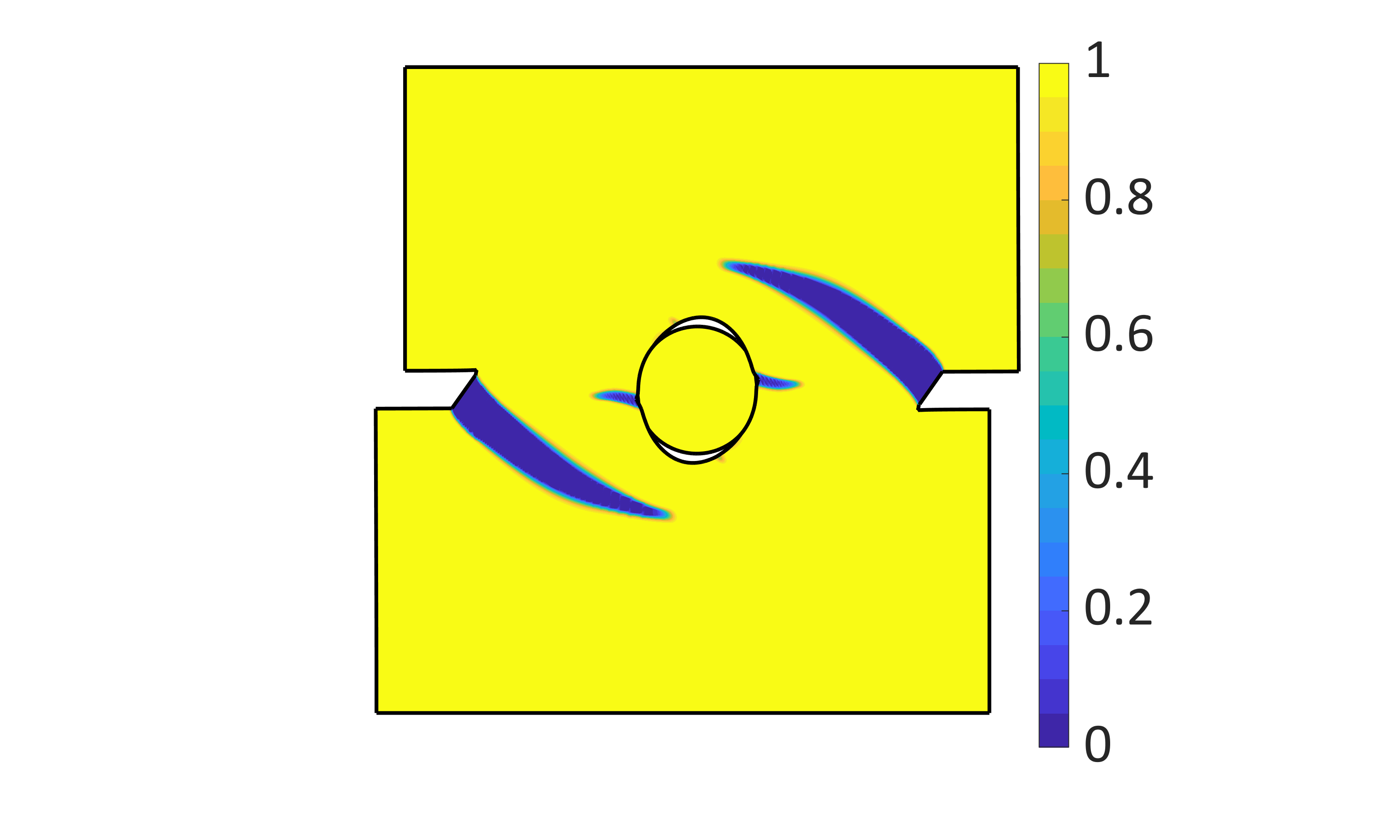}{$t=0.2$ s}
\Pvar{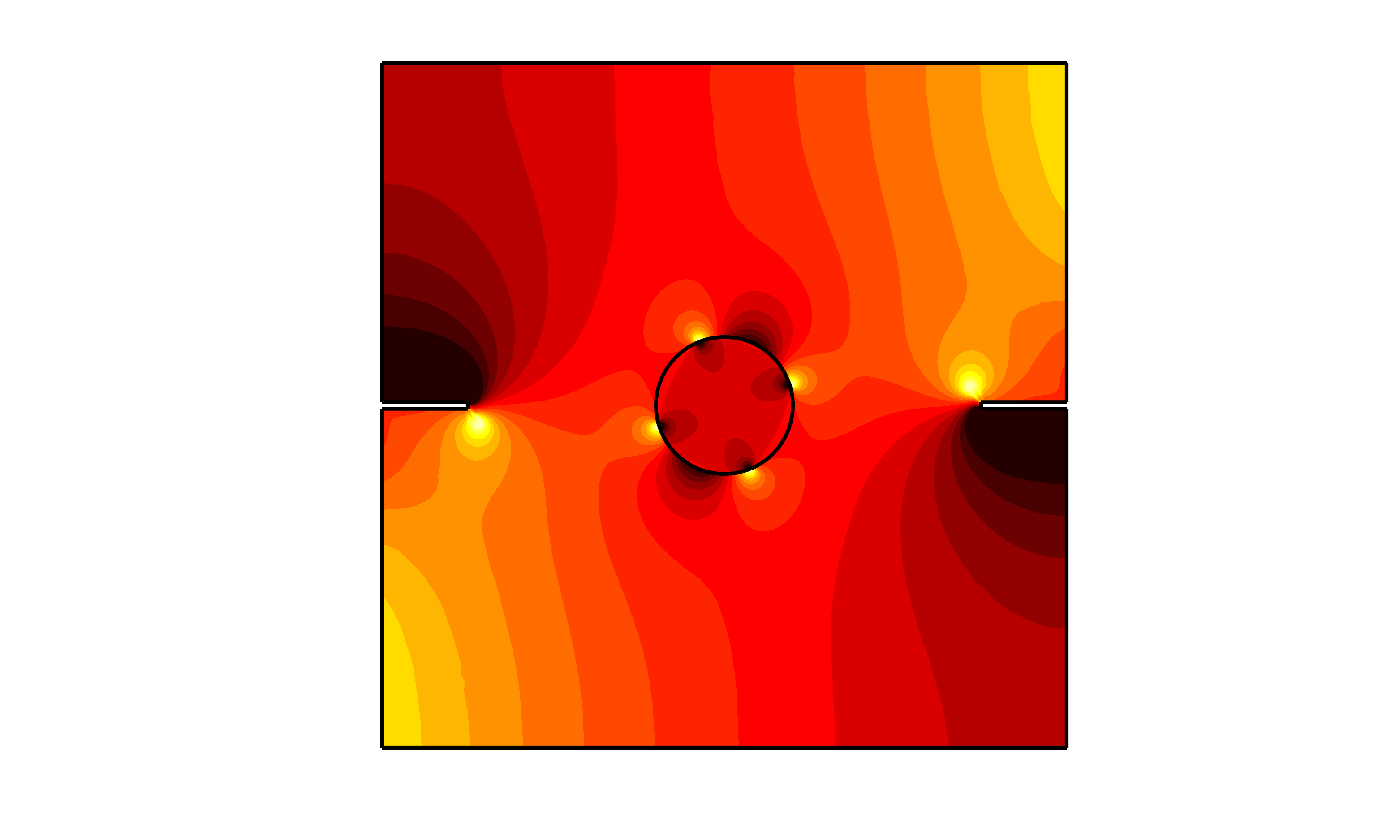}{Stress trace}{530}
\CB[720]{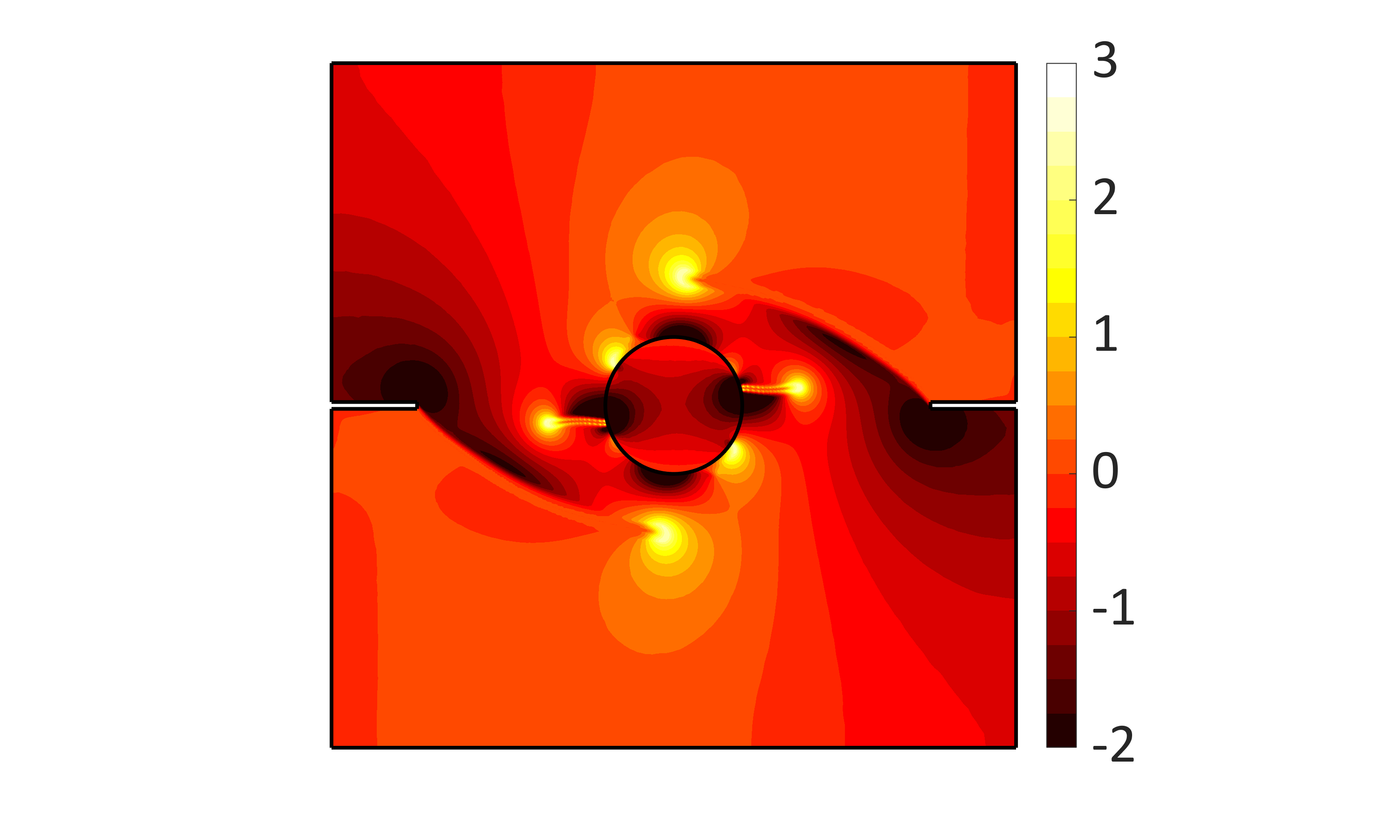}
\PE{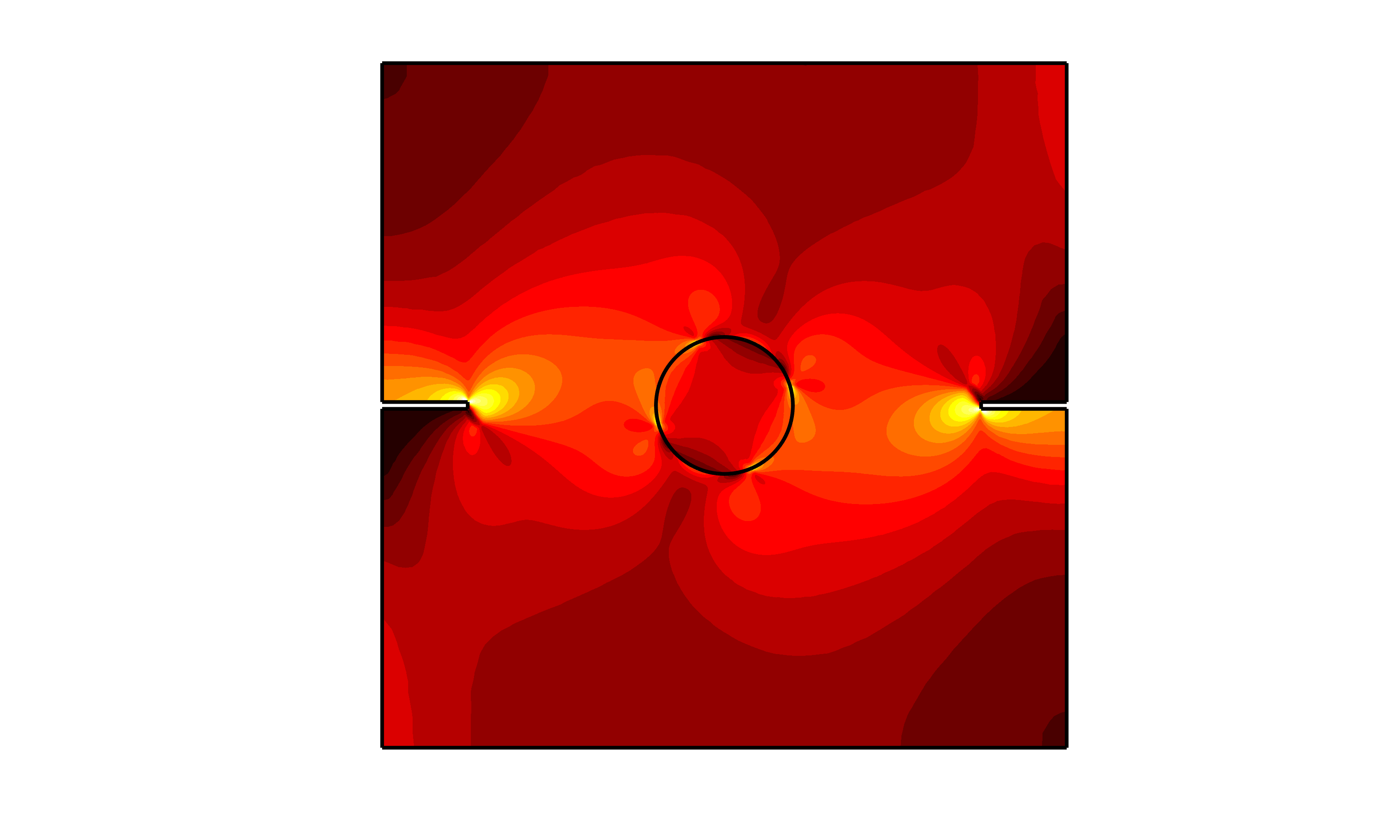}{Deviatoric stress}
\CB[955]{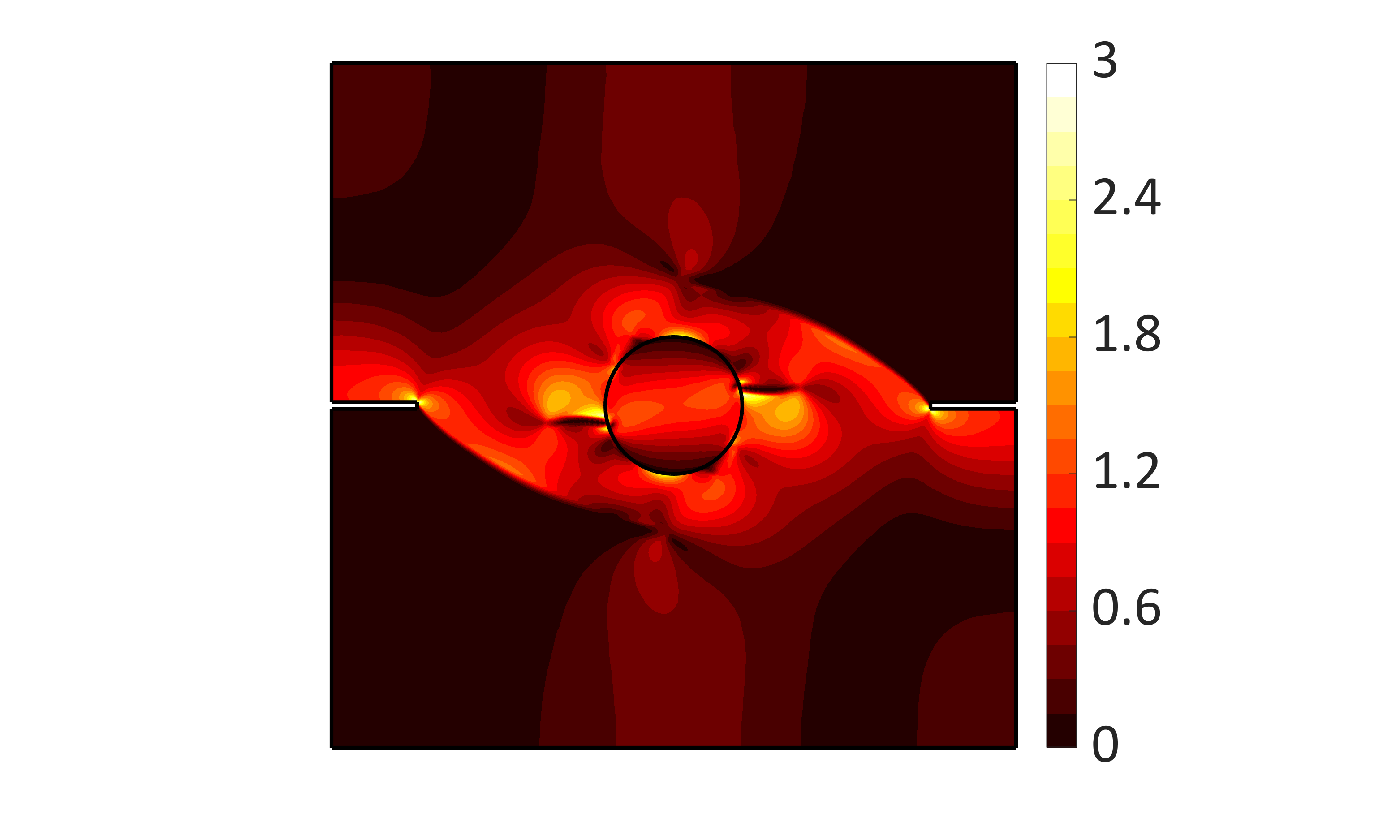}
\end{picture}
\caption{${\Kp}_\text{incl}=3.1$ GPa, $\Gc[iI]=0.01$ Jm$^{-2}$}\label{Fig_NoMoB_EF2_Gm5}
\end{subfigure}
\begin{subfigure}{0.99\textwidth}
\begin{picture}(0.97,0.22)
\PA{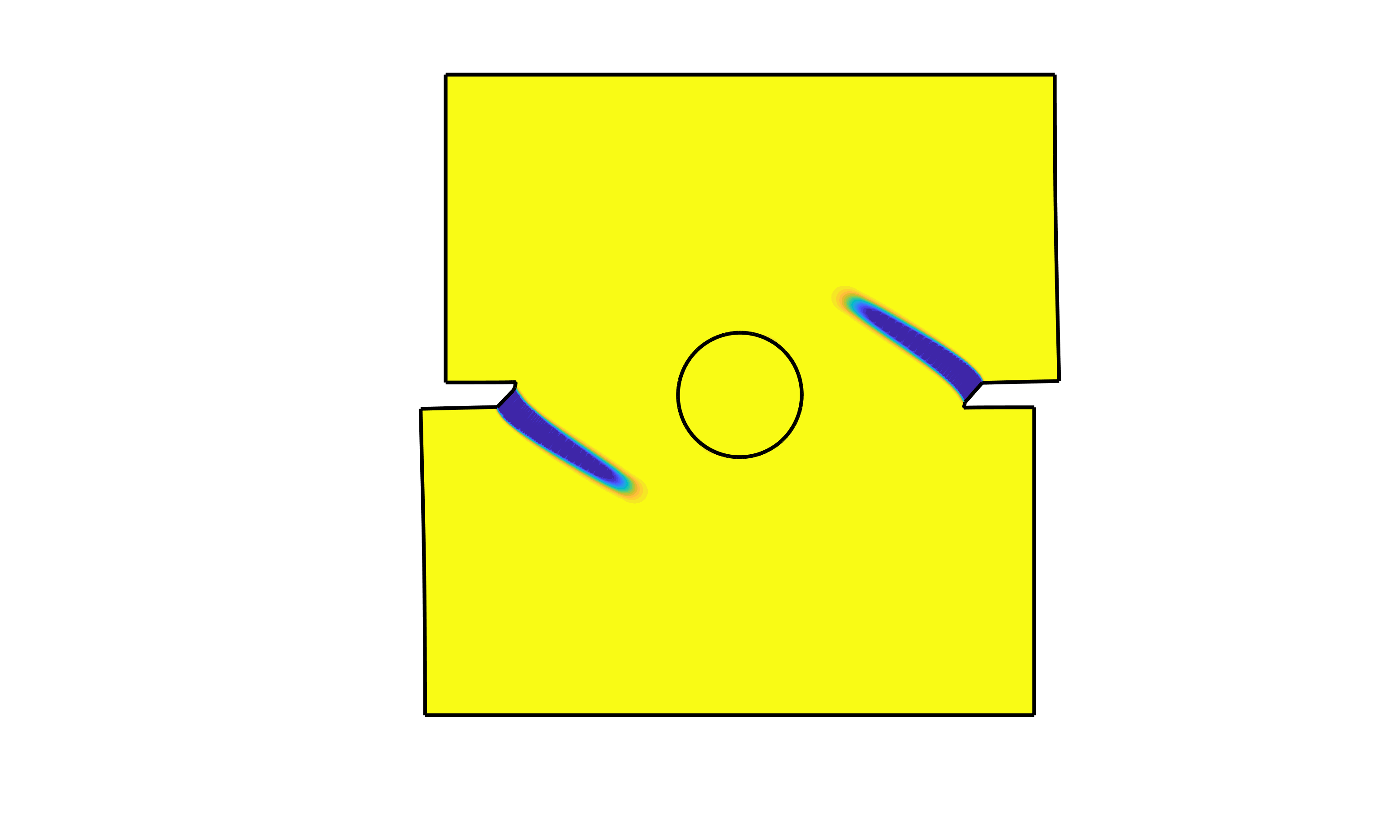}{$t=0.168$ s}
\PBF{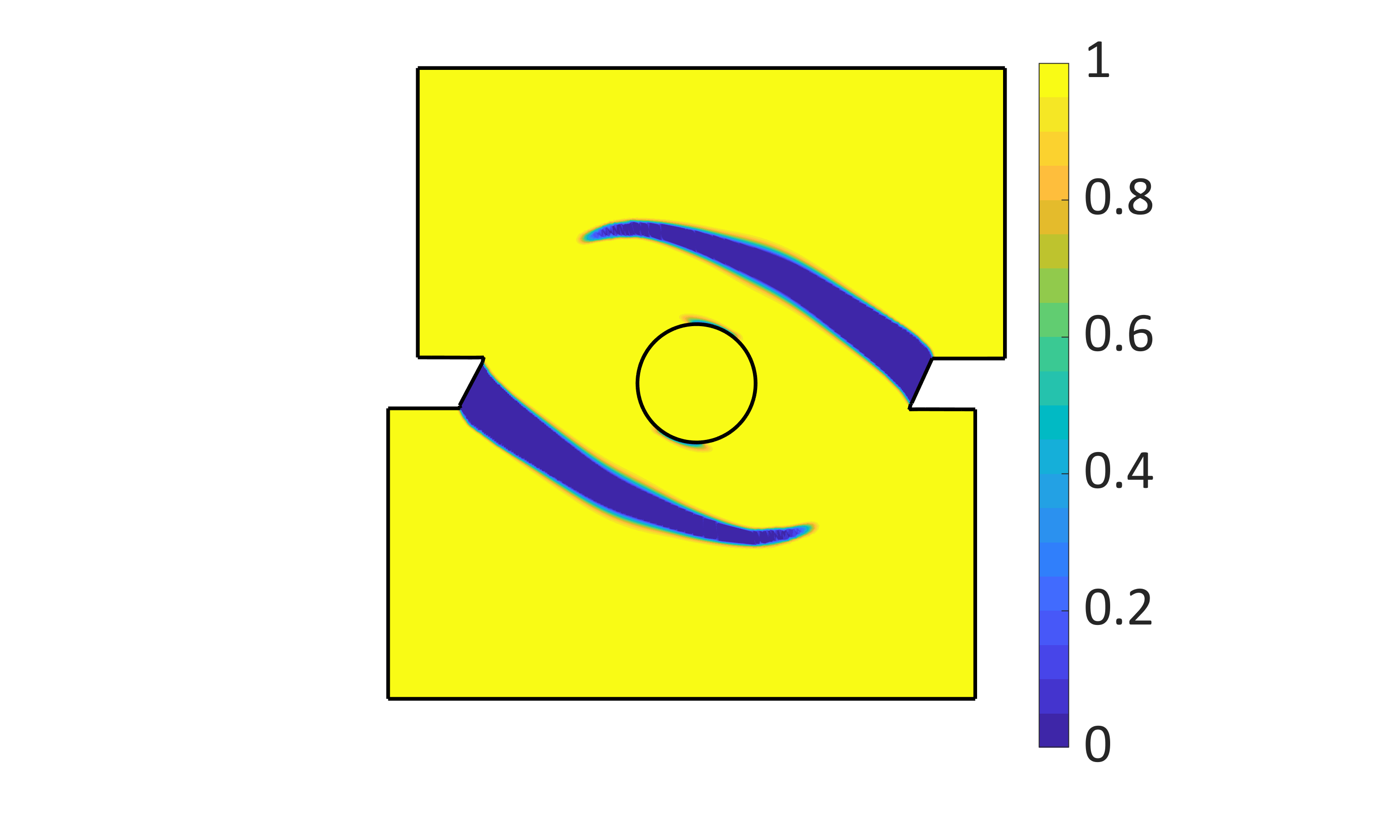}{$t=0.25$ s}
\Pvar{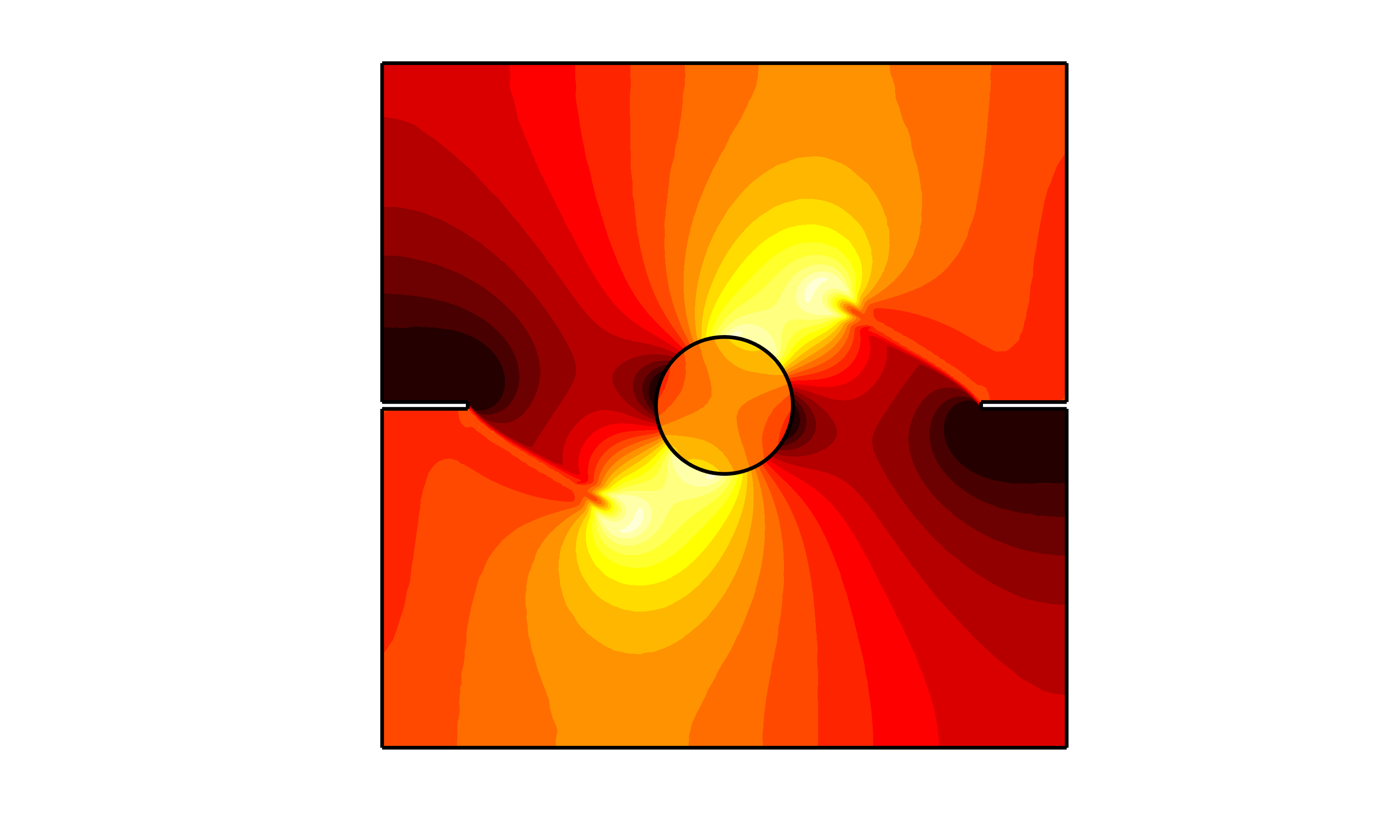}{Stress trace}{530}
\CB[720]{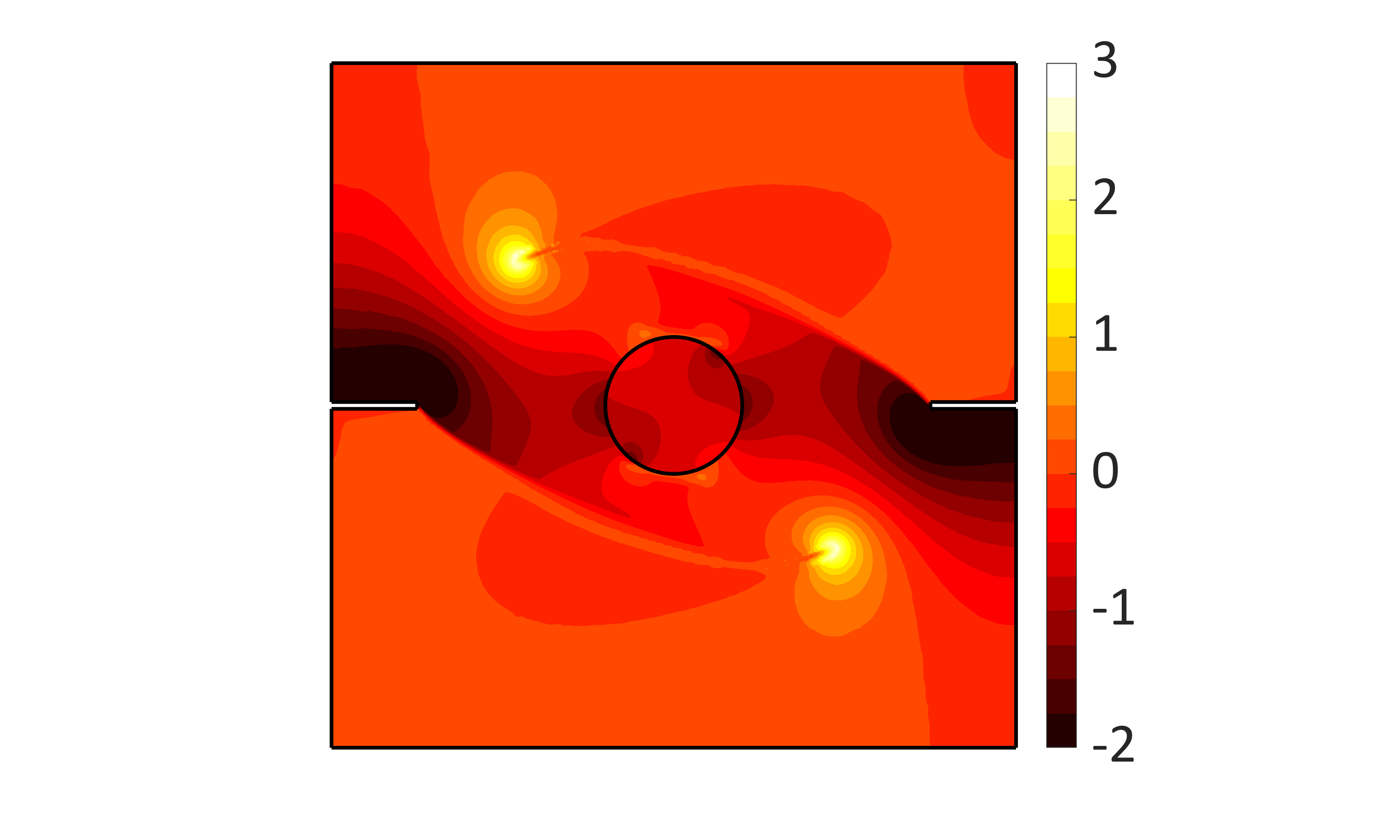}
\PE{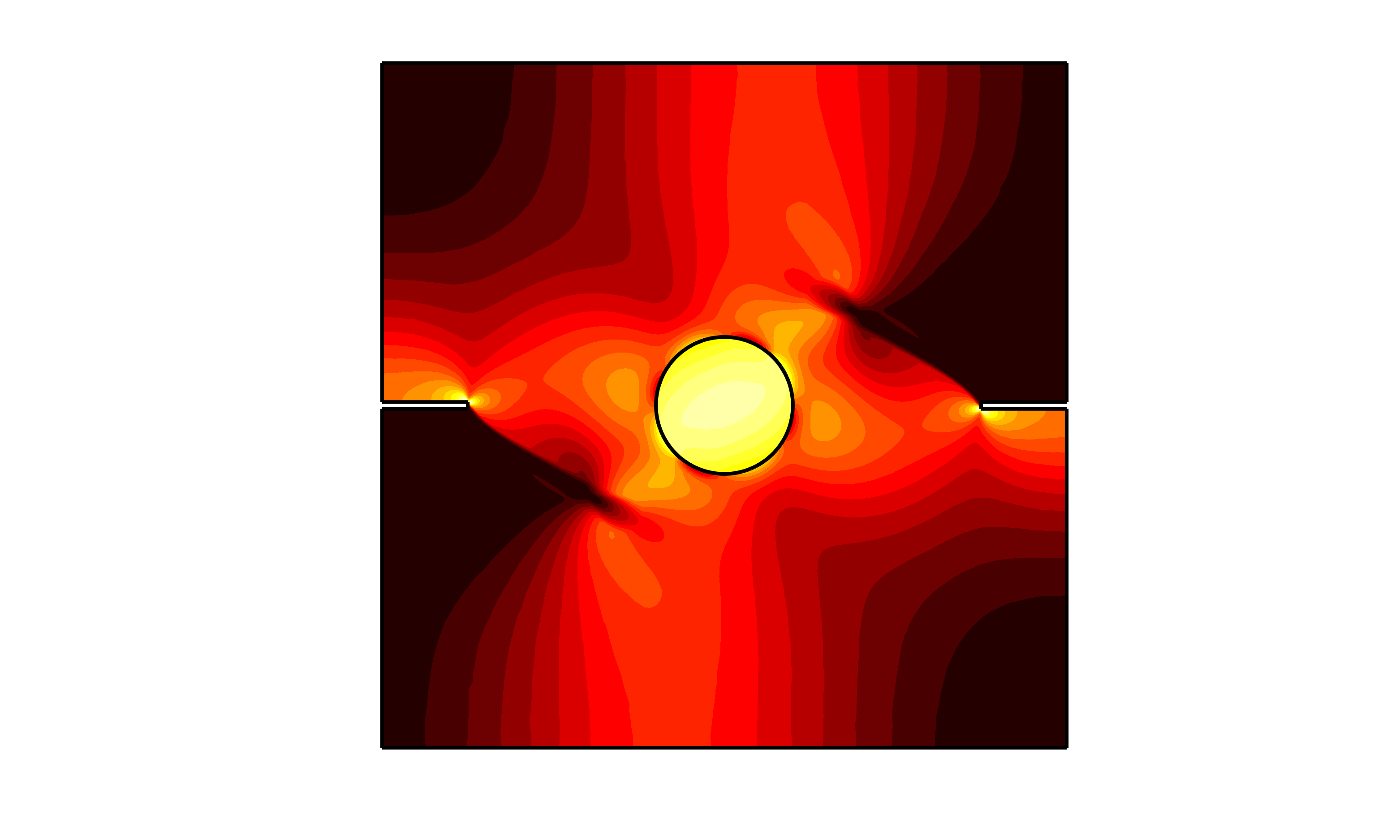}{Deviatoric stress}
\CB[955]{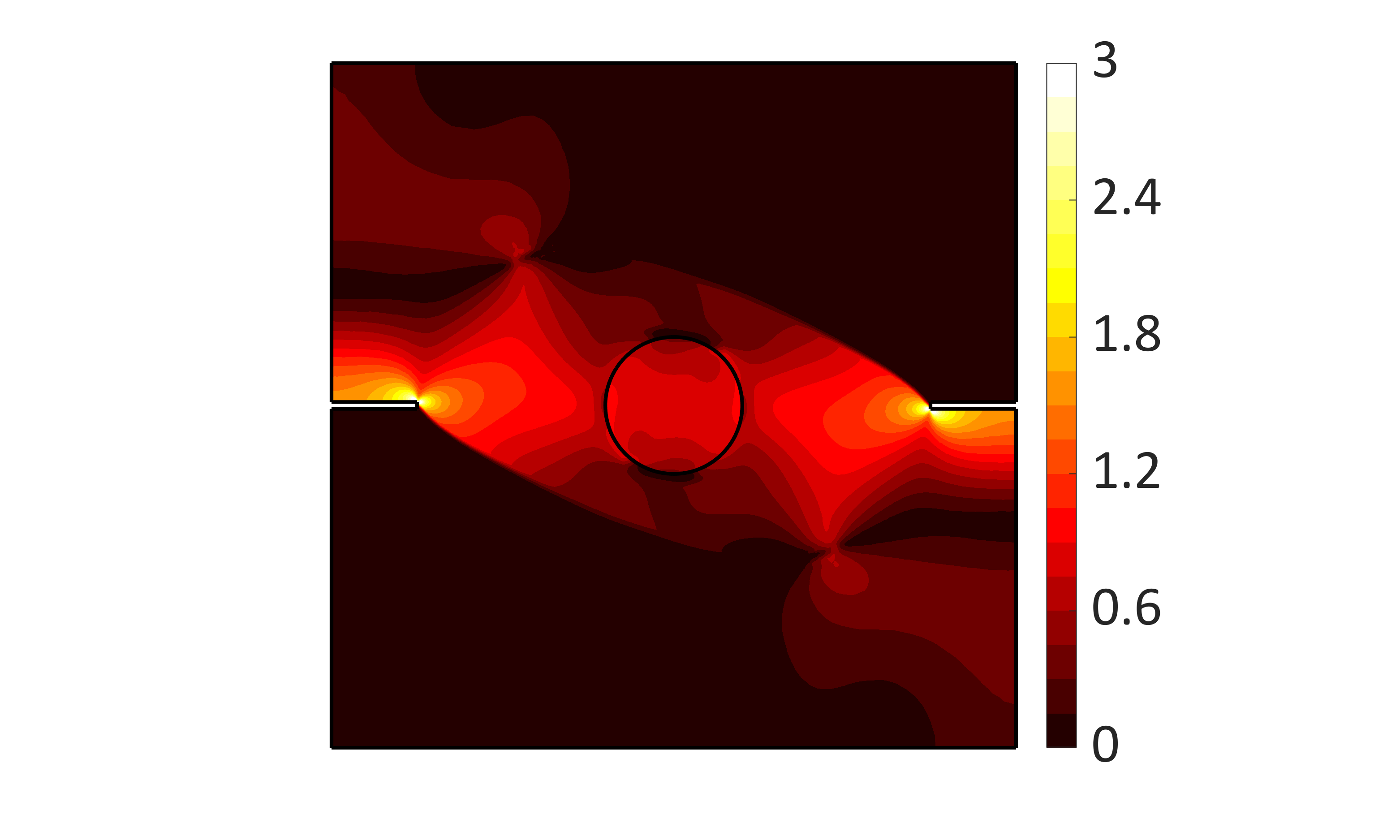}
\end{picture}
\caption{${\Kp}_\text{incl}=51.8$ GPa, $\Gc[iI]=100$ Jm$^{-2}$}\label{Fig_NoMoB_EF70_Gm1}
\end{subfigure}
\begin{subfigure}{0.99\textwidth}
\begin{picture}(0.97,0.22)
\PA{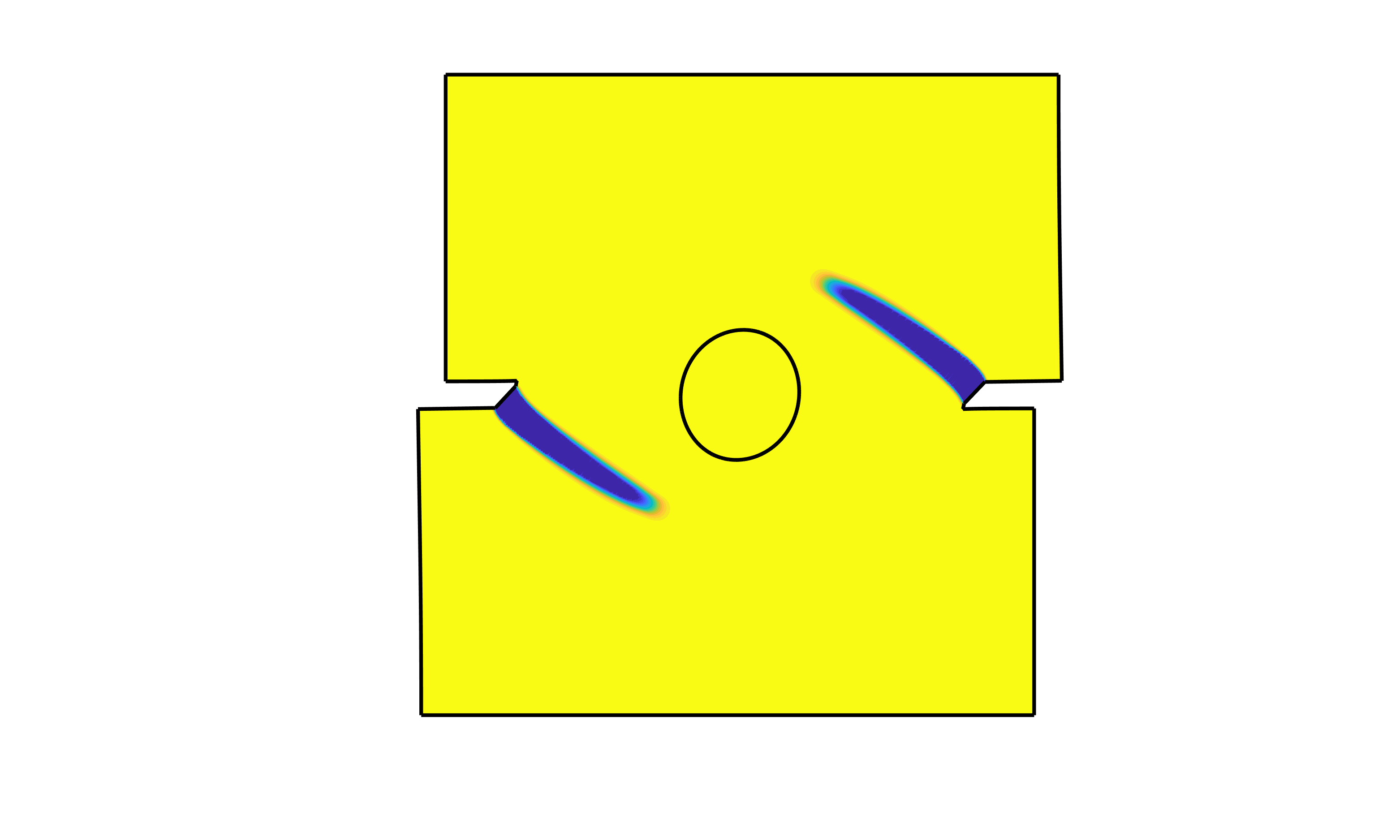}{$t=0.168$ s}
\PBF{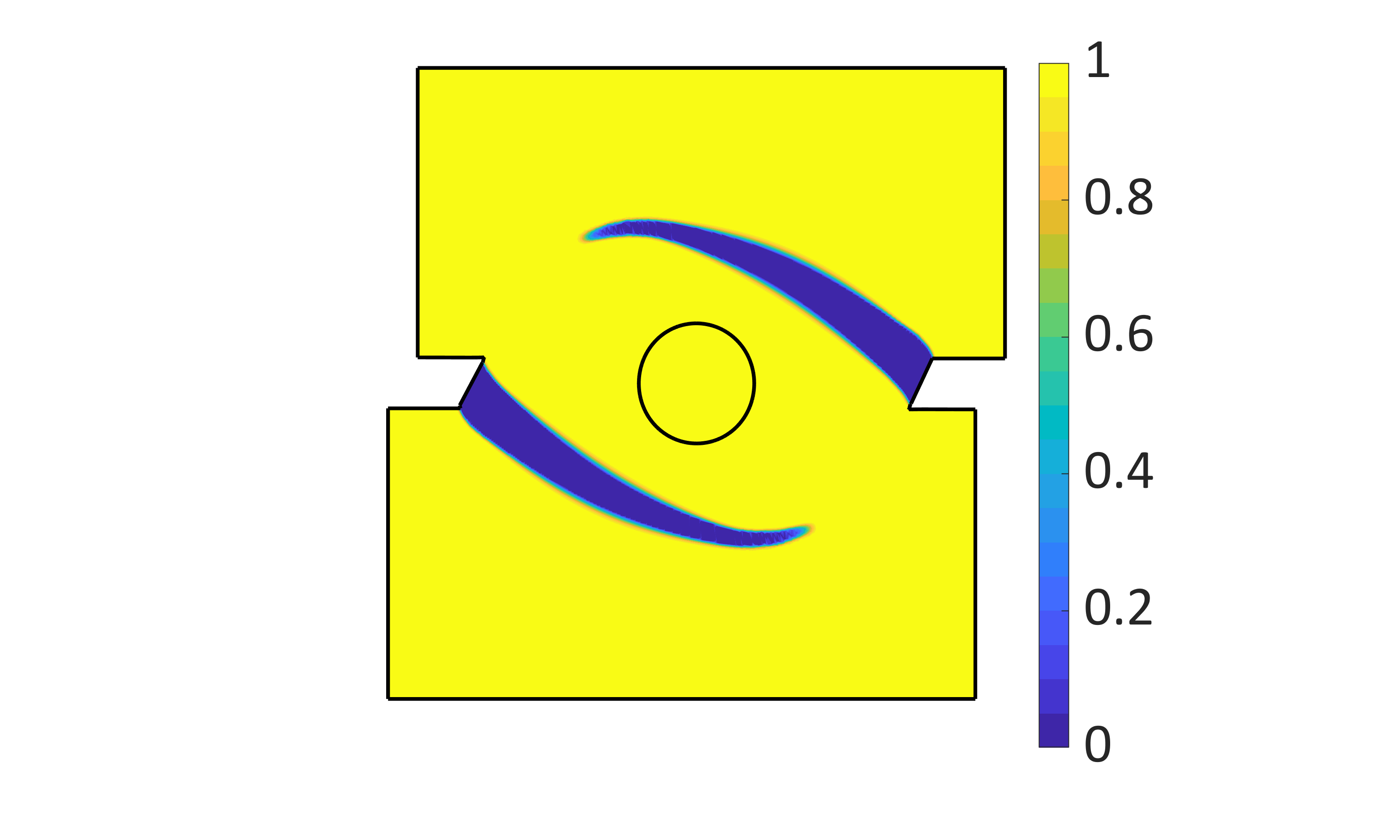}{$t=0.25$ s}
\Pvar{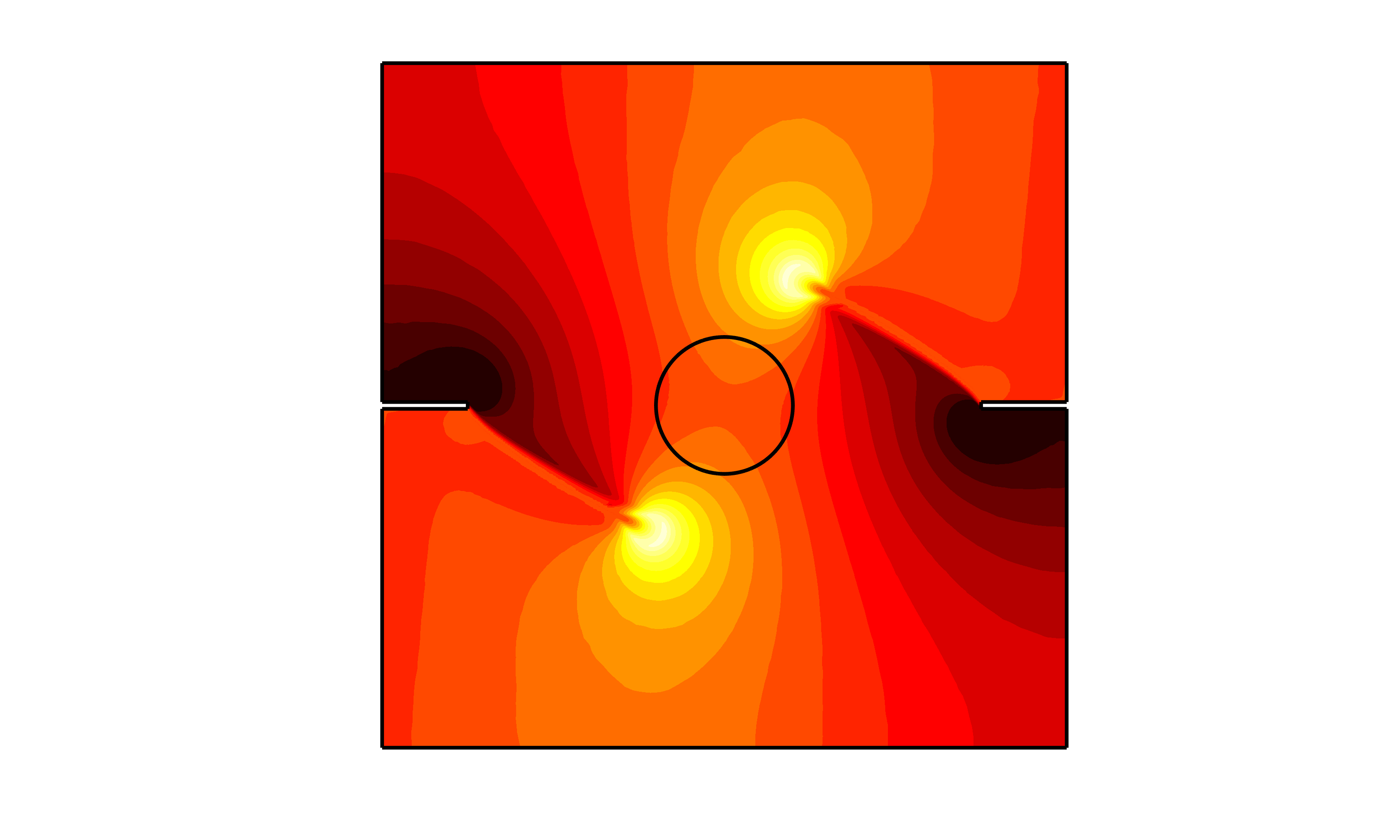}{Stress trace}{530}
\CB[720]{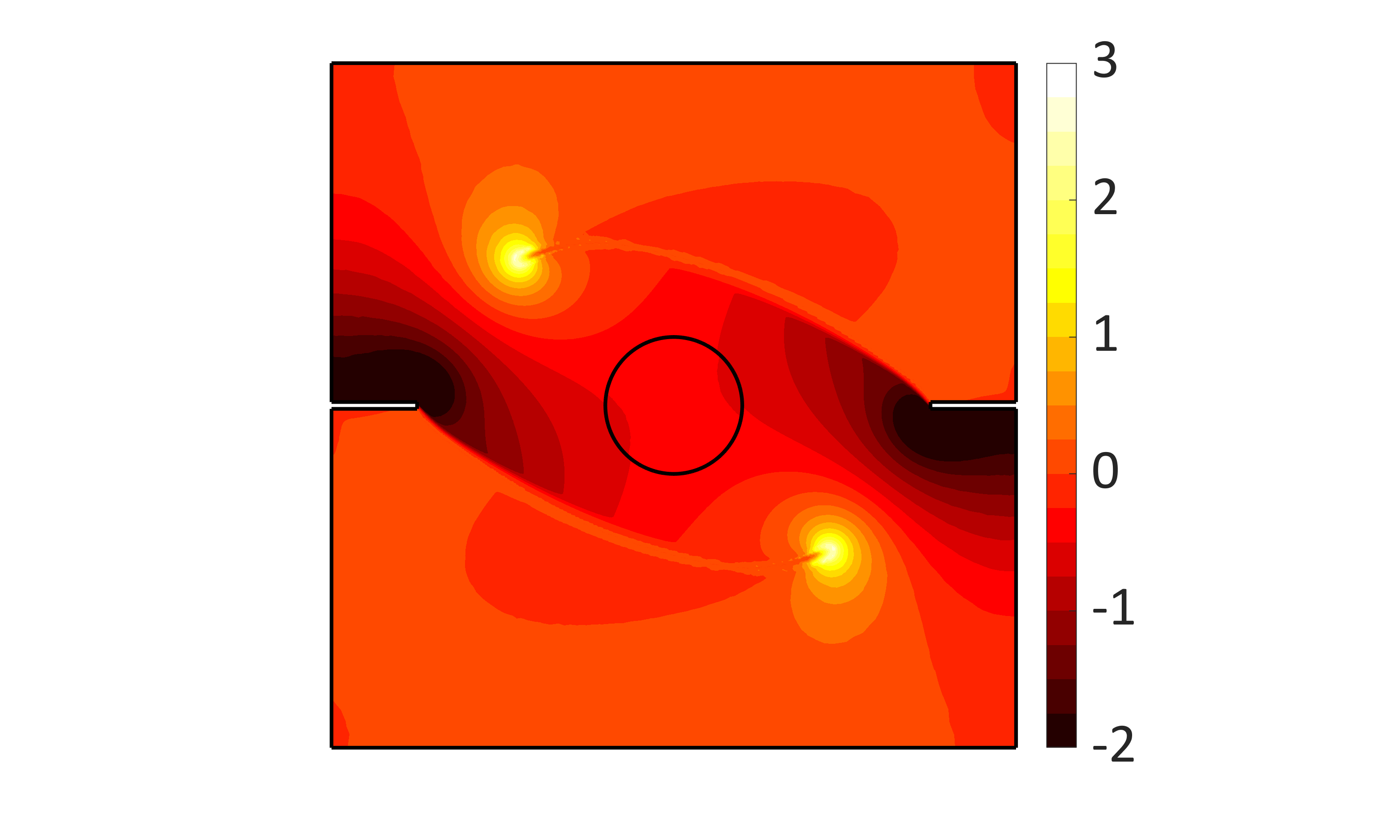}
\PE{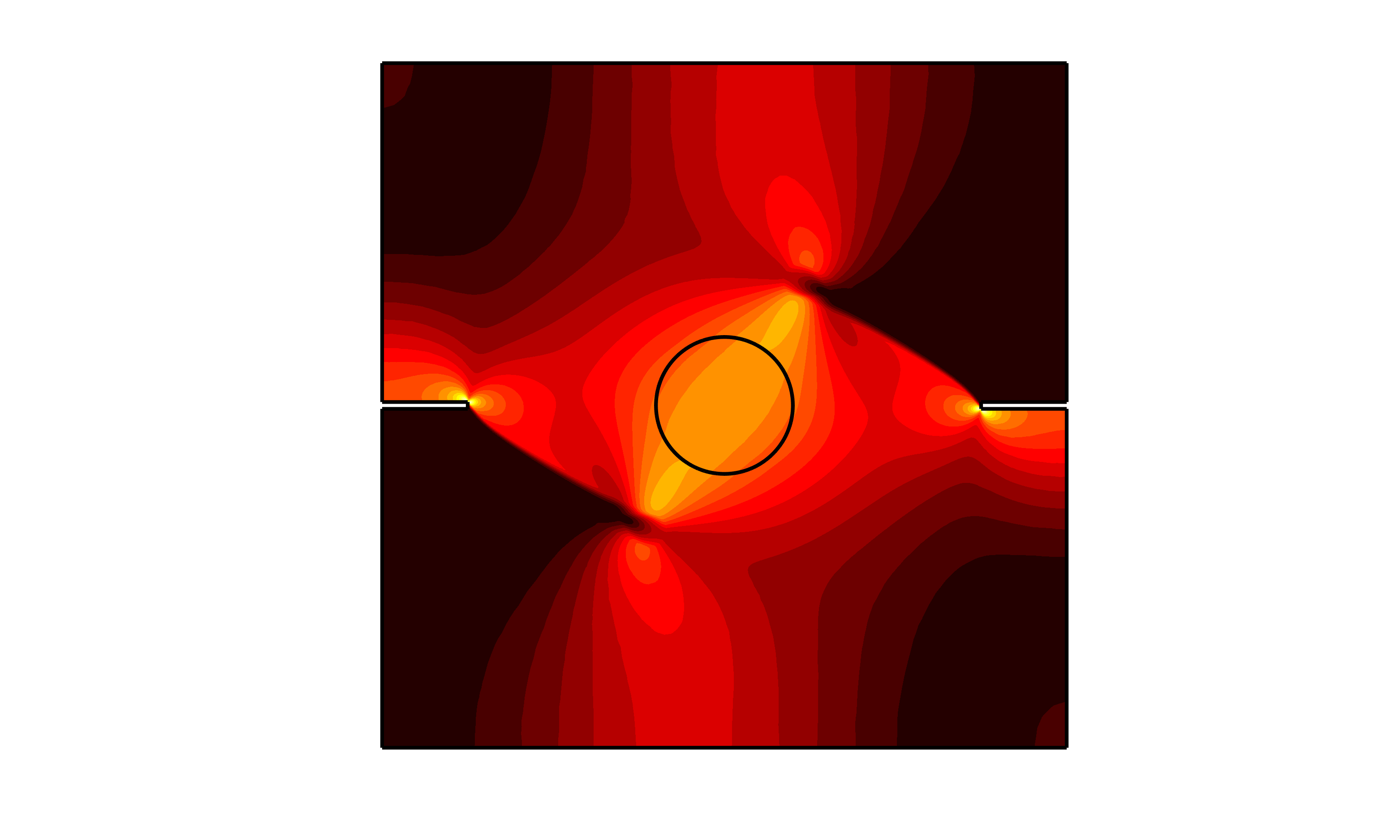}{Deviatoric stress}
\CB[955]{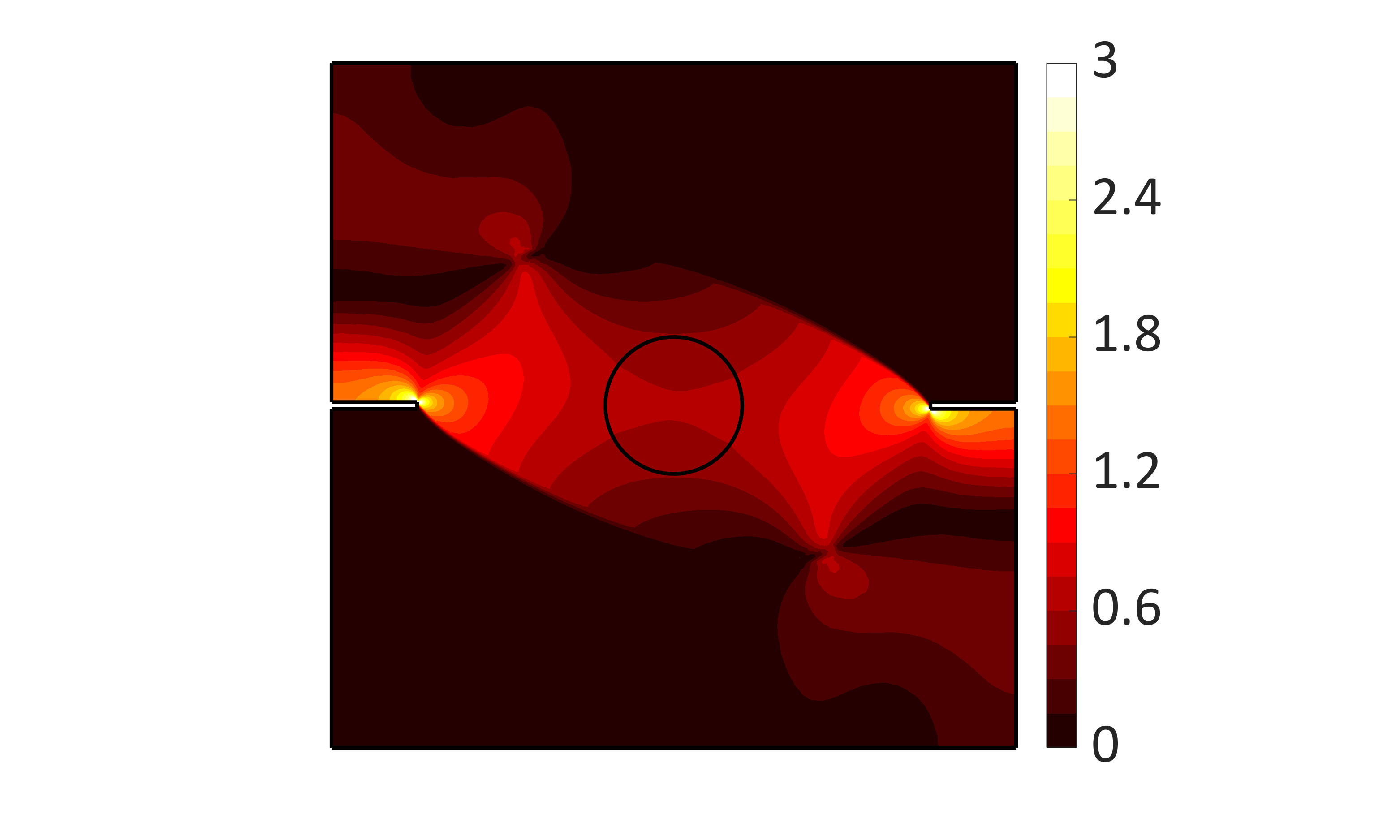}
\end{picture}
\caption{${\Kp}_\text{incl}=3.1$ GPa, $\Gc[iI]=100$ Jm$^{-2}$}\label{Fig_NoMoB_EF2_Gm1}
\end{subfigure}
\caption{Distribution of the phase-field variable  documenting crack propagation (left) and stress values [MPa] (right) at selected time instants for the mixed load case. 
Stresses belong to the first used instant  and the deformation in the crack plot is $100\times$ magnified.}\label{Fig_NoMoB_DS}
\end{figure}

For the cases with smaller interface fracture energy,  an interface crack appears  in accordance with Fig.~\ref{Fig_NoMoB_IF}.
Different debonding angles visible in Figs.~\ref{Fig_NoMoB_EF70_Gm5} and~\ref{Fig_NoMoB_EF2_Gm5} are a consequence of different elastic parameters.
Here, the initiation of the crack in the matrix is shown.
The stress concentration related to both opening and shear stress make the crack to develop in  an oblique angle which corresponds to the direction of maximal normal stress.
The mixity mode parameters  do not allow the shear state to control crack propagation as it could have happen if the ration between $\Gc[II]$ and $\Gc[I]$  was smaller as it occurred in~\cite{vodicka22A1}.
In the present model, the crack starts to propagate when the stress condition~\eqref{Eq_StressCrit_Amod} is satisfied.
The aforementioned values pertinent to the present case can be identified in  the stress graphs  of the referenced subfigures.

The situation relative to the interface is different when the interface damage is eliminated by setting the interface fracture energy to the larger value in Figs.~\ref{Fig_NoMoB_EF70_Gm1} and~\ref{Fig_NoMoB_EF2_Gm1}. 
Here, the first used instant  pertains to the state  where the crack tip in the matrix is affected by the interface.
The opening stress exhibits a concentration of the stresses  in the direction to the interface.
Anyhow, if the materials are the same the interface does not contribute to the stress state, while  if the material characteristics are different, it may cause  a damage process  in the weaker material  so that  a crack nucleation zone appears  in the vicinity of the interface in Fig.~\ref{Fig_NoMoB_EF70_Gm1}, which was also observed by decrease of the interface stresses in Fig.~\ref{Fig_NoMoB_EF70_Gm1_IF_2}.

All the variety of the results which are possible to be obtained document versatility of the proposed computational approach.

%%%%%%%%%%%%%%
\section{Conclusion}\label{Sec_Concl}
%%%%%%%%%%%%%%%

Formation of cracks in compound materials is a complicated process which was intended to be solved by the developed computational approach. 
 Special attention was dedicated on arising of such cracks at interfaces between inhomogeneities  and matrix, and on how they may affect crack formation process  inside the materials.
In any of the situations, the approach is capable of considering the process to depend on the stress state which may result on the form of the emerged cracks.
There may be a tensile stress state which leads to opening of the cracks.
Nevertheless,  if a kind of compressive loading is also applied, the resulting stress state may lead to  mixed mode cracking. Especially, if only compressive load is applied it may by important to identify domains at which  an opening crack appears locally  in combination with damage  in shear.
All such situations have been captured by the provided computational results, which show the effect of changes by modifying the basic characteristic: the fracture energy and its dependence on the mode of the crack.
In this sense, the proposed approach  was assessed and found to be  an useful tool for fracture mechanics.

It is, of course,  clear that similar computational models need some characteristics related the various stages of the crack formation and propagation processes. 
The values of the characteristics may modify form of  damage  processes in materials and along interfaces between material components.
It surely requires comparison with experimental measurements in order to appropriately set these characteristics.
Such adjustments and comparisons with the present computing approach are planned in the near future.

Adaptability of the developed computational code to the changes which may be caused by such experimental observations is guaranteed by  a detailed control  on the computing processes as it is developed by the author in MATLAB.
It naturally utilises discretisation techniques of FEM, of discrete time-stepping method,  and of sequential quadratic programming algorithms and provides a control over almost all details of the solution process.
Anyhow, the computed data can be considered in a good agreement with expectations, so
that the developed computational methodology will be successfully implemented into other complex engineering calculations.\\[1em]
{\bfseries Acknowledgement.} The author acknowledges support from The Ministry of Education, Science, Research and Sport of the Slovak Republic by the grants VEGA 1/0307/23 and VEGA 1/0363/21.

\bibliographystyle{model1-num-names}
\setlength{\bibsep}{0pt}
\bibliography{PFM}

\end{document}